\documentclass[12pt]{article}
\usepackage{amsfonts}
\usepackage{amssymb}
\usepackage{oldgerm}
\def \h {{\mbox{\textgoth{h}}}}
\newread\epsffilein    
\newif\ifepsffileok    
\newif\ifepsfbbfound   
\newif\ifepsfverbose   
\newdimen\epsfxsize    
\newdimen\epsfysize    
\newdimen\epsftsize    
\newdimen\epsfrsize    
\newdimen\epsftmp      
\newdimen\pspoints     
\def\epsfbox#1{\global\def\epsfllx{72}\global\def\epsflly{72}%
   \global\def\epsfurx{540}\global\def\epsfury{720}%
   \def\lbracket{[}\def\testit{#1}\ifx\testit\lbracket
   \let\next=\epsfgetlitbb\else\let\next=\epsfnormal\fi\next{#1}}%
\def\epsfgetlitbb#1#2 #3 #4 #5]#6{\epsfgrab #2 #3 #4 #5 .\\%
   \epsfsetgraph{#6}}%
\def\epsfnormal#1{\epsfgetbb{#1}\epsfsetgraph{#1}}%
\def\epsfgetbb#1{%
\openin\epsffilein=#1
\ifeof\epsffilein\errmessage{I couldn't open #1, will ignore it}\else
   {\epsffileoktrue \chardef\other=12
    \def\do##1{\catcode`##1=\other}\dospecials \catcode`\ =10
    \loop
       \read\epsffilein to \epsffileline
       \ifeof\epsffilein\epsffileokfalse\else
          \expandafter\epsfaux\epsffileline:. \\%
       \fi
   \ifepsffileok\repeat
   \ifepsfbbfound\else
    \ifepsfverbose\message{No bounding box comment in #1; using defaults}\fi\fi
   }\closein\epsffilein\fi}%
\def\epsfclipstring{}
\def\epsfsetgraph#1{%
   \epsfrsize=\epsfury\pspoints
   \advance\epsfrsize by-\epsflly\pspoints
   \epsftsize=\epsfurx\pspoints
   \advance\epsftsize by-\epsfllx\pspoints
   \epsfxsize\epsfsize\epsftsize\epsfrsize
   \ifnum\epsfxsize=0 \ifnum\epsfysize=0
      \epsfxsize=\epsftsize \epsfysize=\epsfrsize
      \epsfrsize=0pt
     \else\epsftmp=\epsftsize \divide\epsftmp\epsfrsize
       \epsfxsize=\epsfysize \multiply\epsfxsize\epsftmp
       \multiply\epsftmp\epsfrsize \advance\epsftsize-\epsftmp
       \epsftmp=\epsfysize
       \loop \advance\epsftsize\epsftsize \divide\epsftmp 2
       \ifnum\epsftmp>0
          \ifnum\epsftsize<\epsfrsize\else
             \advance\epsftsize-\epsfrsize \advance\epsfxsize\epsftmp \fi
       \repeat
       \epsfrsize=0pt
     \fi
   \else \ifnum\epsfysize=0
     \epsftmp=\epsfrsize \divide\epsftmp\epsftsize
     \epsfysize=\epsfxsize \multiply\epsfysize\epsftmp   
     \multiply\epsftmp\epsftsize \advance\epsfrsize-\epsftmp
     \epsftmp=\epsfxsize
     \loop \advance\epsfrsize\epsfrsize \divide\epsftmp 2
     \ifnum\epsftmp>0
        \ifnum\epsfrsize<\epsftsize\else
           \advance\epsfrsize-\epsftsize \advance\epsfysize\epsftmp \fi
     \repeat
     \epsfrsize=0pt
    \else
     \epsfrsize=\epsfysize
    \fi
   \fi
   \ifepsfverbose\message{#1: width=\the\epsfxsize, height=\the\epsfysize}\fi
   \epsftmp=10\epsfxsize \divide\epsftmp\pspoints
   \vbox to\epsfysize{\vfil\hbox to\epsfxsize{%
      \ifnum\epsfrsize=0\relax
        \includegraphics{#1}%
      \else
        \epsfrsize=10\epsfysize \divide\epsfrsize\pspoints
        \includegraphics{#1}%
      \fi
      \hfil}}%
\global\epsfxsize=0pt\global\epsfysize=0pt}%
{\catcode`\%=12 \global\let\epsfpercent=
\long\def\epsfaux#1#2:#3\\{\ifx#1\epsfpercent
   \def\testit{#2}\ifx\testit\epsfbblit
      \epsfgrab #3 . . . \\%
      \epsffileokfalse
      \global\epsfbbfoundtrue
   \fi\else\ifx#1\par\else\epsffileokfalse\fi\fi}%
\def\epsfempty{}%
\def\epsfgrab #1 #2 #3 #4 #5\\{%
\global\def\epsfllx{#1}\ifx\epsfllx\epsfempty
      \epsfgrab #2 #3 #4 #5 .\\\else
   \global\def\epsflly{#2}%
   \global\def\epsfurx{#3}\global\def\epsfury{#4}\fi}%
\def\epsfsize#1#2{\epsfxsize}
\textwidth16cm
\textheight24cm
\addtolength{\evensidemargin}{-1.5cm}
\addtolength{\oddsidemargin}{-1.5cm}
\addtolength{\topmargin}{-1in}
\def \ub {\underline}
\def \id {{\bf 1}}
\def \dd {{\rm d\!I}}

\def \aa {{\mbox{\cmu a}}}
\def \bint {\int\!\!\!\!\!\!\!-}
\def \tr {{\rm tr}}   \def \str {{\rm str}}
\def \lra {\longrightarrow}
\def \utD {{\mathop{{\mathcal D}}_{\widetilde{}}}}
\def \d {{\rm d}}
\def \od {{\displaystyle\mathop{{\mathcal D}}^{(-)}}} 
\def \og {{\displaystyle\mathop{\gamma}^{(-)}}}       
\begin{document}
\newfont{\cmu}{cmu10 scaled 1600}
\setcounter{page}{0}
\pagestyle{empty}
\renewcommand{\theequation}{{\thesection{.}}\arabic{equation}}
\leftline{ETH-TH/97-19\ \hfill hep-th/9706132}
\vspace{2.5cm}
\centerline{\huge{\bf
{Supersymmetric Quantum Theory,}}}
\bigskip\medskip
\centerline{\huge{\bf
{Non-Commutative Geometry,}}}
\bigskip\medskip
\centerline{\huge{\bf
{and}}}
\bigskip\medskip
\centerline{\huge{\bf
{Gravitation}}}
\bigskip\medskip
\bigskip\bigskip
\bigskip\bigskip
\centerline{\huge{
{Lecture Notes}}}
\bigskip\medskip
\centerline{\huge{
{Les Houches 1995}}}

\vspace{2cm}

\large

\centerline{\bf J\"urg Fr\"ohlich}
\smallskip
\centerline{and}
\smallskip
\centerline{\bf Olivier Grandjean, Andreas Recknagel}
\medskip
\bigskip
\centerline{Institut f\"ur Theoretische Physik}

\centerline{ETH-H\"onggerberg}

\centerline{CH--8093 \ Z\"urich}

\vspace{4cm}
\vfil
\normalsize
\centerline{e-mail addresses: juerg, grandj and  anderl@itp.phys.ethz.ch}

\eject
\phantom{xxx}
\eject


\noindent{\Large {\bf Contents}}

\vspace{1cm}

\begin{description}
\item[]\hskip-.2cm Preface
\item[1.] The classical theory of gravitation
\item[2.] (Non-relativistic) quantum theory
\item[3.] Reconciling quantum theory with general relativity: quantum
  space-time-matter
\item[4.] Classical differential topology and -geometry and
  supersymmetric quantum theory
\begin{description}
\item[4.1] Pauli's electron
\item[4.2] The special case where $M$ is a Lie group
\item[4.3] Supersymmetric quantum theory and geometry put into perspective
\end{description}
\item[5.] Supersymmetry and non-commutative geometry
\begin{description}
\item[5.1] Spin$^c$ non-commutative geometry
\item[5.2] Non-commutative Riemannian geometry
\item[5.3] Reparametrization invariance, BRST-cohomology, and
  target-space \hfil\break {super}\-{symmetry}
\end{description}
\item[6.] The non-commutative torus
\begin{description}
\item[6.1] Spin geometry $(N=1)$
\item[6.2] Riemannian geometry $(N=\overline{(1,1)})$
\item[6.3] K\"ahler geometry $(N=\overline{(2,2)})$
\end{description}
\item[7.] Applications of non-commutative geometry to quantum theories
  of gravitation
\begin{description}
\item[7.1] {}From point-particles to strings
\item[7.2] A Schwinger-Dyson equation for string Green functions from
  reparametrization invariance and world sheet supersymmetry
\item[7.3] Some remarks on $M{\rm (atrix)}$ models
\item[7.4] Two-dimensional conformal field theories
\item[7.5] Reconstruction of (non-commutative) target spaces from 
conformal field theory
\item[7.6] Superconformal field theory, and the topology of target spaces
\end{description}
\item[8.] Conclusions
\end{description}

\newpage{\phantom{xxx}}
\eject

\pagestyle{plain}
\pagenumbering{roman}

\noindent {\Large {\bf Preface}}

\vspace{.5cm}

\noindent
These are notes to a course taught by the under-signed at a Les Houches
summer school organized by A.\ Connes and K.\ Gaw\c edzki, in
1995. They follow the program of the lectures presented at Les Houches
and of the notes written there, but they are considerably more
detailed than the lectures and those notes. In working out the
details, I received very valuable help from my two co-authors. Our
work led us to finding some new results which will, in part, be
published as research papers and, in part, are described in these
notes. Thus, these notes contain a mixture of review of very
well-known and less well-known matters and of recent or new results by
many authors (including ourselves).

The writing of these notes was not exactly a leisurely hike. It more
resembled an excursion to the top of Mont Blanc; (I hasten to confess
that I have actually never made it to the top of Mont Blanc, in
reality; but I have some idea of how that would feel from other
experiences in the mountains): One starts to climb the foothills
following very well-known (and, perhaps, not entirely well-known)
tracks --- Sections 1--3 --- until one reaches a refuge -- Section
4 -- where one takes a rest. The following day, one starts to continue
the ascent, very early in the day, through more difficult terrain; the
air is getting thinner, and one discovers unfamiliar tracks ---
Sections 5 and 6. Finally, one approaches the top, along one of the
standard routes, feeling somewhat exhausted --- Section~7, Sects.~7.1
and 7.2. Having reached the top, one is a little out of breath and
decides to admire the view --- an unwritten section. After a good
while, and with some new energy, one starts the descent, choosing 
a new, and somewhat unsafe, route --- Sect.~{7.3}. Fortunately, 
more familiar looking foothills come into sight, soon --- Sect.~{7.4}. 
As one reaches them, one starts to feel ones fatigue --- Sect.~{7.5} 
--- and one decides to take another short rest. Finally, one undertakes 
the last part of the way back to the valley (loosing the canonical 
path) --- Sect.~{7.6}. One is longing for drinks and a good night's
sleep --- Section~{8} and preface. 

Perhaps, the analogy sketched here is not entirely compelling, but it
isn't misleading. 

\medskip

It is probably superfluous to enter into a detailed description of the
various chapters of these notes --- the table of contents speaks for
itself. But a few comments may be helpful.

\smallskip

Sections 1 and 2: Standard stuff --- the experts should skip them (and
reach the foothills by helicopter).

\smallskip

Section 3: An attempt to formulate some constraints on a tentative 
reconciliation of quantum theory with general relativity; (draws on 
ideas mostly due to other people). Reading recommended.

\smallskip

Sect.~4.1: An introduction to differential geometry for readers who
are familiar with Pauli's quantum mechanics of the non-relativistic,
spinning electron. Hopefully useful.

\smallskip

Sect.~4.2: Good, old Lie group theory (put in clothes that physicists
may like). Experts should skip it.

\smallskip

Sect.~4.3: A brief ``tour d'horizon'' of quantum theory, supersymmetry 
and geometry; (``global supersymmetry \dots
is just another name for the differential topology and geometry of
\dots spaces''). Should be clarifying.

\smallskip

Sect.~5.1: Some basic material on non-commutative geometry, according
to Connes.

\smallskip

Sect.~5.2: The Riemannian aspect of non-commutative geometry and
connections to global supersymmetry. A classification of geometries in
terms of supersymmetries and broken supersymmetries. Some algebraic 
topology. 

\smallskip

Sect.~5.3: Group actions on geometric spaces, BRST cohomology and
target space supersymmetry; (``the air is getting thinner'').

\smallskip

Section 6: Analyzing some examples is all-important, in order to
understand the general theory. (The non-commutative torus is, perhaps,
the simplest non-trivial example; further examples appear in Sects.~{7.5} 
and 7.6.)

\smallskip

Sect.~7.1: See Green-Schwarz-Witten, volume I. 

\smallskip

Sect.~7.2: Some sections from Green-Schwarz-Witten, narrated in a,
perhaps, somewhat personal way. Connections between the material in
Section 5 and superstring theory are described. (Emphasis on 
Schwinger-Dyson equations for string Green functions.)

\smallskip

Sect.~7.3: A brief look into the future; (``anything goes'' --- Paul
Feyerabend).

\smallskip

Sect.~7.4: Ground states (``vacua'') of superstring theory are 
described by certain two-dimensional (super-)conformal field theories. 
This section provides a short introduction to two-dimensional, local 
quantum field theory and (super-)conformal field theory and 
explains connections between conformal field theory and group theory. 

\smallskip

Sect.~7.5: Reflections on the question what a conformal field theory 
describing a ground state of string theory teaches us about the geometry 
of space-time (more precisely, of ``internal space''). An attempt to 
view conformal field theories as quantum theories describing ``loop 
space geometry''. 

\smallskip

Sect.~7.6: Tools to explore the topology and geometry of target spaces of
superconformal field theories; an example (WZNW). 

Sects.\ 7.4 -- 7.6 could (or, perhaps, should) have been the core sections of
these notes had the authors not started to feel their fatigue ---
nevertheless hopefully useful reading; (and then it will be time for the 
drinks). 

\smallskip

Section 8: Conclusions are mostly left to the reader. 

\bigskip

I wish to apologize for the shortcomings and imperfections of these
notes and the many typos that may have escaped our attention. But we
were really running out of time.

Much of the material in these notes is inspired by the work of
A.\ Connes and the work of E.\ Witten and of their followers. Our
efforts have been motivated by our desire to try to understand some of
their work and to point to connections between their approaches. We
have drawn on results and ideas of many other colleagues --- too many
to mention them all. Our line of thought is somewhat related to that of
A.\ Jaffe and collaborators. 

We should like to explicitly acknowledge our indebtedness to our
collaborators, A.H.\ Chamseddine, G.\ Felder and K.\ Gaw\c edzki. Had we
not had the privilege of their cooperation and help we would have
little to report here!

We are also grateful to many colleagues and participants of the school
for most valuable discussions. We acknowledge
countless lively discussions with S.\ Mukhanov.

We are indebted to A.\ Connes for sending us much of his work prior to
publication and for various useful comments.

We thank A.\ Schultze for expert typing of the manuscript. 

I am very grateful to A.\ Connes and K.\ Gaw\c edzki  for having invited
me to participate as a student and to 
lecture at their school and for their most generous patience.
\bigskip\medskip

\rightline{J\"urg Fr\"ohlich, \ May 1997\qquad\quad}

\newpage


\pagenumbering{arabic}
\setcounter{page}{1}
\setcounter{equation}{0}%

\section{The classical theory of gravitation}

In this section, we present a brief summary of how classical physics
treats space and time, space-time geometry and its interrelation with 
gravitation. In physics, the geometry of space-time is an object of
experimental exploration and physical modeling. Points in space-time
are identified with (the location of) events, time-like curves with
the world lines of observers or material objects. An observer gathers
information about events or objects by recording light signals emitted
from such events or objects and reaching her from her past light
cone. Information between distant observers is exchanged with the help
of signals consisting of electromagnetic waves. The dynamics of such 
signals is described by classical Maxwell theory. In relativistic
physics it is assumed that information can never be exchanged with a
velocity exceeding the velocity of light. Thus, it is assumed to be
impossible to exchange information between space-like separated
observers. This feature implies a {\it fundamental unpredictability of
  the future} in classical relativistic physics (observers can
receive information, at best, about events inside their past light
cone and hence, for most cosmologies, they can never gather complete
information about ``initial conditions'' prescribed on some space-like
Cauchy surface, because there do not usually exist Cauchy surfaces
completely contained inside the past light cone of any observer. As a
consequence, the maximal amount of information available, in
principle, to an observer does {\it not} enable her to predict her own
future with certainty).

Within {\it classical physics}, space-time is described as a {\it
  four-dimensional Lorentzian manifold} with certain good properties:
It should admit a causal orientation (global distinction between the
past and the future is possible); there should not exist any closed
time-like geodesics (no ``G\"odel universes''); space-time
singularities should be hidden behind horizons (``cosmic censorship
hypothesis'').

In exploring space-time geometry, one assumes that one can detect
signals causing arbitrarily small perturbations of the energy-momentum
tensor (the ``recoil'' of signals on space-time geometry can be
neglected). Among various forms of matter, one assumes the existence
of {\it point particles}. An event is the emission or reflection of an
electromagnetic wave by a point particle. One assumes that, with the
help of {\it arbitrarily weak} signals, one can determine the state of
a point particle {\it arbitrarily precisely}. This is based on the
assumptions that arbitrarily precise watches are available and that
the wavelength of an approximately monochromatic electromagnetic wave
train can be measured arbitrarily precisely (within a space-time
region so small that the deviation of its geometry from Minkowski
space geometry is insignificant). 

All these (fundamentally unrealistic) idealizations lead to the
concept of space-time as a Lorentzian manifold with properties
as described above.

According to the {\it principle of general covariance}, fundamental
laws of nature should take the form of equations between tensor fields
on the space-time manifold. According to the {\it equivalence
  principle}, it must be possible, 
locally in a small vicinity of a space-time point $p$, to construct
coordinate functions $x^\mu$ vanishing at $p$ and such that the
space-time metric $g_{\mu\nu} (p)$ at the point $p$ is given by the
Minkowski metric
\[
\left( \eta_{\mu\nu}\right) \ = \ 
\left( \begin{array}{cccc}
1 &  &	  &\\
  &-1 &	  &\\
  &   &-1 &\\
&   &	&-1
\end{array} \right)
\]
and such that the Christoffel symbols vanish at $p$. In these normal
coordinates, the differential laws describing the dynamics of matter
and radiation in an infinitesimal neighborhood of the point $p$ are
assumed to have the form known from special-relativistic physics. 

Let $T_{\mu\nu}$ denote the energy-momentum tensor of matter
(including point particles, the electromagnetic field, etc.). Let
$R^\mu_{\ \nu\lambda\sigma}$ denote the Riemann curvature tensor,
$R_{\mu\nu} = R_{\ \mu\lambda\nu}^\lambda$ the Ricci tensor, and
$r=R^\mu_{\ \mu}$ the curvature scalar. The Einstein tensor is defined
by 
\begin{equation}
G_{\mu\nu} \ = \ R_{\mu\nu} - \frac 1 2 \ g_{\mu\nu} \ r \ .
\end{equation}
As a consequence of the Bianchi identities, the covariant divergence
of $G_{\mu\nu}$ vanishes. The covariant divergence of the
energy-momentum tensor $T_{\mu\nu}$ vanishes, too (for Lagrangian
models of matter). Thus, the Einstein-Hilbert field equation
\begin{equation}
G_{\mu\nu} \ = \ \kappa \ T_{\mu\nu}
\end{equation}
is meaningful; $\kappa$ is Newton's constant and we choose units such
that $\hbar = c = 1$. Simple dimensional analysis shows that $\kappa$
has dimensions of length$^{d-2}$, where $d$ is the dimension of
space-time: $\kappa = l_{\rm P}^{d-2}$, where $l_{\rm P}$ is the Planck length
$( l_{\rm P} \approx$ 10$^{-33}$ cm, or $l_{\rm P}^{-1} \approx$ 10$^{19}$~GeV).

It is well known that, in the limit of 
weak gravitational fields and for material objects with relative
velocities small compared to the velocity of light, eqs.~(1.2)
formally reproduce the Newtonian theory of gravitation.

While, for suitably chosen initial conditions, eqs.~(1.2) may have
global solutions (see~[1] for an important example) and hence they may
express {\it deterministic} laws of nature, this does {\it not} imply
that a localized observer can gather enough information to predict the
future (as discussed above). This is a basic difference between
non-relativistic and relativistic physics. 

If $T_{\mu\nu}$ is calculated for a gas of very light point
particles then the equation $\nabla_\mu T^{\mu\nu}=0$ implies an
equation of motion for point particles: The world lines of point
particles are geodesics for the space-time metric
$g_{\mu\nu}$. Parametrizing the world line of a massive particle by its
arc length (proper time) $\tau$, the differential equation for a geodesic is
\begin{equation}
\frac{d^2x^\mu(\tau)}{d\tau^2}\ + \ \Gamma_{\nu\lambda}^\mu \left(
  x\left(\tau\right)\right) \ \frac{dx^\nu (\tau)}{d\tau} \
\frac{dx^\lambda (\tau)}{d\tau} \ = \ 0\ ,
\end{equation}
where
\[
\Gamma_{\nu\lambda}^\mu (x) \ = \ \frac 1 2 \ g^{\mu\sigma} (x) 
\left( \frac{\partial g_{\sigma\lambda}(x)}{\partial x^\nu} \ + \
  \frac{\partial g_{\nu\sigma}(x)}{\partial x^\lambda} \ - \
    \frac{\partial g_{\nu\lambda}(x)}{\partial x^\sigma}\right)
\]
are the Christoffel symbols.

Eqs.~(1.3) can be derived from an action principle. The action is
given by
\begin{equation}
S\left(x\left(\cdot\right)\right) \ = \ \frac 1 2 \ \int g_{\mu\nu}
\left(x\left(\tau\right)\right) \ \frac{dx^\mu(\tau)}{d\tau} \cdot
\frac{dx^\nu(\tau)}{d\tau} d\tau 
\end{equation}
with the constraint
\begin{equation}
g_{\mu\nu} \left(x\left(\tau\right)\right) \ \frac{dx^\mu(\tau)}{d\tau} \
\frac{dx^\nu(\tau)}{d\tau} \ = \ 1\ ,
\end{equation}
(for massive test particles).

Unfortunately, the action (1.4) is not
reparametrization-invariant $(\tau$ is arc length). This problem can
be cured by considering the actions

\medskip

\[ 
\!\!\!\!\!\!\!\!\!\!\!\!\!\!\!\!\!\!\!\!\!\!\!{\rm i)}\quad\qquad 
S_{\rm NG}\left(x\left(\cdot\right)\right) \ = \ \int
ds \ = \ \int \sqrt{g_{\mu\nu}\left(x\left(\tau\right)\right)
  \frac{dx^\mu(\tau)}{d\tau}\ \frac{dx^\nu(\tau)}{d\tau}}\  d\tau\ ,
\]
where ``NG'' stands for Nambu-Goto, or

\begin{eqnarray}
\!\!\!\!\!\!\!\!\!\!\!\!\!\!\!\!{\rm ii)} \quad\qquad  S_{\rm P} \left(x\left(\cdot\right), 
h\left(\cdot\right)\right) 
&=& \frac 1 2 \ \int g_{\mu\nu} \left(x\left(\tau\right)\right) \
\frac{dx^\mu(\tau)}{d\tau} \ \frac{dx^\nu(\tau)}{d\tau} \
h(\tau)^{-1/2} d\tau  \nonumber \\
&+& \frac{\mu^2}{2} \int h (\tau)^{1/2} d\tau \ ,
\end{eqnarray}
where $h(\tau)d\tau^2$ is an arbitrary metric on parameter space,
i.e., on an interval $I \subset {\mathbb R}$, and $\mu > 0$ is a
parameter. Here ``P'' stands for 
(Deser-Zumino-)Polyakov. Minimizing $S_{\rm P}$ with respect to
$h(\tau)$ yields the Euler-Lagrange equation
\begin{equation}
h(\tau) \ = \ \frac{1}{\mu^2} \ g_{\mu\nu}
\left(x\left(\tau\right)\right) \ \frac{dx^\mu(\tau)}{d\tau} \cdot
\frac{dx^\nu(\tau)}{d\tau} \ ,
\end{equation}
or, for $h(\tau) \equiv 1$, 
\[
\tau \ = \ \frac{1}{\mu^2} \ \times \ {\rm arc \ length} \ .
\]
Upon using (1.7), the Euler-Lagrange equations obtained by minimizing 
(1.6) with respect to $x(\tau)$  reproduce equation (1.3). The
action~(1.4) can be obtained from (1.6) by ``{\it fixing the gauge}''
$h (\tau) \equiv 1$ (which destroys reparametrization invariance). 

By observing the motion of a gas of test particles of very small mass
with the help of electromagnetic waves, an observer can reconstruct
the geometry of space-time regions contained in his past light
cone. For example, he can measure the components $R_{\ 00\nu}^\mu$ of the
Riemann curvature tensor by studying a correspondence of geodesics
(world lines of a gas of test particles). The Jacobi field $n^\mu$
pointing from one geodesic in the correspondence to an infinitely
close one satisfies the differential equation
\[
\frac{d^2 n^\mu (\tau)}{d\tau^2} \ = \ R_{\ 00\nu}^\mu \left(
  x\left(\tau\right)\right) \;n^\nu\left(\tau\right) \ .
\]
The r.s.\ describes tidal forces whose observation apparently permits
to measure $R_{\ 00\nu}^\mu$.

It is well known how to extend the theory to systems of charged point
particles moving 
through an external electromagnetic field.

All this is very beautiful; but the theory is plagued with
inconsistencies. For example, it turns out to be impossible to take
the concept of a point particle of positive mass (and non-zero charge)
literally. It would lead to divergences and a-causal behaviour. But
quite apart from such mathematical inconsistencies, it is {\it
  impossible} to describe matter and radiation by {\it classical}
physical theories when ``{\it microscopic scales}'' are involved
because, on such scales, their quantum mechanical nature becomes
apparent. (When the action of the trajectory of a point particle is
comparable to Planck's constant $\hbar$, its quantum-mechanical
nature cannot be neglected.)

A fundamental problem of present-day physical theory is to reconcile
(some form of) the quantum theory of matter and radiation with (some
form of) Einstein's relativistic theory of space-time and
gravitation. To understand what the problem is, we shall briefly
recapitulate some basic features of (non-relativistic) quantum theory
and then explain in which way they are incompatible with general
relativity. 

\vspace{1cm}

\setcounter{equation}{0}%
\section{(Non-relativistic) quantum theory}

Quantum theory was born from the study of systems of harmonic
oscillators. The first such system was black-body radiation
corresponding to harmonic oscillations of {\it electromagnetic waves}
in a cavity. Planck and Einstein found the rules of
``quantization''. In his theory of the specific heat of crystals,
Einstein showed that the {\it same rules} of quantization must also be
applied to harmonic oscillations of {\it material} media, in order to
reach agreement with experimental data. De Broglie extended these
ideas to material particles by assuming that such particles have
wave-like properties. The rules 
of quantization were eventually extended to apply to a rather general
class of {\it Hamiltonian systems} with finitely many degrees of
freedom and to systems of infinitely many oscillators with very weak
anharmonicity. There is hardly any doubt that they apply to small
(essentially harmonic) oscillations of the {\it gravitational field}
(space-time metric) around a classical background field. In every
example where a non-linear Hamiltonian system with infinitely many
degrees of freedom is quantized, according to the standard rules,
mathematical difficulties in the quantum theory can be traced to the
fact that an {\it arbitrary number of degrees of freedom} can be
localized in 
an {\it arbitrarily small space-time region}; or, in other words, that
the number of possible events localized in an arbitrarily small
space-time region is arbitrarily large. This suggests that there may
be something wrong with the idea of {\it space-time} as a {\it
  classical, smooth Lorentzian manifold} when it comes to describing
quantum mechanical events in very tiny regions of space-time. A
considerable part of these notes is devoted to trying to find out what
may go wrong and what might be done to cure the problem. (Of course,
it won't be solved here!) 

We continue with a short and standard recapitulation of quantum
mechanics. We 
start with a simple physical system consisting of a charged,
non-relativistic particle interacting with the electromagnetic
field. Historically, the physics of this system was explored on the
basis of the following two ingredients:

\medskip

(A) \ Classical, Newtonian mechanics of the particle and Maxwell's
theory of the electromagnetic field (which, together, form an
infinite-dimensional Hamiltonian system).

(B) \ The theory of photons, due to Planck and Einstein, with the
relations
\begin{equation}
E \ = \ h\nu, \qquad p \ = \ h/\lambda \ ,
\end{equation}
where $h$ is Planck's constant, $E$ and $p$ denote the energy and the
momentum of a photon, i.e., 
of an electromagnetic field oscillator of frequency $\nu$ and wave
length $\lambda=c/\nu$.

\medskip

Unfortunately, these two ingredients are incompatible. Here is what
goes wrong when one tries to combine (A) and (B) naively: The state
of a charged particle at a given time is described by its position
$\vec{x}$ and its momentum $\vec{p}$, the one of the electromagnetic
field by specifying the magnetic and the electric field on a space-like
slice corresponding to, for example, a fixed time (in the rest frame
of the particle). We may attempt to measure the state $\left( \vec{x},
  \vec{p}\right)$ of the particle as follows:

\medskip

(1) \ We turn on a homogeneous magnetic 
field in a region of space where we suspect to find the particle. Due
to the Lorentz force, the trajectory of the particle is bent. If we
know the electric charge of the particle and the velocity of light and
have measured the strength of the magnetic field, we can (according to
(A)) determine the momentum $\vec{p}$ of the particle by measuring the
curvature radius of its trajectory (which, incidentally, necessitates
measuring the position of the particle at at least three distinct times,
{\it or} measuring the electromagnetic radiation caused by the
accelerated motion of the particle). 

(2) \ We measure the position, $\vec{x}$, of the particle by
e.g.~shining light into the region where we suspect to find the
particle and detect light scattered by the particle with the help of a
``Heisenberg microscope''. In studying the interaction of the particle
with a light wave we use relations (B) and the conservation of total
energy and momentum. 

\medskip

Let us suppose that, after having performed measurements (1) and then
(2), we know the position $\vec{x}$ and the momentum $\vec{p}$ within a
precision of $\triangle x^j, \triangle p_j, j=1,2,3$; $(x^j$ is the $j^{\rm
  th}$ component of $\vec{x}$ in a Cartesian coordinate system). Then
$\triangle x^j$ and $\triangle p_j$ are constrained by Heisenberg's
uncertainty relations
\begin{equation}
\triangle x^j \ \triangle p_j \ \geq \ \frac{\hbar}{2}, \ j=1,2,3,
\end{equation}
as some exceedingly well-known arguments show.
Many different gedanken-- and real experiments teach us that (2.2) is
valid {\it independently} of what the tools used to measure
$\vec{x}$ and $\vec{p}$ are. Similarly, when one attempts to measure the
electric and magnetic field averaged over a small region ${\mathcal
  O}$ of space-time by studying e.g.~their influence on the motion of
charged particles, whose positions and momenta are known only to an
accuracy constrained by (2.2), one finds, according to Bohr and
Rosenfeld, that the accuracies, $\triangle E_{{\mathcal O}}, \triangle
B_{{\mathcal O}}$, of these field measurements are constrained by
\begin{equation}
\triangle B_{{\mathcal O}} \cdot \triangle E_{{\mathcal O}} \ \geq \
\hbar\	{\rm const}_{{\mathcal O}} \ .
\end{equation}
Measuring an electromagnetic field in a space-time region through
which a charged particle travels will thus create an uncertainty in
the force acting on the charged particle.

Quantum mechanics is developed on the basis of the postulate that the
uncertainty relations (2.2) and (2.3) are fundamental and hold {\it
  independently} of how the state of the system is measured. 

It is useful to recast the discussion just presented in a more
abstract context: We consider a classical Hamiltonian system,
conveniently one with only finitely many degrees of freedom. We
suppose that the phase space $\Gamma$ of the system is the cotangent
bundle $T^*M$ of a smooth manifold $M$, interpreted as the
configuration space of the system. Phase space $\Gamma$ is equipped
with a symplectic 2-form $\omega$. If $U$ is an open subset of $M$
over which the cotangent bundle is trivial, $T^*U \simeq U \times
{\mathbb R}^n$ (where $n$ is the dimension of $M$), then one can
choose coordinates $q^1, \ldots, q^n$ in $U$ and extend them to
Darboux coordinates $q^1,\ldots,q^n,p_1,\ldots,p_n$, for $T^*U$ such
that
\begin{equation}
\omega \ = \ \sum_{j=1}^n \ dp_j \ \wedge \ dq^j \ .
\end{equation}
A state of the system in $T^*U$ is a point $(q,p) \in T^*U$, where
$q=(q^1,\ldots,q^n)$ is interpreted as a configuration space position
and $p=(p_1,\ldots,p_n)$ as momentum. The symplectic form $\omega$ is
left invariant by symplectic diffeomorphisms of $\Gamma$.
The dynamics of the system is specified by a Hamiltonian vector field
$X_H$, where $H$ is a function on $\Gamma$; $X_H$ is defined by
setting $\omega (X_H, Y) = Y(H) \equiv dH(Y)$, where $Y$ is an
arbitrary vector field. The push forward of a Hamiltonian vector field
under a symplectic diffeomorphism of $\Gamma$ is again a Hamiltonian
vector field. ``Observables'' of the system are real-valued,
continuous (or smooth) functions, $F$, on $\Gamma$. Every observable
$F$ determines a Hamiltonian vector field $X_F$ (as above) and hence
(locally) a one parameter group of symplectic diffeomorphisms. The
algebra of observables with support in a closed subset $\Omega$ of
$\Gamma$ is denoted by ${\mathcal F}(\Omega)$; and ${\mathcal F}:=
{\mathcal F}(\Gamma)$.

In passing from classical Hamiltonian dynamics to quantum theory, one
supposes that, in any {\it real measurement} of the state $(q,p)$ of
the system that determines $q$ up to a precision of $\triangle q$ and
$p$ up to a precision $\triangle p$, the uncertainty relations
\begin{equation}
\triangle q^j \cdot \triangle p_j \ \geq \ \frac{\hbar}{2}, \ j=1,\ldots,n,
\end{equation} 
must hold. One concludes that it is impossible to determine the state
of the system precisely and that, therefore, the {\it classical
  concept} of a {\it state} is {\it not strictly meaningful!} It
follows that the elements of ${\mathcal F}$ {\it cannot} be the
observables of the system, because they separate points of $\Gamma$. 
Furthermore, one concludes that if $\Omega$ is a subset of $\Gamma$ of
finite symplectic volume, vol$_\omega (\Omega) < \infty$, then, by the
uncertainty relations (2.5), the number $N_\Omega$ of states of the
system ``located'' in $\Omega$ that can be resolved by {\it real}
measurements must be bounded by
\begin{equation}
N_\Omega \ \lesssim \ {\rm vol}_\omega (\Omega) / h^n \ .
\end{equation}
Inequality (2.6) suggests that observables measurable in $\Omega$
should form an ``essentially finite-dimensional'' algebra, and hence
${\mathcal F}(\Omega)$ must be deformed to an algebra ${\mathcal
  F}_{\hbar}(\Omega)$ with this property. 

An admissible quantization of the system must respect inequalities
(2.5) and (2.6). In order to construct a quantization, one chooses 
an integrable polarization of $\omega$. A natural choice, in our 
context, is the vertical polarization for which configuration space 
is given by $M$.  In the following,
we only consider this choice. In order not to get stuck with
technicalities, we temporarily assume that $M$ is smooth, compact,
connected and simply connected. (For example in connection with {\it
  quantum statistics}, $\theta$-{\it vacua}, {\it winding modes}, etc.,
it is actually important to consider configuration spaces $M$ which
are {\it not simply connected} or not even connected. But we postpone
this issue.)

Next, we choose a Riemannian metric $g=\left( g_{\mu\nu}\right)$ on
$M$. We denote by $d$vol$_g$ the Riemannian volume form and by
$\triangle_g$ the Laplace-Beltrami operator on smooth functions on $M$
associated with the metric $g$. We define
\begin{equation}
{\mathcal H} \ = \ L^2 \left( M, d {\rm vol}_g\right) 
\end{equation}
to be the Hilbert space of square-integrable functions on $M$. The
operator $\triangle_g$ is essentially self-adjoint on the dense
subspace of smooth functions in ${\mathcal H}$.

We also define a deformation ${\mathcal F}_\hbar$ of the algebra
${\mathcal F}$ of functions on $\Gamma$ as follows: Let $f \in
C_0^\infty ({\mathbb R})$ be an arbitrary, smooth function on
${\mathbb R}$ of compact support. Since $-\triangle_g$ defines a
positive, self-adjoint operator, $f (-\triangle_g)$ is well defined (by
the functional calculus). The operator $-\triangle_g$ defines a one
parameter group, $\alpha_\tau$, $\tau \in {\mathbb R}$, of
$^*$-automorphisms of the {\it algebra} $B({\mathcal H})$ {\it of all
  bounded operators} on ${\mathcal H}$ by setting 
\begin{equation}
\alpha_\tau (A) \ := \ e^{-\,i\tau \hbar \triangle_g} \ A \ e^{i\tau
  \hbar \triangle_g} \ .
\end{equation}
A reasonable definition of a {\it deformation} ${\mathcal F}_\hbar$ 
of the algebra ${\mathcal F}$ of continuous functions on $\Gamma$ is
to define ${\mathcal F}_\hbar$ to be 
the smallest $C^*$-algebra generated by 
\begin{equation}
\left\{ \alpha_\tau (a), \ f\left( -\triangle_g\right)\	 \bigm| \ a \in C
  (M),\	 \tau \in {\mathbb R},\	 f \in C_0^\infty ({\mathbb R})\right\} \ ,
\end{equation}
{\it invariant} under the $^*$-automorphism group $\alpha_\tau$. (Here
and in the following, algebras of operators are defined over the field
of {\it complex numbers}, unless stated otherwise. ``Observables''
will always correspond to {\it self-adjoint} elements of certain
operator algebras.) The algebra ${\mathcal F}_\hbar$ can be thought of
as the {\it ``algebra of functions over quantum phase space''.} It
contains the algebra
\begin{equation}
{\mathcal A} \ = \ C(M) 
\end{equation}
of complex-valued, continuous functions of $M$ as a maximal abelian
$C^*$-subalgebra. For some class of compact regions $\Omega \subset
\Gamma$, one can define algebras ${\mathcal F}_\hbar (\Omega)$
satisfying a suitable variant of the bound (2.6) in an obvious way.

It is useful to describe a second approach to constructing ${\mathcal
  F}_\hbar$: Let the Hilbert space ${\mathcal H}$ be as in
(2.7). Given a diffeomorphism $\varphi$ of $M$, we define a unitary
operator $U_\varphi$ on ${\mathcal H}$ by setting
\[
\left( U_\varphi \psi\right)\left( x\right) \ := \
\sqrt{\frac{\varphi^* d {\rm vol}_g (x)}{d {\rm vol}_g (x)}} \ \psi
\left( \varphi^{-1} (x)\right) \ .
\]
The unitarity of $U_\varphi$ follows from the invertibility of
$\varphi$ and the quasi-invariance of $d$vol$_g$ under
diffeomorphisms. We define
\[
{\mathcal U} \ := \ \left\{ U_\varphi \ \bigm| \ \varphi\;\in\;{\rm
    Diff} \ M\right\} 
\]
and choose ${\mathcal F}_\hbar$ to be the smallest $C^*$-algebra
generated by ${\mathcal U}$ and by $C(M)$. If $M$ consists of a finite number, 
$n$, of points one easily shows that ${\mathcal F}_\hbar = {\mathbb
  M}_n ({\mathbb C})$. 

A {\it quantization} of a classical Hamiltonian system with phase
space $\Gamma = T^*M$, ($M$ compact, smooth, connected and simply
connected) satisfying the requirements expressed in inequalities (2.5)
and (2.6) consists in choosing {\it spectral data} of e.g.~the form
\begin{equation}
\left(\, {\mathcal F}_\hbar,\ {\mathcal A}= C(M),\ {\mathcal H}=
  L^2 \left( M, d {\rm vol}_g\right),\ \triangle := -
  \triangle_g\,\right) \ .
\end{equation}
{\it Dynamics} is specified by choosing a self-adjoint operator $H$
densely defined on ${\mathcal H}$ with the properties:
\medskip
\[
\hskip-4cm\qquad{\rm (i)}\qquad e^{i t H/\hbar} \ a\; e^{-i t H/\hbar} \ \in \; 
{\mathcal F}_\hbar 
\quad {\rm for}\ a\;\in\;{\mathcal F}_\hbar\ \ \ {\rm and\  
for\ all}\ \ t\;\in\;{\mathbb R}\,,  
\]
i.e., the unitary operators \
$e^{itH/\hbar}$ \ determine a one-parameter group of $^*$-automorphisms
of ${\mathcal F}_\hbar$; and 

\smallskip

(ii) \ \ \ in the limit $\hbar \searrow 0$, some classical Hamiltonian
dynamics on $\Gamma = T^*M$ is recovered (in a standard sense that we
do not make precise here; ``quantization'' is a ``deformation'' of
classical, Hamiltonian mechanics).

\medskip

Since nature is intrinsically quantum-mechanical, it is, perhaps, more
interesting to ask how, from spectral data of quantum mechanics, one
can reconstruct e.g.~the topology and geometry of configuration space
$M$, rather than to pursue the question what we mean by the
quantization of a classical Hamiltonian system. (Of course,
reconstructing $M$ may not necessarily directly teach us anything
about {\it physical space}, but something about {\it configuration 
space}.) It is interesting to ask which data of quantum theory
suffice to reconstruct configuration space $M$, together with a
Riemannian metric $g$ on $M$. The complete data are $({\mathcal F},
{\mathcal A}, {\mathcal H}, \triangle)$, as in (2.11), where
${\mathcal F} \equiv {\mathcal F}_\hbar$, and the subscript
``$\hbar$'' will be {\it omitted} from now on. Of course, these data
are redundant, because, knowing ${\mathcal A}, {\mathcal H}$ and
$\triangle, {\mathcal F}$ is determined (by the construction described
above). We propose to view 
$({\mathcal F}, {\mathcal A}, {\mathcal H}, \triangle)$
as {\it abstract spectral data}, where 

\begin{itemize}
\item[(1)] \ ${\mathcal H}$ is a separable Hilbert space;
\item[(2)] \ $\triangle$ is a positive, self-adjoint operator on ${\mathcal
  H}$;
\item[(3)] \ ${\mathcal A}$ is an abstract {\it abelian} $C^*$-{\it algebra}
with the following properties:

\begin{itemize}
\item[(a)] ${\mathcal A}$ has a faithful $^*$-representation, $\pi$, on
${\mathcal H}$;
\item[(b)] ${\mathcal A}$ contains a subalgebra
$\displaystyle\mathop{{\mathcal A}}^{\,\;\rm o}$ dense in ${\mathcal A}$
in the $C^*$-norm of ${\mathcal A}$ such that the operator
\begin{equation}
\frac 1 2 \, \left( \triangle \pi(a)^2\;+\;\pi(a)^2 \triangle\right) \ - \
\pi(a)\,\triangle\,\pi(a) 
\end{equation}
is bounded in norm for arbitrary
$a\;\in\;\displaystyle\mathop{{\mathcal A}}^{\;\,\rm o}$; 
\item[(c)] ${\mathcal F}$ is constructed from ${\mathcal A}$ and $\triangle$
as in (2.9), and ${\mathcal A}$ is maximal abelian in ${\mathcal F}$. 
\end{itemize}
\end{itemize}

Given spectral data satisfying these properties, we may ask the
following questions:

\medskip

$(\alpha)$ \ Does the pair $({\mathcal H},\, \triangle = -\triangle_g)$
determine the manifold $M$ and its geometry? The answer is, in
general, {\it no}: one cannot ``hear the shape of a drum''
[2]. However, certain properties of $M$ {\it are} determined by
$({\mathcal H}, \triangle)$. For example, according to a celebrated
result due to H.~Weyl, the (asymptotics of the) spectrum of
$\triangle$ determines the dimension of $M$ and its Riemannian
volume. Furthermore, using ideas that physicists know from the theory
of {\it quantum-mechanical tunneling}, one can show [3] that 
\[
\frac 1 4 \ h_M^2 \ \leq \ \lambda_1 (M) \ \leq \ C \left( \delta \;
  h_M + h_M^2\right) \ ,
\]
where $h_M=h_M(g)$ is Cheeger's isoperimetric constant of $(M,g)$, and
$\lambda_1 (M) = \lambda_1 (M, g)$ is the smallest non-zero eigenvalue
of $-\triangle_g$, $C$ is a universal constant, and $\delta$ is a
constant depending on the diameter of $M$. Yet, even when \ dim $M=2$,
there are isospectral manifolds that are {\it not} isometric [2].

$(\beta)$ \ Does ${\mathcal A}$ determine $M$ and its geometry? A famous 
theorem, due to Gel'fand (see e.g.\ [4]), says that the space of characters
of an abelian unital $C^*$-algebra ${\mathcal A}$ (the ``spectrum'' of
${\mathcal A}$) is a compact Hausdorff space $K$ with the property
that ${\mathcal A} = C(K)$ \ (the algebra of complex-valued,
continuous functions on $K$). The space $K$ encodes the properties of
$M$ when $M$ is viewed as a compact Hausdorff space, but it does not tell
us anything about e.g.~a differentiable structure on $M$, a geodesic
distance on $M$, etc.

$(\gamma)$ \ Do the data $ ({\mathcal A}, {\mathcal H}, \triangle)$
determine the topology and Riemannian geometry of $M$? Here the answer
is {\it yes}: ${\mathcal A}$ determines $M$ as a compact Hausdorff
space. The algebra
\[
\displaystyle\mathop{{\mathcal A}}^{\;\,\rm o}\ = \ \left\{ a \in
  {\mathcal A} \bigm|\; \Big\| \ \frac 1 2 \, \left( \triangle a^2 + a^2
    \triangle\right) - \; a\triangle a\, \Big\| < \infty \right\} \ ,
\]
which is a norm-dense subalgebra of ${\mathcal A}$, by assumption (b),
can be interpreted as the algebra of {\it Lipshitz-continuous
  functions} on $M$: If ${\mathcal A} = C(M)$ and $\triangle =
-\triangle_g$ then 
\begin{equation}
\frac 1 2 \, \left(\triangle a^2+a^2\triangle\right)\ -a\triangle a \ =
\ -g^{\mu\nu} \left(\partial_\mu a\right)\left(\partial_\nu a\right)\,,
\end{equation}
see [24]. A variant of an argument due to Connes [5] enables us to
reconstruct the {\it geodesic distance} on $M$ determined by $g$:
Given two points, $x$ and $y$, in $M$ (i.e., two characters of
${\mathcal A}$), the geodesic distance between $x$ and $y$ is given by
\begin{equation}
{\rm dist}_g \left(x,y\right)\ = \ {\rm sup}\ |a(x)-a(y)| \ ,
\end{equation}
where the supremum on the r.s.\ of (2.14) is taken over all elements $a$ of
$\displaystyle\mathop{{\mathcal A}}^{\;\,\rm o}$ with the property that
\begin{equation}
\Big\Vert\, \frac 1 2 \, \left( \triangle a^2+a^2\triangle\right) \
-a\triangle a\,\Big\Vert \ \leq \ 1 \ .
\end{equation}
It is not hard to introduce higher-order polynomials in $\triangle$
and elements of ${\mathcal A}$, in order
to test whether $M$ is smooth. (It is not surprising that, on a
classical manifold, one can usually define a notion of
Lipshitz-continuous functions, because, according to a theorem due to
D.~Sullivan, a topological manifold of dimension $\neq 4$ is
automatically Lipshitz.)

Connes' theory of non-commutative geometry [5] starts from the idea
(among other ideas) to view abstract spectral data  
$({\mathcal A}, {\mathcal H}, \triangle)$, where ${\mathcal A}$ is a
$^*$-algebra of bounded operators on the separable Hilbert space
${\mathcal H}$, and $({\mathcal A}, {\mathcal H}, \triangle)$ 
have properties $(a)$ and $(b)$, as a 
starting point for ``non-commutative geometry''.

In the study of quantum-mechanical systems with {\it infinitely many
  degrees of freedom} (quantum field theory) one often encounters a
problem in trying to make sense of ${\mathcal A}$; but an analogue of
the {\it non-commutative} algebra ${\mathcal F}$ is still
meaningful. Recall that ${\mathcal F}$ may be interpreted as the
``algebra of functions'' on ``quantum phase space''; historically, the 
first (and a paradigmatic) example of a {\it non-commutative space}. 
This motivates us to ask the question:

$(\delta)$ \ Under which additional hypotheses do the data 
$({\mathcal F}, {\mathcal H}, \triangle)$, where ${\mathcal F}$ is a
$C^*$-algebra faithfully represented on the separable Hilbert space
${\mathcal H}$ and {\it invariant} under the \ $^*$-auto\-morphism group
$\alpha_\tau$ defined by
\[
\alpha_\tau (A) \ := \ e^{i\tau\triangle} \ A\; e^{-i\tau\triangle} \ ,
\]
$A\in B ({\mathcal H}), \tau \in {\mathbb R}$, determine the topology
and Riemannian geometry of a classical manifold $M$? Suppose we know
that $({\mathcal F}, {\mathcal H}, \triangle)$ are as in (2.7--9)
(i.e., they come from a classical Riemannian manifold $M$, and the
operator $\triangle$ is identified with $-\triangle_g$), how do we
reconstruct $M$ and $g$ just from 
$({\mathcal F}, {\mathcal H}, \triangle)$? 

\medskip

These are important questions to which we don't know complete answers,
at present. They deserve the attention of mathematicians, as will
become apparent in subsequent sections. We shall describe some tools
that may lead to satisfactory answers. But we anticipate one important
observation: In general, topologically distinct classical
configuration spaces $M_1, M_2, \ldots$ may lead to the same spectral
data $({\mathcal F}, {\mathcal H}, \triangle)$ -- ``$T$-duality''.

\medskip

Let us briefly summarize the main points of this section.

\begin{itemize}
\item[(I)] Nature is quantum-mechanical. Quantum theory, described in
  terms of spectral data such as $({\mathcal A}, {\mathcal H},
  \triangle)$ in (1)-(3) above, enables
  one to reconstruct a manifold $M$, interpreted as configuration
  space, equipped with a Lipshitz structure and with a {\it Riemannian
    metric} $g$. The manifold $M$ has, in general, nothing to do with
  (a Cartesian product of several copies of) physical space, {\it
    unless} we study systems of {\it non-relativistic point
    particles}. 
\item[(II)] ``Observables'' of a quantum-mechanical system are
  self-adjoint elements of the algebra ${\mathcal F}$ (of ``functions''
  on ``quantum phase space''). Whereas in classical Hamiltonian
  mechanics observables generate one-parameter groups of symplectic
  diffeomorphisms, ``observables'' of a quantum-mechanical system
  generate one-parameter groups of {\it unitary operators} on the
  Hilbert space ${\mathcal H}$. Pure states in classical mechanics are
  points in phase space. In quantum theory, they are {\it unit rays}
  in ${\mathcal H}$ (i.e., points in a usually infinite-dimensional
  complex projective space).
\item[(III)] In Hamiltonian mechanics, ``symmetries'' are symplectic
  diffeomorphisms of phase space; in quantum theory ``symmetries'' are
  unitary operators on a Hilbert space ${\mathcal H}$ (defining
  $^*$-automorphisms of the algebra ${\mathcal F}$). Dynamics is
  specified by a Hamiltonian vector field in classical mechanics; in
  quantum theory, it is specified by a self-adjoint operator $H$, the
  Hamiltonian (with the property that the unitary operators
  $e^{itH/\hbar}$, $t\in {\mathbb R}$, determine a one-parameter group
  of $^*$-automorphisms of ${\mathcal F}$). 
\item[(IV)] Let ${\mathcal H}$ be the Hilbert space of state vectors of
  a quantum-mechanical system, and let $H$ denote its Hamiltonian. For
  a unit ray $\psi \in {\mathcal H}$ $\left( \Vert \psi \Vert =
    1\right)$, we define
\begin{equation}
\triangle E_\psi \ \stackrel{\rm \; e.g.}{:=} \ 
\Vert \left( H-\langle \psi, H\psi\rangle\right) \ \psi \Vert
\end{equation}
to be the ``energy-uncertainty'' in the state vector $\psi$. One can
show that the time $\triangle t_\psi$ it takes	before the time
evolution $e^{-itH/\hbar} \psi$ of $\psi$ ``deviates substantially''
from $\psi$ satisfies the uncertainty relation
\begin{equation}
\triangle\;E_\psi \ \cdot \ \triangle t_\psi \ \gtrsim \ \hbar \ .
\end{equation}
For a precise version of (2.17) and generalizations and applications
thereof (e.g.~to the theory of {\it resonances}) see [6]. One may
interpret $\triangle t_\psi$ as the {\it ``life time''} of the state
described by $\psi$.
\end{itemize}

Note that, in the present section, we have reviewed some aspects of
non-relativistic quantum theory (leaving aside notions like that of a
particle, of its quantum statistics, etc.) {\it without} taking into
account gravitational effects predicted by the general theory of
relativity. In the next section, we shall sketch some general
conclusions drawn from an attempt to combine quantum theory with
general relativity. Subsequently, we shall turn off the coupling
between matter and the gravitational field again and ask what the
quantum theory of {\it non-relativistic point particles with spin}
(non-relativistic electrons, positrons and positronium) teaches us
about differential topology and -geometry of manifolds. 

\vspace{1cm}

\setcounter{equation}{0}%
\section{Reconciling quantum theory with general\\
 relativity: quantum space-time-matter}
  
Physics as we know it, at present, is founded on two pillars:

(A') \ The analysis of causal sequences of events in a classical
space-time, where classical space-time is a Lorentzian manifold (with
properties as described in Section~{\bf 1}), and the geometry of space-time
coupled to matter is described by Einstein's field equations
\[
G_{\mu\nu} \ = \ \kappa \ T_{\mu\nu} \ .
\]

(B') \ Quantum Theory: Localized events in space-time are caused by
the radiative decay of unstable, localized states of matter. 
Matter and radiation are quantum-mechanical. (The ``rules of
quantization'' must be applied to all degrees of freedom of a physical
system evolving according to some Hamiltonian dynamics: particles,
vibrations of material media, field oscillators, \dots, gravitational
waves, and hence, ultimately, to space-time geometry.)

In their present form, these two pillars are
incompatible. Fortunately, space-time appears to be classical down to
distance scales of the order of the Planck scale, $l_{\rm P}$. {}From a
purely pragmatic point of view, merging space-time structure with
quantum theory in a consistent, unified theory is therefore {\it not}
an urgent task. However, it {\it is} an important task from the point
of view of logical consistency of physical theory.

To anticipate one conclusion of our analysis, we propose --- and we do
not claim to be original in this --- that space-time should be viewed
as a {\it secondary}, or {\it derived structure}, one that emerges
from an underlying fundamental quantum theory of natural phenomena
that treats space, time and matter on an equal footing and that, a
priori, does not talk about space and time. Space-time is expected to
be a feature of such a theory that only emerges in certain limiting
regimes --- just like classical physics can emerge 
from quantum physics in certain limiting regimes.

In attempting to fill this proposal with substance, one will observe
that the concept of point-like events and point-particles is untenable
and that, as a consequence, one needs a generalization of the notion
of a classical space and of classical differential topology and
-geometry. At this point, Connes' theory of non-commutative geometry
is suggestive of what we might want to look for, mathematically.

The question addressed in this section is: Why does a combination of
quantum theory and general relativity force us to modify the concept
of space-time as a classical Lorentzian manifold; what goes wrong with
the idea of point particles and point-like 
events (i.e., ones localized in an arbitrarily small region of
space-time)? In the following, we sketch some crude answer to this
question, following ref.~[7]; see also [8] for various details. 

What we perceive as a localized event in space-time is always the
decay of an unstable, localized state of a physical subsystem that is
inherently quantum-mechanical.

\hbox{\hskip1.2cm\epsfbox{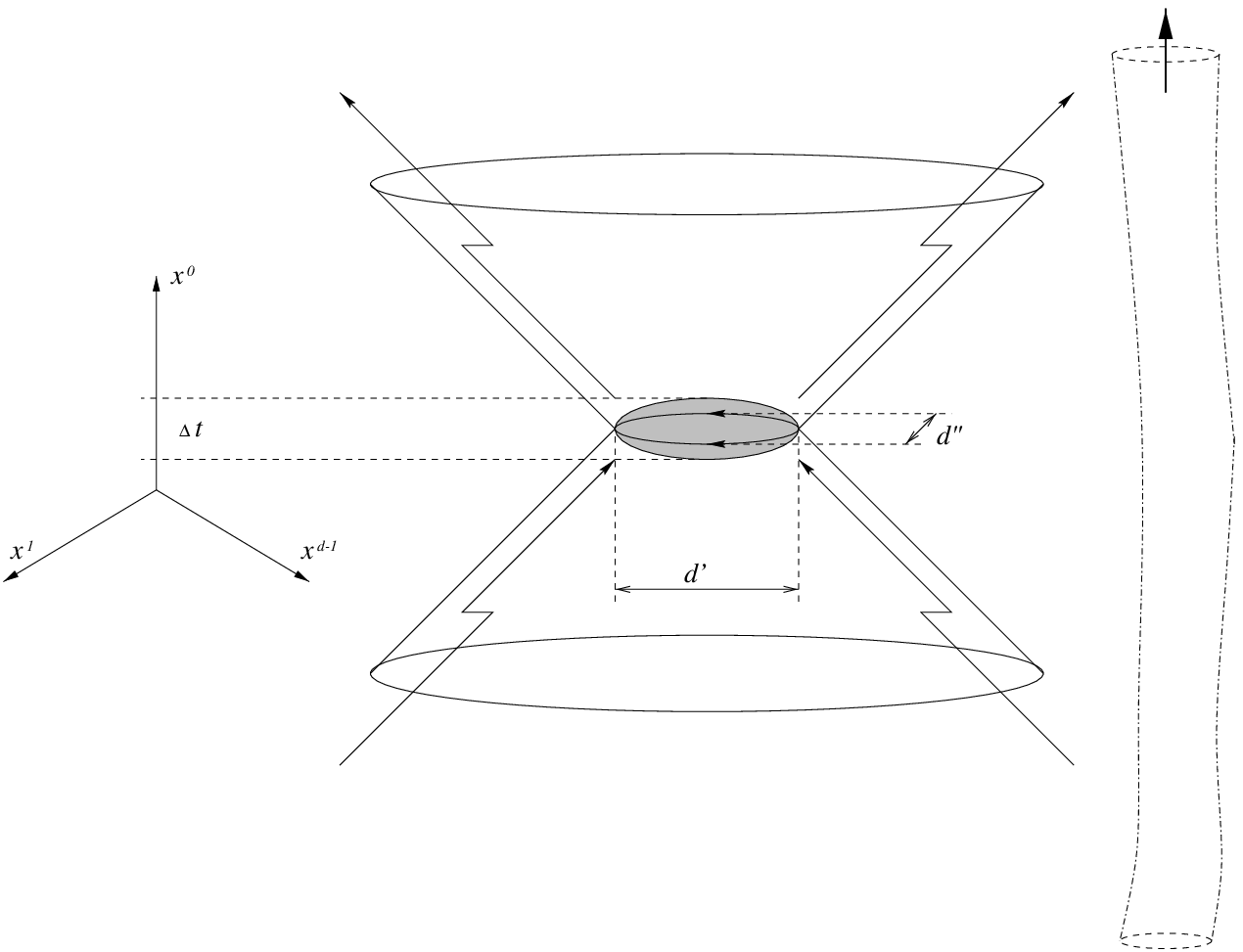}\hskip.3cm}
\vskip-2.3cm
\hbox{\hskip7cm$\scriptstyle{\rm event}$}
\vskip.5cm\hbox{\hskip11.7cm{$\scriptstyle{\rm observer}$}}
\vskip.7cm
\bigskip

\noindent 
The temporal duration of the event is denoted $\triangle t$, $d'$
denotes its maximal and $d''$ its minimal spatial extension. An
observer is located in the forward light cone of the event (spatially
far separated from the event). To be able to give meaning to the
quantities $\triangle t$, $d'$ and $d''$, space-time (outside the
event and around the observer) must be equipped with a {\it
  metric}. The observer applies the usual Heisenberg uncertainty
relations of quantum theory to interpret the observed event.

\medskip

(a) \ The energy uncertainty $\triangle E$ (defined e.g.~as in (2.16))
is bounded below by
\begin{equation}
\triangle E  \ \gtrsim \ \frac{1}{\triangle t} \ ,
\end{equation} 
in units where $\hbar = 1$ and $c=1$.
We also invoke the standard uncertainty relation
\begin{equation}
|\,\triangle\,\vec{p}\,| \ \gtrsim \ \frac{1}{d''}\ .
\end{equation}
Assuming that the motion of the center of mass of the event obeys the
laws of the special theory of relativity, we conclude that if $d''
\lesssim \frac{1}{M}$, where $M$ is the rest mass of the event then
\begin{equation}
\triangle E \ \gtrsim \ \frac{1}{d''} \ , 
\end{equation}
which improves (3.1) if $d'' \ \ll \ \triangle t\,$. 

(b) \ Suppose that $\triangle t \gg d' \approx d''$. Then the metric
well outside the region where the event is localized is given,
approximately, by the Schwarzschild metric (or a Schwarzschild-Newman
metric if the event carries electric charge). The Schwarzschild radius
(radius of the event horizon) is bounded by
\begin{equation}
r_S \ \gtrsim \ \triangle E \cdot l_P^2 \ .
\end{equation}
Next, suppose that $\triangle t \gg d' \gg d''$. Then the metric
outside the event is given, approximately, by the Kerr(-Newman) metric
describing an object with non-vanishing angular momentum $l$. Let $r'$
and $r''$ be the maximal and minimal spatial extensions of the event
horizon, and let $l \lesssim \triangle E r'$. Then $\triangle E
\lesssim \frac{r''}{l_P^2}$ \ and
\begin{equation}
r'\ \cdot r'' \ \approx \ l_P^2 \ .
\end{equation}

(c) \ Suppose that Hawking is wrong, and a black hole is really
black. Then decay products of the event will reach the observer only
if
\begin{equation}
d' \ \approx \ d'' \ \gtrsim \ r_S \ \gtrsim \ \triangle E \cdot l_P^2
\ ,
\end{equation}
for the Schwarzschild metric. Combining (3.6) with (3.3), we conclude
that
\begin{equation}
d' \ \approx \ d'' \ \gtrsim \ l_P \ .
\end{equation}
Combining (3.6) with (3.1), we get that 
\begin{equation}
d' \ \triangle t \ \gtrsim \ l_P^2 \ .
\end{equation}
For the Kerr metric we use that $\triangle E \lesssim
\frac{r''}{l_P^2}$ \ and hence, using (3.3) and the inequalities $d'
\gg d'' \gtrsim r''$, we conclude that
\[
\frac{1}{d'} \ \lesssim \ \triangle E \ \lesssim \ \frac{d''}{l_P^2} \ ,
\]
hence
\begin{equation}
d' \ d'' \ \gtrsim \ l_P^2 \ ,
\end{equation}
and, using that $\triangle E \gtrsim \frac{1}{\triangle t}$ and $d' >
d''$, it follows that
\begin{equation}
d' \cdot \triangle t \ \gtrsim \ l_P^2 \ .
\end{equation}
In all cases, we appear to conclude that if an event is {\it not}
encased in a black hole then
\begin{equation}
d' \cdot d'' \ \gtrsim \ l_P^2 \quad {\rm and} \quad d' \cdot \triangle t
\ \gtrsim \ l_P^2 \ .
\end{equation}
These are the uncertainty relations first proposed in [7]. 

Next, we assume that Hawking's laws of black hole evaporation are
right. If $d'$ and $d''$ denote the maximal and minimal spatial
extension of a black hole then its mass is
\begin{equation}
M \ \gtrsim \ \frac{d''}{l_P^2} \ .
\end{equation}
The Hawking temperature of the black hole is [9] 
\begin{equation}
kT \ = \ \frac{1}{8 \pi l_P^2 M} \ .
\end{equation}
Elementary thermodynamics then implies that
\begin{eqnarray}
- \ \frac{dM}{dt} &\approx& \gamma \left( \frac{1}{d''}\right)^4 \cdot
\left( d'\right)^2 \ = \ \frac{\gamma}{(d'')^2} \left(
  \frac{d'}{d''}\right)^2 \nonumber \\
&\approx& \frac{\gamma}{l_P^4} \ M^{-2} \ \left(
  \frac{d'}{d''}\right)^2 
\end{eqnarray}
for a dimensionless constant $\gamma$. {}From (3.14) we obtain a bound 
on the {\it life time} of a radiating black hole
\begin{equation}
\triangle t \ \approx \ l_P^4 \ M^2 \ \left( \frac{d''}{d'}\right)^2 \
\triangle E \ ,
\end{equation}
with $\triangle E = \triangle M \approx M$ (provided the initial mass
of the black hole is large compared to $\frac{1}{l_P}$ ). 
Using that $\triangle E \gtrsim \frac{1}{\triangle t}$ and $M \approx
\frac{d''}{l_P^2}$ , we conclude that
\[
\left( \triangle t\right)^2 \ \left( \frac{d'}{d''}\right)^2 \ \gtrsim
\ \left( d''\right)^2 \ ,
\]
hence 
\begin{equation}
\phantom{( \triangle t)^2 \ } \triangle t \cdot d' \ \gtrsim \ \left(
  d''\right)^2 \ .
\end{equation}
Furthermore, by (3.1), (3.15) and since $d'' < d'$, 
\[
1 \ \lesssim \ \triangle t \cdot \triangle E \ \approx \ l_P^4 \ M^2
\left( \triangle E\right)^2 \ \approx \ l_P^4 \ M^4
\]
and hence, using (3.12),
\begin{equation}
d'' \ \gtrsim \ l_P \ .
\end{equation}
In conclusion, we find that if a localized event can be interpreted as
the evaporation of a black hole then, again,
\begin{equation}
d'\cdot d'' \ \gtrsim \ l_P^2 \quad {\rm and} \quad \triangle t \cdot
d' \ \gtrsim \ l_P^2 \ ;
\end{equation}
see [8]. {}From (3.12), (3.13) and (3.17) we also derive that
\begin{equation}
k T \ \lesssim \ \frac{1}{l_P} 
\end{equation}
(an upper bound for the Hawking temperature of a black hole). We also
recall the expression, due to Bekenstein and Hawking, for the
entropy of a black hole (in four space-time dimensions)
\begin{equation}
S =  A/4l_P^{2} \ ,
\end{equation}
where $A$ is the area of the horizon of the black hole. This
expression suggests that the number $N_A$ of distinct states of a
black hole is bounded by
\begin{equation}
N_A \ \lesssim \ {\rm exp}\left( {\rm const}\;A/l_P^{d-2}\right) \ .
\end{equation}
In all these formulas, we do not pay attention to values of various
dimensionless constants. 

We should emphasize that our analysis is based on the assumption that
uncertainty relations and Einstein's field equations are valid down to
scales comparable to the Planck scale. It can certainly not be
excluded that quantum theory and the general theory of relativity are
modified, in a more fundamental theory, in such a way that our
analysis is {\it invalidated}. (This might be the case if one
succeeded in constructing some asymptotically free quantum field
theory of matter and the gravitational field; but there is no
evidence, at present, that such a theory can be constructed.)
Keeping the above warning in mind, we shall take the point of view
that the bounds (3.18) on the extension of events in space and time,
the bound (3.19) on the temperature of events and the bound (3.21) on
the number of distinct states of a black hole are fundamental. Our
derivation of these bounds follows [7,8]; but see also [9,10,11]. A
basic result in ref.~[7] (the work that partly motivated~[8]) is that
the {\it uncertainty relations} (3.18) {\it are compatible with the
  special theory of relativity} (Poincar\'e covariance) on large
scales. 

On the basis of inequalities (3.18), one may argue that the number
$N_{{\mathcal O}}$ of events or ``excited modes'' of matter localized
inside an open region ${\mathcal O}$ of finite ({\it metric}) volume
vol$_g({\mathcal O})$ that can be distinguished, in principle,
experimentally is bounded by
\begin{equation}
N_{{\mathcal O}} \ \lesssim \ \frac{{\rm vol}_g({\mathcal O})}{l_P^d}
\ ,
\end{equation}
where $d$ is the dimension of space-time (and $d>2$). Combining this
bound with (3.21), we are tempted to conclude that the total number of
distinct observations of events localized inside some open space-time
region ${\mathcal O}$, with vol$_g({\mathcal O}) < \infty$, is
essentially bounded by \ exp(const$\,$vol$_g({\mathcal O})/l_P^d$). If
${\mathcal A}_{l_P}({\mathcal O})$ 
denotes the algebra of observables
localized in ${\mathcal O}$, and \ vol$_g({\mathcal O}) < \infty$, we
may then argue that ${\mathcal A}_{l_P}({\mathcal O})$ is {\it
  ``essentially finite-dimensional''}:
\begin{equation}
{\rm dim} \left( {\mathcal A}_{l_P}\left({\mathcal O}\right) \right) \ 
\lesssim \ {\rm exp} \left( {\rm const}\cdot{{\rm vol}_g({\mathcal O})}
  \big/ l_P^d \right) \ .
\end{equation}
In particular, if space-time is foliated in {\it compact} space-like
hypersurfaces of codimension 1 one might want to predict that the
algebra of {\it all} local observables is (essentially) {\it finite
  dimensional}.

These conclusions are highly plausible, unless the coupling of modes
of very high energies (comparable to or higher than the Planck
energy) to the gravitational field becomes weak and tends to 0 as the
energy increases to infinity. We propose to take them seriously as
long as modes of energies \ $\gtrsim \ \frac{1}{l_P}$ \ remain
unexcited.

In local relativistic quantum field theory [58], the local algebras
${\mathcal A}({\mathcal O})$ are von Neumann factors of type III$_1$ (see
e.g.~[12]) and hence are genuinely {\it infinite-dimensional} (if
${\mathcal O}$ is, for example, a bounded open double cone). As a
consequence, the bounds (3.22) and (3.23) are {\it violated}. However,
if a local, relativistic quantum field theory describes a finite
number of species of massive asymptotic particles then the number of
states localized in an open region ${\mathcal O}$ of physical space,
with \ vol$_g({\mathcal O}) < \infty$, and with an energy \ $\leq
\varepsilon\; {\rm vol}_g({\mathcal O})$, for an arbitrary \
$\varepsilon < \infty$, is expected to be bounded by \
exp(const$_\varepsilon {\rm vol}_g({\mathcal O})$); see [13] and
refs.~given there. The problem is that \ const$_\varepsilon \to
\infty$, \ as $\varepsilon \to \infty$.

It is sometimes argued that, on the r.s.\ of (3.22) and (3.23), the
dimensionless volume of ${\mathcal O}$ can be replaced by the
dimensionless area of the boundary of ${\mathcal O}$ (``holographic
principle'' [14]). But the plausibility of this prediction is very
limited (unless one considers a single black hole).

We draw the reader's attention to the similarity between the
bound (2.6) and the bound (3.22).

The bounds (3.18), (3.22) and (3.23) say that it is impossible to
determine the location of an event or of some excited modes of matter
in space and time arbitrarily precisely and suggest that, as a
consequence, the concept of a {\it space-time continuum} is {\it not
  strictly meaningful}. The (universal) algebras ${\mathcal
  A}({\mathcal O})$ of observables measurable in a bounded, open
space-time region ${\mathcal O}$ provided by local relativistic
quantum field theory {\it cannot}, ultimately, be the true local
algebras. They must be {\it deformed} to algebras ${\mathcal
  A}_{l_P}({\mathcal O})$ \ satisfying inequality (3.23) -- just like
the algebras \ ${\mathcal F}(\Omega), \ \Omega \subset \Gamma$, of
``observables'' of classical mechanics must be deformed to the
algebras ${\mathcal F}_{\hbar} (\Omega)$ of ``observables'' of quantum
mechanics in such a way that the bound (2.6) holds. The deformation
of classical physics to a fundamental theory of space, time and matter
thus involves {\it two deformation parameters}, $\hbar$ and $l_P$. The
two deformations, in $\hbar$ and in $l_P$, are expected to solve the
problem of {\it singularities} of space-time. For example, the
evaporation of a black hole is not expected to result in a
singularity, because space-time is non-commutative at the Planck
scale.

We have reached satisfactory understanding of what is going on on the
$\hbar$-axis (at $l_P=0$) -- quantum theory of matter in a classical
space-time background -- and on the $l_P$-axis (at $\hbar=0$) --
classical general relativity. But nature is found in the region $\hbar
>0$, $l_P>0$. String perturbation theory is an attempt to understand
something about nature by constructing an expansion in powers of some
function of $l_P$ about $l_P=0$. String theory is clever in that it
treats matter quantum-mechanically in such a way that, on the tree
level, some understanding of what is going on on the
$l_P$-axis is automatically built into the theory.

In conventional quantum gravity, one attempts to construct an
expansion in powers of $\hbar$ about $\hbar=0$ (the $l_P$-axis), in
the following way: One starts by viewing space-time as a classical
manifold equipped with a metric, $g_{\mu\nu}^c$, that is a solution of
the Einstein equation $G_{\mu\nu}^c = \kappa T_{\mu\nu}^c$ (the
superscript ``c'' stands for ``classical''). One then attempts to
include small quantum corrections by setting \
$T_{\mu\nu}=T_{\mu\nu}^c + \triangle T_{\mu\nu}$, where $\triangle
T_{\mu\nu}$ is a ``small'' {\it operator-valued} field. This forces
one to also interpret $G_{\mu\nu}$ as {\it operator-valued}, which, in
turn, entails that the coefficients of the connection $\nabla_\mu$
on the (co-)tangent bundle of the space-time manifold must be
operator-valued, as well. But then vector fields and differential
forms over space-time must be operator-valued, too.
Hence the pairing of a one-form with a vector field, which should
yield a {\it function} on space-time, will in general yield an {\it
  operator-valued} function. This suggests that space-time {\it
  cannot} be viewed as a classical manifold!

In trying to maintain locality (Einstein causality) of the quantum
theory, one would like to imagine that the causal structure on
space-time inherited from the metric $g_{\mu\nu}^c$ solving the {\it
  classical} equations $G_{\mu\nu}^c = \kappa T_{\mu\nu}^c$,
provides an appropriate notion of locality for the quantum
theory. This would be o.k. if the ``quantized'' metric $g_{\mu\nu}$
were conformally equivalent to $g_{\mu\nu}^c$ --- i.e.\ $g_{\mu\nu}= 
{\rm exp}(\phi)g_{\mu\nu}^c$ for some quantum field $\phi$ ---, which, 
in general, {\it  cannot} be true if $d>2$. 
The only way out appears to be to give up the idea of space-time as a
classical manifold! In other words, the problem of quantum gravity is
not, actually, a problem of calculating perturbations in quantum
theory arising from gravitational interactions, or perturbations in
general relativity arising from quantum-mechanical fluctuations, but
to construct a {\it two-parameter deformation}, in $\hbar$ and $l_P$,
of the laws of classical physics resulting in a {\it non-commutative
  generalization of geometry}. 

Our discussion can be summarized by postulating that {\it real
  microscopes cannot} resolve a number of distinct events located in
an open region of space-time of finite volume that would exceed the
bound given in (3.22), and that they cannot be used to determine the
location of an event in space-time with an accuracy violating
(3.18). The bounds (3.18), (3.22) and (3.23) are assumed to be valid
{\it independently} of how such an ``Einstein-Heisenberg microscope''
is built and operated. Intuitively, one would expect that a theory
compatible with (3.18) and (3.22,23) had better be a {\it quantum
  theory} of {\it ``extended objects''} (corresponding to the two
deformation parameters $\hbar$ and $l_P$) that treats space, time and
matter on an equal footing.

Before we try to describe aspects of such a theory, we shall return to
the study of (non-relativistic) quantum theory of point particles with
spin, setting $l_P=0$. We propose to find out what it teaches us about
the geometry of {\it physical} (Newtonian) {\it space}.

\vspace{1cm}

\setcounter{equation}{0}%

\section{Classical differential topology and -geometry\\
and supersymmetric quantum theory}

In this section we describe an approach to differential topology and
-geometry based on the quantum theory of non-relativistic point
particles with spin, {\it Pauli's electron, positron} and {\it
  bound states} thereof. The quantum theory of these particles {\it
 exhibits  supersymmetries}. We show how the classification of different 
types of differential geometries can be derived from the classification 
of supersymmetries. Our approach is inspired by ideas in [15,16,17,5] 
and has appeared in [18,19]; another useful reference is [20]. 

Throughout this section, the recoil of matter on the gravitational
field is neglected, and matter is thought to consist of
non-relativistic point particles. We want to clarify what the quantum
theory of such particles teaches us about the geometry of {\it
  physical space} (time is a parameter). Our presentation follows the
general ideas described in Section~{\bf 2}. But we shall consider the quantum
theory of a single particle with {\it spin}, and spin will turn out to
play a fundamental role. The results of this section set the stage for
a generalization of topology and geometry that enables us to study
non-commutative spaces, as pioneered by Connes~[5]. That
generalization will be described in the next section. The tools 
described there are likely to be useful in exploring aspects of a
theory, yet to be found, that unifies the quantum theory of matter
with gravitation.

We start this section with a recapitulation of Pauli's quantum theory
of the non-relativistic electron with spin, generalized to arbitrary
space dimension.

\vspace{.5cm}

\subsection{Pauli's electron}

Physical space is chosen to be a smooth, orientable, Riemannian
spin$^c$ manifold $(M,g)$ of dimension $n$, where $(g_{ij})$ denotes
the metric on the tangent bundle $TM$ and $(g^{ij})$ the (inverse)
metric on the cotangent bundle $T^*M$. Let $\Lambda^{\,^\bullet} M =
\displaystyle\mathop{\oplus}_k (T^*M)^{\wedge k}$ denote the bundle
of completely anti-symmetric covariant tensors over $M$. Let
$\Omega^{\,^\bullet} (M)$ be the space of smooth sections of
$\Lambda^{\,^\bullet} 
M$, i.e., of smooth differential forms on $M$, and $\Omega_{{\mathbb
    C}}^{\,^\bullet} (M) = \Omega^{\,^\bullet} (M) \otimes {\mathbb C}$ its
complexification. Since we are given a Riemannian metric on $M$,
$\Omega_{{\mathbb C}}^{\,^\bullet} (M)$ is equipped with a Hermitian
structure which, together with	the Riemannian volume
  element $d$vol$_g$, determines a scalar product $(\cdot, \cdot)_g$ on
  $\Omega_{{\mathbb C}}^{\,^\bullet} (M)$. Let ${\mathcal H}_{e-p}$
  denote the Hilbert space completion of $\Omega_{{\mathbb
      C}}^{\,^\bullet} (M)$ in the norm determined by
  $(\cdot,\cdot)_g$. Thus ${\mathcal H}_{e-p}$ is the
  space of complex-valued, square-integrable differential forms on
  $M$. This Hilbert space is 
${\mathbb Z}$-graded, 
\begin{equation}
{\mathcal H}_{e-p} \ = \ \displaystyle\bigoplus_{k=0}^n \
{\mathcal H}_{e-p}^{(k)} \ , 
\end{equation}
where ${\mathcal H}_{e-p}^{(k)}$ is the subspace of square-integrable
differential forms of degree $k$. 

Given a one-form $\xi \in \Omega^1 (M)$, let $X$ be the vector field
corresponding to $\xi$ by the equation 
\begin{equation}
\xi (Y) \ = \ g(X,Y) \ ,
\end{equation}
for any smooth vector field $Y$. For every \ $\xi \in \Omega_{{\mathbb
    C}}^1 (M)$, we define two operators on ${\mathcal H}_{e-p}$:
\begin{equation}
a^* (\xi) \psi \ = \ \xi \wedge \psi 
\end{equation}
and
\begin{equation}
a (\xi) \psi \ = \ X \rightharpoonup \ \psi \ ,
\end{equation}
for all $\psi \in {\mathcal H}_{e-p}$. In (4.4), $\rightharpoonup$
denotes interior multiplication. One easily checks that $a^*(\xi)$ is
the adjoint of $a(\bar{\xi})$ in the scalar product of ${\mathcal
  H}_{e-p}$. Furthermore, one verifies that, for arbitrary $\xi$ and
$\eta$ in $\Omega_{{\mathbb C}}^1 (M)$, 
\begin{eqnarray}
\left\{ a(\xi), \ a(\eta)\right\} &=&\left\{ a^*(\xi), \
  a^*(\eta)\right\} = 0\,, \nonumber \\
\left\{ a(\xi), \ a^*(\eta)\right\} &=& g \left( \xi, \eta\right) \ ,
\end{eqnarray}
where $\{ A,B\} := AB + BA$ denotes the anti-commutator of $A$ and
$B$, and we use the symbol $g$ to denote the (inverse) metric on
$T^*M$. Eqs.~(4.5) are called canonical anti-commutation relations and
are basic in the description of fermions in physics.

Next, for every {\it real} \ $\xi \in \Omega^1 (M)$, we define two
anti-commuting anti-selfadjoint operators \ $\Gamma(\xi)$ and
$\bar{\Gamma} (\xi)$ on ${\mathcal H}_{e-p}$ by
\begin{eqnarray}
\Gamma (\xi) &=& a^*(\xi) \ - \ a(\xi), \\
\bar{\Gamma} (\xi) &=& i \left( a^*\left(\xi\right) \ + \
  a\left(\xi\right)\right) \ .
\end{eqnarray}
One checks that 
\begin{eqnarray}
\left\{ \Gamma (\xi), \Gamma (\eta) \right\} &=& \left\{ \bar{\Gamma}
    (\xi), \bar{\Gamma}(\eta) \right\} \ = \ -
    \,2g\left( \xi,\eta \right)\ , \\
\left\{ \Gamma (\xi), \bar{\Gamma}(\eta)\right\} &=& 0 \ ,
\end{eqnarray}
for arbitrary $\xi$ and $\eta$ in $\Omega^1(M)$. Thus $\Gamma(\xi)$
and $\bar{\Gamma}(\xi)$, $\xi \in \Omega^1(M)$, are anti-commuting
sections of two isomorphic {\it Clifford bundles}, $Cl(M)$, over $M$. 

An $n$-dimensional Riemannian manifold $(M,g)$ is a {\it spin$^c$
  manifold} if and only if $M$ is oriented and there exists a complex
Hermitian vector bundle $S$ of rank $2^{[\frac n 2]}$ over $M$
(where $[k]$ denotes the integer part of $k\in{\mathbb R}$) and a
bundle homomorphism $c$: $T^*M \lra$ End$(S)$ such that
\begin{eqnarray}
c (\xi) \;+\;c^*(\xi) &=& 0 \\
c^* (\xi)\; c(\xi) &=& g\;(\xi,\xi)\ ,
\end{eqnarray}
for arbitrary $\xi \in \Omega^1(M)$. The adjoint $c^*(\xi)$ of
$c(\xi)$ is defined (pointwise) with respect to the Hermitian
structure $\langle \cdot, \cdot\rangle_S$ on $S$. The Hermitian
structure $\langle \cdot, \cdot\rangle_S$ and the Riemannian volume
form \ $d\,{\rm vol}_g$ \ determine a scalar product, $(\cdot, \cdot)_S$,
on the space $\Gamma (S)$ of sections of $S$. The completion of
$\Gamma(S)$ in the norm determined by the scalar product $(\cdot,
\cdot)_S$ is a Hilbert space denoted by ${\mathcal H}_e$, the Hilbert
space of square-integrable Pauli-Dirac spinors on $M$. The
homomorphism $c$ extends uniquely to an irreducible $^*$-representation
of the Clifford algebra over $T_x^* M$ on the fibre $S_x$ of $S$ over
$x$, for all $x\in M$. 

If $M$ is an {\it even-dimensional} spin$^c$ manifold then there is a
section $\sigma\neq 0$ of the Clifford bundle generated by the
operators $c(\xi)$, $\xi \in \Omega^1 (M)$, which anti-commutes with
every $c(\xi)$ and satisfies $\sigma^2=\id$, ($\sigma$ corresponds to
the Riemannian volume form on $M$), and there is an isomorphism
\begin{equation}
i\;:\;\Omega_{{\mathbb C}}^{\,^\bullet} \;(M) \ \lra \ \Gamma (\bar{S})
\;\otimes_{{\mathcal A}}\; \Gamma(S) \ , 
\end{equation}
where ${\mathcal A} = C (M)$, and where $\bar{S}$ is the
``charge-conjugate'' bundle associated to $S$, obtained from $S$ by
complex conjugation of the transition functions of $S$. The bundle
$\bar{S}$ inherits a natural Clifford action $\bar{c}\;:\;T^* M \;\lra$
End$(\bar{S})$ from the Clifford action $c$ on $S$, and the
isomorphism $i$ is an intertwiner satisfying 
\begin{eqnarray}
i\;\circ\;\Gamma(\xi) &=& \left( \id\;\otimes\;c
  \left(\xi\right)\right)\;\circ\;i \ ,\\
i\;\circ\;\bar{\Gamma} \left(\xi\right) &=& \left( \bar{c}
  \left(\xi\right)\;\otimes\;\sigma\right)\;\circ\;i \ ,
\end{eqnarray}
for all $\xi \in\Omega^1(M)$. The element $\sigma$ is inserted on the
r.s.\ of (4.14) to ensure that the Clifford actions $\Gamma$ and
$\bar{\Gamma}$ anti-commute, as required in (4.9). 

If $M$ is an {\it odd-dimensional} spin$^c$ manifold then the Clifford
algebra associated with a cotangent space $T_x^*M$ contains a {\it
  central} element, $\sigma$, corresponding to parity. There is then
an isomorphism
\begin{equation}
i\;:\;\Omega_{{\mathbb C}}^{\,^\bullet} (M) \ \lra \ \Gamma
(\bar{S})\otimes_{{\mathcal A}} \ \Gamma (S) \otimes {\mathbb C}^2 
\end{equation}
such that
\begin{eqnarray}
i\;\circ\;\Gamma(\xi) &=& \left( \id\;\otimes\;c
  \left(\xi\right)\;\otimes\;\tau_3\right)\;\circ\;i \ ,\nonumber\\
i\;\circ\;\bar{\Gamma}(\xi) &=& \left( \bar{c}
  \left(\xi\right)\;\otimes\;\id\;\otimes\;\tau_1\right) \;\circ\;i \ ,
\end{eqnarray}
where \ $\tau_1 = {0 \ 1 \choose 1 \ 0}$ \ and \ $\tau_3 = {1 \quad 0
  \choose 0 \ -1}$  .

A connection $\nabla_S$ on $S$ is called a {\it spin$^c$ connection} 
iff it satisfies the ``Leibniz rule''
\begin{equation}
\nabla_X^S \left( c(\eta)\psi\right) \ = \ c \left( \nabla_X
  \eta\right) \psi \ + \ c(\eta) \nabla_X^S \psi \ ,
\end{equation}
for arbitrary vector fields $X$, one-forms $\eta$ and 
sections $\psi \in \Gamma (S)$, where $\nabla$ is a connection on
$T^*M$. We say that $\nabla^S$ is compatible with the Levi-Civita
connection iff, in (4.17), $\nabla = \nabla^{L.C.}$ .

If $\nabla_1^S$ and $\nabla_2^S$ are two Hermitian spin$^c$
connections compatible with the {\it same} connection $\nabla$ on
$T^*M$ then
\begin{equation}
\left( \nabla_1^S \ - \ \nabla_2^S\right)\;\psi \ = \
i\;\alpha\;\otimes\;\psi 
\end{equation}
for some real one-form \ $\alpha\in\Omega^1(M)$. The physical
interpretation of $\alpha$ is that it is the difference of two {\it
  electromagnetic vector potentials}. If $R_{\nabla^S}$ denotes the
curvature of a spin$^c$ connection $\nabla^S$ then
\begin{equation}
2^{-[\frac n 2]}\;\tr\,\left( R_{\nabla^S} \left( X,Y\right)\right) \ =
\ F_A \left( X,Y\right) \ ,
\end{equation}
for arbitrary vector fields $X,Y$, where $F_{2A}$ is the curvature
(``electromagnetic field\break 
strength'') of a U(1)-connection $2A$
(``vector potential'') on a line bundle canonically associated to $S$;
$A$ itself defines a ``virtual U(1)-connection''. See [20] and [18] for
details. 

The Pauli-Dirac operator associated with a spin$^c$ connection
$\nabla^S$ is defined by
\begin{equation}
D_A \ = \ c\;\circ\;\nabla^S \ .
\end{equation}
We are now prepared to describe Pauli's quantum theory of the
non-relativistic electron. The Hilbert space of pure state vectors of
a one-electron system is chosen to be ${\mathcal H}_e$, the space of
square-integrable Pauli-Dirac spinors. The dynamics of an electron,
with gyromagnetic factor $g$ measuring the strength of the magnetic
moment of the electron set equal to 2, is given by the Hamiltonian
\begin{equation}
H_A\ = \ \frac{\hbar^2}{2m} \ D_A^2\;+\;v\ = \
\frac{\hbar^2}{2m}\;\left( - \triangle_A^S\;+\;\frac r 4\; +\;c \left(
    F_A\right)\right) \;+\;v \ ,
\end{equation}
where $m$ is the mass of the electron, $v$ is the scalar
(``electro-static'') potential, which is a function on $M$, $r$
denotes the scalar curvature, $\triangle_A^S$ is the ``Lichnerowicz
(covariant) Laplacian'', and $c(F_A)$ denotes Clifford multiplication
by the 2-form $F_A$. For conditions ensuring that $H_A$ is bounded
from below and self-adjoint see [17,20,21]. 

Considering position measurements as fundamental, one chooses
${\mathcal A} := C(M)$ as an algebra of observables. {}From the point of
view of quantum physics it is, however, more natural to choose the
algebra ${\mathcal F}$ of ``functions on {\it quantum phase
space}'' as an algebra of observables. The (non-commutative) algebra
${\mathcal F}$ is defined to be the smallest $C^*$-algebra generated
by 
\[
\left\{ \alpha_\tau (a), \ f\left( H_A^0\right) \bigm| a \in C(M),\ 
  \tau \in {\mathbb R},\ f \in C_0^\infty ({\mathbb R})\right\}
\]
where $H_A^0 = \frac{\hbar^2}{2m} \ D_A^2$ \ and
\[
\alpha_\tau (B) \ := \ e^{i\tau H_A^0/\hbar} \ B\;e^{-\,i\tau
  H_A^0/\hbar} \ ,
\]
for any $B\in B({\mathcal H}_e)$.

Connes has shown that the spectral triple \ $\left( {\mathcal A},
  {\mathcal H}_e, D_A\right)$ encodes the topology and Riemannian
geometry of $M$ completely; see [5]. It is less clear how
much information about $M$ is encoded into the data 
$\left( {\mathcal F}, {\mathcal H}_e, D_A\right)$, viewed as abstract
spectral data, and some interesting mathematical questions remain to
be solved.

If $v=0$ then
\begin{equation}
H_A^0 \ = \ D^2
\end{equation}
where \ $D\;:=\;\frac{\hbar}{\sqrt{2m}} \; D_A$, ($D$ is self-adjoint),
i.e., $H_A^0$ is the square of a {\it ``supercharge''}. If $M$ is
even-dimensional then, as discussed above, the Clifford bundle over
$M$ has a section $\sigma$ which is a unitary involution on ${\mathcal
  H}_e$ with the property that
\[
\left[ \sigma, a\right] \ = \ 0 \quad {\rm for \ all} \quad a \in
{\mathcal A} \quad (a \in {\mathcal F}) \ ,
\]
but
\begin{equation}
\left\{ \sigma, D\right\} \ = \ 0 \ .
\end{equation}
Then $\sigma$ defines a ${\mathbb Z}_2$-grading of ${\mathcal
  H}_e$. The data $\left( {\mathcal A}, {\mathcal H}_e, D,
  \sigma\right)$ yield an example of $N=1$ {\it supersymmetric quantum
  mechanics}\renewcommand{\thefootnote}{\fnsymbol{footnote}}\footnote[1]{Our 
nomenclature {\it deviates} from the one used in the older literature, e.g.~in
  [22]!}. An important
topological invariant of $M$ provided by Pauli's supersymmetric
quantum mechanics of a non-relativistic electron is given by
the index of $D$, 
\begin{equation}
\str\,\left( e^{-\;\beta H_A^0}\right) \ := \tr\,\left(
  \sigma\,e^{-\;\beta H_A^0}\right) 
\end{equation} 
which is easily seen to be {\it independent} of $\beta$. 
Using a fairly standard {\it Feynman-Kac formula} to express
(4.24) as a functional integral and studying the small $\beta$
asymptotics of this integral, Alvarez-Gaum\'e has been able to
rederive the $\hat{A}$ genus and the index density for the Dirac
operator $D$ in a simple manner; see~[22] (and [17]). We shall not
pursue this theme here. 

In order to describe the twin of Pauli's electron, the non-relativistic 
positron, we replace the bundle $S$ by the charge-conjugate spinor 
bundle $\bar{S}$. A Hermitian spin connection
$\nabla^S$ on $S$ uniquely determines a spin connection \ 
$\nabla^{\bar{S}}$ on $\bar{S}$, \ by setting \break
 $\nabla_X^{\bar{S}} = C
\nabla_X^S C^{-1}$, where $C:S\lra \bar{S}$ is charge conjugation, and
$X$ is an arbitrary {\it real} vector field. The space of square
integrable sections of $\bar{S}$ is denoted by ${\mathcal H}_p$;
${\mathcal H}_p$ is a {\it right} module, while ${\mathcal H}_e$ is a
{\it left} module for ${\mathcal A}$ and $Cl(M)$. One defines 
\[
\bar{D}_A \ = \ \bar{c}\;\circ\;\nabla^{\bar{S}}
\]     
and sets
\begin{equation}
\bar{H}_A \ = \ \frac{\hbar^2}{2m} \ \bar{D}_A^2 \ - v \ .
\end{equation}
The physical interpretation of these changes is simply that the sign
of the electric charge of the particle is reversed, replacing $A$ by
$-A$ (and $v$ by $-v$), keeping everything else, such as its mass $m$,
unchanged. 

The third character of the play is the (non-relativistic) {\it
  positronium}, the particle corresponding to a {\it ground state} of a
bound pair of an electron and a positron. As an algebra of
``observables'' we continue to use e.g.~${\mathcal A} = C(M)$, as for
the electron and the positron. The Hilbert space of pure state vectors
of positronium is
\begin{equation}
{\mathcal H}_{e-p} \ = {\mathcal H}_p\;\otimes_{{\mathcal A}}\;
{\mathcal H}_e\ ,\quad\	 {\rm dim}\,M \ {\rm even} \ ,
\end{equation}
and
\begin{eqnarray}
{\mathcal H}_{e-p} &=& \left( {\mathcal H}_p \otimes_{{\mathcal A}}
  {\mathcal H}_e\right)_+ \ \oplus \  \left( {\mathcal H}_p \otimes
  {\mathcal H}_e\right)_- \nonumber \\
&\cong& \left( {\mathcal H}_p \otimes_{{\mathcal A}} {\mathcal H}_e
\right) \ \otimes \ {\mathbb C}^2\ , \quad\ {\rm dim}\,M \ {\rm odd} \
.
\end{eqnarray}
Elements in $({\mathcal H}_p \otimes_{{\mathcal A}} {\mathcal H}_e)_+$
are even, elements $({\mathcal H}_p \otimes_{{\mathcal A}} {\mathcal
  H}_e)_-$ are odd under space reflection. By (4.12) and (4.15), the
Hilbert space ${\mathcal H}_{e-p}$ is the Hilbert space (4.1) of
square-integrable differential forms on $M$. A connection
$\widetilde{\nabla}$ on ${\mathcal H}_{e-p}$ can be defined as follows: If
$\phi \in {\mathcal H}_{e-p}$ is given by $\phi = \psi_1
\otimes_{{\mathcal A}} \psi_2 (\otimes u)$, with $\psi_1\in {\mathcal
  H}_p$, $\psi_2 \in {\mathcal H}_e$, $(u \in {\mathbb C}^2)$, we set
\begin{equation}
\widetilde{\nabla} \phi \ = \ \left( \nabla^{\bar{S}}
  \psi_1\right)\;\otimes_{{\mathcal A}}\;\psi_2\;\left( \otimes
  u\right) \ + \ \psi_1\;\otimes_{{\mathcal A}}\; \nabla^S \psi_2
\;(\otimes u) \ .
\end{equation}
If $\nabla^S$ is compatible with the connection $\nabla$ on $T^*M$
then $\widetilde{\nabla}$ is the connection on $\Lambda^{\,^\bullet}
M$ determined by $\nabla$. We note that $\widetilde{\nabla}$ is {\it
  independent} of the electromagnetic vector potential $A$ (the
virtual U(1)-connection), which, physically,
comes from the fact that the electric charge of positronium is zero.

We define two first-order differential operators on ${\mathcal
  H}_{e-p}$ by
\begin{equation}
{\mathcal D} \ = \ \Gamma\;\circ\;\widetilde{\nabla}\ ,\ \ \ 
\bar{{\mathcal D}} \ = \ \bar{\Gamma}\;\circ\;\widetilde{\nabla} \ ,
\end{equation} 
with $\Gamma$ and $\bar{\Gamma}$ defined as in (4.13), (4.14), (4.16)
(see also (4.6,7)). {\it If} $\nabla^S$ {\it is compatible with
  the Levi-Civita connection} $\nabla^{L.C.}$ then ${\mathcal D}$ and
$\bar{{\mathcal D}}$ satisfy the algebra
\begin{equation}
\left\{ {\mathcal D}, \bar{{\mathcal D}} \right\}\ = \ 0, \quad {\mathcal D}^2
  \ = \ \bar{{\mathcal D}}^2
\end{equation}
defining $N=(1,1)$ {\it supersymmetry}.

The quantum theory of non-relativistic positronium is formulated in
terms of the $N=(1,1)$ spectral data
\begin{equation}
\left( {\mathcal A}, {\mathcal H}_{e-p}, {\mathcal D}, \bar{{\mathcal
      D}} \right) \ ,
\end{equation}
and its dynamics is determined by the Hamiltonian 
\begin{equation}
H \ = \ \frac{\hbar^2}{2\mu} \ {\mathcal D}^2 \ = \
\frac{\hbar^2}{2\mu} \ \bar{{\mathcal D}}^2 \ ,
\end{equation}
where $\mu=2m$ is the mass of positronium. Such data are meaningful even 
for manifolds that are {\it not} spin${}^c$; in physics jargon, one could say 
that, on manifolds which do not carry a spinor bundle $S$, an electron 
and a positron are ``permanently confined'' to a positronium bound state. 
 
The Weitzenb\"ock formula
says that
\begin{equation}
H \ = \ \frac{\hbar^2}{2\mu} \ \left( - \triangle\;+\;\frac r 4
  \;-\;\frac 1 8 \;R_{ijkl} \bar{\Gamma}^i \bar{\Gamma}^j
    \Gamma^k \Gamma^l\right) \ ,
\end{equation}
where \ $-\triangle = \nabla_i^* g^{ij} \nabla_j = - g^{ij} \left(
  \nabla_i \nabla_j - \Gamma_{ij}^k \nabla_k\right)$ \ is the Bochner
Laplacian, $\Gamma_{ij}^k$ are the Christoffel symbols of the
Levi-Civita connection, $r$ is scalar curvature, and $R_{ijkl}$ are
the components of the Riemann curvature tensor, all in local
coordinates \ $q^j, j=1,\ldots, n$, on $M$; finally \ $\Gamma^j=\Gamma
(dq^j)$, $\bar{\Gamma}^j = \bar{\Gamma}(dq^j)$, \ and the summation
convention is used in (4.33). One recognizes the r.s.\ of (4.33) to be
proportional to the 
Laplacian on the space of differential forms. This is not surprising:
We introduce two operators $\dd$ and $\dd^*$ by
\begin{equation}
\dd \ = \ \frac 1 2 \ \left( {\mathcal D} - i \bar{{\mathcal
      D}}\right)\ , \quad \dd^* \ = \ \frac 1 2 \ \left( {\mathcal D}
  + i \bar{{\mathcal D}}\right) \ .
\end{equation}
Then the relations (4.30) show that 
\begin{equation}
\dd^2 \ = \ \left( \dd^*\right)^2 \ = \ 0, \quad H \ = \
\frac{\hbar^2}{2\mu}\;\left( \dd \dd^* + \dd^* \dd \right) \ .
\end{equation}
Using (4.6), (4.7) and (4.29), (4.34), one easily verifies that 
\begin{equation}
\dd \ = \ a^*\ \circ \ \widetilde{\nabla} \ = \ \aa \ \circ
\widetilde{\nabla}
\end{equation}
where $a^*$ is defined in (4.3), and $\aa$ denotes
anti-symmetrization; in local coordinates, $\dd =  a^* (dq^j)
\widetilde{\nabla}_j$. Since the torsion $T(\widetilde{\nabla})$ of a
connection $\widetilde{\nabla}$ on $\Omega^{\,^\bullet}(M)$ is defined
by 
\begin{equation}
T(\widetilde{\nabla}) \ = \ d\;-\;\aa
\;\circ\;\widetilde{\nabla} \ ,
\end{equation}
where $d$ denotes {\it exterior differentiation}, we conclude that
\[
\dd \ = d \ \ \Longleftrightarrow \ \ T (\widetilde{\nabla}) = 0 \ \ 
\Longleftrightarrow \ \ (4.30)\ {\rm holds} \ ,
\]
assuming that, in (4.30), ${\mathcal D}$ and $\bar{{\mathcal D}}$ are
  self-adjoint operators on ${\mathcal H}_{e-p}$, which is implied,
  formally, by the assumption that $\widetilde{\nabla}$ is a Hermitian
  connection. Thus $\dd = d$ is exterior differentiation precisely if
  $\widetilde{\nabla}$ is the Levi-Civita connection on ${\mathcal
    H}_{e-p}$.

It follows that the $N=(1,1)$ supersymmetric quantum theory of
non-relativistic positronium can be formulated on general, orientable
Riemannian manifolds $(M,g)$ which need {\it not} be spin$^c$. 

If $\gamma$ is the operator on ${\mathcal H}_{e-p}$ with eigenvalue
$+1$ on forms of even degree and $-1$ on forms of odd degree then
\begin{equation}
\left\{ \gamma, \dd\right\} \ = \ \left\{ \gamma, \dd^*\right\}\;=\;0\,,
\quad\ \left[ \gamma, a\right]\;=\;0\ , 
\end{equation}
for all $a\in{\mathcal A}$.

An algebra ${\mathcal F}$ of ``functions on quantum phase
space'' can be defined as in (2.8), (2.9): \ ${\mathcal F}$ is
generated by
\begin{equation}
\left\{ \alpha_\tau (a), f (H) \bigm| a \in {\mathcal A}, \tau
  \,\in\,{\mathbb R}, \ f\,\in\,C_0^\infty ({\mathbb R})\right\} \ ,
\end{equation}
with
\[
\alpha_\tau (A) \ = \ e^{i\tau	H/\hbar} \ A\; e^{-\,i\tau H/\hbar} \ ,
\]
for all $A \in B ({\mathcal H}_{e-p})$, $\tau \in {\mathbb R}$.

The spectral data \ $({\mathcal A}, {\mathcal H}_{e-p}, \dd, \dd^*,
\gamma)$, or $({\mathcal F}, {\mathcal H}_{e-p}, \dd, \dd^*, \gamma)$, 
define an example of $N=(1,1)$ supersymmetric quantum mechanics: There
are two supercharges $\dd$ and $\dd^*$ (or ${\mathcal D}$ and
$\bar{{\mathcal D}}$) satisfying the algebra (4.35) (or (4.30)). When
$\dd = d$ (exterior differentiation) the ${\mathbb Z}_2$-grading
$\gamma$ can be replaced by a ${\mathbb Z}$-grading $T$ counting the
degree of a differential form. Furthermore, if $M$ is orientable one
can define a unitary Hodge involution, $*$, on ${\mathcal H}_{e-p}$
such that $* d *^{-1} = \zeta d^*$, $|\zeta|=1$, and $*a*^{-1} = a$, 
for all $a\in {\mathcal A}$ (or $a \in {\mathcal F}$). If $M$ is 
even-dimensional then $*$
can be constructed from the element $\sigma$ anti-commuting with $i
\circ \Gamma (\xi) \circ i^{-1}$ \ and commuting with $i \circ
\bar{\Gamma} (\xi) \circ i^{-1}$, for all $\xi \in \Omega^1 (M)$:
\begin{equation}
* \ = \ \sigma \ .
\end{equation}

$N=(1,1)$ supersymmetric quantum theory yields topological invariants
for $M$ if $M$ is even-dimensional:

\noindent the {\it Euler number}
\begin{equation}
\chi (M) \ = \tr\,\left( \gamma\;e^{-\beta\,H}\right) \ ,
\end{equation}
and the {\it Hirzebruch signature}
\begin{equation}
\tau (M) \ = \tr\,\left( *\ e^{-\beta\,H} \right) \ .
\end{equation}
Since $\gamma$ and $*=\sigma$ anti-commute with ${\mathcal D}$, the
r.s.\ of (4.41) and (4.42) are easily seen to be independent of $\beta$
and of the metric on $M$. Using a path integral representation of the
r.s.\ of (4.41), one derives the Gauss-Bonnet
formula. Similarly, (4.42) can be evaluated in terms of the Hirzebruch
polynomial; see [22,17].

In Section~{\bf 1}, we have considered the equations of motion for a
classical, relativistic scalar particle and have derived them from an
action principle with an action $S$ given in eq.~(1.4). In this
section, we study {\it quantum-mechanical, non-relativistic} particles
with {\it spin}. Space-time is given by $N := M\times {\mathbb R}$,
where $M$ is space, $x^0 = \tau \in {\mathbb R}$, and
\[
g_{\mu\nu} \ = \ \left( \begin{array}{c|ccc}
1  &0 &\cdots &0\\ \hline
0 &&&\\
\vdots &&- g_{ij} &\\
0 &&& 
\end{array}\right) \ ,
\]
where $g=(g_{ij})$ is the Riemannian metric on $M$. In this situation
the action of a scalar particle with mass $\mu$ is given by
\begin{equation}
S\left( x\left(\cdot\right)\right) \ = \ -\;\frac \mu 2 \int g_{ij}
\left(x\left(\tau\right)\right) \ \frac{dx^i(\tau)}{d\tau} \cdot
\frac{dx^j(\tau)}{d\tau} \ d\tau \ ,
\end{equation}
where, now, $\tau$ is {\it time}. Quantum-mechanically, the
Hamiltonian of a scalar particle is given by
\[
H \ = \ - \ \frac{\hbar^2}{2\mu} \ \triangle_g \ ,
\]
where $\triangle_g$ is the Laplace-Beltrami operator acting on
\[
{\mathcal H} \ = \ L^2 \left( M, d\,{\rm vol}_g \right) \ .
\]
According  to {\it Feynman} and {\it Kac}, the heat kernel $\left(
  e^{-\;\beta H/\hbar}\right) (x,y)$, $x, y \in M$, is given by
\begin{equation}
\left( e^{-\;\beta H/\hbar}\right) (x,y) \ = \
\int\limits_{{x(0) = x \atop x (\beta) = y}}
e^{\frac 1 \hbar \ S_\beta \left( x (\cdot)\right)
} \
\prod_{\tau\,\in\,[0,\beta]} \ dx (\tau) \ ,
\end{equation}
where $S_\beta$ is given by (4.43) with the $\tau$-integration
extending over the interval $[0,\beta]$, and $dx(\tau) := d{\rm
  vol}_g(x(\tau))$. The mathematical interpretation of the integrand
on the r.s.\ of (4.44) is that it is given by the Wiener
\def\btimes{{\sf X}}
measure on path space \
$\displaystyle\mathop{\btimes}_{\tau\,\in\,[0,\beta]} \ M_\tau$ , where
$M_\tau$ is a copy of $M$ (with $M$ compact) and the Cartesian product
is equipped with the Tychonov topology; see e.g.~[23].

In order to evaluate expressions like (4.24), (4.41) or (4.42), we
require a generalization of formulae (4.43,44) for particles with
spin. People who know their path integral formulation of
non-relativistic many-body theory will have little difficulty in
finding such a generalization (see e.g.~[22,24], and [25] for some
details). As an example, we consider the Hamiltonian $H$ given in
(4.33) and we propose to derive a path integral representation for the
heat kernel corresponding to $H$. In order to be explicit, we work in
a local coordinate patch of $M$, with coordinate functions now denoted
by $x^1,\ldots, x^n$. \ It is advantageous to reexpress the r.s.\ of
(4.33) in terms of the creation and annihilation operators $a^{*j} :=
a^* (dx^j)$, $a^j := a(dx^j) \equiv g^{jl}\,a (\partial_l)$ defined in
(4.3), (4.4), respectively. Then 
\[
H \ = \ \frac{\hbar^2}{2\mu} \ \left( - \triangle - R_{ijkl} \ a^{*i}
  \ a^j \ a^{*k} \ a^l \right) 
\]
with
\begin{equation}
- \triangle \ = \ \nabla_i^*\;g^{ij}\;\nabla_j\; , \quad \nabla_j \ =
\ \partial_j \ - \ \Gamma_{jl}^k\;a^{*l}\;a_k \ ,
\end{equation}
where $a_k = g_{km} a^m$. As usual in the functional integral
formulation of non-relativistic fermions (see e.g.~[25]), we now
associate Grassmann variables $\psi^{*j} (\tau)$ with $a^{*j}$ and
Grassmann variables $\psi^j (\tau)$ with $a^j$, $\tau \in {\mathbb
  R}$, such that 
\begin{equation}
\left\{ \psi^i (\tau),\;\psi^j(\tau')\right\} \ = \ 
\left\{ \psi^i (\tau),\;\psi^{*j}(\tau')\right\} \ = \ 
\left\{ \psi^{*i} (\tau),\; \psi^{*j} (\tau')\right\} \ = \ 0 \ .
\end{equation}
The action corresponding to (4.45) is then given by
\begin{eqnarray}
&& S_\beta \left( x,\psi, \psi^*\right) \ = \ -\;\mu
\int\limits_0^\beta \left[ \frac 1 2 \ g_{jk}
  \left( x\left( \tau\right)\right) \ \frac{dx^j(\tau)}{d\tau} \cdot
  \frac{dx^k (\tau)}{d\tau}\right. \nonumber \\
&& \qquad +\ i\;g_{jk} \left( x\left( \tau \right)\right) \ \psi^{*j}
(\tau) \;D_\tau\;\psi^k (\tau) \\
&& \qquad \left. -\;\frac 1 2 \ \left( \frac \hbar \mu \right)^2 \ R_{ijkl}
\left( x\left( \tau\right)\right) \ \psi^{*i} (\tau)\; \psi^{j}
(\tau) \;\psi^{*k} (\tau)\; \psi^l (\tau)\right] \ d\tau \ , \nonumber
\end{eqnarray}
where
\[
D_\tau\;\psi^k (\tau) \ = \ \frac{d\;\psi^k (\tau)}{d\tau} \ + \
\Gamma_{lm}^k \ \frac{d\;x^l (\tau)}{d\tau} \ \psi^m (\tau) \ .
\]
Then 
\begin{eqnarray}
\chi (M) &=& \tr\,\left( \gamma \; e^{- \beta H}\right) \nonumber \\
&=& \int e^{\frac 1 \hbar \ S_\beta \left( x,\psi,\psi^*\right)} 
\prod_{\tau\,\in\,[0,\beta]} dx (\tau) \prod_{j=1}^n d\;\psi^j (\tau)
\ d\;\psi^{*j} (\tau) \ ,
\end{eqnarray}
and {\it periodic} boundary conditions are imposed at $\tau =
0,\beta$. Hence, for very small $\beta$, the constant 
modes dominate the functional integral (4.48). It is then easy to
evaluate it (asymptotically, as $\beta \to 0$) using the saddle point
method. The result is the general Gauss-Bonnet formula. The
calculations of $\tau (M)$ and of (4.24) (index of $D$) are a little
harder, although the basic ideas are the same; see [22], and [17] for
rigorous proofs.

We do not want to enter into more detail, but rather continue our
journey through non-relativistic quantum theory. Below (4.39), we have
identified the spectral data
\begin{equation}
\left( {\mathcal A},\  {\mathcal H}, \ \dd,\  \dd^*,\ \gamma \right) \
,
\end{equation}
with relations (4.35), (4.38) of $N=(1,1)$ supersymmetric quantum
theory of which non-relativistic positronium is an example if one takes
${\mathcal A} = C(M)$, ${\mathcal H} = {\mathcal H}_{e-p}$, $\dd^\# =
d^\#$ ($x^\#$ denotes $x$ or $x^*$), and for $\gamma$ the operator 
detecting the
parity of a differential form. In this example, the data (4.49)
completely encode the differential topology and Riemannian geometry of
$(M,g)$. Furthermore, they can be completed to
\begin{equation}
\left( {\mathcal A}, \ {\mathcal H}, \ \dd, \ \dd^*, \ T, \ * \
\right)
\end{equation}
where  $T$ counts the degree of a differential form, and $*$ is
the Hodge operator, with \ $*\dd*^{-1} = \zeta \dd^*$ , $|\zeta |=1$ , and
$*a*^{-1}=a$ , for all \ $a\,\in\,{\mathcal A}$. We say that the data
(4.50) define some $N=\overline{(1,1)}$ \ {\it supersymmetric quantum theory}.
It is important to distinguish $N=(1,1)$ from $N=\overline{(1,1)}$ 
supersymmetry: Every $N=\overline{(1,1)}$ supersymmetry is an $N=(1,1)$ 
supersymmetry, but it may turn out to be impossible to enlarge $N=(1,1)$
to $N=\overline{(1,1)}$ supersymmetry,
even in the context of quantum theory on {\it classical} manifolds
$(M,g)$. An example is provided by choosing a connection $\nabla$ on
$T^*M$ with non-vanishing torsion $(T(\nabla)\neq 0)$. We assume that
the torsion of $\nabla$ defines a {\it closed} three-form, denoted
$\vartheta$. \ {\it Locally}, in some coordinate patch of $M$, we can
then construct a 2-form $\beta$ with $d\beta = \vartheta$ and define
an operator \ $B := \beta\wedge\, = \beta_{ij} a^{*i} a^{*j}$ . We then
define a new ``exterior derivative''
\[
d_\lambda \ := \ e^{\lambda B} \ d\;e^{-\;\lambda B} \ .
\]
Clearly $d_\lambda^2 = (d_\lambda^*)^2 = 0$ . \ Two Pauli--Dirac
operators can now be defined by
\[
{\mathcal D} \ :=\ d_\lambda\;+\;d_\lambda^* \ , \quad \bar{{\mathcal
    D}} \ := \ i\;\left( d_\lambda - d_\lambda^*\right) \ ,
\]
for arbitrary real $\lambda$. Since $d_\lambda^2 = (d_\lambda^*)^2 =
0$, ${\mathcal D}$ and $\bar{{\mathcal D}}$ obey the $N=(1,1)$
algebra (see also [24] where a specific choice for $\lambda$ is
made). Of course, there is no natural ${\mathbb Z}$-grading operator
$T$ in this example. If the form $\vartheta$ is {\it not exact}
(i.e., there does not exist a globally defined 2-form $\beta$ with
$d\beta=\vartheta$) then $d_\lambda$, for $\lambda \neq 0$, and
$d=d_{\lambda=0}$ may give rise to different cohomologies; see [24] for
examples and expressions for the action functionals corresponding to
${\mathcal D}^2$. We shall return to these issues in Sects.~{\bf 4.2} and
{\bf 5}.
\smallskip

Our findings can be summarized as follows:
Pauli's quantum theory of a non-rel\-at\-iv\-ist\-ic electron, such as
described by the $N=1$ spectral data $({\mathcal A}, \ {\mathcal
 H}_{e-p}, \ D_A)$, with ${\mathcal A} = C(M)$, or of positronium,
such as described by the $N=\overline{(1,1)}$ spectral data \ $({\mathcal A}, \
{\mathcal H}_{e-p}, \ d, d^*, T, * \ )$, {\it completely encode} the
{\it topology} and {\it geometry} of the Riemannian manifold
$(M,g)$. When ${\mathcal A}$ is replaced by an algebra ${\mathcal
  F}$ of ``functions over quantum phase space'', there remain
interesting mathematical problems to be reckoned with, which we plan
to discuss in future work --- see also Section~{\bf 5}.

\medskip

Readers not interested in quantum physics may ask what one gains by
reformulating differential topology and geometry in terms of spectral
data, such as those provided by $N=1$ (electron) or $N=\overline{(1,1)}$
(positronium) supersymmetric quantum mechanics, beyond a slick
algebraic reformulation. The answer ---	 as emphasized by Connes --
is{ {\it generality!} Supersymmetric quantum mechanics enables us to
  study {\it highly singular spaces} or {\it discrete objects}, like
  graphs, lattices and aperiodic tilings (see e.g.~[5]), and also {\it
    non-commutative spaces}, like quantum groups, as {\it geometric}
  spaces, and to extend standard constructions and tools of algebraic
  topology or of differential geometry to this more general context,
  so as to yield non-trivial results. Moreover, as we have argued in
  Section~{\bf 3}, quantum physics ultimately {\it forces} us to generalize
  the basic notions and concepts of geometry.

The principle that the time evolution of a quantum mechanical system
is a one-parameter {\it unitary group} on a Hilbert space, whose
generator is the Hamiltonian of the system  
(a {\it self-adjoint} operator), entails that the study of
supersymmetric quantum mechanics is the study of {\it metric}
geometry. Let us ask then how we would study manifolds like {\it
  symplectic} manifolds that are, a priori, {\it not} endowed with a
metric. The example of symplectic manifolds is instructive, so we
sketch what one does (see [18]). 

Let $(M,\omega)$ be a symplectic manifold. The symplectic form $\omega$
is a globally defined closed 2-form. It is known that every symplectic
manifold can be equipped with an almost complex structure $J$ such
that the tensor $g$ defined by
\begin{equation}
g (X,Y) \ = \ -\;\omega\,(JX,Y) \ ,
\end{equation}
for all vector fields $X,Y$, is a Riemannian metric on $M$. Thus, we
can study the Riemannian manifold $(M,g)$ with $g$ from (4.51) by
exploring the quantum mechanical propagation of e.g.~positronium on
$M$, using the spectral data \ $({\mathcal A}, {\mathcal H}_{e-p}, d,
d^*, T, * )$ \ of \ $N=\overline{(1,1)}$ supersymmetric quantum mechanics, with
${\mathcal A} = C(M)$. We must ask how these data ``know'' that $M$
is symplectic. The answer is as follows: We can view the ${\mathbb
  Z}$-grading $T$ as the generator of a U(1)-symmetry (a ``global
gauge symmetry'') of the system. It may happen that this symmetry can
be enlarged to an SU(2)-symmetry, with generators $L^1, L^2, L^3$
acting on ${\mathcal H}_{e-p}$ such that they commute with all
elements of ${\mathcal A}$ and have the following additional
properties:

\quad i) $L^3 = T\;-\;\frac n 2 $ \ with $n = {\rm dim}\;M$.\hfill\break
Defining $L^\pm = L^1 \pm i L^2$, \ the structure equations of su(2) =
Lie(SU(2)) imply that

\quad ii) $[L^3, L^\pm] \ = \ \pm\;2 L^\pm$ , $[L^+, L^- ] \ = \
  L^3$,\hfill\break
and, since in quantum mechanics symmetries are represented unitarily,

\quad iii) $(L^3)^* \ = \ L^3$ , \ $(L^\pm)^* \ = \ L^\mp$ .\hfill\break
We also assume that

\quad iv) $[L^+,d] \ = \ 0$ , \hfill\break
hence $L^-$ commutes with $d^*$ by property iii). Next we define an
operator $\tilde{d}^*$ by 
\begin{equation}
\tilde{d}^* \ = \ [L^-, d] \ ;
\end{equation}
it satisfies $[L^+, \tilde{d}^*] = d$ \ because of ii) and iii), and
also
\[
\{ \tilde{d}^*, d\} \ = \ 0
\]
since $d$ is nilpotent. Assuming, moreover, that 

\quad v) $[L^-, \tilde{d}^*] = 0$\hfill\break
we find that $(d,\tilde{d}^*)$ transforms as a doublet under the
adjoint action of $L^3, L^+, L^-$ and that $\tilde{d}^*$ is
{\it nilpotent}. Thus, $(\tilde{d}, -d^*)$ with $\tilde{d} =
(\tilde{d}^*)^*$ is an SU(2)-doublet, too, and $\tilde{d}^2 = 0$. 

\bigskip

\noindent
The theorem is that the spectral data
\begin{equation}
\left( {\mathcal A}, \ {\mathcal H}_{e-p}, \ d, \ d^*, \ \left\{ L^3,
    L^+, L^-\right\}, \ *\;\right) \ ,
\end{equation}
with properties \ i) - v) assumed to be valid, encode the geometry of
a symplectic manifold $(M,\omega)$ equipped with the metric $g$
defined in (4.51). The identifications are as follows:
\begin{eqnarray}
&& L^3 \ = \ T \ - \ \frac n 2 \ , \ L^+ \ = \ \omega\wedge \ = \
\frac 1 2 \ \omega_{ij} \ a^{*i} \ a^{*j} \ , \nonumber\\
&& L^- \ = \ (L^+)^* \ = \ \frac 1 2 \ (\omega^{-1})^{ij} \ a_i\; a_j
\ .\nonumber
\end{eqnarray}
Assumption iv) is equivalent to $d\omega=0$. Further details can be
found in [18]. 

We say that the spectral data (4.53) define $N=4^s$ supersymmetric
quantum mechanics, because there are four ``supersymmetry generators''
$d, \tilde{d}^*, \tilde{d}, d^*$; the superscript $s$ stands for
``symplectic''.

Note that we are {\it not} claiming that
\begin{equation}
\{\, d, \tilde{d}\,\} \ = \ 0
\end{equation}
because this equation does, in general, {\it not} hold. However, {\it
  if} it holds then $(M,\omega)$ is in fact a {\it K\"ahler manifold},
with the $J$ from eq.~(4.51) as its complex structure and $\omega$ as
its K\"ahler form. Defining
\begin{equation}
\partial \ = \ \frac 1 2 \ (d-i\tilde{d}), \quad \bar{\partial} \ = \
\frac 1 2 \ (d+i\tilde{d}) \ ,
\end{equation} 
one finds that, thanks to eqs.~(4.52, 4.54) and because $d$ and
$\tilde{d}$ are nilpotent,
\begin{equation}
\partial^2 \ = \ \bar{\partial}^2 \ = \ 0\,, \quad \{ \partial,
\partial^*\} \ = \ \{ \bar{\partial}, \bar{\partial}^* \} \ .
\end{equation}
There is a useful alternative way of saying what it is that identifies
a symplectic manifold $(M,\omega)$ as a K\"ahler manifold: Eq.~(4.54)
is a consequence of the assumption that an $N=4^s$ supersymmetric
quantum mechanical model has an {\it additional} U(1)-symmetry ---
which, in physics jargon, one is tempted to call a ``global chiral
U(1)-gauge symmetry'': We define
\begin{eqnarray}
d_\theta &=& {\rm cos}\,\theta\, d \ + \ {\rm sin}\,\theta\,\tilde{d}
\ , \nonumber\\
\tilde{d}_\theta &=& -{\rm sin}\,\theta\, d \ + \ {\rm
  cos}\,\theta\,\tilde{d} \ ,
\end{eqnarray}
and assume that $(d_\theta, \tilde{d}_\theta^*)$ and
$(\tilde{d}_\theta, -d_\theta^*)$ are again SU(2)-doublets with the
same properties as $(d, \tilde{d}^*)$ and $(\tilde{d}, - d^*)$, for
all real angles $\theta$. Then the nilpotency of $d, \tilde{d}$ and of
$\tilde{d}_\theta$ for all $\theta$ implies eq.~(4.54). Furthermore
\[
\partial_\theta \ = \ \frac 1 2 \ (d_\theta - i\tilde{d}_\theta) \ =
\ e^{i\theta} \partial, \quad 
\bar{\partial}_\theta \ = \ \frac 1 2 \ (d_\theta + i\tilde{d}_\theta)
\ = \ e^{- i\theta} \;\bar{\partial} \ .
\]
Assuming that the symmetry (4.57) is implemented by a one-parameter
unitary group on ${\mathcal H}_{e-p}$ with an infinitesimal generator
denoted by $J_0$, we find that
\begin{equation}
[ J_0, d] \ = \ -\;i\,\tilde{d}, \quad [ J_0,\tilde{d}] \ = \ i\,d \ .
\end{equation}
Geometrically, $J_0$ can be expressed in terms of the {\it complex
  structure} $J$ on a K\"ahler manifold ---  it is bilinear in $a^*$
and $a$ with coefficients given by $J$. Defining
\begin{equation}
{\mathcal T} \ := \ \frac 1 2 \ (L^3+J_0) , \quad \
\overline{\mathcal T} \ := \ \frac 1 2 \ (L^3-J_0) \ ,
\end{equation}
one checks that 
\begin{eqnarray}
&& [\,{\mathcal T}, \partial\,] \ = \partial, 
\qquad [\,{\mathcal T}, \bar{\partial}\,] \ = \ 0 \ , \nonumber\\
&& [\,\overline{{\mathcal T}},\partial\,] \ = \ 0, 
\qquad [\,\overline{{\mathcal T}},\bar{\partial}\,] \
= \ \bar{\partial} \ .
\end{eqnarray}
Thus ${\mathcal T}$ is the holomorphic and $\overline{{\mathcal T}}$ the 
anti-holomorphic ${\mathbb Z}$-grading of complex differential forms. 
The spectral data
\begin{equation}
\left( {\mathcal A}, \ {\mathcal H}_{e-p}, \ d, \ d^*\;, \ \left\{
    L^3, L^+, L^-\right\}, \ J_0, \ *\right)
\end{equation}
belong to $N=4^+$ supersymmetric quantum mechanics. We have seen 
that they contain the spectral data
\begin{equation}
\left( {\mathcal A}, \ {\mathcal H}_{e-p}, \ \partial, \ \partial^*, \
  \bar{\partial}, \ \bar{\partial}^*, \ {\mathcal T}, \ 
 \overline{{\mathcal T}}, \ *\right)
\end{equation}
characterizing {\it K\"ahler manifolds}. We say that these define
$N=\overline{(2,2)}$ supersymmetric quantum mechanics. If one drops the
requirement that $\partial$ anti-commutes with $\bar{\partial}^*$
(amounting to the breaking of the SU(2) symmetry generated by
$L^3,L^+, L^-$) \ the data (4.62) characterize {\it complex Hermitian
  manifolds}, see [18].

Alternatively, complex Hermitian manifolds can be described by $N=(1,1)$
spectral data, as in eq.~(4.50), with an additional U(1) symmetry
generated by a self-adjoint operator $J_0$ with the property that
$\tilde{d} := i[J_0,d]$ is nilpotent, and different from $d$. Then
$\tilde{d}$ and $d$ anti-commute, and one may define $\partial$ and
$\bar{\partial}$ through eqs.~(4.55). One verifies that 
\[
\partial^2 \ = \ \bar{\partial}^2 \ = \ 0 \quad {\rm and} \quad 
\{ \partial, \bar{\partial}\} \ = \ 0 \ .
\]
Having proceeded thus far, one might think that on certain K\"ahler
manifolds with special properties the U(1) symmetries generated by
${\mathcal T}$ and $\overline{{\mathcal T}}$ are embedded into SU(2) 
symmetries with generators ${\mathcal T}^3 = {\mathcal T}, 
{\mathcal T}^+, {\mathcal T}^-$ (analogously for the anti-holomorphic 
generators) which satisfy properties i) through v) from above, 
with $d$ and $d^*$ replaced by $\partial$ and $\partial^*$, 
and such that $\tilde{\partial}^* = [{\mathcal T}^-,\partial]$ ---
as well as analogous relations for the anti-holomorphic generators.
Alternatively, one might assume that, besides the SU(2) symmetry
generated by $L^3, L^+, L^-$ there are actually {\it two} ``chiral''
U(1)-symmetries with generators $I_0$ and $J_0$, enlarging the
original U(1) symmetry.

Indeed, this kind of symmetry enhancement can happen,\	and what one
finds are spectral data characterizing {\it Hyperk\"ahler
  manifolds}. The two ways of enlarging the \break 
SU(2)$\times$U(1) symmetry
of K\"ahler manifolds to larger symmetry groups characteristic of
Hyperk\"ahler manifolds are {\it equivalent} by a theorem of
Beauville; see e.g.~[26]. The resulting spectral data define what is
called $N=(4,4)$ supersymmetric quantum mechanics, having two sets of
four supercharges, $\{ \partial, \tilde{\partial}^*, \bar{\partial}^*,
{\stackrel{\simeq}{\partial}} \}$ and $\{ \tilde{\partial}, \partial^*,
{\stackrel{\simeq}{\partial^*}},\bar{\partial} \}$, with the property
that each set transforms in the fundamental representation of Sp(4) --- 
see [18] for the details. This yields the data of $N=8$
supersymmetric quantum mechanics --- from which we can climb on to
$N=(8,8)$ or $N=16$ supersymmetric quantum mechanics and enter the
realm of very rigid geometries of symmetric spaces with special
holonomy groups [27,26].

Of course, the operators
\begin{equation}
I \ := \ {\rm exp}\,(- i\pi\,I_0), \quad J \ := \ {\rm exp}\,(-
i\pi\,J_0), \quad K \ :=\ IJ
\end{equation}
in the group of ``chiral symmetries'' of the spectral data of $N=(4,4)$
supersymmetric quantum mechanics correspond to the three complex
structures of Hyperk\"ahler geometry.
One may then try to go ahead and enlarge these ``chiral'' symmetries
by adding further complex structures. This leads to the study of
hypercomplex manifolds with many complex structures. 
See e.g.~[28] and references
therein for some formal considerations in this direction, and also 
Section~{\bf 5}.

We could now do our journey through the land of geometry and
supersymmetric quantum mechanics in reverse and pass from special
(rigid) geometries, i.e., supersymmetric quantum mechanics with high
symmetry, to more general ones by reducing the supersymmetry
algebra. The passage from special to more general geometries then
appears in the form of {\it supersymmetry breaking} in supersymmetric
quantum mechanics (in a way that is apparent from our previous
discussion). The symmetry generators in the formulation of geometry as
supersymmetric quantum mechanics are bilinear expressions in the
creation and annihilation operators $a^*$ and $a$ from eqs.~(4.3,4) 
with coefficients that are tensors of rank two. It is quite
straightforward to find conditions that guarantee that such tensors
generate symmetries and hence to understand what kind of {\it
  deformations} of geometry {\it preserve} or {\it break} the
symmetries. Furthermore, the general transformation theory of quantum
mechanics enables us to describe the deformation theory of the
supersymmetry generators $(D_A; {\mathcal D}, \bar{{\mathcal D}};$ or
$d, d^*)$ including isospectral deformations (as unitary
transformations). Deformations of $d$ and $d^*$ played an important
role in Witten's proof of the Morse inequalities [16] and in exploring
geometries involving anti-symmetric tensor fields such as torsion ---
recall the example described above --- which are important in conformal 
field theory.

\medskip

We hope we have made our main point clear: {\it Pauli's quantum
  mechanics of the non-relativistic electron and of positronium on a
  general manifold (along with its internal symmetries) neatly encodes
  and classifies all types of differential geometry}. See [18] for
details. 

\vfil\eject
\subsection{The special case where $M$ is a Lie group}

What is special if physical space $M$ is a {\it Lie group} $G$? We
briefly discuss this special case, because it will help us to
understand conformal field theory and the operator formalism for BRST
cohomology. For simplicity, we only consider finite-dimensional,
compact, connected, semi-simple Lie groups. The group is denoted by
$G$, and {\tt g} denotes its Lie algebra. For each $g\in G$, we denoted
by $L_g$ the left action of $g$ on $G$. The tangent maps of $L_g$ are
denoted by $D_g$. The Lie algebra ${\tt g}$ of $G$ can be viewed as the
space of left-invariant vector fields: Let $\varphi$ be an arbitrary
function on $G$ and let $X\in \Gamma(TG)$ be a vector field on
$G$. Then $X$ is left-invariant if
\begin{equation}
D_g\;X (\varphi)(h) \ = \ X\;(\varphi)(g^{-1} h)
\end{equation}
for all $h$ and $g$ in $G$. The space of left-invariant vector fields
is canonically isomorphic to the tangent space $T_e G={\tt g}$ at the unit
element $e\in G$. The Lie algebra ${\tt g}$ acts on itself by
\begin{equation}
ad_X (Y) \ = \ [X,Y], \quad X, Y \;\in\;{\tt g} \ ;
\end{equation}
(adjoint representation). We define a symmetric, ${\tt g}$-invariant
Killing form $\langle\cdot,\cdot\rangle$ on ${\tt g}\times{\tt g}$ by 
\begin{equation}
\langle X,Y\rangle \ := \tr\,\left( ad_X \cdot ad_Y\right)\,,\quad \ X,
Y\;\in\;{\tt g} \ ,
\end{equation}
which is non-degenerate (if $G$ is semi-simple) and
negative-definite. The Killing form defines a Riemannian metric $g$ on
$TG$ by setting
\begin{equation}
g (X,Y) \ := \ -\;\langle D_{h^{-1}} X, D_{h^{-1}} Y\rangle \ , 
\end{equation}
for arbitrary $X$ and $Y$ in $T_hG$.

The Haar measure $dg$ on $G$ corresponds to the volume form
associated with (4.67), normalized such that $\int_G dg=1$. For
compact Lie groups $dg$ is invariant under the left action $L_h$ {\it
  and} under the right action $R_h$ of $h \in G$ on
$G$. Corresponding to the right action $R$ of $G$ on $G$, one can
define right-invariant vector fields on $G$. The left (or right)
action of $G$ on $G$ can be used to show that the tangent bundle $TG$
is parallelizable and it determines a flat connection $\nabla_L$
(or $\nabla_R$, respectively) with non-vanishing torsion. By $\nabla$ we
denote the Levi-Civita connection corresponding to the metric (4.67).

Obviously we have all the data necessary to define a model of $N=1$
(electron) or $N=(1,1)$ (positronium) {\it supersymmetric quantum
  theory} of particle motion on a compact, connected, semi-simple Lie
group $G$.

Let $\{ T_i\}_{i=1}^n$, where $n={\rm dim}\;G$, be a basis of
left-invariant vector fields on $G$, and let $\{
\vartheta^i\}_{i=1}^n$ be the dual basis of 1-forms. The structure
constants $f_{ij}^k$ of ${\tt g}$ in the basis $\{ T_i\}_{i=1}^n$ are
defined by
\begin{equation}
\left[ T_i, T_j\right] \ = \ f_{ij}^k \ T_k \ .
\end{equation}
The coefficients of the metric $g$ in (4.67) in this basis are given
by
\begin{equation}
g_{ij} \ = \ g\left( T_i, T_j\right) \ =  -\;f_{il}^k\;f_{jk}^l \
.
\end{equation}
Using the $G$-invariance of the Killing form,
\[
\langle \left[ X,Y\right], Z\rangle \ + \ \langle Y,\left[
  X,Z\right]\rangle \ = \ 0 \ ,
\]
for $X,Y,Z \in {\tt g}$, one shows that metricity and vanishing torsion
yield the following formula for the Levi-Civita connection on $TG$:
\begin{equation}
\nabla T_i \ = \ \frac 1 2 \ f_{ki}^j\ \vartheta^k\, \otimes \, T_j \ .
\end{equation}
As in (4.3) and (4.4), we define creation- and annihilation operators
$a^*$ and $a$ by setting
\begin{equation}
c^j \ \equiv \ a^{*j} \ := \ \vartheta^j\wedge\,, \quad 
b_j \ \equiv \ a_j \ := \ T_j \rightharpoonup \ .
\end{equation}
They satisfy the canonical anti-commutation relations
\begin{equation}
\left\{ c^i, c^j \right\} \ = \ \left\{ b_i, b_j\right\} \ = \ 0,
\quad \left\{ c^i, b_j\right\} \ = \ \delta_j^i \ .
\end{equation}
Then the Levi-Civita connection on $T^*G$ is given by
\begin{equation}
\nabla \ = \ \vartheta^j \, \otimes \, \Bigl( T_j\;-\;\frac 1 2 \
  f_{ji}^k\;c^i\;b_k \Bigr) \ ,
\end{equation}
where $T_j (\varphi)$ is the directional derivative of a function
$\varphi$ on $G$ in the direction of $T_j$.
Since the torsion of $\nabla$ vanishes, exterior differentiation on
$G$ is given by
\begin{equation}
d\ = \ a^* \circ \nabla \ = \ c^j\,T_j\;-\;\frac 1 2 \ 
f_{ij}^k\; c^i\;c^j\;b_k \ , 
\end{equation}
with $d^2=0$. To physicists $d$ is known under the name of {\it BRST 
  charge} (see e.g.~[29]).

Since, by definition,
\[
\nabla_L T_i \ = \ 0, \quad \nabla_L \vartheta^i \ = \ 0 \,
\]
we find that
\begin{eqnarray}
T(\nabla_L) &=& d\;-\;a^*\circ \nabla_L\;=\;d\;-\;a^*\circ
\nabla\;+\;a^* \circ (\nabla-\nabla_L) \nonumber\\
&=& a^*\circ (\nabla-\nabla_L)\;=\;- \frac 1 2 \ f_{ij}^k\;c^i\;c^j\;b_k
\ .
\end{eqnarray}
The corresponding 3-form
\[
\theta \ = \ -\;\frac 1 2 \ f_{ijk} c^i\,c^j\,c^k \ = \ \frac 4 3 \
tr\;(g^{-1} dg)^{\wedge 3} 
\]
is closed (the Jacobi identity implies $d\theta = 0$) but not
exact. Locally, we can choose a 2-form $\beta$ with
$d\beta=\theta$. Setting $B := \beta\wedge$, we can consider the
deformed exterior derivatives $d_\lambda := e^{\lambda B} \ d\;
e^{-\lambda B}$ \ and define \ ${\mathcal D} := d_\lambda +
d_\lambda^*$ and $\overline{{\mathcal D}} := i (d_\lambda -
d_\lambda^*)$. Since $d_\lambda^2 = (d_\lambda^*)^2 = 0$, it follows
that  $\{ {\mathcal D}, \overline{{\mathcal D}}\} = 0$ and ${\mathcal D}^2
= \overline{{\mathcal D}}{}^2$. For a suitable choice of $\lambda$ (see [24])
one finds that
\begin{eqnarray}
{\mathcal D} &=& \Gamma^i \Bigl(\, T_i\; -\; \frac{1}{12} \ f_{ijk}\; \Gamma^j
  \Gamma^k \,\Bigr) \ , \nonumber \\
\overline{{\mathcal D}} &=& \bar{\Gamma}^i \Bigl(\, \bar{T}_i\;-\;
  \frac{i}{12} \ f_{ijk} \; \bar{\Gamma}^j \bar{\Gamma}^k \,\Bigr) \ ,
\end{eqnarray}
where $\bar{T}_i$ is the directional derivative defined by a
right-invariant vector field, also denoted by $\bar{T}_i$, 
$i=1,\ldots,n$, and $\bar{\vartheta}^1,\ldots,\bar{\vartheta}^n$ is
the dual basis of 1-forms. Moreover $\Gamma^i = \Gamma(\vartheta^i)$,
$\bar{\Gamma}^i = \bar{\Gamma} (\bar{\vartheta}^i)$.  One finds that
\begin{equation}
{\mathcal D}^2\;=\;\overline{{\mathcal D}}{}^2\;=\;g^{ij} T_iT_j\ + \
\frac{g^{\scriptscriptstyle\vee} n}{24} \ \geq \ 
\frac{g^{\scriptscriptstyle\vee} n}{24} \ , 
\end{equation}
where $g^\vee$ is the dual Coxeter number of $G$. The lower bound
(4.77) proves that the Hilbert space ${\mathcal H}_{e-p} = L^2
(\Lambda^{\,^\bullet} G, d{\rm vol}_g)$ does not contain any
$d_\lambda$-closed vectors that are not $d_\lambda$-exact; the physicists 
call this phenomenon {\it spontaneously broken supersymmetry}. It does,
however, contain $d$-closed vectors that are not $d$-exact. This
proves that $\vartheta$ is {\it not} exact, hence $H^3 (G) \neq 0$, 
while $H^2 (G) = H^4 (G) = 0$ under our hypotheses on $G$ (see
e.g.~[30]).

We proceed towards reviewing some standard facts of representation theory
like the Peter-Weyl theorem etc., the reason being that notions very
similar to those encountered in group representation theory will
appear again in the study of two-dimensional conformal field theory!

Let $G$ be as above. By {\sl 1} we denote the trivial representation and by
${\mathcal R} = \{\hbox{{\sl 1}}, I, J, \ldots \}$ the complete list of
irreducible representations of $G$. Since $G$ has been assumed to be
compact, its irreducible representations are all finite-dimensional
and unitarizable. Given $I\in{\mathcal R}$, let
$I^{\vee}$ denote the representation
contragredient to $I$, which can be defined by the property that the
tensor product representation $I\otimes I^{\vee}$
  contains the trivial representation precisely once.

Besides the Hilbert spaces ${\mathcal H}_e$ and ${\mathcal H}_{e-p}$,
we define the Hilbert space
\begin{equation}
{\mathcal H} \ = \ L^2\,\left( G, dg\right) \ .
\end{equation}
This space is a bi-module for the group algebra ${\mathbb C} [G]$: It
carries the left-regular and the right-regular representation which
commute with each other. The dense subspace ${\mathcal S} \subset
{\mathcal H}$ of smooth functions on $G$ carries the left-regular
representation of left-invariant vector fields and the right-regular
representation of right-invariant vector fields, which commute with
each other. The Peter--Weyl theorem says that
\begin{equation}
{\mathcal H} \ = \ {\bigoplus}_{I\,\in\,{\mathcal R}}
\ W_I \, \otimes \, W_{I^{\vee}} \ , 
\end{equation}
 where $W_I$ is the representation space for the 
representation $I\in{\mathcal R}$, which carries a canonical
scalar product with respect to which $I$ is unitary. Choosing
orthonormal bases in the spaces $W_I$, the Peter--Weyl theorem can be
formulated as saying that the matrix elements $I_{ij} (g)$,
$I\in{\mathcal R}, g\in G$, of irreducible representations form
an orthonormal basis of ${\mathcal H}$. 

Since $G$ is compact, every representation of $G$ is a direct sum of
irreducible representations. Given $I$ and $J$ in ${\mathcal R}$, we
can form the tensor product representation $I\otimes J$ and consider
its decomposition into irreducible representations (Clebsch--Gordan series)
\begin{equation}
I \;\otimes\;J \ = \ \displaystyle\bigoplus_{K\;\in\;{\mathcal
    R}} \ N_{IJ}^K \ K \ ,
\end{equation} 
where $N_{IJ}^K$ is the multiplicity of $K$
in $I\otimes J$, which the physicists call {\it fusion rule}. The
fusion rule $N_{IJ}^K$ is equal to the number of distinct intertwiners
\begin{equation}
V_\alpha \left( I, J | K\right) \ : \ W_K\;\lra\;W_I \otimes W_J \ ,
\end{equation}
$\alpha = 1,\ldots,N_{IJ}^K$, which are called {\it Clebsch--Gordan
  matrices}. These matrices are isometries satisfying
\begin{eqnarray}
&& V_\beta^* \left( I,J|K\right) \ V_\alpha \left( I,J|K\right) \ = \
\delta_{\alpha\beta} \ \id\bigm|_{W_K} \nonumber \\
&& V_\alpha \left( I,J|K\right) \ V_\beta^* \left( I,J|K\right) \ = \
\delta_{\alpha\beta} \ P_{W_K^\alpha} \ , \nonumber
\end{eqnarray}
where $P_{W_K^\alpha}$ is the orthogonal projection onto the $\alpha^{\rm
  th}$ copy, $W_K^\alpha$, of $W_K$ appearing in $W_I \otimes W_J$. 

We define
\begin{equation}
C_{ikn} \left( I,J|K\right)_{jlm} \ := \ \int\limits_G I_{ij} (g) \
J_{kl} (g) \ \overline{K_{mn} (g)} \ dg\ ,
\end{equation}
where $\overline{K_{mn} (g)} \;=\;\left( K(g)^*\right)_{nm}\;=\;K_{nm}
(g^{-1})$, \ because all irreducible representations are unitary. 
Then, using the left- and right-invariance of the Haar measure
$(dg = d(hg) = d(gh)$ for any $h \in G$), one verifies immediately
that 
\[
 \sum_{i',k'} I_{ii'} (h)\; J_{kk'} (h)\; C_{i'k'n} (I,J|K)_{jlm}
= \sum_{m'} \; C_{ikn} (I,J|K)_{jlm'} K(h)_{m'm} \ ,
\]
and
\[
\sum_{j',l'} C_{ikn} (I,J|K)_{j'l'm}\; I_{j'j} (h) J_{l'l} (h)
= \sum_{n'} K(h)_{nn'}\;C_{ikn'} (I,J|K)_{jlm} \ .
\]
One concludes without difficulty that 
\begin{equation}
C(I,J|K) \ = \ \sum_\alpha V_\alpha (I,J|K) \otimes V_\alpha^* (I,J|K)
\ .
\end{equation}
It follows from the definition of the constants	 \ $C_{ikn}
(I,J|K)_{jlm}$ \ that they are the {\it structure constants} of the
abelian $C^*$-algebra ${\mathcal A} = C(G)$:
Choosing the functions $I_{ij} (\cdot)$, $I\in{\mathcal R}$, as generators
of ${\mathcal A}$, we conclude from (4.82) that
\begin{equation}
I_{ij} (g)\; J_{kl} (g) \ = \ \sum_{K,m,n} C_{ikn}\; (I,J|K)_{jlm}\;
K_{mn} (g) \ . 
\end{equation}
Put differently, $C_{ikn} (I,J|K)_{jlm}$ is the matrix element of the
operator $J_{kl} (\cdot) \in {\mathcal A}$ between the vectors $I_{ij}
(\cdot) \in {\mathcal H}$ and $K_{mn} (\cdot) \in {\mathcal H}$. 
\smallskip

The group $G$ has a left- and a right-representation on the space
${\mathcal H}_{e-p}$ of square-integrable differential forms over $G$:
If $L_g$ (resp.\ $R_g\,$) is the diffeomorphism of $G$ determined by left 
(resp.\	 right) multiplication by $g$ and $\alpha$ is a differential form on
$G$ then
\begin{equation}
\lambda(g)\,\alpha \ := \ L_g^* \alpha\ ,\quad \ \rho(g)\,\alpha \ := \ R_g^*
\alpha \ , 
\end{equation}
where $\varphi^*\alpha$ denotes the pull back of $\alpha$ under the
diffeomorphism $\varphi$, define the left (resp.\ right) representation of
$G$ on ${\mathcal H}_{e-p}$. Let ${\mathcal H}_{e-p}^I$ denote the
subspace of differential forms with the property that
$\lambda\bigm|_{{\mathcal H}_{e-p}^I} \cong I$, where $I$ is some
representation of $G$, not necessarily irreducible. Then, for an 
arbitrary $\alpha \in {\mathcal
  H}_{e-p}^I$, 
\[
d\,\alpha \ = \ Q_I\,\alpha \ ,
\]
where 
\begin{equation}
Q_I \ = \ c^j\;i(T_j)\;-\;\frac 1 2 \ f_{mn}^k\;c^m\;c^n\;b_k \ ,
\end{equation}
and $i=dI$ is the representation of the Lie algebra ${\tt g}$ of $G$
corresponding to $I$. The ``BRST operator'' $Q_I$ is nilpotent:
\begin{equation}
Q_I^2 \ = \ 0  
\end{equation}
If $T$ is defined by
\begin{equation}
T \ = \ \sum_j \ c^j\,b_j 
\end{equation}
then the eigenspace of $T$ corresponding to an eigenvalue
$p=0,1,2,\ldots, {\rm dim}G$ \ consists precisely of all square-integrable
$p$-forms on $G$. Physicists call the grading operator $T$ the ``ghost
number operator''. The $k^{\rm th}$ ``{\it BRST cohomology group}'' of
$Q_I$ is given by
\begin{eqnarray}
H_I^k &=& {\rm ker}\;Q_I\biggm|_{{\mathcal C}_I^k} \ \Bigg/ \ {\rm im}\;
Q_I\biggm|_{{\mathcal C}_I^{k-1}} \nonumber \\
&=& H^k \left( {\tt g}, dI\right) \ ,
\end{eqnarray}
where ${\mathcal C}_I^k$ is the eigenspace of $T\bigm|_{{\mathcal
    H}_{e-p}^I}$ corresponding to the eigenvalue $k$. It is easy to
check that $H_I^0$ consists of all functions (0-forms) in ${\mathcal
  H}_{e-p}^I$ {\it invariant} under $L_g, g\in G$.

It is known that the notions introduced here are meaningful under much
weaker assumptions on $G$ (or ${\tt g}$).

\medskip

We have reviewed all these ``elementary'' notions and results in Lie
group theory in more detail than may be bearable, because analogous
notions and results will turn up in conformal field theory. 

\vspace{.5cm}

\subsection{Supersymmetric quantum theory and geometry put into
  perspective}

What we call $N=1$ supersymmetric quantum mechanics (Pauli's electron)
is a structure that has played and important role in Connes'
exploration of the ``metric aspect'' of non-commutative geometry 
[5]. For the formulation of non-commutative geometry, it is a useful
exercise to translate our discussion of the passage from $N=(1,1)$ to
$N=(2,2)$ to $N=(4,4)$ etc. supersymmetric quantum mechanics,
accomplished by adding symmetries, into the language of $N=1$
supersymmetric quantum mechanics, encoding spin$^c$--geometry enriched
by additional symmetries.

In a more rudimentary, but concrete form, the $N=1$ spectral data have
been studied for a long time, by physicists and mathematicians
alike. The work of Lichnerowicz on the Dirac operator (e.g.~formula
(4.21), and also the realization that for a compact spin manifold $M$
without boundary $r>0$ implies that the index of $D=D_{A=0}$ vanishes)
and also index theory (see [17] and refs.~given there) can be viewed as the
study of Pauli's electron on a general manifold and hence of $N=1$
supersymmetric quantum mechanics. In particular, the study of zero
modes of $D_A$ has turned out to be important in topology. 

Zero modes of $D_A$ play an important role in a concrete physics
context as well. Electrons are {\it fermions}. The Hilbert space of
pure state vectors of a system describing $N$ non-relativistic
electrons which move on a manifold $M$ (identified with physical
space) is thus given by
\begin{equation}
{\mathcal H}^{(N)} \ = \ {\mathcal H}_e^{\wedge N} \ , 
\end{equation}
where ${\mathcal H}_e$ is the one-electron Hilbert space introduced
above after eq.~(4.11).

Let us suppose that $K$ static nuclei of atomic numbers
$Z_1,\ldots,Z_K$, with $Z_j \leq Z_* < \infty$ for all
$j=1,\ldots,K$, are present at the points $y_1,\ldots,y_K$ of physical
space $M$. We also fix an arbitrary (virtual) U(1)--connection $A$
with \ div $A=0$ ($A$ is the electromagnetic vector potential in the
Coulomb gauge). Let $v(x,y)$ denote the Green function of the scalar
Laplacian on $M$. The Hamiltonian of the system is defined by
\begin{equation}
H^{(N)} \ = \ \sum_j D_{A(x_j)}^2 \ + \ V_C\left( x_1,\ldots,x_N;\;
  y_1,\ldots,y_K\right) \ ,
\end{equation}
where
\[
V_C\left( x_1,\ldots,x_N;y_1,\ldots,y_K\right) = \sum_{1\leq i<j\leq
  N}v\left( x_i,y_j\right) - \sum_{i=1,\ldots,N \atop l=1,\ldots,K}
Z_l v\left( x_i,v_l\right) + \sum_{1\leq l < k \leq K} Z_l Z_k v\left(
  y_l, y_k\right) \ .
\]
Let 
\begin{equation}
{\mathcal E}_{\rm field} (A) \ = \ \Gamma \int\limits_M \| F_A \|^2 \
d\,{\rm vol}_g \ ,
\end{equation}
where $F_A$ is the curvature of $A$ and $\Gamma > 0$ is a
constant. Physically, ${\mathcal E}_{\rm field} (A)$ is the energy of the
  magnetic field described by $A$ (for \ dim $M=3$).

The problem of stability of non-relativistic matter is the problem of
showing that the ``energy functional''
\begin{equation}
{\mathcal E} (\psi,A) \ = \ \langle \psi, H^{(N)}
\psi\rangle_{{\mathcal H}^{(N)}} \ + \ {\mathcal E}_{\rm field} (A) \
,
\end{equation}
where $\psi \in {\mathcal H}^{(N)}$ has norm 1, is bounded from below
by
\begin{equation}
{\mathcal E} (\psi,A) \ \geq \ -\; C (N+K) 
\end{equation}
for a finite constant $C$ only depending on $\Gamma$ and $Z^*$, but
independent of $N$ and $K$, provided $\Gamma$ is large enough.

If \ dim $M\equiv n \leq 2$ \ stability of matter is not a
particularly challenging problem; see [31]. If $n\geq 4$ it cannot be
valid, as simple scaling arguments show: \ $\left[ {\mathcal E}_{\rm
    field}\right] = {\rm length}^{n-4}, \ [V_C] = {\rm
  length}^{2-n}$. \ If $n=3$, which is the dimension of physical
space, then \ $[{\mathcal E}_{\rm field}] = [V_C]$, and the problem of
proving (4.94) has a physically interesting and mathematically
non-trivial solution; see [32] (and refs.~given there). In this case
there is a critical value, $\Gamma_c$, of $\Gamma$ such that (4.94)
holds for $\Gamma > \Gamma_c$ and $Z_*$ small enough (depending on
$\Gamma$), but fails for $\Gamma < \Gamma_c$.

The reason why $\Gamma$ must be large enough for (4.94) to hold in
$n=3$ dimensions is that $[{\mathcal E}_{\rm field}]=[V_C]$ and, for a
large class of connections $A$, the Pauli-Dirac operator $D_A$ has
{\it zero-modes}. For a class of connections $A$ with ${\mathcal
  E}_{\rm field} (A) <\infty$, such zero-modes were constructed by
Loss and Yau, [33]. For this purpose, these authors studied the
equations
\begin{equation}
D_A \psi \ = \ 0\; , \quad c\,(F_A) \ = \ \kappa\,(\psi\psi^*)_0 \ ,
\end{equation} 
where $\psi \in {\mathcal H}_e$, $c$ denotes Clifford multiplication,
see eqs.~(4.10,11,12), $(\psi\psi^*)_0$ is the traceless
part of the matrix $\psi\psi^*$, and $\kappa$ is constant. Using a
clever ansatz for $\psi$ and $A$, Loss and Yau exhibited solutions of
these equations.

Of course, eqs.~(4.95) are related to the famous Seiberg-Witten
equations [34],which Witten discovered from the study of
supersymmetric Yang-Mills theory on {\it four-dim\-ens\-ion\-al}, compact,
orientable, smooth Riemannian manifolds (which are automatically
spin$^c$). They have invigorated four-dimensional differential
topology. Apparently, they also emerge from problems of stability in
the quantum mechanics of non-relativistic electrons (at least when \
dim $M=3$)!

As we have seen in Sect.~{\bf 4.1}, the study of the quantum theory of
non-relativistic electrons and positrons leads to the discovery of
{\it gauge symmetries of the second kind} and of {\it
  supersymmetry}. Evidently, the gauge group underlying Pauli's
quantum mechanics of the electron on an $n$-dimensional manifold is
Spin$^c(n)$, which locally is isomorphic to \ Spin$(n) \times $ U(1);
\ Spin$(n)$ is the group of rotations in ``spin space'', while U(1) is
the group of electromagnetic phase transformations. Local U(1)-gauge
invariance has been used and studied ever since Fock and Weyl
discovered it. Invariance of Pauli's quantum mechanics under local
Spin$(n)$ rotations has escaped the attention of physicists until
recently [35]. Yet, it has important qualitative implications
(description of spin-orbit interactions as torsion in the spin
connection, quantum mechanical Larmor theorem, Barnett-Einstein-de
Haas effect, etc.).

Supersymmetric quantum mechanics and its uses in algebraic topology
have been studied extensively [16,22,36]. But the fact that the form
of supersymmetric quantum mechanics which the mathematicians have
explored so successfully is really Pauli's quantum theory of the
non-relativistic electron with spin and of positronium and the use of
supersymmetry arguments in non-relativistic quantum physics have
apparently been somewhat under-emphasized. 

The example of electrodynamics with non-relativistic, quantum
mechanical matter considered above is much more instructive than the
reader may have realized. It was recognized by Jordan and Dirac in the
late twenties that, in order to construct a theory correctly
describing radiation of atoms and molecules and the quantum
mechanics of the radiation field (which stood at the origin of quantum
theory), one must {\it quantize} the electromagnetic field and
interpret the U(1)-connection $A$ (the electromagnetic vector
potential) as an {\it operator-valued distribution} on a Hilbert space
${\mathcal F}$ (the Fock space) of pure state vectors describing
photon configurations. The resulting quantum theory consists of the
following data:
\begin{itemize}
\item[(i)] Its Hilbert space of pure state vectors is given by
\[
{\mathcal H}_{\rm tot} \ = \ {\mathcal H}^{(N)} \ \otimes \ {\mathcal
F} \ , 
\]
where, recall, $N$ is the number of electrons in the system, which in
this theory is conserved.
\item[(ii)] The time evolution is generated by (a renormalized version
  of) the Hamiltonian
\[
H_{\rm tot} = \ H^{(N)} \ + \ \id \ \otimes \ H_{\rm field} \ ,
\]
where $H_{\rm field}$ is the usual free-photon Hamiltonian on
${\mathcal F}$.
\item[(iii)] The algebra of observables ${\mathcal A}_{\rm tot}$ of the
  system contains the algebra
\[
{\mathcal A}^{(N)} \ \otimes \ {\mathcal B} \ ,
\]
where ${\mathcal A}^{(N)}$ is the algebra of smooth functions on
$M^{\times N}$, and ${\mathcal B}$ is an algebra of bounded functions
of the quantized magnetic field (smeared out with test functions). 
\end{itemize}

\noindent For the sake of backing up this tale with mathematically precise 
results --- see e.g.~[37] --- one must assume that \ dim $M\leq 3$ (although,
even under this hypothesis, there remain open problems).
\smallskip

Our purpose in mentioning this example is {\it not} to now engage in a
discussion of the beautiful and very rich physics described by quantum
electrodynamics with non-relativistic matter, as defined in (i)
through (iii) above. We just want to make the following crucial point:
As we have learned in Sect.~{\bf 4.1}, the electromagnetic vector
potential $A$ (the ``virtual'' U(1)-connection) is {\it part of the
  geometric data} (associated with spin$^c$ manifolds) on which the
quantum mechanics of non-relativistic electrons hinges. Apparently,
the {\it physics} of natural phenomena forces us to view the
connection $A$ as {\it operator-valued}, and to subject its curvature
$F_A$ describing the electric field $E$ (time-derivative of $A$) and
the magnetic field $B$ (spatial derivatives of $A$) to the uncertainty
relations (2.3) of Section~{\bf 2}. It is the quantization of the
electromagnetic field that makes excited atoms emit light
spontaneously. The model of quantum electrodynamics with
non-relativistic matter described in (i--iii) above enables us to
describe such (and other) phenomena in a physically acceptable and
mathematically rigorous way (at least if the Hamiltonian of the theory
is regularized at short distances). The example is interesting in a
second respect: Using an operator-theoretic version of renormalization
group methods, see [37], one can make precise the claim that classical
behaviour, in particular classical geometry, reappears at very large
distance (and long time) scales.

Let $E$ be a Hermitian vector bundle over $M$ associated to a
principal $G$-bundle, where $G$ is a compact Lie group (e.g.~SU(2),
SU(3)). Let $V$ denote the fiber of $E$. We can equip $E$ with a
connection $\nabla^E$ inherited from a connection on the principal
$G$-bundle over $M$. A heavy quark in a hadron can be viewed as a
variant of a non-relativistic electron described by data consisting of
${\mathcal H}_q$, the space of square-integrable sections of a
Hermitian vector bundle over $M$ with fiber isomorphic to $W\otimes
V$, where $W\cong {\mathbb C}^{[\frac n 2 ]}$. This bundle is
equipped with a connection determined by $\nabla^S$ and
$\nabla^E$. This connection determines a Pauli-Dirac operator $D$, the
square of which gives rise to a Hamiltonian $H$ as in eq.~(4.21). In
order to arrive at a correct quantum mechanical description of heavy
quarks bound in a hadron, one must generalize the theory to encompass
$N=1,2,3,\ldots$ quarks, {\it and one must quantize} $\nabla^E$: The
components of $\nabla^E$ are interpreted as operator-valued
distributions. The resulting ``quantum chromodynamics'' (with
non-relativistic quarks) is not so well understood, yet.

Our discussion makes it tempting to imagine that, in nature, {\it all}
the data characterizing the $K$-theory and geometry of physical
space-(time) must be quantized, {\it including the spin connection and
  the Riemannian metric}. This is the topic of {\it quantum gravity},
a quantum theory that, to date, is not well understood and hence is
really not a theory in the mathematical sense, yet. But we are able to
guess some of its crude features, and these compel one to generalize
classical to non-commutative geometry, as described in Section~{\bf 5}.
\medskip

Our discussion in Sects.~{\bf 4.1} and {\bf 4.2} has made it clear
that, in the 
context of finite-dimensional classical manifolds, (globally)
supersymmetric quantum mechanics --- as it emerges from the study of
non-relativistic electrons and positrons --- is just another name for
classical differential topology and geometry. Actually, this is a
general fact: {\it Global supersymmetry, whether in quantum mechanics
  or in quantum field theory, is just another name for the
  differential topology and geometry of (certain) spaces.}

In the following, we indicate why globally supersymmetric quantum
{\it field theory}, too, is nothing more than geometry of
infinite-dimensional spaces. However, once we pass from supersymmetric
quantum mechanics to quantum field theory, there are surprises.
\smallskip

A non-linear $\sigma$-model is a field theory of maps from a parameter
``space-time'' $\Sigma$ to a target space $M$. Under suitable
conditions on $M$, a non-linear $\sigma$-model can be extended to a
supersymmetric theory, (see e.g.~[38]). One tends to imagine that
such models can be quantized. When $\Sigma$ is the real line ${\mathbb
  R}$, this is indeed possible, and one recovers supersymmetric
quantum mechanics in the sense explained in the previous
sections. When $\Sigma = S^1 \times {\mathbb R}$, there is hope that
quantization is possible, and one obtains an analytic tool to explore
the infinite-dimensional geometry of loop space $M^{S^1}$. When
$\Sigma = L \times {\mathbb R}$ with \ dim $L\geq 2$, the situation is
far less clear, but a supersymmetric non-linear $\sigma$-model with
parameter space-time $L\times{\mathbb R}$ could be used to explore the
geometry of $M^L$ --- i.e., of poorly understood infinite-dimensional
manifolds.

A (quantized) supersymmetric $\sigma$-model with $n$ global
supersymmetries and with parameter space $\Sigma = {\mathbb T}^d
\times {\mathbb R}$, where ${\mathbb T}^d$ is a $d$-dimensional torus,
formally determines a model of supersymmetric quantum mechanics by
{\it dimensional reduction}: The supersymmetry algebra of the
non-linear $\sigma$-model with this $\Sigma$ contains the algebra of
infinitesimal translations on $\Sigma$; by restricting the theory to
the Hilbert sub-space that carries the {\it trivial} representation of
the group of translations of ${\mathbb T}^d$ one obtains a model of
supersymmetric quantum mechanics. The supersymmetry algebra can be
reduced to this ``zero-momentum'' subspace, and the restricted algebra
is of the form discussed in Sect.~{\bf 4.1}. Starting from $\hat{n}$
supersymmetries and a $d+1$--dimensional parameter space-time
$\Sigma$, one ends up with a model of $N=(n,n)$ supersymmetric quantum
mechanics where $n=\hat{n}$ for $d=1,2,n=2\hat{n}$ for
$d=3,4,n=4\hat{n}$ for $d=5,6,$ and $n=8\hat{n}$ for $d=7,8$ (see
e.g.~[39]). The resulting supersymmetric quantum mechanics (when
restricted to an even smaller subspace of ``zero modes'') is expected
to encode the geometry of the target space $M$ --- in, roughly speaking, 
the sense outlined in Sect.~{\bf 4.1}. {}From what we have learned
there, it follows at once that {\it target spaces} of $\sigma$-models
with {\it many supersymmetries} or with a {\it high-dimensional
  parameter space-time} must have {\it very special geometries}.

This insight is not new. It has been gained in a number of papers,
starting in the early eighties with work of Alvarez-Gaum\'e and
Freedman, see [38]. Supersymmetric quantum mechanics is a rather old
idea, too, beginning with papers by Witten [15,16] --- which, as is
well known, had a lot of impact on mathematics. In later works,
``supersymmetry proofs'' of the index theorem were given [22]. The
reader may find comments on the history of global supersymmetry
e.g.~in [40].

A (quantum) field theory of Bose fields can always be thought of as a
$\sigma$-{\it model} (linear  or non-linear), i.e., as a theory of
maps from a parameter ``space-time'' $\Sigma = L \times {\mathbb R}$
to a target space $M$. At the level of classical field theory, we may
attempt to render such a model globally supersymmetric, and then to
quantize it. As we have discussed above, the resulting quantum field
theory --- if it exists ---  provides us with the {\it spectral data}
to explore the geometry of what one might conjecture to be some
version of the formal infinite-dimensional manifold $M^L$. It may
happen that the quantum field theory exhibits some form of invariance
under re-parameterizations of parameter space-time $\Sigma$ (though
such an invariance can be destroyed by {\it anomalies}, even if
present at the classical level). However, when $\Sigma = {\mathbb R}$,
re-parameterization invariance can be imposed; when e.g.~$\Sigma = S^1
\times {\mathbb R}$, it leads us to the tree-level formulation of
first-quantized string theory. Let us be content with ``conformal
invariance'' and study (supersymmetric) {\it conformal}
$\sigma$-models of maps from parameter space-time $\Sigma = S^1 \times
{\mathbb R}$ to a target space $M$ which, for concreteness, we choose
to be a smooth, compact Riemannian manifold without boundary, at the
classical level. An example would be $M=G$, some compact (simply
laced) Lie group. The corresponding field theory is the supersymmetric
Wess-Zumino-Witten model [41], which is surprisingly well
understood. Thanks to the theory of Kac-Moody algebras and centrally
extended loop groups, see e.g.~[42], its quantum theory	 is under
fairly complete mathematical control and provides us with the 
spectral data of some $N=(1,1)$ supersymmetric quantum theory 
(related to the example discussed in Sect.~{\bf 4.2} with 
spontaneously broken supersymmetry; see e.g.~[24]). We might
expect that these spectral data encode the geometry of the loop space
$G^{S^1}$ over $G$. The {\it surprise} is, though, that they encode
the geometry of loop space over a {\it quantum deformation} of $G$
(where the deformation parameter depends on the {\it level} $k$ of the
Wess-Zumino-Witten model in such a way that, formally, $k\to\infty$
corresponds to the classical limit). We expect that this is an example
of a {\it general phenomenon}: Quantized supersymmetric
$\sigma$-models with parameter space-time $\Sigma$ of dimension $d\geq
2$ --- assuming that they exist --- tend to provide us with the
spectral data of a supersymmetric quantum theory which encodes the
geometry of a {\it ``quantum deformation''} of the target space of the
underlying classical $\sigma$-model.
\smallskip

This is one reason why quantum physics forces us to go beyond
classical differential geometry. A second, more fundamental topic
which calls for ``quantum geometry'' is the outstanding problem of
unifying the quantum theory of matter with the theory of gravity
within a theory of {\it quantum gravity}, as discussed in Section~{\bf 3} 
above.

\def \dd {{d}}
\setcounter{equation}{0}%

\section{Supersymmetry and non-commutative geometry}

In this section, we attempt to describe the geometry of generalized
spaces, such as discrete sets, graphs, quantum phase spaces, and more
general non-commutative spaces, in such a way that the spin$^{\rm c}$,
Riemannian, complex, etc.\ geometries of classical manifolds emerge as
special cases. This is the subject that Connes calls non-commutative
geometry (NCG), see [5,43].

Our approach to NCG is inspired by [15,16,5,43]; we follow the
presentation in [18], where the reader finds further details (in
particular on classical geometry). The emphasis is put on the general
structure of the theory and on key ideas, rather than on technical
details.

NCG is not a particularly well developed theory, yet, and the number
of well understood examples is quite limited. Typically, they 
involve discrete sets, the quantum phase spaces
over discrete sets, and deformation quantizations of K\"ahler
manifolds.
\smallskip

There are three starting points for generalizing classical geometry to
NCG:

\begin{itemize}
\item[(1)] {\it Geometry of non-commutative metric spaces}. Here we
  start from spectral data \ $\left( {\mathcal A}, {\mathcal H},
    \triangle\right)$  where
\begin{itemize}
\item[(i)] ${\mathcal H}$ is a separable Hilbert space;
\item[(ii)] ${\mathcal A}$ is a $C^*$--algebra faithfully represented
  on ${\mathcal H}$;
\item[(iii)] $\triangle$ is a self-adjoint operator on ${\mathcal H}$
  such that \ exp$\,(-\varepsilon \triangle)$ is trace class, for
  arbitrary $\varepsilon >0$; there exists a norm-dense subalgebra
  $\stackrel{\ \circ}{{\mathcal A}}$ of ${\mathcal A}$ such that the operator
\[
\frac 1 2 \ \left( \triangle\,a^2 \ + \ a^2\,\triangle\right) \ - \
a\,\triangle\, a
\]
is bounded  for an arbitrary $a \in
\stackrel{\;\;\circ}{{\mathcal A}}$.
\end{itemize}
This structure has been described and studied in [24]; see also
Section {\bf 2}. We shall not pursue it here, although this approach
leads to interesting mathematical problems.
\item[(2)] {\it Spin$^{\rm c}$ geometry}, cast in the form of $N=1$
  supersymmetric quantum theory, which is inspired by Pauli's quantum
  mechanics of the non-relativistic electron with spin. This is
  Connes' starting point [5].
\item[(3)] {\it Riemannian geometry}, cast in the form of $N=(1,1)$ 
  or $N=\overline{(1,1)}$ supersymmetric quantum theory, which is
  inspired by Pauli's quantum mechanics of non-relativistic
  positronium. This approach is described in some detail in [18].
\end{itemize}

In classical spin$^{\rm c}$ geometry, we can always pass from (2) to (3) by
considering the tensor product bundle of the spinor- and the
charge-conjugate spinor bundle; in NCG it may not always be possible to
pass from (2) to (3), and this justifies that we describe both approaches. 

An example of NCG (the non-commutative torus) will be described in 
Section~{\bf 6} and 
applications to string- and membrane theory in Section~{\bf 7}.  

\subsection{Spin$^{\bf c}$ non-commutative geometry}

Our starting point is a natural generalization of the $N=1$
supersymmetric quantum theory of a non-relativistic electron described
in Sect.~{\bf 4.1}.

\bigskip

\noindent 1) \ub{The spectral data of spin$^{\rm c}$ NCG.}
\bigskip

\noindent {\bf Definition A.} A {\it spin$^{\rm c}$ 
non-commutative space} is described by $N=1$ spectral data
$({\mathcal A},{\mathcal H},D,\gamma)$ with the following properties:
\begin{itemize}
\item[(1)] ${\mathcal H}$ is a separable Hilbert space;
\item[(2)] ${\mathcal A}$ is a unital $^*$--algebra faithfully
  represented on ${\mathcal H}$;
\item[(3)] $D$ is a self-adjoint operator on ${\mathcal H}$ such that
\begin{itemize}
\item[i)] for each $a\in{\mathcal A}$, the commutator $[D,a]$ defines
  a {\it bounded} operator on ${\mathcal H}$,
\item[ii)] the operator \ exp$ (-\varepsilon D^2)$ is trace class for
  all $\varepsilon >0$;
\end{itemize}
\item[(4)] $\gamma$ is a ${\mathbb Z}_2$--grading on ${\mathcal H}$,
  i.e., $\gamma=\gamma^*=\gamma^{-1}$, such that
\[
\{ \gamma, D\}\ = \ 0, \quad [\gamma, a] \ = \ 0, \quad {\rm for \
  all} \quad a \in {\mathcal A} \ .
\]
\end{itemize}

In NCG, ${\mathcal A}$ plays the role of the ``algebra of functions
over a non-commutative space''. The existence of a unit in ${\mathcal
  A}$ and property 3)~ii) mean that we are only considering
``compact'' non-commutative spaces.

Note that if the Hilbert space ${\mathcal H}$ is infinite-dimensional,
condition 3)~ii) implies that the operator $D$ is unbounded. By
analogy with classical differential geometry, $D$ is interpreted as a
(generalized) Dirac operator.

Also note that the fourth condition in Definition~A does not impose
any restriction on $N=1$ spectral data: In fact, given a triple
$(\tilde{{\mathcal A}}, \tilde{{\mathcal H}}, \tilde{D})$ satisfying
properties (1--3) above, we can define a set of $N=1$ even
spectral data $({\mathcal A}, {\mathcal H}, D, \gamma)$ by setting
\begin{eqnarray}
&& {\mathcal H}\;=\;\tilde{{\mathcal H}} \otimes {\mathbb C}^2 \ ,
\qquad {\mathcal A}\;=\;\tilde{{\mathcal A}} \otimes \id_2 \ ,
\nonumber \\
&& D\;=\;\tilde{D} \otimes \tau_1 \ , 
\qquad \  \gamma\;=\;\id_{\tilde{{\mathcal H}}} \otimes \tau_3 \ ,
\nonumber 
\end{eqnarray}
where $\tau_i$ are the Pauli matrices acting on ${\mathbb C}^2$. 
See [5] for further background and motivation.

\bigskip

\noindent 2) \ub{Differential forms}.

\medskip

\noindent Given a unital $^*$--algebra ${\mathcal A}$, as in 1), let
$\Omega^{\,^\bullet} ({\mathcal A})$ denote the universal unital,
graded, differential algebra of ``forms'' constructed by Connes and
Karoubi [44], which can be described as follows: 
\[
\Omega^{\,^\bullet} ({\mathcal A}) \ = \
\displaystyle\bigoplus_{n=0}^\infty \ 
\Omega^n ({\mathcal A}) \ ,
\]
where $\Omega^n ({\mathcal A})$ is spanned by elements $\alpha$ of the
form
\begin{equation}
\alpha \ = \ \sum_j a_j^0\;\delta a_j^1 \cdots \delta a_j^n \ ,
\end{equation}
with $a_j^i \in {\mathcal A}$ for all $i$ and $j$, and the
``derivation'' $\delta$ has the following properties:
\begin{itemize}
\item[(i)] $\delta$ is linear, and $(\delta a)^* := - \delta a^*$,
  which makes $\Omega^{\,^\bullet} ({\mathcal A})$ into a $^*$--algebra;
\item[(ii)] $\delta$ satisfies the Leibniz rule
\[
\delta\,(ab)\ = \ (\delta a)\,b \ + \ a\,(\delta b) \ ,
\]
for all $a,b$ in ${\mathcal A}$; in particular $\delta\;1=0$, where 1
is the unit element in ${\mathcal A}$;
\item[(iii)] $\delta^2 = 0$ .
\end{itemize}

Given spectral data $({\mathcal A}, {\mathcal H}, D, \gamma)$, as in
Definition~A, we define a $^*$--homomorphism $\pi$ from 
$\Omega^{\,^\bullet} ({\mathcal A})$ 
to $ B({\mathcal H})$ by setting 
\[
\pi\,(a) \ := \ a , \quad \pi\,(\delta a) \ := \ [D,a] \ . 
\]
A graded $^*$--ideal $J$ of $\Omega^{\,^\bullet} ({\mathcal A})$ 
is defined by
\begin{equation}
J \ = \ \displaystyle\bigoplus_{n=0}^\infty \ J^n \ ,\quad \ J^n \ :=
\ {\rm ker}\;\pi \bigm|_{\Omega^n({\mathcal A})} \ .
\end{equation}
Since in general $J$ is not a {\it differential} ideal, the graded
quotient $\Omega^{\,^\bullet} ({\mathcal A})/J$  does not define a
differential algebra. However, it is easy to show [5] that the
graded sub-complex
\[
J\;+\;\delta\,J \ := \ \displaystyle\bigoplus_{n=0}^\infty \
\left( J^n\;+\;\delta\,J^{n-1}\right) \ ,
\]
with $J^{-1} := \{ 0\}$, is a two-sided graded differential $^*$--ideal
of $\Omega^{\,^\bullet} ({\mathcal A})$.
(This follows from $\delta^2=0$ and from the Leibniz rule.)

The unital, graded, differential $^*$--algebra of differential forms,
$\Omega_D^{\,^\bullet} ({\mathcal A})$, is defined by 
\[
\Omega_D^{\,^\bullet} ({\mathcal A}) \ = \
\displaystyle\bigoplus_{n=0}^\infty \ \Omega_D^n \;({\mathcal
  A}) \ ,
\] 
where 
\begin{equation}
\Omega_D^n ({\mathcal A}) \ := \ \Omega^n ({\mathcal A}) \big/ \left(
  J^n + \delta\,J^{n-1}\right) \ .
\end{equation}
Each subspace $\Omega_D^n({\mathcal A})$ is a bi-module over
${\mathcal A} = \Omega_D^0 ({\mathcal A})$. 
\smallskip

An $n$--form on ${\mathcal H}$ is an {\it equivalence class} of bounded
operators on ${\mathcal H}$: For \ $[\alpha] \in \Omega_D^n ({\mathcal A})$,
\[
\pi([\alpha]) \ = \ \pi(\alpha) \ + \ \pi
\left( \delta J^{n-1} \right) \ .
\]
The image of $\Omega_D^{\,^\bullet} ({\mathcal A})$
under $\pi$ is ${\mathbb Z}_2$--graded:
\[
\pi\left( \Omega_D^{\,^\bullet} ({\mathcal A}) \right) \ = \
\pi\left( \displaystyle\bigoplus_{n=0}^\infty \ 
\Omega_D^{2n} ({\mathcal A}) \right) \ \oplus \ \pi \left(
\displaystyle\bigoplus_{n=0}^\infty \ 
\Omega_D^{2n+1} ({\mathcal A}) \right) \ ,  
\]
where elements of the first summand on the r.s.\ commute with $\gamma$,
while elements of the second summand anti-commute with $\gamma$ (where 
$\gamma$ is the ${\mathbb Z}_2$-grading of Definition~A).

\bigskip

\noindent 3) \ub{Integration}
\medskip

\noindent Property 3)~ii) of the Dirac operator in Definition~A allows us
to define a notion of integration over a non-commutative space in the
same way as in the classical case. Note that, for certain sets of
$N=1$ spectral data, we could use the Dixmier trace, as Connes
originally proposed; but the definition given below, first introduced
in [45], works in greater generality. Moreover, it is closer to
constructions in quantum field theory.

\bigskip

\noindent {\bf Definition B.} The {\it integral} over the
non-commutative space described by the $N=1$ spectral data $({\mathcal
  A}, {\mathcal H}, D, \gamma)$ is a state $\int\!\!\!\!\!-$ \ on \ $\pi 
(\Omega^{\,^\bullet} ({\mathcal A}))$ \ defined by 
\[
\int\!\!\!\!\!\!\!- \ : \ \left\{ \begin{array}{ccl}
\pi \left( \Omega^{\,^\bullet} ({\mathcal A}) \right) &\lra
&{\mathbb C} \\
\omega &\longmapsto &\int\!\!\!\!\!- \
\omega\;:=\;\displaystyle\mathop{\rm Lim}_{\varepsilon\to 0^+} \ 
\frac{{\rm Tr}_{{\mathcal H}}\left( \omega\, e^{-\varepsilon D^2}\right)}{
  {\rm Tr}_{{\mathcal H}} \left( e^{-\varepsilon D^2}\right)} \ , 
\end{array}
\right.
\]
where $\displaystyle\mathop{\rm Lim}_{\varepsilon\to 0^+}$ denotes
some limiting procedure making the functional $\int\!\!\!\!\!-$ linear
and positive semi-definite; existence of such  a procedure can be
shown analogously to [5,46], where the Dixmier trace is discussed.
\medskip

For the integral $\int\!\!\!\!\!-$ to be a useful tool, we need an
additional property that must be checked in each example:
\medskip

\noindent {\bf Assumption A.} The state $\int\!\!\!\!\!-$ on $\pi(
\Omega^{\,^\bullet} ({\mathcal A}))$ is {\it cyclic}, i.e., 
\[
\bint \ \omega \eta^* \ = \ \bint \ \eta^* \omega
\]
for all $\omega,\eta \in \pi (\Omega^{\,^\bullet} ({\mathcal A}))$
. (A weaker form is to only assume that $\int\!\!\!\!\!- \
\omega\,a\;=\; \int\!\!\!\!\!- \ a\, \omega$, for all $a\in{\mathcal
  A}$, $\omega \in \pi \left( \Omega^{\,^\bullet}\left( {\mathcal
      A}\right)\right)$.) 
\medskip

The state $\int\!\!\!\!\!-$ determines a positive semi-definite
sesqui-linear form on $\Omega^{\,^\bullet} ({\mathcal A})$ by setting
\begin{equation}
(\omega,\eta) \ := \ \bint \ \pi(\omega)\;\pi(\eta)^*
\end{equation}   
for all $\omega,\eta \in \Omega^{\,^\bullet} ({\mathcal A})$. In the
formulas below, we will often drop the representation symbol $\pi$
under the integral, as there is no danger of confusion.

Note that the commutation relations of the grading $\gamma$ with the
Dirac operator imply that forms of odd degree are orthogonal to those
of even degree with respect to $(\cdot,\cdot)$.
\smallskip

By $K^k$ we denote the kernel of this sesqui-linear form restricted to
$\Omega^k({\mathcal A})$. More precisely, we set
\begin{equation}
K \ := \ \displaystyle\bigoplus_{k=0}^\infty \ K^k\ , \quad K^k
\ := \ \left\{ \omega \in \Omega^k ({\mathcal A})\, |\, (\omega,\omega)
  = 0 \right\} \ .
\end{equation}
Obviously, $K^k$ contains the ideal $J^k$ defined in eq.~(5.2); in the
classical case they coincide. Assumption A is needed to show that $K$
is a two-sided graded $^*$--ideal of the algebra of universal forms,
too, so that we can pass to the quotient algebra, see~[18]. We now define 
\begin{equation}
\tilde{\Omega}^{\,^\bullet} ({\mathcal A}) \ := \ 
\displaystyle\bigoplus_{k=0}^\infty \ \tilde{\Omega}^k
({\mathcal A})\; , \quad \tilde{\Omega}^k ({\mathcal A}) \ 
:= \ \Omega^k ({\mathcal A}) \big/ K^k \ .
\end{equation}
The sesqui-linear form $(\cdot,\cdot)$ defines a positive definite
scalar product on $\tilde{\Omega}^k({\mathcal A})$, and we denote by
$\tilde{{\mathcal H}}^k$ the Hilbert space completion of this space
with respect to the scalar product, 
\begin{equation}
\tilde{{\mathcal H}}^{\,^\bullet} \ := \
\displaystyle\bigoplus_{k=0}^\infty  \ \tilde{{\mathcal H}}^k \
, \quad \tilde{{\mathcal H}}^k \ := \ \overline{\tilde{\Omega}^k
  ({\mathcal A})}^{(\cdot,\cdot)} \ .
\end{equation}
$\tilde{{\mathcal H}}^k$ is to be interpreted as the {\it space of
square-integrable $k$--forms}. Note that $\tilde{{\mathcal
    H}}^{\,^\bullet}$ does not quite coincide with the Hilbert space
that would arise from a GNS construction using the state
$\int\!\!\!\!\!-$ on $\tilde{\Omega}^{\,^\bullet} ({\mathcal A})$:
Whereas in $\tilde{{\mathcal H}}^{\,^\bullet}$, orthogonality of forms
of different degree is installed by definition, there may occur
mixings among forms of even degrees (or among odd forms) in the GNS
Hilbert space.

One now shows that the space $\tilde{\Omega}^{\,^\bullet} ({\mathcal A})$ is a
unital graded $^*$--algebra. For any $\omega \in \tilde{\Omega}^k
({\mathcal A})$, the left and right actions of $\omega$ on
$\tilde{\Omega}^p ({\mathcal A})$, with values in
$\tilde{\Omega}^{p+k} ({\mathcal A})$, 
\[
m_L (\omega)\eta \ := \ \omega\eta\; , \quad m_R (\omega)\eta \ := \
\eta\omega \ ,
\] 
are continuous in the norm given by $(\cdot,\cdot)$. 

Since the algebra $\tilde{\Omega}^{\,^\bullet} ({\mathcal A})$ may
fail to be differential, we introduce the unital graded differential
$^*$--algebra of square-integrable differential 
forms $\tilde{\Omega}_D^{\,^\bullet} ({\mathcal A})$ as the graded
quotient of $\Omega^{\,^\bullet} ({\mathcal A})$ by $K+\delta K$, 
\begin{equation}
\tilde{\Omega}_D^{\,^\bullet} ({\mathcal A})\;:=\;
\displaystyle\bigoplus_{k=0}^\infty\; \tilde{\Omega}_D^k
({\mathcal A}), \ \tilde{\Omega}_D^k ({\mathcal A}) \;:=\; \Omega^k
({\mathcal A}) \big/ \left( K^k + \delta K^{k-1}\right)\; \cong\;
\tilde{\Omega}^k ({\mathcal A}) / \delta K^{k-1} \ .
\end{equation}  
Note that we can regard the ${\mathcal A}$--bi-module
$\tilde{\Omega}_D^{\,^\bullet} ({\mathcal A})$ 
as a ``smaller version'' of $\Omega_D^{\,^\bullet} ({\mathcal A})$, in
the sense that there exists a projection from the latter onto the
former. 

In the classical case, differential forms can be identified with the
orthogonal complement of $Cl^{(k-2)}$ within $Cl^{(k)}$, where $Cl^{(k)}$
denotes the $k\,$th subspace in the filtration of the space of sections 
of the Clifford bundle, see [5,18]. Now, we use the scalar 
product $(\cdot,\cdot)$ on
$\tilde{{\mathcal H}}^k$ to introduce, for each $k\geq 1$, the
orthogonal projection
\begin{equation}
P_{\delta K^{k-1}} \ : \ \tilde{{\mathcal H}}^k \; \lra\;
\tilde{{\mathcal H}}^k
\end{equation}     
onto the image of $\delta K^{k-1}$ in $\tilde{{\mathcal H}}^k$, and we
set 
\begin{equation}
\omega^\perp \ := \ \left( 1 - P_{\delta K^{k-1}} \right)\omega\ \ \  \in
\tilde{{\mathcal H}}^k
\end{equation}
for each element $[\omega] \in \tilde{\Omega}_D^k ({\mathcal A})$.
This allows us to define a positive definite scalar product on 
$\tilde{\Omega}_D^k ({\mathcal A})$ via the representative
$\omega^\perp$ :
\begin{equation}
\left(\left[ \omega\right], \left[ \eta\right]\right) \ := \ \left(
  \omega^\perp, \eta^\perp\right)
\end{equation}
for all $[\omega],[\eta] \in \tilde{\Omega}_D^k ({\mathcal A})$. In
the classical case, this is just the usual inner product on the space
of square-integrable $k$--forms. 

\bigskip

\noindent 4) \ub{Vector bundles and Hermitian structures}

\medskip

\noindent We follow the algebraic formulation of classical differential
geometry, in order to generalize the notion of a vector bundle to the
non-commutative case.

\bigskip

\noindent {\bf Definition C.} [5] A {\it vector bundle} ${\mathcal E}$
over the non-commutative space described by the $N=1$ spectral data 
$({\mathcal A}, {\mathcal H}, D, \gamma)$ is a finitely generated,
projective left ${\mathcal A}$--module. 
\medskip

Recall that a module ${\mathcal E}$ is {\it projective} if there exists
another module ${\mathcal F}$ such that the direct sum ${\mathcal E}
\oplus {\mathcal F}$ is {\it free}, i.e., ${\mathcal E} \oplus
{\mathcal F} \cong {\mathcal A}^n$ as left ${\mathcal A}$--modules, for
some $n\in{\mathbb N}$. Since ${\mathcal A}$ is an algebra, every
${\mathcal A}$--module is a vector space; therefore, left ${\mathcal
  A}$--modules are representations of the algebra ${\mathcal A}$, and
${\mathcal E}$ is projective iff there exists a module ${\mathcal F}$
such that ${\mathcal E} \oplus {\mathcal F}$ is isomorphic to a
multiple of the left-regular representation.
\smallskip

By Swan's Lemma [47], a finitely generated, projective left module
corresponds, in the commutative case, to the space of sections of a
vector bundle. 

It is straightforward to define the
notion of a Hermitian structure over a vector bundle:

\bigskip

\noindent {\bf Definition D.} [5] A {\it Hermitian structure} over a
vector bundle ${\mathcal E}$ is a sesqui-linear map (linear in the
first argument)
\[
\langle \cdot, \cdot\rangle \ : \ {\mathcal E} \times {\mathcal E}
\lra {\mathcal A}
\]
such that for all $a, b \in {\mathcal A}$ and all $s,t \in {\mathcal
  E}$
\begin{itemize}
\item[1)] $\langle as,bt\rangle \ = \ a \langle s,t\rangle b^*$\,;
\item[2)] $\langle s,s\rangle \ \geq \ 0$ ;
\item[3)] the ${\mathcal A}$--linear map
\[
g\;: \ \left\{ \begin{array}{lll}
{\mathcal E} &\lra &{\mathcal E}_R^* \\
s &\longmapsto &\langle s,\cdot\,\rangle \ ,
\end{array} \right.
\]
where ${\mathcal E}_R^* := \{ \phi \in {\rm Hom} ({\mathcal E},
  {\mathcal A}) | \phi (as) = \phi(s) a^*\}$, is an isomorphism of
  left ${\mathcal A}$--modules, i.e., $g$ can be regarded as a metric
  on ${\mathcal E}$.
\end{itemize}

In the second condition, the notion of positivity in ${\mathcal A}$ is
simply inherited from the algebra ${\mathcal B} ({\mathcal H})$ of all
bounded operators on the Hilbert space ${\mathcal H}$.

\bigskip

\noindent 5) \ub{Generalized Hermitian structure on
  $\tilde{\Omega}^k({\mathcal A})$} 

\medskip

\noindent It turns out that the ${\mathcal A}$--bi-modules
$\tilde{\Omega}^k({\mathcal A})$ carry Hermitian structures in a
slightly generalized sense. Let ${\mathcal A}''$ be the weak
closure of the algebra ${\mathcal A}$ acting on $\tilde{{\mathcal
    H}}^0$, i.e., ${\mathcal A}''$ is the von Neumann algebra
generated by $\tilde{\Omega}^0 ({\mathcal A})$ acting on the Hilbert
space $\tilde{{\mathcal H}}^0$.

\bigskip

\noindent {\bf Theorem} [45,18]. There is a canonically defined
sesqui-linear map
\[
\langle \cdot, \cdot \rangle_D \ : \ \tilde{\Omega}^k ({\mathcal A})
\times \tilde{\Omega}^k ({\mathcal A}) \lra {{\mathcal A}}''
\]
such that, for all $a,b \in {\mathcal A}$ and all $\omega,\eta \in
\tilde{\Omega}^k ({\mathcal A})$, 
\begin{itemize}
\item[1)] $\langle a \omega, b\eta \rangle_D \ = \ a \langle
  \omega,\eta\rangle_D\, b^*$;
\item[2)] $\langle \omega,\omega\rangle_D \ \geq \ 0$;
\item[3)] $\langle \omega a, \eta\rangle_D \ = \ \langle \omega, \eta
  a^*\rangle_D$ .
\end{itemize}
We call $\langle \cdot, \cdot \rangle_D$ a {\it generalized Hermitian
  structure} on $\tilde{\Omega}^k ({\mathcal A})$. It is the
non-commutative analogue of the {\it Riemannian metric on the bundle
  of differential forms}. Note that $\langle \cdot, \cdot\rangle_D$
takes values in ${{\mathcal A}}''\,$, and thus property 3) of
Definition~D is not directly applicable.
For the proof of the theorem see [45,18].

\bigskip

\noindent 6) \ub{Connections}

\medskip

\noindent {\bf Definition E.} A {\it connection} $\nabla$ on a vector
bundle ${\mathcal E}$ over a non-commutative space is a ${\mathbb
  C}$--linear map
\[
\nabla \ : \ {\mathcal E} \ \lra \ \tilde{\Omega}_D^1
({\mathcal A}) \otimes_{{\mathcal A}} {\mathcal E} 
\]
such that
\[
\nabla (as) \ = \ \delta a \otimes s + a \nabla s
\]
for all $a \in {\mathcal A}$ and all $s \in {\mathcal E}$.
\medskip

Given a vector bundle ${\mathcal E}$, we define a space of ${\mathcal
  E}$--valued differential forms by
\[
\tilde{\Omega}_D^{\,^\bullet} ({\mathcal E}) \ := \
\tilde{\Omega}_D^{\,^\bullet} ({\mathcal A}) \otimes_{{\mathcal A}}
{\mathcal E} \ ;
\]
if $\nabla$ is a connection on ${\mathcal E}$ then it extends
uniquely to a ${\mathbb C}$--linear map, again denoted $\nabla$,
\begin{equation}
\nabla \ : \ \tilde{\Omega}_D^{\,^\bullet} ({\mathcal E}) \
\lra \ \tilde{\Omega}_D^{\,^{\bullet+1}}\ ({\mathcal E})
\end{equation}
such that
\begin{equation}
\nabla (\omega s) \ = \ (\delta\,\omega)\, s \ + \ (-1)^k\; \omega \nabla
s
\end{equation}
for all $\omega \in \tilde{\Omega}_D^k ({\mathcal A})$ and all $s \in 
\tilde{\Omega}_D^{\,^\bullet} ({\mathcal E})$.

\bigskip

\noindent {\bf Definition F.} The {\it curvature} of a connection $\nabla$
on a vector bundle ${\mathcal E}$ is given by
\[
R (\nabla) \ = \ -\; \nabla^2\;:\; {\mathcal E} \ \lra \
\tilde{\Omega}_D^2 \;({\mathcal A}) \otimes_{{\mathcal A}} {\mathcal
  E} \ .
\]
Note that the curvature extends to a map
\[
R (\nabla) \ : \ \tilde{\Omega}_D^{\,^\bullet} ({\mathcal E}) \
\lra \ \tilde{\Omega}_D^{\,^{\bullet+2}} ({\mathcal E})
\]
which is left ${\mathcal A}$--linear, as follows from eq.~(5.13) and
Definition E.

\bigskip

\noindent {\bf Definition G.} A connection $\nabla$ on a Hermitian
vector bundle $({\mathcal E}, \langle\cdot,\cdot\rangle)$ is called
{\it unitary} if
\[
\delta\,\langle s,t\rangle \ = \ \langle\, \nabla s,t\,\rangle \ - \
\langle \,s,\,\nabla t\,\rangle 
\]
for all $s, t \in {\mathcal E}$, where the r.s.\ of this equation is
defined by
\begin{equation}
\langle \omega \otimes s, t\rangle \ = \ \omega\,\langle s,t\rangle \
, \quad \langle s,\eta \otimes t\rangle \ = \ \langle s,t \rangle \,
\eta^*
\end{equation}
for all $\omega, \eta \in \tilde{\Omega}_D^1 ({\mathcal A})$ and all
$s, t \in {\mathcal E}$. 

\bigskip

\noindent 7) \ub{Riemannian curvature and torsion}

\medskip

\noindent Throughout this subsection, we make three additional 
assumptions which limit the generality of our results, but turn 
out to be fulfilled in interesting examples.

\bigskip

\noindent {\bf Assumption B.} We assume that the $N=1$ spectral data
under consideration have the following additional properties:
\begin{itemize}
\item[1)] $K^0=0$. (This implies that $\tilde{\Omega}_D^0 ({\mathcal
    A}) = {\mathcal A}$ \ and \ $\tilde{\Omega}_D^1 ({\mathcal A}) =
  \tilde{\Omega}^1 ({\mathcal A})$; \ thus $\tilde{\Omega}_D^1
  ({\mathcal A})$ carries a generalized Hermitian structure.)
\item[2)] $\tilde{\Omega}_D^1 ({\mathcal A})$ is a vector bundle,
  called the {\it cotangent bundle over} ${\mathcal
    A}$. $(\tilde{\Omega}_D^1 ({\mathcal A})$ is always a left
  ${\mathcal A}$--module. Here, we assume, in addition, that it is {\it
    finitely generated} and {\it projective}.)
\item[3)] The generalized metric $\langle\cdot,\cdot\rangle_D$ on
  $\tilde{\Omega}_D^1 ({\mathcal A})$ defines an isomorphism of left
  ${\mathcal A}$--modules between $\tilde{\Omega}_D^1 ({\mathcal A})$
  and the space of ${\mathcal A}$--anti-linear maps from
  $\tilde{\Omega}_D^1 ({\mathcal A})$ to ${\mathcal A}$, i.e., for
  each ${\mathcal A}$--anti-linear map
\[
\phi \ : \ \tilde{\Omega}_D^1 ({\mathcal A})
\;\lra\;{\mathcal A}
\]
with $\phi (a \omega) = \phi (\omega) a^*$, for all $\omega \in
\tilde{\Omega}_D^1 ({\mathcal A})$ and all $a \in {\mathcal A}$, there
is a unique $\eta_\phi \in \tilde{\Omega}_D^1 ({\mathcal A})$ with
\[
\phi (\omega) \ = \ \langle \eta_\phi, \omega\rangle_D \ .
\]
\end{itemize}

If $N=1$ spectral data $({\mathcal A}, {\mathcal H}, D, \gamma)$
satisfy these assumptions, we are able to define non-commutative
generalizations of classical notions like {\it curvature} and {\it
  torsion}. Whereas torsion and Riemann curvature can be introduced
whenever $\tilde{\Omega}_D^1 ({\mathcal A})$ is a vector bundle, the
third assumption above will provide a substitute for the 
procedure of ``contracting indices'' leading to {\it Ricci} and {\it 
scalar  curvature}. 

\bigskip

\noindent {\bf Definition H.} Let $\nabla$ be a connection on the
cotangent bundle $\tilde{\Omega}_D^1 ({\mathcal A})$ over a
non-commutative space $({\mathcal A}, {\mathcal H}, D, \gamma)$
satisfying Assumption B. The {\it torsion} of $\nabla$ is the
${\mathcal A}$--linear map 
\[
T (\nabla) \ := \ \delta\,-\,m \circ \nabla \ : \ \tilde{\Omega}_D^1
({\mathcal A}) \;\lra \; \tilde{\Omega}_D^2 ({\mathcal A})
\]
where $m\;:\;\tilde{\Omega}_D^1 ({\mathcal A}) \otimes_{{\mathcal A}}
\tilde{\Omega}_D^1 ({\mathcal A}) \lra \tilde{\Omega}_D^2
({\mathcal A})$ denotes the product of 1-forms in
$\tilde{\Omega}_D^{\,^\bullet} ({\mathcal A})$. 

\medskip

Using the definition of a connection, ${\mathcal A}$--linearity of
torsion is easy to verify. In analogy to the classical case, a unitary
connection $\nabla$ with $T(\nabla)=0$ is called a {\it Levi-Civita
  connection}. Note, however, that for a given set of non-commutative
spectral data, there may be several Levi-Civita connections --- or
none at all. 
\smallskip

Since we assume that $\tilde{\Omega}_D^1 ({\mathcal A})$ is a vector
bundle, we can define the {\it Riemannian curvature} of a connection
$\nabla$ on the cotangent bundle as a specialization of
Definition~F. To proceed further, we make use of part 2) of
Assumption~B, which implies that there exists a finite set of
generators $\{ E^{{\mathcal A}}\}$ of $\tilde{\Omega}_D^1 ({\mathcal
  A})$ and an associated ``dual basis'' $\{ \varepsilon_A\} \subset
\tilde{\Omega}_D^1 ({\mathcal A})^*$,
\[
\tilde{\Omega}_D^1 ({\mathcal A})^* \;:=\; \left\{
  \phi\,:\,\tilde{\Omega}_D^1 ({\mathcal A})\,\lra\,{\mathcal A}\ \Big|\ 
  \phi (a\omega) = a\phi(\omega)\quad {\rm for \ all} \ a \in
  {\mathcal A},\; \omega \in \tilde{\Omega}_D^1 ({\mathcal A}) \right\}
\ ,
\]
such that each $\omega \in \tilde{\Omega}_D^1 ({\mathcal A})$ can be
written as $\omega = \varepsilon_A (\omega) E^A$, see e.g.\ [48].   
Since the curvature is ${\mathcal A}$--linear, there
is a family of elements $\{ R_{\; B}^A\} \subset \tilde{\Omega}_D^2
({\mathcal A})$ with 
\begin{equation}
R (\nabla) \ = \ \varepsilon_A \otimes R_{\; B}^A \otimes E^B \ ;
\end{equation}
here and in the following the summation convention is used. Put
differently, we have applied the canonical isomorphism of vector
spaces
\[
{\rm Hom}_{{\mathcal A}} \left( \tilde{\Omega}_D^1 ({\mathcal A}), \
  \tilde{\Omega}_D^2 ({\mathcal A}) \otimes_{{\mathcal A}}
  \tilde{\Omega}_D^1 ({\mathcal A})\right) \ \cong \ 
\tilde{\Omega}_D^1 ({\mathcal A})^* \otimes_{{\mathcal A}}
\tilde{\Omega}_D^2 ({\mathcal A}) \otimes_{{\mathcal A}}
\tilde{\Omega}_D^1 ({\mathcal A}) 
\]
--- which exists because $\tilde{\Omega}_D^1 ({\mathcal A})$ is
projective --- and chosen explicit generators $E^A,
\varepsilon_A$. Then we have that $R(\nabla)\omega = \varepsilon_A
(\omega) R_{\;B}^A \otimes E^B$ for any 1-form $\omega \in
\tilde{\Omega}_D^1 ({\mathcal A})$.

Note that although the components $R_{\;B}^A$ need not be unique, the
tensor on the r.s.~of eq.~(5.15) is well-defined. Likewise, the Ricci
and the scalar curvature, to be introduced below, will be {\it
  invariant} combinations of those components, as long as we make sure
that all maps we use have the correct ``tensorial properties'' with
respect to the ${\mathcal A}$--action.
\smallskip

The last part of Assumption~B guarantees that to each
$\varepsilon_A$ there exists a unique 1-form $e_A \in
\tilde{\Omega}_D^1 ({\mathcal A})$ such that
\[
\varepsilon_A (\omega) \ = \ \langle \omega, e_A\rangle_D
\]
for all $\omega \in \tilde{\Omega}_D^1 ({\mathcal A})$. Every such
$e_A$ determines a bounded operator $m_L (e_A)\;:\;\tilde{{\mathcal
    H}}^1 \lra \tilde{{\mathcal H}}^2$ acting on $\tilde{{\mathcal
    H}}^1$ by left multiplication with $e_A$. The adjoint of this
operator w.r.t.~the scalar product $(\cdot,\cdot)$ on $\tilde{{\mathcal
    H}}^{\,^\bullet}$ is denoted by
\begin{equation}
e_A^{\rm ad} \ : \ \tilde{{\mathcal H}}^2 \;\lra\;\tilde{{\mathcal
    H}}^1 \ .
\end{equation} 
$e_A^{\rm ad}$ is a map of right ${\mathcal A}$--modules, and it is
easy to see that the correspondence $\varepsilon_A\;\longmapsto\; e_A^{\rm
  ad}$ is right ${\mathcal A}$--linear: For all $b \in {\mathcal A}$,
$\omega \in \tilde{\Omega}_D^1 ({\mathcal A})$, we have that
\[
(\varepsilon_A \cdot b)(\omega) \ = \ \varepsilon_A(\omega) \cdot b \
= \ \langle \omega, e_A\rangle\, b\ = \ \langle \omega, b^* e_A\rangle
\ ,
\]
and, furthermore, for all \ $\xi_1 \in \tilde{{\mathcal H}}^1$, $\xi_2
\in \tilde{{\mathcal H}}^2$,
\[
\left( b^* e_A \left( \xi_1\right), \xi_2\right) \ = \ \left(
  e_A\left(\xi_1\right), b\,\xi_2\right) \ = \ \left( \xi_1, e_A^{\rm
    ad} \left(b\,\xi_2\right)\right) \ ,
\]
where scalar products have to be taken in the appropriate spaces
$\tilde{{\mathcal H}}^k$. Altogether, the asserted right ${\mathcal
  A}$--linearity follows. Therefore the map
\[
\varepsilon_A \otimes R_{\; B}^A \otimes E^B \;\longmapsto \; e_A^{\rm ad}
\otimes R_{\; B}^A \otimes E^B
\]
is well-defined and has the desired tensorial properties. 

The definition of Ricci curvature involves another operation which we
require to be similarly well-behaved:
The orthogonal projections $P_{\delta K^{k-1}}$ on $\tilde{{\mathcal
    H}}^k$, see eq.~(5.9), satisfy
\[
P_{\delta K^{k-1}} (axb) \ = \ a\,P_{\delta K^{k-1}} (x)\,b 
\]
for all $a, b \in {\mathcal A}$ and all $x \in \tilde{{\mathcal
    H}}^k$.
For a proof see~[18].
This shows that projecting onto the ``2--form part'' of $R_{\; B}^A$
is an ${{\mathcal A}}$--bi-module map, i.e., we may apply
\[
e_A^{\rm ad} \otimes R_{\; B}^A \otimes E^B \longmapsto e_A^{\rm ad}
\otimes \left( R_{\; B}^A\right)^\perp \otimes E^B
\]
with $(R_{\; B}^A)^\perp = (1-P_{\delta K^1}) R_{\; B}^A$ as in
eq.~(5.10). 

Altogether, we arrive at the following definition of the
{\it Ricci curvature},
\[
{\rm Ric}\,(\nabla) \ = \ e_A^{\rm ad}\,\left( \left( R_{\; B}^A
  \right)^\perp \right) \;\otimes\; E^B \in \tilde{{\mathcal H}}^1\;
\otimes_{{\mathcal A}} \tilde{\Omega}_D^1 ({\mathcal A}) \ ,
\]
which turns out to be {\it independent} of any choices. In the following, we will
also use the abbreviation
\[
{\rm Ric}_B \ := \ e_A^{\rm ad}\,\left( \left(
    R_{\;B}^A\right)^\perp\right)
\]
for the components (which, again, are not uniquely defined).
\smallskip

{}From the components \ ${\rm Ric}_B$ we can pass to scalar
curvature. Again, we have to make sure that all maps occurring in this
process are ${\mathcal A}$--equivariant so as to obtain an invariant
definition. For any 1-form $\omega \in \tilde{\Omega}_D^1 ({\mathcal
  A})$, right-multiplication on $\tilde{{\mathcal H}}^0$ with $\omega$
defines a bounded operator $m_R(\omega)\,:\,\tilde{{\mathcal
    H}}^0\;\lra\;\tilde{{\mathcal H}}^1$, and we denote by
\begin{equation}
\omega_R^{\rm ad} \ : \ \tilde{{\mathcal H}}^1 \;\lra\;\tilde{{\mathcal
    H}}^0 
\end{equation}
the adjoint of this operator. In a similar fashion as above, one
establishes that
\[
(\omega\,a)_R^{\rm ad}\,(x) \ = \ \omega_R^{\rm ad}\,(x\,a^*)
\]
for all $x\in\tilde{{\mathcal H}}^1$ and $a \in {\mathcal A}$. This
makes it possible to define the scalar curvature $r\,(\nabla)$ of a
connection $\nabla$ as
\[
r\,(\nabla) \ = \ \left( E^{B\,*}\right)_R^{\rm ad}\;\left({\rm
    Ric}_B\right)\;\in\;\tilde{{\mathcal H}}^0 \ .
\]
As was the case for the Ricci tensor, acting with the adjoint of
$m_R(E^{B^*})$ serves as a substitute for the ``contraction of
indices''. We summarize our results in the following

\bigskip

\noindent {\bf Definition I.} Let $\nabla$ be a connection on the cotangent
bundle $\tilde{\Omega}_D^1 ({\mathcal A})$ over a non-commutative space
$({\mathcal A}, {\mathcal H}, D, \gamma)$ satisfying Assumption~B. The
{\it Riemannian curvature} $R(\nabla)$ is the left ${\mathcal
  A}$--linear map
\begin{equation}
R\,(\nabla) \ = \ -\;\nabla^2\;:\; \tilde{\Omega}_D^1 ({\mathcal A})
\;\lra\;\tilde{\Omega}_D^2 ({\mathcal A})\;\otimes_{{\mathcal
    A}}\;\tilde{\Omega}_D^1 ({\mathcal A}) \ .
\end{equation}
Choosing a set of generators $E^A$ of $\tilde{\Omega}_D^1 ({\mathcal
  A})$ and dual generators $\varepsilon_A$ of $\tilde{\Omega}_D^1
({\mathcal A})^*$, and writing $R\,(\nabla) = \varepsilon_A \otimes
R_{\; B}^A \otimes E^B$ as above, the {\it Ricci tensor} ${\rm Ric}
(\nabla)$ is given by
\begin{equation}
{\rm Ric}\,(\nabla) \ = \ {\rm Ric}_B\;\otimes\;
E^B\;\in\;\tilde{{\mathcal H}}^1 \;\otimes_{{\mathcal A}}\; \Omega_D^1
({\mathcal A}) \ ,
\end{equation}
where ${\rm Ric}_B := e_A^{\rm ad} \left(\left(
    R_{\;B}^A\right)^\perp\right)$, see eqs.~(5.10) and
(5.16). Finally, the {\it scalar curvature} $r\,(\nabla)$ of the
connection $\nabla$ is defined as
\begin{equation}
r\,(\nabla) \ = \ \left( E^{B*}\right)_R^{\rm ad} \;\left({\rm
  Ric}_B\right)\;\in\;\tilde{{\mathcal H}}^0 \ ,
\end{equation} 
with the notation of eq.~(5.17). The tensors ${\rm Ric}(\nabla)$ and
$r\,(\nabla)$ do not depend on the choice of generators.
\medskip

In [18] it is shown how to derive {\it Cartan structure equations} for
$\nabla, R (\nabla)$ and $T(\nabla)$ in NCG, in full generality. As in
classical geometry, these equations are useful for explicit
calculations. In [45,49] they have been exploited to study explicit
examples of non-commutative spaces arising in the Connes-Lott
formulation [5,50] of the standard model. 

\bigskip

\noindent 8) \ub{Generalized K\"ahler non-commutative geometry and
  higher supersymmetry}

\medskip

\noindent In this subsection we return to basics. Recall that a 
spin$^{\rm c}$ non-commutative space is described by some $N=1$ 
supersymmetric quantum theory, formulated in terms of spectral 
data $({\mathcal A}, {\mathcal H}, D, \gamma)$ with the properties 
specified in Definition A of subsection~1).

As in Sect.~{\bf 4.1}, we may ask what it is that characterizes
$({\mathcal A}, {\mathcal H}, D, \gamma)$ as the analogue of a
non-commutative K\"ahler (or Hyperk\"ahler, etc.) space. For a
classical K\"ahler manifold $M$ of even real dimension $n$, the
K\"ahler form enables one to define {\it two} Dirac operators, $D_1 :=
D$ and $D_2$, on the space of square-integrable
sections of the bundle $S$ of Pauli-Dirac spinors, which satisfy 
$D_1^2 = D_2^2$ and 
anti-commute with each other. (Likewise, one finds two Dirac
operators, $\bar{D}_1$ and $\bar{D}_2$, on the space ${\mathcal
  H}_p$ of charge-conjugate Pauli-Dirac spinors which anti-commute
and are transformed into each other by the K\"ahler form.) There is an
isomorphism from $S$ to
$\displaystyle\mathop{\oplus}_{p=0}^n\;\Lambda^{p,0} (M)$, the bundle
of holomorphic forms, under which $D_1$ is mapped to $\partial +
\partial^*$ and $D_2$ is mapped to $i\,(\partial - \partial^*)$. Hence
$\partial$ corresponds to $D_1 - i\,D_2$ and $\partial^*$ to 
$D_1+i\,D_2$. Since $\{ D_1, D_2\} = 0$ and $D_1^2 =
D_2^2$, the operators $D_1 \pm i\,D_2$ are indeed nilpotent.

Apparently, K\"ahler geometry can be characterized by the existence of
{\it two} supersymmetry generators $D_1$ and $D_2$ on ${\mathcal
  H}_e$ which anti-commute with each other and with the ${\mathbb
  Z}_2$-grading $\sigma$ (see Sect.~{\bf 4.1}) and whose squares are
equal to each other. In quantum mechanics
\begin{equation}
H \ = \ D_1^2 \ = \ D_2^2
\end{equation}
is interpreted as the Hamiltonian (generator of the time evolution) of
the system. 

The structure described here can be extended to non-commutative
geometry in a straightforward way: Consider $N=1$ spectral data
$({\mathcal A}, {\mathcal H}, D, \gamma)$, as described in 
Definition~A of subsection~1). We propose to explore the consequences of
the assumption that, besides $D_1 := D$, the operator $H := D^2$ has
further {\it self-adjoint} square roots, $D_2,\cdots,D_n$, such that
\begin{equation}
\left\{ \gamma, D_i\right\} \ = \ 0, \quad \left\{ D_i, D_j\right\} \
= \ 2\;\delta_{ij}\,H \ ,
\end{equation}
for all $i,j=1,\cdots,n$. Obviously, $H$ is a central element of the
algebra generated by $D_1, \cdots, D_n$. Thus, on every eigenspace of
$H$ corresponding to a non-zero eigenvalue $e \neq 0$, the operators 
$i\,D_1,\cdots,i\,D_n$ define a representation of the {\it Clifford
  algebra} over ${\mathbb R}^n$, while for $e=0$ we have $D_1=\dots=D_n=0$ 
(as follows from~(5.22)). If $n$ is odd the existence of $\gamma$
implies that this Clifford representation is necessarily reducible. In
classical geometry, one has that $n=1$, or $n$ even.

The group of $^*$--automorphisms of (5.22) is SO($n$). There is a
unitary representation $\rho$ of the group \ Spin($n$) on ${\mathcal
  H}$ such that
\begin{equation}
\rho\,(g)\,\underline{D} \cdot \underline{\xi}\;\rho\,(g^{-1}) \ = \
\underline{D} \cdot R\,(g)\,\underline{\xi} \ ,
\end{equation}
for all $g \in {\rm Spin} (n)$, where $\underline{\xi} \in {\mathbb
  R}^n$, $\underline{D} \cdot \underline{\xi} =
\mathop{\sum}_{j=1}^n \; D_j \xi^j$, and $R : g \in
{\rm Spin}(n) \longmapsto R(g) \in {\rm SO}(n)$ is the canonical
homomorphism from \ Spin($n$) to SO($n$). Assuming that, for $n\geq
2$, the ${\mathbb Z}_2$--grading $\gamma$ belongs to $\rho\left({\rm
    Spin}\left(n\right)\right)$, we conclude that $n$ must be {\it
  even}.

In spin geometry of classical manifolds, $\rho$ commutes with the
action of ${\mathcal A}$ on ${\mathcal H}$. We say that the spectral
data
\begin{equation}
\left( {\mathcal A}, {\mathcal H}, D_1,\cdots, D_n, \gamma\right)
\end{equation} 
exhibit $N=\bar{n}$ {\it supersymmetry} iff the representation $\rho$
of ${\rm Spin}\,(n)$ on ${\mathcal H}$ {\it commutes} with the
representation of ${\mathcal A}$ on ${\mathcal H}$.
\smallskip

As an example, we consider $N=\bar{2}$ supersymmetric spectral data
$({\mathcal A}, {\mathcal H}, D_1, D_2, \gamma)$. We define
\[
\partial\;=\;D_1\;-\;i\,D_2 \ , \quad \partial^*\;=\; D_1\;+\;i\,D_2 \ .
\]
By (5.22), 
\begin{equation}
\partial^2\;=\;(\partial^*)^2\;=\;0 \quad {\rm and}\quad \{ \partial,
\partial^*\}\;=\; 4 H \ .
\end{equation}
If ${\mathcal T}$ denotes the generator of the representation $\rho$
of ${\rm spin}(2) \cong {\mathbb R}$ on ${\mathcal H}$ then, for a
suitable normalization of ${\mathcal T}$, 
\begin{equation}
[\,{\mathcal T}, \partial\,]\;=\;\partial\ , \quad [\,{\mathcal
  T},\partial^*\,]\;=\;-\,\partial^* \ .
\end{equation}
The eigenvalues of ${\mathcal T}$ (which are, in general, neither
integer nor half-integer) thus correspond to ``degrees'' of
``holomorphic forms''. Thanks to (5.25), (5.26), the Hilbert space
${\mathcal H}$ can be interpreted as a direct sum of ${\mathbb
Z}$--graded complexes for $\partial$.

This structure reminds us of classical K\"ahler geometry, with
\begin{eqnarray}
&&{\mathcal A} \;=\; C\,(M)\;, \quad {\mathcal H} \ = \ L^2\left(
   \Lambda^{\,^{\bullet,0}} (M)\,, d{\rm vol}_g\right) \nonumber\\
&& \quad D_1\;=\; \partial\;+\;\partial^*\;, \quad D_2\;=\; i\,\left(
  \partial\;-\;\partial^*\right) \ ,\nonumber
\end{eqnarray}
${\mathcal T}$ counts the degree of a holomorphic form, and $\gamma =
(-1)^{{\mathcal T}}$. 

All this suggests to say that $N=\bar{2}$ supersymmetric spectral
data describe {\it ``non-commutative K\"ahler spaces''}. Similarly,
one may view $N=\bar{4}$ supersymmetric spectral data as encoding the
geometry of {\it ``non-commutative Hyperk\"ahler spaces''}. The
non-commutative torus, see [5,51,52] and Section~{\bf6}, is an example 
of a non-commutative K\"ahler space. 

\bigskip

\noindent 9) \ub{Aspects of the algebraic topology of $N=n$
  supersymmetric spectral data}.

\medskip

\noindent An obvious topological invariant is the index of $D$,  
see [15,22,17]:
\begin{equation}
{\rm Ind}\,(D) \ = \ {\rm tr}\,\left( \gamma\;e^{-\,\beta H}\right) \ ,
\end{equation}
where $H=D^2$; compare to eq.~(4.24).

With the help of the functional ${\rm tr}\,(\gamma\,e^{-\,\beta H}
(\cdot))$ one is able  to construct an analogue of the Chern character
and cyclic cocycles (e.g., the JLO cocycles) for the algebra
${\mathcal A}$; see [5,52].

For spectral data with $N=\bar{n}$ supersymmetry and $n \geq 2$, one
can construct an abstract analogue of Dolbeault-Hodge theory (see
subsections {\bf 5.2}, 7), 9)), whose precise relationship to
cyclic cohomology remains to be elucidated. 

\vspace{.5cm}

\subsection{Non-commutative Riemannian geometry}

A notion conspicuously absent from our discussion in Sect.~{\bf 5.1}
is that of {\it reality}. The structure introduced there does not
enable us to construct combinations of Dirac operators that are {\it
  real} operators. Of course, the bundle of Pauli-Dirac spinors is a
{\it complex} Hermitian vector bundle, and the Dirac operator $D_A$ of
Sect.~{\bf 4.1} is not, in general, a real operator. However, the
bundle of differential forms {\it is} a {\it real} vector bundle, and
exterior differentiation and its adjoint are real operators. Our
discussion of non-relativistic positronium in Sect.~{\bf 4.1} suggests
a way to introduce a notion of reality: To $N=\bar{n}$ spectral data
$({\mathcal A}, {\mathcal H}, \{ D_i\}_{i=1}^n , \gamma)$, one tries
to associate {\it ``charge-conjugate''} data $({\mathcal
  A},\bar{{\mathcal H}}, \{ \bar{D}_i\}_{i=1}^n, \bar{\gamma})$ (this
corresponds to replacing the electron by the positron, as in Sect.~{\bf
  4.1}) and then to construct a ``tensor product'' of these data
(corresponding to positronium, see Sect.~{\bf 4.1}), yielding {\it ``real''
spectral data} (corresponding to the fact that the electric charge of
positronium is zero). The definition of ``charge conjugation'', for
the example of classical spin$^{\rm c}$ manifolds, can be inferred
from many text books on quantum field theory; e.g.~[53]. When
${\mathcal A}$ is a non-commutative $^*$--algebra a proper definition of
charge conjugation requires some care~[54], because one now must 
distinguish between left and right ${\mathcal A}$--modules. It is not
hard to see that one should assume that ${\mathcal H}$ and
$\bar{{\mathcal H}}$ be ${\mathcal A}$--bi-modules. The Hilbert space
of the tensor product theory is then given by ${\mathcal H}
\otimes_{{\mathcal A}} \bar{{\mathcal H}}$. The details of this
construction are described in [54,18] and in subsection~5) 
below. 

In classical geometry, there are of course manifolds that do {\it not} 
admit any spin$^c$ structure and where one has to proceed along a 
different route, leading towards Riemannian geometry. In analogy, we will 
in this section describe supersymmetric spectral 
data which directly provide a notion of {\it non-commutative 
Riemannian geometry}.

\vfil\eject

\noindent 1) \ub{$N=(1,1)$ supersymmetry and Riemannian geometry}

\medskip

\noindent An appropriate definition of $N=(1,1)$ supersymmetry can be 
inferred from our discussion of positronium in Sect.~{\bf 4.1}. 

\bigskip

\noindent {\bf Definition A.} The data $({\mathcal A}, {\mathcal H}, {\mathcal
  D}, \gamma, \bar{{\mathcal D}},\bar{\gamma})$ are called $N=(1,1)$
{\it (supersymmetric) spectral data} iff
\begin{itemize}
\item[(1)] ${\mathcal H}$ is a separable Hilbert space;
\item[(2)] ${\mathcal A}$ is a unital $^*$--algebra faithfully
  represented on ${\mathcal H}$ by bounded operators;
\item[(3)] ${\mathcal D}$ and $\bar{{\mathcal D}}$ are operators that
  are essentially self-adjoint on a common dense domain in ${\mathcal
    H}$ and such that 
\begin{itemize}
\item[(i)] $\{ {\mathcal D}, \bar{{\mathcal D}}\} \;=\;0\,, \quad
  {\mathcal D}^2\;=\;\bar{{\mathcal D}}^2\;=:\;H\,$;
\item[(ii)] for each $a \in {\mathcal A}$, the commutators $[{\mathcal
    D},a]$ and $[\bar{{\mathcal D}}, a]$ extend to bounded operators
  on ${\mathcal H}$;
\item[(iii)] ${\rm exp}\,(-\,\varepsilon H)$ is trace class for
  arbitrary $\varepsilon > 0$;
\end{itemize}
\item[(4)] $\gamma$ and $\bar{\gamma}$ are ${\mathbb Z}_2$--gradings on
  ${\mathcal H}$ such that
\begin{itemize}
\item[(i)] $[\gamma,a]\;=\;[\bar{\gamma},a]\;=\;0$, for all $a \in
  {\mathcal A}$,
\item[(ii)] $\{ \gamma, {\mathcal D}\}\;=\;[\bar{\gamma},{\mathcal
    D}]\;=\;0$, \ $\{ \bar{\gamma}, \bar{{\mathcal D}}
  \}\;=\;[\gamma, \bar{{\mathcal D}}]\;=\;0\,.$ 
\end{itemize}
\end{itemize}

\noindent {\bf Remarks.} (a) As for $N=1$ supersymmetric spectral data,
${\mathbb Z}_2$--gradings $\gamma$ and $\bar{\gamma}$ may always be
introduced ``by hand'' if not given at the beginning:
\begin{eqnarray}
&& {\mathcal H}\;\lra\;{\mathcal H}\;\otimes\; {\mathbb
  C}^2\;\otimes\;{\mathbb C}^2\ , \quad 
  {\mathcal D}\;\lra\;{\mathcal
  D}\;\otimes\; \tau_1\;\otimes\; \id\ ,\quad
  \bar{{\mathcal D}}\;\lra\;\bar{{\mathcal D}}\;\otimes\; \id\;\otimes\;
  \tau_1\ ,\nonumber \\
&& \gamma\;=\;\id\;\otimes\; \tau_3\;\otimes\;\id\ ,
  \quad \bar{\gamma}\;=\;\id\;\otimes\;\id\;\otimes\;\tau_3\;,
  \nonumber 
\end{eqnarray}
where $\tau_1, \tau_2$ and $\tau_3$ denote the usual Pauli matrices. 

\medskip

(b) Setting
\[
d\;=\;{\mathcal D}\;-\;i\,\bar{{\mathcal D}}\;, \quad 
  d^*\;=\;{\mathcal D}\;+\;i\,\bar{{\mathcal D}} \ ,
\]
the relations in point (3) (i) of the Definition imply that 
\begin{equation}
d^2\ = \ (d^*)^2\;=\;0 \ , \quad d\,d^*\;+\;d^*\,d\ = \
4\,H \ .
\end{equation}
Thus $d$ plays the role of {\it exterior differentiation}.

\medskip

(c) Setting $\tilde{\gamma} = \gamma \bar{\gamma}$, $D_1 = {\mathcal
  D}$, $D_2 := \bar{{\mathcal D}}$, the data $({\mathcal A}, {\mathcal
  H}, D_1, D_2, \tilde{\gamma})$ define $N=2$ spectral data. Let $T$
denote the generator of the representation $\rho$ of ${\rm spin}(2)
\cong {\mathbb R}$ on ${\mathcal H}$ that implements the group SO(2)
$\cong$ U(1) of $^*$--automorphisms of the Clifford algebra generated
by $D_1$ and $D_2$, as discussed in subsection 8) of Sect.~{\bf
  5.1}. Then
\[
[\,T, d\,] \ = \ d \ , \quad [\,T, d^*\,] \ = \ -\,d^* \ .
\]
Thus $T$ counts the ``degree of differential forms''.\\
{\it If}
\begin{equation}
[T,a]\ =\ 0\  \quad {\rm for \ all} \ a \in {\mathcal A} \ ,
\end{equation}
we say that the {\it data} $({\mathcal A}, {\mathcal H}, {\mathcal D},
\gamma, \bar{{\mathcal D}}, \bar{\gamma}, T)$ {\it exhibit}
$N=\overline{(1,1)}$ {\it supersymmetry}. 

\medskip

(d) A {\it Hodge $*$ operator} can be defined by setting $* := \gamma$. Then
we find that 
\begin{equation}
*\;d\  =\  -\, d^*\; *\ , \quad [*,a] \ =\  0, \quad {\rm for \
  all} \ a\ \in \ {\mathcal A} \ .
\end{equation}
(Alternatively, one could choose $* := \bar{\gamma}$, with $*
\;d\;=\;d^* \ *$.)
On a classical manifold $M$, a Hodge operator with the above properties
exists whenever $M$ is compact, orientable, and of even dimension. For
a slightly more general definition of $*$, applicable e.g.\ when $M$ 
is odd-dimensional, see [18].

\medskip

In conclusion, $N=(1,1)$ spectral data can also be described in terms
of $({\mathcal A}, {\mathcal H}, d, \tilde{\gamma}, *)$, with
properties as in (b) -- (d) above, and we call them
$N=\overline{(1,1)}$ spectral data if the operator $T$ from remark (c)
commutes with ${\mathcal A}$.

We now start to explore the mathematical structure described by
$N=\overline{(1,1)}$ (supersymmetric) spectral data.

\bigskip

\noindent 2) \ub{Differential forms}

\medskip

\noindent Recall that the $N =\ \stackrel{{\scriptscriptstyle(-)}}{n}$ 
spectral data discussed in
Sect.~{\bf 5.1} do not enable one to introduce any notion of reality,
or, equivalently, of {\it complex conjugation}. We must show that,
starting from $N=(1,1)$ data, one can introduce a complex conjugation
with the property that $d$ is a {\it real} operator.

We first introduce an involution $\natural$, called complex
conjugation, on the universal graded, differential algebra
$\Omega^{\,^\bullet}({\mathcal A})$ of forms defined in subsection 2)
of Sect.~{\bf 5.1} ({\it without} the assumption that $(\delta a)^* =
- \delta a $ !):
\[
\natural \ : \ \Omega^{\,^\bullet} ({\mathcal A}) \lra \
\Omega^{\,^\bullet} ({\mathcal A}) 
\]
is the unique ${\mathbb C}$--anti-linear anti-automorphism such that
\begin{equation}
\natural\, (a) \;\equiv\; a^\natural\;:=\;a^*\ , \quad 
\natural\, (\delta a)\;\equiv\;(\delta a)^\natural\;:=\;\delta\,(a^*)
\end{equation}
for all $a \in {\mathcal A}$. If we write $\hat{\gamma}$ for the mod 2
reduction of the canonical ${\mathbb Z}$--grading on
$\Omega^{\,^\bullet} ({\mathcal A})$, we have
\begin{equation}
\delta\,\natural\,\hat{\gamma} \ = \ \natural\,\delta \ .
\end{equation}
We define a representation of $\Omega^{\,^\bullet} ({\mathcal A})$ on
${\mathcal H}$, again denoted by $\pi$, by setting
\begin{equation}
\pi\,(a) \ := \ a\;, \quad \pi\,(\delta a)\;:=\;[ d, a] 
\end{equation}
for all $a \in {\mathcal A}$. The map $\pi$ is a ${\mathbb Z}_2$--graded
representation in the sense that 
\[
\pi\,(\hat{\gamma} \omega \hat{\gamma}) \ = \ \gamma\,
\pi\,(\omega)\,\gamma
\]
for all $\omega \in \Omega^{\,^\bullet} ({\mathcal A})$ .

Although the abstract algebra of universal forms is the same as in the
$N=1$
 setting, the interpretation of the universal differential
$\delta$ has changed: In the $N=(1,1)$ framework, it is represented on
${\mathcal H}$ by the {\it nilpotent} operator $d$, instead of the
self-adjoint Dirac operator $D$, as in Sect.~{\bf 5.1}. This implies
that
\begin{equation}
\pi\,(\delta\omega) \ = \ \left[ d, \pi\left( \omega\right)\right]_g
\end{equation}
for all $\omega \in \Omega^{\,^\bullet} ({\mathcal A})$, where
$[\cdot, \cdot]_g$ denotes the graded commutator (defined with the
${\mathbb Z}_2$--grading on $\pi (\Omega^{\,^\bullet}({\mathcal A}))$
{}from above). The validity of eq.~(5.34) is the main difference
between the $N=(1,1)$ and the $N=1$ formalism. It ensures that there
are no forms $\omega \in \Omega^p ({\mathcal A})$ with $\pi
(\omega)=0$ but $\pi (\delta \omega) \neq 0$. 

\medskip

\noindent {\bf Proposition A.} [18] The graded vector space
\[
J \ = \ \displaystyle\bigoplus_{k=0}^\infty \ J^k \ , \quad J^k
\ :=\ {\rm ker}\;\pi \Bigm|_{\Omega^k ({\mathcal A})}
\]
with $\pi$ defined in (5.33) is a two-sided, graded, {\it
  differential} $^\natural$--ideal of $\Omega^{\,^\bullet} ({\mathcal
  A})$. 
\medskip

As a consequence of this proposition, the algebra of differential
forms
\begin{equation}
\Omega_d^{\,^\bullet} ({\mathcal A})\ := \ 
\displaystyle\bigoplus_{k=0}^\infty \ \Omega_d^k ({\mathcal
A})\, , \quad \Omega_d^k ({\mathcal A}) \;:=\; \Omega^k ({\mathcal A})
/ J^k \ ,
\end{equation}
is represented on the Hilbert space ${\mathcal H}$ via $\pi$. For
later purposes, we will also need an involution on
$\Omega_d^{\,^\bullet} ({\mathcal A})$, and, according to Proposition
A, it is given by the anti-linear map $\natural$ of (5.31). Note that
the ``natural'' involution $\omega\mapsto\omega^*$, which is inherited
from ${\mathcal H}$ and was used in the $N=1$ case, is no longer
available here: The space $\pi (\Omega^k ({\mathcal A}))$ is {\it not}
closed under taking adjoints, simply because $d$ is not self-adjoint.
However, it {\it is} closed under complex conjugation $\natural$,
which is implemented on ${\mathcal H}$ by 
\begin{equation}
\pi\,(\natural \omega) \ = \ *\,\pi\,(\omega)^*\,* \ , \quad\ \omega \in 
\Omega_d^{\,^\bullet} ({\mathcal A}) \ ,
\end{equation}
where $*$ is the Hodge operator.

\bigskip
\noindent 3) \ub{Integration}
\medskip

\noindent The integration theory follows the same lines as in the
$N=1$ case. The state $\int\!\!\!\!\!-$ is given as in Definition B of 
Sect.~{\bf 5.1}, with $4H=4D^2$ written as $\triangle = d d^* + d^*d$, see 
eq.\ (5.28). Again, 
we require Assumption~A of subsection {\bf 5.1}, 3) about the cyclicity of the
integral. This yields a sesqui-linear form on $\Omega_d^{\,^\bullet}
({\mathcal A})$ as before:
\begin{equation}
(\omega, \eta) \ = \bint \ \pi(\omega)\,\pi(\eta)^*
\end{equation}
for all $\omega, \eta \in \Omega_d^{\,^\bullet} ({\mathcal A})$. 

Because of the presence of the Hodge $*$--operator, the form
$(\cdot,\cdot)$ has an additional feature in the $N=(1,1)$ setting,
namely the inner product defined in eq.~(5.37) behaves like a {\it
  real} functional with respect to the involution $\natural$: For 
$\omega, \eta \in \Omega_d^{\,^\bullet} ({\mathcal A})$ we have
that 
\begin{equation}
\left( \omega^\natural , \eta^\natural \right) \ = \ 
\overline{ \left( \omega, \eta \right)}
\end{equation}
where the bar denotes ordinary complex conjugation. This is proven in
[18].

Note that, in examples, $p$-- and $q$--forms for $p\neq q$ are often
orthogonal w.r.t.\ the inner product $(\cdot,\cdot)$. (This also implies
eq.~(5.38).)

Since $\Omega_d^{\,^\bullet} ({\mathcal A})$ is a $^\natural$--, and not
a $^*$--algebra, the statement that the ideal $K$ defined in (5.5) is a
two-sided, graded $^*$-ideal of $\Omega^{\,^\bullet} ({\mathcal A})$ is
replaced by 
\medskip

\noindent {\bf Proposition~B.} [18] The graded kernel $K$, see eq.~(5.5), of
the sesqui-linear form $(\cdot,\cdot)$ is a two-sided, graded
$^\natural$-ideal of $\Omega_d^{\,^\bullet} ({\mathcal A})$.
\medskip

The remainder of subsection {\bf 5.1}, 3) carries over to the $N=(1,1)$ case,
with the only differences that $\tilde{\Omega}^{\,^\bullet} ({\mathcal
  A})$ is a $^\natural$--algebra and that the quotients $\Omega^k
({\mathcal A}) / (K^k + \delta K^{k-1}) \cong\tilde{\Omega}^k
({\mathcal A}) / \delta K^{k-1}$ are denoted by $\tilde{\Omega}_d^k
({\mathcal A})$. 

Upon passing from $\Omega_d^{\,^\bullet} ({\mathcal A})$ to the
algebra of {\it square-integrable forms} $\tilde{\Omega}_d^{\,^\bullet}
({\mathcal A})$, one might, however, lose the advantage of working
with differential ideals: Whereas $J$ has this property in the
$N=(1,1)$ setting, there may exist $\omega \in K^{k-1}$ with $\delta
\omega \notin K^k$. But it turns out that $K$ {\it vanishes} in many
interesting examples, and, for these, we have a representation of the
algebra $\tilde{\Omega}_d^{\,^\bullet} ({\mathcal A})$ of square-integrable
forms on $\tilde{{\mathcal H}}^{\,^\bullet}$. 

\bigskip
\noindent 4) \ub{Unitary connections and scalar curvature}
\medskip

\noindent Except for the notions of unitary connections and scalar
curvature, all definitions and results of subsections {\bf 5.1}, 4--8)
literally carry over to the $N=(1,1)$ case. The two exceptions
explicitly involve the $^*$-involution on the algebra of differential
forms, which is no longer available now. Therefore, we have to modify
the definitions for $N=(1,1)$ non-commutative geometry as follows:

\bigskip

\noindent {\bf Definition B.} \ A connection $\nabla$ on a Hermitian
vector bundle $({\mathcal E}, \langle \cdot, \cdot\rangle )$ over an
$N=(1,1)$ non-commutative space is called {\it unitary} if
\[
\lbrack\,d,\langle s, t \rangle\,\rbrack \ = \ \langle \nabla s, t \rangle \ + \
\langle s, \nabla t\rangle 
\]
for all $s,t \in {\mathcal E}$; since in general $\langle s, t \rangle 
\in {\mathcal A}''$, this equality is taken on the Hilbert space. 
The Hermitian structure on the r.s.\ is
extended to ${\mathcal E}$--valued differential forms by
\[
\langle \omega \otimes s, t\rangle \ = \ \omega \langle s,t\rangle \ ,
\quad \langle s, \eta \otimes t \rangle \ = \ \langle s,t \rangle\,
\eta^\natural
\]
for all $\omega, \eta \in \tilde{\Omega}_d^{\,^\bullet} ({\mathcal
  A})$ and $s, t \in {\mathcal E}$.

\bigskip

\noindent {\bf Definition C.} \ The scalar curvature of a connection
$\nabla$ on $\tilde{\Omega}_d^1 ({\mathcal A})$ is defined by
\begin{equation}
r\,(\nabla) \ = \ \left( E^{B\,\natural}\right)_R^{\rm ad} \; \left(
  {\rm Ric}_B\right) \;\in\; \tilde{{\mathcal H}}_0 \ .
\end{equation} 

\bigskip
\noindent 5) \ub{Remarks on the relation between $N=1$ 
 and $N=(1,1)$ spectral data}
\medskip

\noindent The definitions of $N=1$ and $N=(1,1)$ non-commutative
spectral data provide two different generalizations of classical
Riemannian differential geometry. In  classical geometry, one can
always find an $N=(1,1)$ description of a manifold originally given by
an $N=1$ set of data, whereas a non-commutative $N=(1,1)$ set of
spectral data appears to define a different mathematical structure
than a spectral triple, because of the additional generalized Dirac
operator which must be given on the Hilbert space. Thus, it is a
natural and important question under which conditions on an $N=1$
spectral triple $({\mathcal A}, {\mathcal H}, D)$ there exists an
associated $N=(1,1)$ set of data $({\mathcal A}, \tilde{{\mathcal H}},
d, *)$ over the same non-commutative space ${\mathcal A}$.

We have not been able, yet, to answer the question of how to pass from
$N=1$ to $N=(1,1)$ data in full generality; but in the following we propose 
one construction. Our guideline is the classical case, where the
main step in passing from $N=1$ to $N=(1,1)$ data is to replace the
Hilbert space ${\mathcal H}=L^2 (S)$ by $\tilde{{\mathcal
    H}}=L^2(\bar{S}) \otimes_{{\mathcal A}} L^2 (S)$ carrying two
actions of the Clifford algebra and therefore two anti-commuting Dirac
operators ${\mathcal D}$ and $\bar{{\mathcal D}}$, with all the
properties required in Definition A of subsection 1).

It is plausible that there are other approaches to this question, in
particular ones of a more operator algebraic nature, e.g.\ using
a ``Kasparov product of spectral triples'', but we will not enter
these matters here.
\smallskip

The first problem one meets when trying to copy the classical step
from $N=1$ to $N=(1,1)$ is that ${\mathcal H}$ should be an ${\mathcal
  A}$--bi-module. To ensure this, we require that the set of $N=1$
(even) spectral data $({\mathcal A}, {\mathcal H}, D, \gamma)$ is
endowed with a {\it real structure} [54], i.e., that there exists an
anti-unitary operator $J$ on ${\mathcal H}$ such that
\[
J^2\;=\;\epsilon\; 1, \quad J\gamma\;=\;\epsilon' \gamma J\; , \quad
JD\;=\;D\,J 
\]
for some (independent) signs $\epsilon, \epsilon' = \pm 1$, and such
that, in addition,
\[
J a J^* \ {\rm commutes \ with} \ b \ {\rm and} \ [D,b]
\ {\rm for \ all} \ a, b \in {\mathcal A} \ .
\]
This definition of a real structure was introduced by Connes in [54];
$J$ is of course related to {\it charge conjugation}, which, in this
context, can be expressed in terms of Tomita's modular conjugation
(see subsection 6) below).
\smallskip

In the present context, $J$  provides a canonical right
${\mathcal A}$--module structure on ${\mathcal H}$ by defining
\[
\xi \cdot a \ :=\ Ja^*J^*\,\xi
\]
for all $a \in {\mathcal A}, \xi \in {\mathcal H}$, see [54]. We can
extend this to a right action of $\Omega_D^1 ({\mathcal A})$ on
${\mathcal H}$ if we set
\[
\xi \cdot \omega \ := \ J\omega^*J^*\,\xi
\]
for all $\omega \in \Omega_D^1 ({\mathcal A})$ and $\xi \in {\mathcal
  H}$; for simplicity, the representation symbol $\pi$ has been
omitted. Note that, by the assumptions on $J$, the right action
commutes with the left action of ${\mathcal A}$. Thus ${\mathcal H}$
is an ${\mathcal A}$--bi-module. Moreover, we can form tensor products
of bi-modules {\it over the algebra} ${\mathcal A}$ just as in the
classical case. If ${\mathcal H}$ carries a Hermitian structure as in
Definition D of Sect.~{\bf5.1}, then ${\mathcal H} \otimes_{{\mathcal A}} 
{\mathcal H}$ is endowed with a natural scalar product. 
\smallskip

The real structure $J$ allows us to define an anti-linear ``flip''
operator
\[
\Psi \ : \ \left\{
\begin{array}{c}
\Omega_D^1 ({\mathcal A}) \otimes_{{\mathcal A}} {\mathcal
  H}\;\lra\;{\mathcal H} \otimes_{{\mathcal A}} \Omega_D^1 ({\mathcal
  A}) \\
\omega \otimes \xi \;\longmapsto\; J \xi \otimes \omega^* 
\end{array} \right. \ .
\]
It is straightforward to verify that $\Psi$ is well-defined and that
it satisfies
\[
\Psi\,(a\,s) \ = \ \Psi\,(s)\, a^*
\]
for all $ a\in {\mathcal A},\  s \in \Omega_D^1 ({\mathcal A})
\otimes_{{\mathcal A}} {\mathcal H}$.

{}From now on, we furthermore assume that ${\mathcal H}$ is a {\it
  projective} left ${\mathcal A}$-module. (In fact, the existence of 
a dense projective left ${\mathcal A}$-module ${\mathcal H}_0$ inside 
${\mathcal H}$ is sufficient for our purposes.) Then ${\mathcal H}$ can be 
equipped with connections
\[
\nabla \ : \ {\mathcal H} \;\lra\; \Omega_D^1 ({\mathcal A})
\otimes_{{\mathcal A}} {\mathcal H} \ ,
\]
i.e., ${\mathbb C}$--linear maps such that
\[
\nabla (a\,\xi) \ = \ \delta\, a \otimes \xi \,+\,a \nabla \xi
\]
for all $a\in {\mathcal A}$ and $\xi \in {\mathcal H}$. For each
connection $\nabla$ on ${\mathcal H}$, there is an ``associated
right-connection'' $\overline{\nabla}$ defined with the help of the flip
$\Psi$:
\[
\overline{\nabla} \ : \ \left\{
\begin{array}{c}
{\mathcal H}\;\lra\; {\mathcal H} \otimes_{{\mathcal A}} \Omega_D^1
({\mathcal A}) \\
\xi\;\longmapsto\; - \Psi\,\left( \nabla J^* \xi\right)
\end{array} \right.
\]
$\overline{\nabla}$ is again ${\mathbb C}$--linear and satisfies
\[
\overline{\nabla}\,(\xi a) \ = \ \xi \otimes \delta a\;+\;\left(
  \overline{\nabla} \xi \right)\, a \ .
\]
A connection $\nabla$ on ${\mathcal H}$, together with its associated
right connection $\overline{\nabla}$, induces a ${\mathbb C}$--linear
``tensor product connection'' $\widetilde{\nabla}$ on ${\mathcal H}
\otimes_{{\mathcal A}} {\mathcal H}$ of the form
\[
\widetilde{\nabla} \ : \ \left\{
\begin{array}{c} 
{\mathcal H} \otimes_{{\mathcal A}} {\mathcal H}\;\lra\;{\mathcal H}
\otimes_{{\mathcal A}} \Omega_D^1 ({\mathcal A}) \otimes_{{\mathcal
    A}} {\mathcal H} \\
\quad\xi_1 \otimes \xi_2 \;\longmapsto\; \overline{\nabla} \xi_1 \otimes \xi_2 +
\xi_1 \otimes \nabla \xi_2
\end{array} \right. \ .
\]
Because of the position of the factor $\Omega_D^1 ({\mathcal A})$,
$\widetilde{\nabla}$ is not quite a connection in the usual sense. 
\smallskip

In the classical case, the last ingredient needed for the definition of the
two Dirac operators  of an $N=(1,1)$ Dirac bundle were the  two
anti-commuting Clifford actions $\Gamma$ and $\overline{\Gamma}$ on
${\mathcal H}$. Their obvious generalizations to the non-commutative
case are the ${\mathbb C}$--linear maps
\begin{equation}
\Gamma \ : \ \left\{
\begin{array}{c}
{\mathcal H} \otimes_{{\mathcal A}} \Omega_D^1 ({\mathcal A})
\otimes_{{\mathcal A}} {\mathcal H}\;\lra\; {\mathcal H}
\otimes_{{\mathcal A}} {\mathcal H} \\
\qquad \ \quad\; \xi_1 \otimes \omega \otimes \xi_2\;\longmapsto\;\xi_1 \otimes
\omega \xi_2
\end{array} \right. 
\end{equation}
and
\begin{equation}
\phantom{xxxx}\overline{\Gamma} \ : \ \left\{
\begin{array}{c}
{\mathcal H} \otimes_{{\mathcal A}} \Omega_D^1 ({\mathcal A})
\otimes_{{\mathcal A}} {\mathcal H}\;\lra\;{\mathcal H}
\otimes_{{\mathcal A}} {\mathcal H} \\
\qquad\qquad\ \ \xi_1 \otimes \omega \otimes \xi_2 \;\longmapsto\; \xi_1
\omega \otimes \gamma\,\xi_2
\end{array} \right. \ .
\end{equation}
With these, we may introduce two operators ${\mathcal D}$ and
$\overline{{\mathcal D}}$ on ${\mathcal H} \otimes_{{\mathcal A}} {\mathcal
  H}$ in analogy to the classical case:
\begin{equation}
{\mathcal D} \ := \ \Gamma \circ \widetilde{\nabla} \ , \quad
\overline{{\mathcal D}} \ := \ \overline{\Gamma} \circ \widetilde{\nabla}\ .
\end{equation}
In order to obtain a set of $N=(1,1)$ spectral data, one has to find a
connection $\nabla$ on ${\mathcal H}$ which makes the operators
${\mathcal D}$ and $\overline{{\mathcal D}}$ self-adjoint and ensures that
the anti-commutation relations in point (3)(i) of Definition A in
subsection 1) are satisfied. 

Although we are not able, in general, to prove the existence of a
connection $\nabla$ on ${\mathcal H}$ which supplies ${\mathcal D}$
and $\overline{{\mathcal D}}$ with the correct algebraic properties, the
naturality of the construction presented above as well as the
similarity with the procedure in Sect.~{\bf 4.1} leads us to expect
that this problem can be solved in many cases of interest. (See 
Section {\bf6} for an example.)

\bigskip
\noindent 6) \ub{Riemannian and spin$^{\bf c}$ ``manifolds'' in
  non-commutative geometry}
\medskip

\noindent In this section, we want to address the following question:
What is the additional structure that makes an $N=(1,1)$
non-commutative space into a {\it non-commutative ``manifold''}, into
a spin$^{\rm c}$ {\it ``manifold''}, or into a {\it quantized phase
  space}? There exists a definition of non-commutative manifolds in
terms of $K$--homology, see [5], but in the formalism introduced
in the present work it is possible to find more direct criteria. In
our search for the characteristic features of non-commutative
manifolds we will, as before, be guided by the classical case and by
the principle that they should be natural from the physics point of
view.

Extrapolating from classical geometry, we are led to the following
requirement an $N=(1,1)$ space $({\mathcal A}, {\mathcal H}, d, \gamma
*)$ should satisfy in order to be a ``manifold''. The data must extend
to a set of $N=\overline{(1,1)}$ {\it spectral data} $({\mathcal A},
{\mathcal H}, d, T, *)$ where $T$ is a self-adjoint operator on
${\mathcal H}$ such that

\quad$\!\!\!\!\!$ i) $\,\ \ [\,T,a\,] = 0$\ for all\ $a \in {\mathcal A}\,$;

\ ii) $\ \ [\,T, d\,] = d\,$;

iii) $\ \ T$ has integral spectrum, and $\gamma$ is the mod 2 reduction of
$T$, i.e., $\gamma = \pm 1$ on ${\mathcal H}_\pm$,\\
\phantom{\quad iiiiiiiiii)}  where
\[
{\mathcal H}_\pm \ = \ {\rm span}\,\left\{ \xi \in {\mathcal H} \bigm|
  T \xi\;=\;n \xi \ {\rm for \ some} \ n \in {\mathbb Z},\ 
  (-1)^n\;=\;\pm 1 \right\}\ .
\]

Before we can formulate other properties characteristic of
non-commutative manifolds, we recall some basic facts about {\it
  Tomita-Takesaki theory}. Let ${\mathcal M}$ be a von Neumann algebra
acting on a separable Hilbert space ${\mathcal H}$, and assume that
$\xi_0 \in {\mathcal H}$ is a cyclic and separating vector for
${\mathcal M}$, i.e., 
\[
\overline{{\mathcal M} \xi_0} \ = \ {\mathcal H} \ 
\]
and
\[
a\, \xi_0 \ = 0 \quad \Longrightarrow \quad a \ = 0 \ 
\]
for any $a \in {\mathcal M}$, respectively. Then we may define an
anti-linear operator $S_0$ on ${\mathcal H}$ by setting
\[
S_0\,a \,\xi_0 \ = \ a^*\, \xi_0
\]
for all $a\in {\mathcal M}$. One can show that $S_0$ is closable, and
we denote its closure by $S$. The polar decomposition of $S$ is
written as
\[
S \ = \ J\;\triangle^{\frac 1 2} 
\]
where $J$ is an anti-unitary involutive operator, referred to as
(Tomita's) {\it modular conjugation}, and the so-called {\it modular
  operator} $\triangle$ is a positive self-adjoint operator on
${\mathcal H}$. The fundamental result of Tomita-Takesaki theory is
the following theorem:
\[
J {\mathcal M} J \ = \ {\mathcal M}'\ , \quad
\triangle^{it} {\mathcal M} \triangle^{-it} \ = \ {\mathcal M}
\]
for all $t\in {\mathbb R}$, where ${\mathcal M}'$ denotes the
commutant of ${\mathcal M}$ on ${\mathcal H}$. Furthermore, the vector
state $\omega_0 (\cdot) := (\xi_0, \cdot\,\xi_0)$ is a {\it KMS}-state for
the automorphism $\sigma_t := {\rm Ad}_{\triangle^{it}}$ of ${\mathcal
  M}$, i.e., 
\[
\omega_0 \left( \sigma_t \left( a\right) b\right) \ = \ \omega_0
\left( b\,\sigma_{t-i}\left( a \right)\right)
\]
for all $a, b \in {\mathcal M}$ and all real $t$.

\medskip

Let $({\mathcal A}, {\mathcal H}, d, T, *)$ be a set of $N =
\overline{(1,1)}$ spectral data. We define the analogue $Cl_{{\mathcal
  D}} ({\mathcal A})$ of the space of sections of the Clifford bundle,
\[
Cl_{{\mathcal D}} ({\mathcal A}) \ = \ \left\{ a_0\,\left[\, {\mathcal
      D}, a_1\,\right] \cdots \left[\, {\mathcal D}, a_k\,\right] \bigm|
  k\in {\mathbb Z}_+,\, a_i \in {\mathcal A} \right\} \ ,
\]
where ${\mathcal D} = d + d^*$, and, corresponding to the second
generalized Dirac operator $\overline{{\mathcal D}} = i (d-d^*)$, 
\[
Cl_{\overline{{\mathcal D}}} ({\mathcal A}) \ = \ \left\{ a_0 \left[\,
    \overline{{\mathcal D}}, a_1\,\right] \cdots \left[\, \overline{{\mathcal D}}, a_k
    \,\right] \bigm| k \in {\mathbb Z}_+\; , \ a_i \in {\mathcal
      A}\right\} \ . 
\]
In the classical setting, $Cl_{{\mathcal D}} ({\mathcal
  A})$ and $Cl_{\overline{{\mathcal D}}} ({\mathcal A})$ operate on
${\mathcal H}$ by the two actions $\Gamma$ and $\overline{\Gamma}$,
respectively; see Sect.~{\bf 4.1}. In the general case, we notice
that, in contrast to the algebras $\Omega_d ({\mathcal A})$ and
$\Omega_{{\mathcal D}} ({\mathcal A})$ introduced before,
$Cl_{{\mathcal D}}({\mathcal A})$ and $Cl_{\overline{{\mathcal D}}}
({\mathcal A})$ form $^*$--algebras of operators on ${\mathcal H}$,
but are neither ${\mathbb Z}$--graded nor differential.

 We want to apply Tomita-Takesaki theory to the von
Neumann algebra 
${\mathcal M} :=$ \break
$ \left( Cl_{{\mathcal D}} \left( {\mathcal
      A}\right)\right)''$. Suppose there exists a vector $\xi_0 \in
{\mathcal H}$ which is cyclic and separating for ${\mathcal M}$, and
let $J$ be the anti-unitary conjugation associated to ${\mathcal M}$
and $\xi_0$. Suppose, moreover, that for all $a\,\in\, ^J\!{\mathcal A}
:= J{\mathcal A}J$ the operator $[\overline{{\mathcal D}}, a]$ uniquely
extends to a bounded operator on ${\mathcal H}$. Then we can form the
algebra of bounded operators $Cl_{\overline{{\mathcal D}}} (^J\!{\mathcal
  A})$ on ${\mathcal H}$ as above. The properties $J{\mathcal A} J \subset 
{\mathcal A}'$ and 
$\{\,{\mathcal D},\overline{\mathcal D} \,\}=0$ imply that 
$Cl_{{\mathcal D}} \left( {\mathcal    A}\right)$
and $Cl_{\overline{{\mathcal D}}} (^J\!{\mathcal  A})$
commute in the graded sense; to arrive at truly commuting
algebras, we first decompose $Cl_{\overline{{\mathcal D}}} (^J\!{\mathcal
  A})$ into a direct sum
\[
Cl_{\overline{{\mathcal D}}} (^J\!{\mathcal A}) \ = \ Cl_{\overline{{\mathcal
    D}}}^+ (^J\!{\mathcal A}) \oplus Cl_{\overline{{\mathcal D}}}^-
(^J\!{\mathcal A})
\]
with
\[
Cl_{\overline{{\mathcal D}}}^\pm (^J\!{\mathcal A}) \ = \ \left\{ \omega \in
  Cl_{\overline{{\mathcal D}}} (^J\!{\mathcal A}) \bigm| \gamma \omega \ = \
  \pm \, \omega \gamma \right\} \ .
\]
Then we define the ``twisted algebra'' $\widetilde{Cl}_{\overline{{\mathcal
      D}}} (^J\!{\mathcal A}) := Cl_{\overline{{\mathcal D}}}^+ (^J\!{\mathcal
  A}) \oplus \gamma\, Cl_{\overline{{\mathcal D}}}^- (^J\!{\mathcal
  A})$. This algebra commutes with $Cl_{{\mathcal D}}({\mathcal A})$.

We propose the following definitions: The $N=\overline{(1,1)}$ spectral data
$({\mathcal A}, {\mathcal H}, d, T, *)$ describe a {\it
  non-commutative manifold} if
\[
\widetilde{Cl}_{\overline{{\mathcal D}}} (^J\!{\mathcal A}) \ = \ J\,
Cl_{{\mathcal D}} ({\mathcal A})\, J \ .
\]
Furthermore, inspired by classical geometry, we say that a
non-commutative manifold $({\mathcal A}, {\mathcal H}, d, T, *,
\xi_0)$ is $spin^c$ if the Hilbert space factorizes as a
$Cl_{{\mathcal D}} ({\mathcal A}) \otimes
\widetilde{Cl}_{\overline{{\mathcal D}}} (^J\!{\mathcal A})$ module in the
form
\[
{\mathcal H} \ = \ {\mathcal H}_{{\mathcal D}} \otimes_{{\mathcal Z}}
{\mathcal H}_{\overline{{\mathcal D}}} 
\]
where ${\mathcal Z}$ denotes the center of ${\mathcal M}$.
\medskip

Next, we introduce a notion of ``quantized phase space''. 
We consider a set of $N=(1,1)$ spectral data $({\mathcal F}, {\mathcal
  H}, d, \gamma, *)$, where we now think of ${\mathcal F}$ as the
algebra of ``phase space functions'' (i.e., of pseudo-differential
operators, in the Schr\"odinger picture of quantum mechanics;
${\mathcal F}$ is constructed as in eq.~(4.39) of Sect.~{\bf 4.1}), 
rather than of functions over configuration space. We are, therefore,
not postulating the existence of a cyclic and separating vector for
the algebra $Cl_{{\mathcal D}} ({\mathcal F})$. Instead, for
each $\beta > 0$, we define the {\it temperature} or {\it KMS state}
\[
\mathop{\bint}\limits_\beta \ : \ \left\{ 
\begin{array}{l}
Cl_{{\mathcal D}} ({\mathcal F}) \ \lra \ {\mathbb C} \\
\qquad \quad  \omega \longmapsto \ \mathop{\int\!\!\!\!\!-}\limits_\beta \;\omega
\ := \ {\displaystyle \frac{{\rm
    Tr}_{{\mathcal H}} (\omega\,e^{-\beta {\mathcal D}^2})}{{\rm
    Tr}_{{\mathcal H}} (e^{-\beta{\mathcal D}^2})}}
\end{array} \right. \ ,
\]
The $\beta$--integral $\mathop{\int\!\!\!\!\!-}\limits_\beta$ clearly 
is a faithful
state, and through the GNS-construction we obtain a faithful
representation of $Cl_{{\mathcal D}} ({\mathcal F})$ on a Hilbert
space ${\mathcal H}_\beta$ with a cyclic and separating vector
$\xi_\beta \in {\mathcal H}_\beta$ for ${\mathcal M}= (Cl_{{\mathcal
    D}} ({\mathcal F}))''$. Each bounded operator $A \in {\mathcal B}
({\mathcal H})$ on ${\mathcal H}$ induces a bounded operator $A_\beta$
on ${\mathcal H}_\beta$; this is easily seen by computing matrix
elements of $A_\beta$,
\[
\langle A_\beta\, x,y\rangle \ = \ \mathop{\bint}\limits_\beta \ A x y^*
\]
for all $x, y \in {\mathcal M} \subset {\mathcal H}_\beta$, and by using
  the explicit form of the $\beta$--integral. We denote the modular
  conjugation and the modular operator on ${\mathcal H}_\beta$ by
  $J_\beta$ and $\triangle_\beta$, respectively, and we assume that,
  for each $a \in {\mathcal M}$, the commutator
\[
\left[\, \overline{{\mathcal D}}, J_\beta \, a J_\beta \,\right] \ = \ \frac 1
i \ \frac{d}{dt} \ \left( \left( e^{it\overline{{\mathcal D}}}\right)_\beta \
    J_\beta\,a J_\beta \left( e^{-it\overline{{\mathcal D}}}
    \right)_\beta\right) \Bigm|_{t=0}
\]
defines a bounded operator on ${\mathcal H}_\beta$.

Then we can define an algebra of bounded operators
$\widetilde{Cl}_{\overline{{\mathcal D}}} (^{J_\beta} {\mathcal F})$ on
${\mathcal H}_\beta$, which is contained in the commutant of
$Cl_{{\mathcal D}} ({\mathcal F})$, and we say that the $N=(1,1)$
spectral data $({\mathcal F}, {\mathcal H}, d, \gamma, *)$ describe a
{\it quantized phase space} if the following equation holds:
\[
J_\beta \, Cl_{{\mathcal D}} ({\mathcal F}) \, J_\beta \ = \
\widetilde{Cl}_{\overline{{\mathcal D}}} \left(^{J_\beta} {\mathcal
    F}\right) \ .
\]

\bigskip
\noindent 7) \ub{Algebraic topology of $N=\overline{(1,1)}$ spectral
  data}
\medskip

\noindent Let $({\mathcal A}, {\mathcal H}, {\mathcal D}, \gamma,
\bar{{\mathcal D}}, \bar{\gamma})$ be some $N=(1,1)$ or
$N=\overline{(1,1)}$ supersymmetric spectral data with all the
properties (1 -- 4) specified in Definition A of subsection
1). We set
\begin{equation}
H \ := \ D^2\; , \ \tilde{\gamma}\;=\;\gamma\,\bar{\gamma}\; , \
  *\;=\;\gamma \ .
\end{equation}
Then we can define the {\it Euler number} $\chi$ and the {\it
  Hirzebruch signature} $\tau$ as in formulae (4.41) and (4.42) of Sect.~{\bf
  4.1}:
\begin{equation}
\chi\;=\;\chi\left( {\mathcal A}, {\mathcal H}, {\mathcal D},
  \gamma, \bar{{\mathcal D}}, \bar{\gamma}\right) \ := \ {\rm
  tr}_{{\mathcal H}} \left( \tilde{\gamma}\, e^{-\beta H}\right) \ ,
\end{equation}
and
\begin{equation}
\tau \ := \ {\rm tr}_{{\mathcal H}} \left( *\, e^{-\beta H}\right) \ .
\end{equation}
They are independent of $\beta$ and define homotopy invariants of the
spectral data.
\smallskip

The data $\left( {\mathcal A}, {\mathcal H}, {\mathcal D},
  \tilde{\gamma}\right)$ permit us to introduce a functional ${\rm
  tr}_{{\mathcal H}} \left( \tilde{\gamma}\,e^{-\beta H}
  (\cdot)\right)$ that gives rise to a JLO cyclic cocycle (the ``Euler
cocycle'') for the algebra ${\mathcal A}$. Likewise, the data
$({\mathcal A}, {\mathcal H}, {\mathcal D}, *)$ yield the functional
${\rm tr}_{{\mathcal H}} (*\,e^{-\beta H} (\cdot))$ and give rise to a
second JLO cyclic cocycle (the ``signature cocycle''). See [5,52,55]
for the construction of such cocycles.
\smallskip

What is, perhaps, more useful is that the $N=(1,1)$ data $({\mathcal
  A}, {\mathcal H}, {\mathcal D}, \gamma, \bar{{\mathcal D}},
\bar{\gamma})$ give rise to a {\it de Rham--Hodge theory} on
${\mathcal H}$. In order not to get lost in somewhat uninteresting
generalities, we only consider $N=\overline{(1,1)}$ spectral data, but
see [55] for more general results. As usual, we introduce exterior
differentiation and its adjoint by setting 
\begin{equation}
\dd \ := \ {\mathcal D}\;- i\,\bar{{\mathcal D}} \ , \quad \dd^* \ = \
{\mathcal D} \;+\;i\,\bar{{\mathcal D}} \ .
\end{equation}
Furthermore, there exists a ${\mathbb Z}$--grading operator $T$ such
that 
\begin{equation}
[T,a]\;=\;0\;,\ \quad [T,\dd ] \;=\; \dd\;, \ \quad [T,\dd^*]\;=\;-\,\dd^*\; ;
\end{equation}
see Remarks (b) and (c) of subsection 1). Let ${\mathcal H}_0
\subseteq {\mathcal H}$ be an arbitrary (e.g.\ minimal, non-zero) subspace of
${\mathcal H}$ {\it invariant} under the $^*$-algebra generated by $\{
{\mathcal A}, {\mathcal D}, \gamma, \bar{{\mathcal D}}, \bar{\gamma}, T\,
\}\,$. One can normalize $T$ such that \ ${\rm
  spec}\,T \subseteq \,{\mathbb Z}$, hence ${\mathcal H}_0$ becomes a
    ${\mathbb Z}$--graded complex:
\begin{equation}
{\mathcal H}_0 \ = \ \displaystyle\bigoplus_{p\,\in\,{\mathbb
    Z}} \ {\mathcal H}_0^p \ ,
\end{equation}
where ${\mathcal H}_0^p$ is the eigenspace of $T$ corresponding 
to the eigenvalue $p \in {\mathbb Z}$. Furthermore, ${\mathcal H}_0^p$
is invariant under ${\mathcal A}$, and (5.47) implies that
\begin{equation}
\dd\;:\;{\mathcal H}_0^p\;\lra\;{\mathcal H}_0^{p+1}\; , \ \quad 
\dd^*\;:\;{\mathcal H}_0^p\;\lra\;{\mathcal H}_0^{p-1}  \ .
\end{equation}
We say that ${\mathcal H}_0^p$ is the subspace of {\it ``vector
  $p$--forms''} and define the $p^{\rm th}$ cohomology space by
\begin{equation}
H_{\dd}^p\ := \ {\rm ker}\,\left( \dd\bigm|_{{\mathcal H}_0^p}\right)
\Big/\, {\rm im}\, \left( \dd\bigm|_{{\mathcal H}_0^{p-1}}\right) \ .
\end{equation}
A {\it harmonic} vector form $\psi \in {\mathcal H}_0$ is one that
satisfies
\begin{equation}
\dd\,\psi \ = \ \dd^*\,\psi \ = \ 0 \ .
\end{equation}
Since $4H = \dd \dd^* + \dd^*\dd$, and $\dd^*$ is the adjoint of $\dd$ 
on ${\mathcal H}$, we conclude that
\begin{equation}
\psi \ {\rm is \ harmonic} \quad \Longleftrightarrow \quad H\,\psi \ = 0 \ .
\end{equation}
Let ${\mathcal H}_h^p \subset{\mathcal H}_0^p$ denote the subspace of 
harmonic vector $p$--forms. Then the usual arguments show that
\begin{equation}
{\mathcal H}_0^p \ = \ {\mathcal H}_h^p \ \oplus \ \dd\,{\mathcal
  H}_0^{p-1} \ \oplus \ \dd^*\, {\mathcal H}_0^{p+1}
\end{equation}
(Hodge decomposition) and
\begin{equation}
H_{\dd}^p \ \cong \ {\mathcal H}_h^p  
\end{equation}
as vector spaces. By (5.52),
\begin{equation}
\left( {\rm ker} \,H\right)^\bot \ = \ \dd\,{\mathcal H}_0 \ \oplus \
\dd^*\,{\mathcal H}_0 \ ,
\end{equation}
and it follows easily that the cohomology of $\dd$ is {\it trivial} on
$({\rm ker}\,H)^\bot$ . In particular, if supersymmetry is spontaneously 
broken, in the sense [15] that \ ${\rm ker}\, H = \{0\}$, then $H_{\dd}^p =
\{0\}$, for all $p$. But as the example in Sect.~{\bf 4.1}, below (4.50),
shows, we should absolutely {\it not} jump to the conclusion that the
data $({\mathcal A}, {\mathcal H}_0, {\mathcal D}, \gamma,
\bar{{\mathcal D}}, \bar{\gamma})$ describe a non-commutative space
with ``trivial homology''. We will expand on this issue below. 
\smallskip

Note that from (5.53--55) we can conclude that
\[
\chi_0 \ := \ {\rm tr}_{{\mathcal H}_0} \left(\left( - 1\right)^T\;
  e^{-\beta H}\right) \ = \ \sum_p (-1)^p \,B_p \ ,
\]
where $B_p={\rm dim}{\mathcal H}_h^p ={\rm dim} H_{\dd}^p$ \ is the
$p^{\rm th}$ Betti number. The absolute convergence of the sum on the
r.s.\ follows from the assumption that $e^{-\beta H}$ is trace
class. Likewise, one can derive a formula for $\tau_0$. 
\smallskip

Let us next examine the cohomology of graded commutation by $\dd$ on
the {\it algebra} $\Omega^{\,^\bullet}_\dd ({\mathcal A})$ of
differential forms. This task is indispensable in view of the above
remark that if supersymmetry is spontaneously broken in the sense
that \ ${\rm ker}H=\{0\}$ then the cohomology spaces $H_{\dd}^p$ are all
trivial. By construction, $\Omega^{\,^\bullet}_\dd ({\mathcal A})$ is
a unital, graded, differential $^\natural$-algebra -- see subsection 2),
(5.35) and (5.36) -- with a faithful $^\natural$-representation $\pi$ on
${\mathcal H}$. In the following, we omit the symbol $\pi$. For
$\alpha \in \Omega^{\,^\bullet}_\dd ({\mathcal A})$, we define
\begin{equation}
\delta\,\alpha \ := \ \left[\, \dd, \alpha \,\right]_g \ ,
\end{equation}
where $[\cdot, \cdot]_g$ is the ${\mathbb Z}_2$--graded commutator,
and 
\begin{equation}
\tau\,\alpha \ := \ \left[\, T, \alpha\,\right]; \ \quad \tau\,\alpha \ = \
n\,\alpha \ \Longleftrightarrow \ \alpha \in \Omega_\dd^n ({\mathcal A}) \
.
\end{equation}
We define cohomology spaces by setting
\begin{equation}
H_\dd^n ({\mathcal A}) \ := \ {\rm ker} \left(
  \delta\bigm|_{\Omega_\dd^n ({\mathcal A})}\right) \Big/\;{\rm im}
\left(\delta \bigm|_{\Omega_\dd^{n-1}({\mathcal A})}\right) \ .
\end{equation}
Thanks to the graded Leibniz rule obeyed by $\delta$, 
\begin{equation}
H_\dd^{\,^\bullet} ({\mathcal A}) \ := \ {\rm ker} \left(
  \delta\bigm|_{\Omega^{\,^\bullet}_\dd ({\mathcal A})}\right)
\Big/\;{\rm im} \left( \delta\bigm|_{\Omega^{\,^\bullet}_\dd ({\mathcal
      A})}\right) 
\end{equation}
is a unital, graded $^\natural$-algebra. 

Let us suppose there is a vector $\varphi_0 \in {\mathcal H}_0$ which
is {\it cyclic} and {\it separating} for $\Omega^{\,^\bullet}_\dd
({\mathcal A})$ and which is closed, i.e., $\dd \varphi_0 = 0$. Then
(under some hypothesis of {\it ``elliptic regularity''}) one finds that
\begin{equation}
H_\dd^p ({\mathcal A}) \ \cong \ H_\dd^p \ , \ {\rm for \ all} \ p \ .
\end{equation}
The situation described here is the one encountered in the de
Rham-Hodge theory of classical, smooth, compact manifolds. But if
supersymmetry is spontaneously broken (i.e., if 
$H$ is strictly positive) then a
vector $\varphi_0$ with the properties assumed above does {\it not}
exist.

{}From the point of view of the theory of $^*$-algebras and of quantum
theory, the formalism developed so far has a drawback: The algebra 
$\Omega^{\,^\bullet}_\dd ({\mathcal A})$ is {\it not} a $^*$-algebra,
because $\dd \neq \dd^*$. Given $a\in {\mathcal A}$, we define
\begin{equation}
\delta^*\,a \ := \ \left[\, a, \dd^*\,\right] \ .
\end{equation}
This gives rise to a graded $^\natural$-algebra (the algebra of {\it
  ``poly-vector fields''})
\[
\Omega^{\,^\bullet}_{\dd^*} ({\mathcal A}) \ = \ \left(
  \Omega^{\,^\bullet}_\dd ({\mathcal A})\right)^* \ 
\]
with a graded differential $\delta^*$ given by 
\begin{equation}
\delta^*\,\alpha \ := \ \left[ \alpha, \dd^*\right]_g \ 
\end{equation}
for all $\alpha \in \Omega_{\dd^*}^{\,^\bullet} ({\mathcal A})$. We
define $\Phi_\dd^{\,^\bullet} ({\mathcal A})$, the {\it ``field
  algebra''}, to be the smallest $^*$-algebra of (generally unbounded)
operators generated by $\Omega_{\dd}^{\,^\bullet} ({\mathcal A})$
and $\Omega_{\dd^*}^{\,^\bullet} ({\mathcal A})$ which is closed under the
action of $\delta$ and $\delta^*$. (Note that the graded commutator of
$\dd$ with the adjoint of a differential form is, in general, an
unbounded operator.) Alternatively, $\Phi_{\dd}^{\,^\bullet}
({\mathcal A})$ can also be defined as the $^*$-algebra generated by
${\mathcal A}$ and by arbitrary multiple graded commutators of
${\mathcal D}$ and $\bar{{\mathcal D}}$ with elements of ${\mathcal
  A}$. $\Phi_\dd^{\,^\bullet} ({\mathcal A})$ 
 is obviously ${\mathbb Z}_2$--graded and, for
$N=\overline{(1,1)}$ spectral data (for which a ${\mathbb Z}$--grading
operator $T$ exists), it is ${\mathbb Z}$--graded:
\begin{eqnarray}
\Phi_{\dd}^{\,^\bullet} ({\mathcal A}) &=&
\displaystyle\bigoplus_{n\,\in\,{\mathbb Z}} \ \Phi_\dd^n
({\mathcal A}) \ , \nonumber\\
\Phi_\dd^n ({\mathcal A}) &=& \left\{ \phi \in
  \Phi_{\dd}^{\,^\bullet} ({\mathcal A}) \bigm| \left[ T,
    \phi\right] \ = \ n\,\phi\right\} \ .
\end{eqnarray}
Operators $\delta$ and $\delta^*$ are defined on
$\Phi_{\dd}^{\,^\bullet} ({\mathcal A})$ as in (5.56), (5.62), and
$\delta : \Phi_\dd^n ({\mathcal A}) \lra \Phi_\dd^{n+1} ({\mathcal
  A})$, $\delta^* : \Phi_\dd^n ({\mathcal A}) \lra \Phi_\dd^{n-1}
({\mathcal A})$, with $\delta^2 = (\delta^*)^2 = 0$. Thus
$\Phi_\dd^{\,^\bullet}({\mathcal A})$ is a graded complex for $\delta$
(and for $\delta^*$). In the situation described above eq.~(5.60), the
study of the complex $(\Phi_\dd^{\,^\bullet} ({\mathcal A}),
\delta^\#)$, $\delta^\# = \delta$ or $\delta^*$, does not yield any
interesting results beyond those of de Rham-Hodge theory. In
general, this complex is not very well understood. It may be useful to
study it in connection with notions of ``diffeomorphisms of
non-commutative spaces'' and with deformation theory (see Sect.~{\bf
  5.3}). 

\bigskip 

There is a theory dual to the cohomology theory for the complexes
$(\Omega_\dd^{\,^\bullet}({\mathcal A}), \delta)$ and 
$(\Phi_\dd^{\,^\bullet} ({\mathcal A}), \delta^\#)$, see [5,52,55]. 
It involves the notions of {\it currents}, which are operators
analogous to currents in de Rham theory. A current $C$ is an arbitrary
(densely defined, closed, \dots) operator on ${\mathcal H}$ commuting
with all elements of the algebra ${\mathcal A}$ and such that
\begin{equation}
\left\{ \tilde{\gamma}, C\right\} = 0 \quad (C \ {\rm is \ {\it odd}}), 
\quad\ {\rm  or} \ 
\quad\left[ \tilde{\gamma}, C\right] = 0 \quad (C \ {\rm is \ {\it
    even}}) \ .
\end{equation}
We say that $C$ is {\it closed} iff
\begin{equation}
\left[ \dd, C\right]_g \ = \ 0 \ .
\end{equation}
Obviously, $C=\id$ (the unit of ${\mathcal A}$) is a closed, even
current, while $C=\tilde{\gamma}$ is an even current which is {\it
  not} closed. Note that {\it closed, even} currents commute with 
$\Omega_\dd^{\,^\bullet} ({\mathcal A})$, while {\it
  closed, odd} currents {\it graded-commute} with
$\Omega_{\dd}^{\,^\bullet} ({\mathcal A})$. 
Given a current $C$, we would like to study functionals ({\it ``signed
  weights''})
\[
{\rm Tr} (\tilde{\gamma}\, C\,(\,\cdot\,)\,) \ ,
\]
where ${\rm Tr}$ is some trace to be specified more precisely. For this
purpose, one can define ``regularized'', multi-linear functionals
\begin{eqnarray}
&& \rho_C^\beta \left( \alpha_1 \left( \tau_1\right), \ldots, \alpha_n
  \left( \tau_n\right)\right) \ := \nonumber \\
&&\qquad {\rm tr}_{{\mathcal H}} \left(
  \tilde{\gamma}\,C\,e^{-(\beta\,-\tau_n + \tau_1)H} \alpha_1\,
  e^{(\tau_1 - \tau_2)H} \alpha_2 \ldots e^{(\tau_{n-1} -\tau_n)H}
  \alpha_n\right) \ ,
\end{eqnarray}
where $0 \leq \tau_1 \leq \tau_2 \leq \ldots \leq \tau_n < \beta\,$, \
$\alpha_1,\ldots, \alpha_n\, \in \Omega_\dd^{\,^\bullet} ({\mathcal
  A})$, (or in $\Phi_\dd^{\,^\bullet} ({\mathcal A})$), $n=0,1,2,\ldots$
. One may then attempt to construct a weight by considering the
limiting functional
\[
\mathop{\bint}\limits_C \ (\,\cdot\,) \ := \ \displaystyle\mathop{{\rm
    Res}}_{\beta\,\to\,0} \ \rho_C^\beta\,(\,\cdot\,) \ ,
\]
where $\displaystyle\mathop{\rm Res}_{\beta \,\to\,0}$ is a
prescription for choosing a residue of $\rho_C^\beta (\cdot)$ when
$\beta \searrow 0$, e.g., $\displaystyle\mathop{\rm
  Lim}_{\beta\,\to\,0}\ Z_\beta^{-1} \rho_C^\beta (\cdot)$, for some
function $Z_\beta$. 

Note that the ordinary integral can be written as 
$\int\!\!\!\!\!- (\,\cdot\,) =
\mathop{\int\!\!\!\!\!-}\limits_{\tilde{\gamma}} (\,\cdot\,)$, with
$Z_\beta = {\rm tr}_{{\mathcal H}} \ (e^{-\beta H})$. The functionals
in (5.66) and their limits as $\beta \searrow 0$ are building blocks
for Hochschild- and cyclic cocycles, see [5,52]. 

One easily verifies that
if $C$ is odd then (5.66) vanishes whenever the form $\alpha_1 \cdots 
\alpha_n$ is even; while if $C$ is even (5.66) vanishes whenever $\alpha_1
\cdots \alpha_n$ is odd.

If $C$ is {\it closed} then, for arbitrary $\alpha_1$ and $\alpha_2$,
\[
\rho_C^\beta \left( \delta\, \alpha_1 (\tau_1), \, \alpha_2
  (\tau_2)\right) \ = \ (-1)^{{\rm deg}\,\alpha_1+1} \ \rho_C^\beta
\left( \alpha_1 (\tau_1), \; \delta\,\alpha_2 (\tau_2)\right) \ ,
\]
(and similarly with $C$ replaced by $C^*$ and $\delta$  by $\delta^*$, 
respectively); in particular, we conclude that 
\[
\rho_C^\beta \left( \delta\,\alpha (\tau)\right) \ = \ \rho_C^\beta
(\delta\,\alpha, \id) \ = \ \rho_C^\beta (\alpha, \delta\,\id) \ = \ 0\ .
\]
Thus, weights corresponding to {\it closed} currents {\it vanish} on
{\it exact} forms. (If, in addition, $C$ is self-adjoint similar
identities hold when $\delta$ is replaced by $\delta^\#=\delta$
{\it or} $=\delta^*$.)
\smallskip

If $C$ is a closed current commuting with $e^{itH}, \ t \in {\mathbb
  R}$, then $\rho_C^\beta$ satisfies the {\it graded KMS condition}
\[
\rho_C^\beta \left( \alpha_1, \alpha_2 (\tau)\right) \ = \ (-1)^{{\rm
    deg}\alpha_1\cdot {\rm deg} \alpha_2}\; \rho_C^\beta \left( \alpha_2,
  \alpha_1 (\beta-\tau)\right) \ .
\]
{\it Formally}, this identity continues to hold for the weight
$\mathop{\int\!\!\!\!\!-}\limits_C (\cdot)$ (with $\tau = 0,\, \beta\searrow
0)$, {\it even} if $C$ does not commute with $e^{itH}, \ t \in
{\mathbb R}$. It should be regarded as a characteristic property of
weights corresponding to closed currents.
\smallskip

Ultimately, one should probably explore the topology of
non-commutative (phase) spaces $({\mathcal A}, \alpha_\tau)$, where
$\alpha_\tau, \tau \in {\mathbb R}$, is a $^*$-automorphism group of
${\mathcal A}$, by studying the theory of {\it ``superselection
  sectors''} [58,59] of $({\mathcal A}, \alpha_\tau)$, i.e., of inequivalent 
irreducible representations of ${\mathcal A}$ with the property that 
$\alpha_\tau$ is implemented by a unitary group $e^{i\tau H}$, 
with $H \geq 0$. 

\vspace{.5cm}
\noindent 8) \ub{Central extensions of supersymmetry, and equivariance}
\medskip

\noindent We consider spectral data 
$({\mathcal A},{\mathcal H},{\mathcal D},
\gamma,\bar{{\mathcal D}},\bar{\gamma})$ 
with all the properties specified in Definition A of subsection 1), 
except that, in point (3) (i), we {\it only} assume that
\begin{equation}
{\rm (3) \ (i')} \phantom{ZeichnungZeichnung} \{ {\mathcal D},
\bar{{\mathcal D}} \} \ = \ 0 \
. \phantom{ZeichnungZeichnungZeichnung}
\end{equation}
We define three operators
\begin{eqnarray}
&& H \ = \ \frac 1 2 \,\left( {\mathcal D}^2 + \bar{{\mathcal D}}^2
\right)\;, \ 
P\ \equiv \ P_1 \ := \ \frac 1 2 \,\left( {\mathcal D}^2 -
  \bar{{\mathcal D}}^2 \right) \ , \nonumber \\
&& \qquad P_2 \ := \ \frac i 2 \,\tilde{\gamma} \,\left( {\mathcal
    D}\bar{{\mathcal D}} - \bar{{\mathcal D}} {\mathcal D}\right) \ .
\end{eqnarray}
If $\tau_0 \;=\; {1 \ 0 \choose 0 \ 1}$, \ 
$\tau_1\;=\; {0 \ \tilde{\gamma} \choose \tilde{\gamma} \ 0}$, \ 
$\tau_2\;=\; {0 \ -i\tilde{\gamma} \choose i\tilde{\gamma} \ \ 0}$, \
$\tau_3\;=\; {1 \ \  0 \choose 0 \ -1}$, \
$\tilde{\gamma} \;=\; \gamma \bar{\gamma}$,
denote the ``graded'' Pauli matrices, and
\[
\utD\ := \ {{\mathcal D} \choose \bar{{\mathcal D}} }
\]
then\[
H\ = \ \frac 1 2 \ \utD^* \tau_0 \utD\,,\quad  P_1 \;=\; \frac 1 2 \ \utD^*
\tau_3 \utD\,, \quad P_2\;=\;\frac 1 2 \ \utD^* \tau_2 \utD \ .
\]
The operator $\frac{\tilde{\gamma}}{2}\,{\displaystyle \utD^* \tau_1 \utD}$ vanishes 
by (5.67). Note that, formally, $H, P_1$ and $P_2$ are {\it commuting,
  self-adjoint} operators on ${\mathcal H}$ with the property that, 
for every unit vector $\underbar{n} \in {\mathbb R}^3$, 
\begin{equation}
H + \underbar{n}\; \underbar{P} \ = \ \frac 1 2  \ \utD^* \left( \id +
  \underline{n} \cdot \underline{\tau}\right) \utD \;\geq\; 0 \ .
\end{equation}
The physicists might want to call this the {\it ``relativistic
  spectrum condition''}. Moreover, formally, ${\mathcal D}$ and
$\bar{{\mathcal D}}$ commute  with $H,P$ and $P_2$. (We do not enter a
discussion of the functional analysis necessary to make these formal
calculations rigorous facts.) Note that $H$ and $P$ are central elements of the graded Lie
algebra spanned by ${\mathcal D}, \bar{{\mathcal D}}, H$ and $P$. 

Defining
\begin{equation}
\dd \ := \ {\mathcal D}\;-\; i\,\bar{{\mathcal D}} \ , \quad 
\dd^*\ = \ {\mathcal D}\;+\; i\,\bar{{\mathcal D}} \ ,
\end{equation}
we find that
\begin{eqnarray}
&&\phantom{ZeichnungZeichnung} \left\{ \dd, \dd\right\} \ = \ 
\left\{ \dd^*, \dd^* \right\} \ = \ 4P \ ,
\phantom{ZeichnungZeichnungZeichnung} \nonumber \\
&&\!\!\!\!\!\!\!\!\!\!{\rm and} \\
&&\phantom{ZeichnungZeichnungZeichnung\ \ } \left\{ \dd, \dd^*\right\} \ =
  \ 4 H \ . \nonumber
\end{eqnarray}
If $P\neq 0$ we say that the relations (5.71) or (5.67) describe {\it
  centrally extended} $N=(1,1)$ supersymmetry, [57]. 
\smallskip

Centrally
extended $N=(1,1)$ supersymmetry is well known to mathematicians: Let
$(M,g)$ be a Riemannian manifold, and let $X$ be a Killing vector
field. If we define
\begin{equation}
\dd \ = \ \d + i\,\lambda\,X\rightharpoonup \ = \ a^*\;\circ\;\nabla 
+ i\,\lambda\,a (\xi) \ ,
\end{equation}
where $\xi$ is the one-form corresponding to $X$ by $\xi (Y) = g
(X,Y)$, and $a^*$ and $a$ are as in Section~{\bf 4}, then
\begin{equation}
\{ \dd, \dd\} \ = \ 
\{ \dd^*, \dd^*\} \ = \
2i\,\lambda\{ \d , X\rightharpoonup \} \ \equiv \ 2i\,\lambda L_X \ ,
\end{equation}
where $L_X$ is the Lie derivative in the direction of $X$. Thus the
operator $P$ in (5.71) plays the r\^ole of a Lie derivative in the
direction of some Killing vector field and thus generates an action of 
$S^1$ (or ${\mathbb R}$) on the non-commutative space described by 
$({\mathcal A},{\mathcal H},{\mathcal D},\gamma,\bar{{\mathcal
    D}},\bar{\gamma})$ .

Centrally extended supersymmetry always appears in {\it quantum field
theory}, where $H$ has an interpretation as Hamiltonian 
(generator of time translations), and $P$ is the 
momentum operator (generator of space translations, 
if space is one-dimensional). 
\smallskip

If we continue to assume that ${\rm exp}(-\varepsilon H)$ is trace
class, for arbitrary $\varepsilon >0$, then the spectrum of $P$ is
discrete. If 0 is in the spectrum of $H$ then it is also in the
spectrum of $P$, because of the ``relativistic spectrum condition''
(5.69), and the subspace
\begin{equation}
{\mathcal H}_0 \ = \ \left\{ \psi\;\in\;{\mathcal H} \bigm| P\,\psi \
  = 0 \ \right\}
\end{equation}
is non-empty. We may then restrict the operators $H, {\mathcal D},
\gamma, \bar{{\mathcal D}}$ and $\bar{\gamma}$ to ${\mathcal H}_0$,
where they generate a standard $N=(1,1)$ supersymmetry algebra of the
type considered in previous sections.

Assuming for simplicity  that $P$ generates an
$S^1$--action and that $e^{i\theta P}, \ \theta \in [0,2\pi)$, defines
a $^*$-automorphism group of ${\mathcal A}$ (i.e.\ $e^{i\theta P}\, a\,
  e^{- i\theta P} \,\in\, {\mathcal A}$, for all \
  $a\in {\mathcal A}\,$), we can define the fixed-point
subalgebra
\begin{equation}
{\mathcal A}_0 \ = \ \Bigl\{\, \frac{1}{2\pi}\,\int\limits_0^{2\pi}
  d\theta \;e^{i\theta P}\, a\,e^{-i \theta P}\ \Bigg|\  a\,\in\,{\mathcal
    A}\,\Bigr\} \ ,
\end{equation}
and the data $({\mathcal A}_0, {\mathcal H}_0, {\mathcal D}_0,
\gamma_0, \bar{{\mathcal D}}_0, \bar{\gamma}_0)$ ,
where $\od_0 = \od\big|_{{\mathcal H}_0} \ , \ \og_0 = \og
\big|_{{\mathcal H}_0}$ , are $N=(1,1)$
spectral data in the sense of subsection 1).
\smallskip

The interesting topological invariants, in the present context, are
\begin{equation}
\chi \ = \ \tr_{{\mathcal H}_0} \left( \tilde{\gamma}_0 \;e^{-\beta
    H_0}\right) \ = \ \tr_{{\mathcal H}} \left( \tilde{\gamma}\;
  e^{-\beta H}\;e^{i\theta P}\right) \ ;
\end{equation}
(the r.s.\ is easily seen to be {\it independent} of $\theta$) and
\begin{equation}
\tau(\theta) \ := \ \tr_{{\mathcal H}} \left( *\,e^{-\beta
    H}\;e^{i\theta P}\right) \ ,
\end{equation}
where $*$ is the Hodge operator introduced in Remark (d) of subsection
1); one easily checks that $\tau (M; \theta)$ is independent of
$\beta$. For classical Riemannian manifolds, one can derive Lefschetz
fixed point formulae for $\chi$ and $\tau (\theta)$ --- as well as for
the $\hat{A}$ genus of $N=1$ spectral data; see [56,22]. The easiest
example is 
\[
\chi (M) \ = \ \sum_i \chi (M_i) \ ,
\]
where $M_1, M_2, \ldots$ are the connected components of the fixed
point set of the Killing vector field $X$ (Lefschetz fixed point
theorem). 

Of course, for the data $({\mathcal A}_0, {\mathcal H}_0, {\mathcal
  D}_0, \gamma_0, \bar{{\mathcal D}}_0, \bar{\gamma}_0)$, all the
results of subsection 7) can be carried over. Here we are entering the
realm of $S^1$--{\it equivariant cohomology}, but we shall not develop
this theme here, beyond saying that the $S^1$--equivariant cohomology
is determined by $H_{\dd_0}^{\,^\bullet} ({\mathcal A}_0)$; see
Sect.~{\bf 5.3}.  An example that is important in the study of two-dimensional
supersymmetric $\sigma$--models is to choose as an algebra ${\mathcal
  A}$ ``something like'' $C(M^{S^1})$, where $M^{S^1}$ is the loop space
over some compact manifold $M$ (interpreted as the {\it target space}
of the $\sigma$--model), and ${\mathcal H}$ is the Hilbert space of
physical state vectors of the $\sigma$--model. The operator $P$ is
chosen to represent the generator of rigid rotations, $\varphi \mapsto
\varphi + \theta$, \ $\theta \in [0, 2\pi)$, of loops in $M^{S^1}$ on
${\mathcal H}$. Considering a $\sigma$--model exhibiting ``unbroken''
$N=(1,1)$ supersymmetry [15], one concludes, {\it formally}, that the
de Rham-Hodge theory for $({\mathcal A}_0, {\mathcal H}_0, {\mathcal
  D}_0, \gamma_0, \bar{{\mathcal D}}_0, \bar{\gamma}_0)$ yields the de
Rham cohomology of $M$. Upon closer examination of the situation, one
finds that the natural $^*$-algebra ${\mathcal A}$ provided by a
quantized supersymmetric $\sigma$--model is really an ``algebra of
functions'' on {\it quantum phase space} over some {\it deformation}
of $M^{S^1}$. The deformation of target space may be ``invisible'' at
the level of de Rham theory (although the algebraic structure of \
$H_{\dd_0}^{\,^\bullet} ({\mathcal A}_0)\;=\;{\rm ker} \left(
  \delta\big|_{\Omega_{\dd_0}^{\,^\bullet} ({\mathcal A}_0)}\right)
  \Big/\; {\rm im} \left( \delta\big|_{\Omega_{\dd_0}^{\,^\bullet} ({\mathcal
     A}_0)}\right) $ \ is generally not that of a graded-commutative 
  algebra); but when one attempts to reconstruct the
 Riemannian geometry of target space form $({\mathcal A}_0,
    {\mathcal H}_0, {\mathcal D}_0, \gamma_0, \bar{{\mathcal D}}_0,
    \bar{\gamma}_0)$ 
one may find that it is surprisingly different form that of $M$. The
example of the supersymmetric WZW model (where $M$ is a semi-simple, 
compact Lie group) is instructive; see [24] and Section~{\bf 7}. 

\vspace{.5cm}
\noindent 9) \ub{$N=(n,n)$ supersymmetry, and supersymmetry breaking}
\medskip

\noindent In this section we describe non-commutative generalizations of
complex Hermitian, sym\-plectic, K\"ahler, hypercomplex, and
Hyperk\"ahler geometry in terms of $N=(n,n)$ supersymmetric spectral
data with partially broken supersymmetry, following the ideas of
subsection 8) of Sect.~{\bf 5.1}.

\bigskip

\noindent {\bf Definition D.} \ The data $({\mathcal A}, {\mathcal H},
\{ {\mathcal D}_i \}_{i=1}^n$ , $\gamma$, \ $ \{ \bar{{\mathcal
      D}}_i\}_{i=1}^n$ , $\bar{\gamma})$ are called $N=(n,n)$
 {\it (supersymmetric) spectral data} iff properties (1) and (2) of 
  Definition  A in subsection 1) hold, and \ 
\begin{itemize} 
\item[(3)] $\{ {\mathcal
    D}_i\}_{i=1}^n$ , $\{ \bar{{\mathcal D}}_i\}_{i=1}^n$ \ are
  essentially self-adjoint on a common dense domain in ${\mathcal H}$
  such that 
\begin{itemize}
\item[(i)] $\{ {\mathcal D}_i, {\mathcal D}_j\} \ = \ \{\bar{{\mathcal
      D}}_i, \bar{{\mathcal D}}_j \} = 0\,$, for all $i\neq j\,$, $\{
  {\mathcal D}_i, \bar{{\mathcal D}}_j \} = 0\,$, for all $i,j$;
\item[(ii)] for each $a \in {\mathcal A}\,$, the commutators
  $[{\mathcal D}_i, a]$ and $[\bar{{\mathcal D}}_i, a]\,$,
  $i=1,\ldots,n\,$, extend to bounded operators on ${\mathcal H}\,$;
\item[(iii)] defining $H := \mathop{\sum}_{i=1}^n
  ({\mathcal D}_i^2 + \bar{{\mathcal D}}_i^2 )\,$, the operator ${\rm
    exp} (- \varepsilon H)$ is required to be trace class, for
  arbitrary $\varepsilon > 0\,$;
\end{itemize}
\item[(4)] $\gamma$ and $\bar{\gamma}$ are ${\mathbb Z}_2$--gradings
  on ${\mathcal H}$ such that 
\begin{itemize}
\item[(i)] $[ \gamma, a] = [\bar{\gamma}, a] = 0\,$, for all $a \in
  {\mathcal A}\,$;
\item[(ii)] $\{ \gamma, {\mathcal D}_i\} = [ \bar{\gamma}, {\mathcal
    D}_i] = 0\,$, $\{\bar{\gamma}, \bar{{\mathcal D}}_i \} = 
  [\bar{\gamma}, {\mathcal D}_i] = 0\,$,
for all $i$.
\end{itemize}
\end{itemize}
\medskip

\noindent The operators 
\begin{equation}
{\mathcal D}_i\,,\ \bar{{\mathcal   D}}_i\,,\ 
L_i \ := \ {\mathcal D}_i^2\,, \ \bar{L}_i \ := \ \bar{{\mathcal
    D}}_i^2 \,,\quad i=1, \ldots, n\,, 
\end{equation}
form a graded Lie algebra with $L_i\,,\bar{L}_i$ as central
elements. The latter operators are 
positive and commute with $H$; thus they have discrete spectrum, by
Definition D, (3 iii). 

On each eigenspace of $\{ L_1,\ldots, L_n,
\bar{L}_1,\ldots, \bar{L}_n \}$, the generalized Dirac operators
${\mathcal D}_1, \ldots,\allowbreak {\mathcal D}_n, \bar{{\mathcal D}}_1, \ldots,
\bar{{\mathcal D}}_n$ form a {\it finite-dimensional} representation
of a Clifford algebra in $m \leq 2n$
dimensions. The automorphism group of the graded Lie algebra generated
by ${\mathcal D}_1, \ldots, {\mathcal D}_n,\allowbreak \bar{{\mathcal D}}_1,
\ldots, \bar{{\mathcal D}}_n,\allowbreak L_1, \ldots, L_n$ , and $\bar{L}_1,
\ldots, \bar{L}_n$ is thus unitarily implemented on ${\mathcal H}$. If
this representation commutes with ${\mathcal A}$ we say that the
spectral data 
$\left( {\mathcal A}, {\mathcal H}, \{ {\mathcal D}_i\}_{i=1}^n\; ,
  \; \gamma , \; \{ \bar{{\mathcal D}}_i\}\; , \bar{\gamma}\right)$ 
are $N=\overline{(n,n)}$ supersymmetric. The general theory of
$N=(n,n)$ and $N=\overline{(n,n)}$ supersymmetric data can now be
developed by combining subsection 8) of Sect.~{\bf 5.1} with
subsections 8) and 7) of Sect.~{\bf 5.2}, above. Apart from a few
details, there is nothing interesting to invent or to check.

We do, however, discover {\it new} geometric structure if we study
various ways of {\it breaking supersymmetry}, see also [18]. In order to
stay as close to notions in classical geometry as possible, it is
useful to reformulate $N=(n,n)$ spectral data in an alternative
way. Since we shall not aim at full generality here, 
we restrict our attention to
$N=(n,n)$ spectral data with a ``charge-conjugation symmetry'':
\begin{equation}
L_i \ = \bar{L}_i \ , \quad i = 1, \ldots, n \ .
\end{equation}
(In the general case, we pass to the subspace ${\mathcal H}_0$ of
${\mathcal H}$ on which (5.79) holds; see subsection 8) above.) 
Thanks to (5.79) we can define $n$ nilpotent differentials
\begin{equation}
\dd_j \ := \ {\mathcal D}_j\;-\;i\,\bar{{\mathcal D}}_j\,,\quad 
j=1, \ldots, n\,,
\end{equation}
with adjoints $\dd_j^*= {\mathcal D}_j\;+\;i\,\bar{{\mathcal D}}_j$. 
We introduce a ${\mathbb Z}_2$--grading $\tilde{\gamma}$ and a Hodge $*$
operator by 
\begin{equation}
\tilde{\gamma} \ := \ \gamma\;\bar{\gamma} \ , \quad * \ = \ \gamma \ .
\end{equation}
Then
\begin{equation}
\dd_j \ = \ (\dd_j^*)^2 = 0\;,\quad \ \{ \tilde{\gamma}, \dd_j\} = 0\; ,\quad \
*\;\dd_j\;* \;=\; -\,\dd_j^* \ ,
\end{equation}
for all $j=1, \ldots, n$.

If the spectral data have $N=\overline{(n,n)}$ supersymmetry, there is
also a ${\mathbb Z}$--grading operator $T$ such that for $j=1,\ldots,n$ 
\begin{equation}
[ T, \dd_j]\;=\; \dd_j\;, \quad \ [T, \dd_j^*]\;=\;-\,\dd_j^* \ .
\end{equation}
The reformulation (5.79--83) makes it clear that
there are {\it two} types of automorphisms of $N=(n,n)$ spectral data,
which we call {\it ``horizontal''} and {\it ``vertical''}. Let $G$ be
the subgroup of $^*$-automorphisms of the graded Lie algebra generated
by $\{ {\mathcal D}_i, \bar{{\mathcal D}}_i, L_i, \bar{L}_i\}_{i=1}^n$
\ that is implemented on ${\mathcal H}$ by a unitary representation
$\pi$ with the property that $\pi (G)$ commutes with ${\mathcal
  A}$. Let ${\mathcal G}$ denote the Lie algebra of $G$. An element $J
\in d \pi ({\mathcal G})$ is said to be the generator of a {\it
  horizontal symmetry} iff
\begin{equation}
[ J,\; \dd_i] \ = \ J_i^{\ j} \ \dd_j \ ,
\end{equation}
for some complex numbers $J_i^{\ j}$ (we are using the summation
convention). A similar equation for $\dd_i^*$ follows by taking the
adjoint of (5.84). An element $\Omega \in d \pi ({\mathcal
  G}^{{\mathbb C}})$ is said to be the generator of a {\it vertical
  symmetry} iff
\begin{equation}
[ \Omega, \; \dd_i^*] \ = \ \Omega_i^{\ j}\; \dd_j \ ,
\end{equation}
for some real (or complex) numbers $\Omega_i^{\ j}$ . Of course, we
also demand that $J$ and $\Omega$ commute with $\tilde{\gamma}$. We
assume that there exists a ${\mathbb Z}$--grading operator $T$. Recall
the (graded) {\it Jacobi identities}:
\begin{eqnarray}
&& [\; [A,B\;],C ] \ + \ [\;[B,C],A\,] \ + \ [\;[ C, A],B\,] \ = \ 0 \ ,
\nonumber\\
&& [\{ A,B\},C] \ - \ \{[B,C],A\} \ + \ \{[C,A],B\} \ = \ 0 \ ,\\
&& [\{ A,B\},C] \ + \ [\{ B,C\}, A] \ + \ [\{ C,A\}, B] \ = \ 0 \
. \nonumber 
\end{eqnarray}
The first identity in (5.86) and eqs.~(5.83,84) imply that
$[T,J]$ commutes with ${\mathcal A}$ and with $\dd_j$, and, since $J$
must be anti-selfadjoint, with $\dd_j^*$, for all $j$. Since $i[T,J]$
belongs to $d\pi ({\mathcal G})$, we conclude that 
\begin{equation}
\left[ T,J \right] \ = \ 0 \ .
\end{equation}
Here we assume that ${\mathcal G}$ acts faithfully on the
differentials and their adjoints. The first identity in (5.86) and
eqs.~(5.83,85) imply that, for some operator $Z \in d \pi
({\mathcal G}^{{\mathbb C}})$ commuting with ${\mathcal A}$ and with
all the $\dd_i^*$, 
\begin{equation}
[ T, \Omega ] \ = \ 2\;\Omega \ + \ Z \ .
\end{equation}
By (5.88) and (5.86),
\begin{equation}
\left[ T, \left[ \Omega, \dd_i\right]\right] = \left[ \Omega, \left[
    T, \dd_i\right]\right] - \left[ \dd_i, \left[ T,
    \Omega\right]\right] = 3 \left[ \Omega, \dd_i\right] + \left[ Z,
  \dd_i\right] \ .
\end{equation}
If $[Z, \dd_i]=0$, for all $i$ then $Z=0$, and it follows that
$\Omega$ has degree 2 and $[\Omega, \dd_i]$ has degree 3.

In the following, we focus our attention on spectral data with the
property that all {\it odd} elements of the graded Lie algebra
generated by $\dd_i, \dd_i^*$ and ${\mathcal G}$ have in fact 
${\mathbb Z}$--degree $\pm 1$. Then it follows from (5.89) and
$Z=0$ that
\begin{equation}
[ \Omega, \dd_i] = 0 \ , \quad i = 1, \ldots, n \ .
\end{equation}
Eqs.~(5.84), (5.87) and eqs.~(5.85), (5.90) and (5.88)
show that, in the context of classical manifold theory, a horizontal
symmetry generator $J$ corresponds to a {\it complex structure}, while
a vertical symmetry operator $\Omega$ corresponds to wedging by a {\it
  symplectic} 2-{\it form}; (eq.~(5.90) says that a symplectic form is
{\it closed}).

It is not hard to elucidate the algebraic structure of differentials
and horizontal and vertical symmetries further by repeated use of
eqs.~(5.83--90); but we refrain from going into further details
here. Instead, we propose to change our point of view: Rather than
starting from $N=(n,n)$ spectral data, we may start from
$N=\overline{(1,1)}$ spectral data, as in subsection 1) of Sect.~{\bf
5.2}, and enrich them by {\it horizontal} and/or {\it vertical}
symmetries.
\smallskip

Thus, let $({\mathcal A}, {\mathcal H}, \dd, \tilde{\gamma}, *)$\ be
some $N=\overline{(1,1)}$ spectral data, and let ${\mathcal G}_h$ be a
Lie algebra of ``horizontal symmetries'' represented on ${\mathcal H}$
by anti-selfadjoint operators  which commute with ${\mathcal A},
\tilde{\gamma}, T$ and $*$. Let $\{ J_1, \ldots, J_{n-1}\}$ be a basis
for ${\mathcal G}_h$ with the property that the operators 
\begin{equation}
\dd_1 \ :=\ \dd, \quad \dd_k \ := \ [J_{k-1}, \dd ], \ k=2,\ldots,n \ , 
\end{equation}
span a ${\mathcal G}_h$--module under the adjoint action. 
The (graded) Jacobi identity (first and second equation in (5.86))
shows that
\begin{equation}
\{ \dd_i, \dd_j\} \ =\ 0\ , \quad {\rm for \ all} \ i \ {\rm and} \ j \
.
\end{equation}
However, the structure described by ${\mathcal G}_h$ and (5.91) does 
{\it not} imply that $\{\dd_i, \dd_j^*\} = 0$ for $i\neq j$, 
as would be the case if the differentials $\dd_i$ were
derived from $N=(n,n)$ spectral data as in eq.~(5.80). 
We may now introduce {\it   Dolbeault differentials}
\begin{equation}
\partial_k \:=\ \dd + i\,\dd_k\,,\quad \ \bar{\partial}_k \:=\ \dd - i\,\dd_k \ ,
\end{equation}
$k=2,\ldots,n$. Assuming that, for all $a \in {\mathcal A}$, the operator
$\{ \partial_k, [\bar{\partial}_k, a]\}$ is bounded, we can introduce a
bi-graded, bi-differential algebra $\Omega_{\partial_k,
  \bar{\partial}_k}^{\,^{\bullet,\bullet}} ({\mathcal A}) \;=\;
  \displaystyle\mathop{\oplus}_{p, q} \; \Omega_{\partial_k,
    \bar{\partial}_k}^{\ p,q} ({\mathcal A})$ \ in the obvious way,
  satisfying
\[
\Omega_\dd^n ({\mathcal A}) \ = \ \displaystyle\bigoplus_{p+q=n}
\ \Omega_{\partial_k,\bar{\partial}_k}^{\ p,q} ({\mathcal A}) \ ;
\]
see [18]. The elements of $\Omega_{\partial_k,
  \bar{\partial}_k}^{\,^{\bullet,\bullet}} ({\mathcal A})$ \ are
called {\it Dolbeault forms}. We can construct an integral 
$\int\!\!\!\!\!- $ and a metric $\langle\!\langle\,\cdot,
\cdot\, \rangle\!\rangle$ on $\Omega_{\partial_k,
  \bar{\partial}_k}^{\,^{\bullet,\bullet}} ({\mathcal
  A})$ (using $H = \frac 1 4 \ (\dd \dd^* + \dd^* \dd)$). Furthermore,
one can define {\it holomorphic vector bundles}, ${\mathcal E}$, as
finitely generated, projective (left-) modules for ${\mathcal A}$
equipped with a connection $\nabla = \nabla^{(1,0)} + \nabla^{(0,1)}$,
where $\nabla^{(p,q)} : {\mathcal E} \;\lra\;\Omega_{\partial_k,
  \bar{\partial}_k}^{\ p,q} ({\mathcal
  A}) \otimes_{{\mathcal A}} {\mathcal E}$ for $p+q=1$, such that
\begin{equation}
\nabla^{(0,1)} \;\circ\;\nabla^{(0,1)} \ = \ 0 \ .
\end{equation}
See [18] for details. Apparently, the structure we are exploring here
resembles the theory of {\it (hyper)complex Hermitian manifolds} in
classical geometry; see [26,27,28]. We therefore say that spectral
data $({\mathcal A}, {\mathcal H}, \dd, \tilde{\gamma}, *, T,
{\mathcal G}_h)$, where ${\mathcal G}_h$ is a Lie algebra of
horizontal symmetries as in (5.91), define a {\it (hyper)complex
  Hermitian non-commutative space}. The special case, where ${\mathcal
  G}_h = {\mathbb R}$, i.e., with only {\it one} complex structure
$J$, characterizes {\it complex non-commutative geometry}; the
special case where ${\mathcal G}_h = {\rm su} (2)$, i.e., with three
complex structures $I, J$ and $K$, is characteristic of standard
hypercomplex geometry (see [26]). Further cases are discussed, for
classical manifolds, in [27,28].
\smallskip

Of course, {\it if} it so happens that 
\begin{equation}
\{ \dd_i, \dd_j^*\} \ = \ 0 \  \quad {\rm for \ all} \ i\neq j \ ,
\end{equation}
then the data $({\mathcal A}, {\mathcal H}, \dd, \tilde{\gamma}, *, T,
{\mathcal G}_h)$ determine $N=(n,n)$ or $N=\overline{(n,n)}$ {\it
  supersymmetric spectral data}, with ${\mathcal D}_i = \dd_i +
\dd_i^*, \bar{{\mathcal D}}_i = i\,(\dd_i - \dd_i^*)$, $\gamma = *$
and $\bar{\gamma} = * \tilde{\gamma}$. Thus, the anti-commutators
\begin{equation}
\{ \dd_i, \dd_j^*\} \ , \quad i\neq j \ ,
\end{equation}
describe an {\it explicit breaking} of $N=(n,n)$ supersymmetry, and,
in geometry, broken supersymmetry is apparently a rather standard
phenomenon.
\smallskip

Besides the algebras
$\Omega_{\partial_k,\bar{\partial}_k}^{\,{\bullet,\bullet}} ({\mathcal
  A})$, we should also consider the bi-graded, differential algebras 
\begin{equation}
\Omega_{\partial_k,\bar{\partial}_k^*}^{\,^{\bullet,\bullet}}
({\mathcal A}) \ = \ \displaystyle\bigoplus_{p,q} \
\Omega_{\partial_k, \bar{\partial}_k^*}^{\ p,q} ({\mathcal A}) \ ,
\end{equation}
which are differential algebras of (generally unbounded) operators for
$\partial_k$ and for $\bar{\partial}_k^*$, but not, usually,
bi-differential algebras, (unless $\{ \partial_k, \bar{\partial}_k^*
\} = 0$). These algebras are defined in an obvious way. The study of
the cohomology of the complex
$(\Omega_{\partial_k,\bar{\partial}_k^*}^{\,{\bullet,\bullet}}
({\mathcal A}), \ \partial_k)$ is important in {\it deformation
  theory}, e.g., for the Kodaira-Spencer theory of deformations of
the complex structure $J_{k-1}$ ; see [60].
\medskip

{\it If}, for some $k \geq 2$ , $\{ \dd, \dd_k^*\} = 0$ and {\it if}\ 
$4L_1 = \dd \dd^* + \dd^*\dd = \dd_k \dd_k^* + \dd_k^* \dd_k = 4 L_k$
\ i.e., if there is an {\it unbroken} $N=(2,2)$ supersymmetry, then
$\Omega_{\partial_k, \bar{\partial}_k^*}^{\,^{\bullet,\bullet}}
({\mathcal A})$ is a bi-differential algebra, and $(\Omega_{\partial_k,
  \bar{\partial}_k^*}^{\,^{\bullet,\bullet}} ({\mathcal A}),
\partial_k, \bar{\partial}_k^* )$ is a {\it bi-complex}. \ {\it Mirror
  symmetry} is a map from spectral data $({\mathcal A}, {\mathcal H},
\partial_k, {\mathcal T}, \bar{\partial}_k, \overline{\mathcal T})$ to data 
\[
\left(\, {\mathcal B},\; {\mathcal H},\; \partial_k'\;:=\; \partial_k,
  \;{\mathcal T}' := {\mathcal T},\; \bar{\partial}_k':= 
\bar{\partial}_k^*, \;  \overline{\mathcal T}':= -\,
\overline{\mathcal T} \,\right)
\]
where ${\mathcal T}$ and $\overline{\mathcal T}$ are the holomorphic 
and anti-holomorphic
${\mathbb Z}$--grading operators, and ${\mathcal B}$ is a second
$^*$-algebra on ${\mathcal H}$ with the same properties as ${\mathcal
  A}$. (One should, perhaps, assume that the phase space algebras
generated by $({\mathcal A}, H)$ and by $({\mathcal B}, H)$, where
$H=L_1=L_k$, {\it coincide}, as is true in $N=(2,2)$ supersymmetric
conformal field theory in two dimensions, see e.g.\ [24].)
\smallskip

Next, we consider $N=\overline{(1,1)}$ spectral data enriched by {\it
  vertical} symmetries. This leads us to a natural notion of {\it
  symplectic} (and ``hyper-symplectic'') {\it non-commutative
  geometry}, [18]. Thus, let $({\mathcal A}, {\mathcal H}, \dd,
\tilde{\gamma}, *)$ be some $N=\overline{(1,1)}$  spectral data, and
let $\Omega_1,\ldots,\Omega_n$ be a basis of {\it vertical
  symmetries}, in the sense that the operators
\begin{equation}
\dd_k \ := \ [ \Omega_k,\;\dd^* ]
\end{equation}
are {\it nilpotent} and
\begin{equation}
[ T, \Omega_k ]\;=\; 2 \Omega_k \ , \quad
[ \dd, \Omega_k ]\;=\;0 \ ,
\end{equation}
for all $k=1,\ldots,n$. It follows from the graded Jacobi identity
(second equation in (5.86)) that 
\begin{equation}
\{ \dd_k,\; \dd^* \} \ = 0 \ ,
\end{equation}
for all $k=1,\ldots,n$. We are however {\it not} claiming that
\begin{equation}
\{ \dd_k,\;\dd\} \ = \ 0 \ ,
\end{equation}
because (5.101) does {\it not} follow from the structure required, so
far, and is not valid in examples. Note that, from (5.98) and its
adjoint, from (5.99) and from the Jacobi identity, it follows that 
\begin{equation}
[\,\Omega_l, \dd_k^*\,]\;=\;[\,\Omega_l, [\,\dd, \Omega_k^*\,]]\;=\;
[[\,\Omega_k^*, \Omega_l\,],\dd\,]
\end{equation}
is an operator of degree 1 which anti-commutes with $\dd$. {\it If}
this operator is {\it nilpotent} then $[\,\Omega_k^*, \Omega_l\,]$ is a
linear combination of $T$ and of {\it horizontal} symmetries, i.e., is
related to complex structures. 

Let us consider the case where $n=1$, setting $\Omega_1 =: \Omega$,
$\dd =: \dd_1$ and $[\,\Omega, \dd^*\,] =: \dd_2$. If $\{\, \dd_1, \dd_2\,\}
\neq 0$ then there {\it cannot} exist a horizontal symmetry relating
$\dd_1$ and $\dd_2$. It follows that {\it either} $i [\,\Omega^*,
\Omega\,]$ is a new horizontal symmetry, {\it or} $[\,\Omega^*, \Omega\,] =
T$ after appropriate normalization. The second alternative describes what 
one might want to call a {\it symplectic non-commutative space}. Apparently,
such a space is characterized by $N=\overline{(1,1)}$ spectral data,
enriched by {\it one} vertical symmetry $\Omega$, i.e., by 
\begin{equation}
\left( {\mathcal A}, {\mathcal H}, \dd_1 \equiv \dd, \tilde{\gamma},
  *, \Omega \right)
\end{equation}
with the properties that
\begin{eqnarray}
&& \dd_2\; :=\; [\,\Omega, \dd^*\,] \quad {\rm is \ {\it nilpotent}}, \ \quad
[\,\dd, \Omega\,]\; =\; 0 \ , \nonumber \\
&& [\,T, \Omega\,]\;=\;2\,\Omega,\quad \ [\,\Omega^*, \Omega\,] \;=\; T \ . 
\end{eqnarray}
Obviously, the operators $\Omega, \Omega^*$ and $T$ determine a
unitary representation of the Lie algebra $sl_2$, and
\[
{\dd_1 \choose \dd_2^*} \quad {\rm and} \quad {\dd_2 \choose \dd_1^*}
\]
form two $sl_2$ doublets under the action of $\Omega^*, \Omega, T$.
\smallskip

{\it If} $\{ \dd_1, \dd_2\}=0$ then, setting ${\mathcal D}_j := \frac
1 2 \ (\dd_j + \dd_j^*)$ , $\bar{{\mathcal D}}_j := \frac i 2 \ (\dd_j
- \dd_j^*)$ , we find that $({\mathcal A}, {\mathcal H}, {\mathcal
  D}_1, {\mathcal D}_2, \gamma=*, \bar{{\mathcal D}}_1, \bar{{\mathcal
    D}}_2, \bar{\gamma} = * \tilde{\gamma})$ are $N=(2,2)$ spectral
data, which give rise to a non-commutative {\it K\"ahler geometry}.

Our analysis suggests to define a {\it mirror map} from a {\it
  symplectic} to a {\it complex-Hermitian} non-commutative space as a
map $m$,
\begin{equation}
m\;:\;\left( {\mathcal A}, {\mathcal H}, \dd_1, \dd_2,
  \tilde{\gamma}, *\right) \;\longmapsto\;
\left( {\mathcal B}, {\mathcal H}, \dd_1' := \dd_1, \dd_2' := \dd_2^*,
  \tilde{\gamma}, * \right) 
\end{equation}
where ${\mathcal B}$ is a second $^*$-algebra on ${\mathcal H}$ with
the same properties as ${\mathcal A}$. The Dolbeault differentials for
${\mathcal B}$ are then given by
\begin{equation}
\partial \ = \ \dd_1 - i \dd_2^*, \ \quad\bar{\partial} = \dd_1 + i \dd_2^*
\ .
\end{equation}
Of course, $m$ cannot preserve the ${\mathbb Z}$--grading $T$. 
\smallskip

Spectral data with several vertical symmetry generators
$\Omega_1,\ldots,\Omega_n$ describe (non-commutative) geometries with
several (at least $n-1$) complex structures corresponding to a {\it
  possibly broken} $N=(n+1, n+1)$ supersymmetry.

In [18] the reader can find additional technical details on complex
Hermitian and symplectic (non-commutative) geometry.
\medskip

In conclusion, we have been able to characterize (non-commutative)
complex-Hermitian and symplectic geometry in terms of spectral data
with {\it broken} $N=(2,2)$ supersymmetry
\[
\{ \dd_1, \dd_2^*\} \neq 0 \quad {\rm or} \quad \{ \dd_1, \dd_2 \} \neq 0 \ ,
\]
respectively. Unbroken $N=(2,2)$ supersymmetry corresponds to
(non-commutative) K\"ahler geometry. In spin$^{\rm c}$ geometry, as
developed in Sect.~{\bf 5.1}, supersymmetry breaking corresponds to
spectral data with Dirac operators $D_1, D_2, \ldots$ such that 
\[
\{ D_1, D_2 \} \ \neq \ 0 \ .
\]
There is some beginning of an understanding under what conditions
supersymmetry can be restored by {\it deforming} the generators $
{\mathcal D}_i, \bar{{\mathcal D}}_i $ \ (or $D_i$, resp.), $i =
2,3,\ldots\,$; see [55]. 

Finally, we note that one can also study {\it asymmetric} $N=(n,m)$ or
$N=\overline{(n,m)}$ \break
supersymmetric spectral data, with $n\neq m$, thus
leaving the realm of real geo\-metry; such data 
appear in superconformal field theory and string theory [29].

\vspace{.5cm}

\subsection{Reparametrization invariance, BRST cohomology, and\\
target space supersymmetry}

\noindent
In this section we study non-commutative spaces  described by $N=n$
or $N=(n,n)$ spectral data which admit some symmetries, called {\it
  reparametrizations}. If a $^*$-algebra ${\mathcal A}$ is interpreted
as the ``algebra of functions'' over some (non-commutative) ``space''
then ``symmetries'' of this space can be described as
$^*$-automorphisms of ${\mathcal A}$. They form a group denoted by
${\rm Aut}({\mathcal A})$. Infinitesimal symmetry transformations are
$^*$-derivations of ${\mathcal A}$ and form a Lie algebra denoted by
${\rm Der}({\mathcal A})$. In classical geometry, where ${\mathcal A}
= C(M)$, with $M$ e.g.\ a smooth, compact 
manifold, ${\rm Der}({\mathcal A})$ consists of (all Lie derivatives
associated with) {\it smooth vector fields}, i.e., $L \in {\rm
  Der}({\mathcal A})$ iff $L = L_X = \{ d, a (\xi)\}$, where $\xi$ is
the 1-form corresponding to a smooth vector field $X$  w.r.t.\ some
Riemannian metric on $M$ (see Sect.~{\bf 4.1}). Thus $[d,L]=0$ for all
$L\in {\rm Der}({\mathcal A})$, and it follows that $L$ is
automatically a derivation of the differential algebra
$\Omega_d^{\,^\bullet}({\mathcal A})$. (It is, however, {\it not}
necessarily a derivation of $\Omega_{d^*}^{\,\,^\bullet} ({\mathcal
  A})$ or of $\Phi_d^{\,^\bullet} ({\mathcal A})$. But this 
holds if $X$ is an isometry, i.e., if $L_X$ commutes with $d^*$.)

In non-commutative geometry, derivations of ${\mathcal A}$ have {\it
  no} reason to commute with the differential $\dd$ of some $N=(1,1)$
supersymmetric
spectral data $({\mathcal A}, {\mathcal H}, \dd, \tilde{\gamma}, 
*)$. In general, they do {\it not} commute with $\dd$ and $\dd^*$ (or
with ${\mathcal D}$ and $\bar{{\mathcal D}}$). One might want to call
  the subgroup of ${\rm Aut}({\mathcal A})$ commuting with $\dd$ the
  {\it diffeomorphism group} of ${\mathcal A}$, ${\rm Diff}_\dd
  ({\mathcal A})$, and the subgroup of ${\rm Aut}({\mathcal A})$
  commuting with $\dd$ {\it and} with $\dd^*$ the group of {\it
    isometries} of ${\mathcal A}$. Our purpose, in this section, is
  not primarily to study diffeomorphisms or isometries of ${\mathcal
    A}$, but certain subalgebras ${\mathcal G} \subseteq {\rm
    Der}({\mathcal A})$, called {\it algebras of infinitesimal
    reparametrizations} of ${\mathcal A}$, with properties described
  presently.

In order not to get lost in generalities, we start from $N=1$
supersymmetric spectral data
\begin{equation}
\left( {\mathcal A}, {\mathcal H}, D, \gamma \right) \ .
\end{equation}
Let ${\mathcal G}$ be some Lie subalgebra of ${\rm Der}({\mathcal
  A})$, with a basis $T_1, \ldots, T_n$. (To be on safe grounds, we
temporarily assume that ${\mathcal G}$ is finite-dimensional, i.e., $n
< \infty$.) 

\bigskip

\noindent{\bf Definition A}. \ ${\mathcal G}$ is a {\it Lie algebra of
  infinitesimal reparametrizations} of $({\mathcal A}, {\mathcal H},
D, \gamma)$ iff 

({i}) \ ${\mathcal G}$ is implemented on ${\mathcal H}$ in a
representation $d\pi$ by anti-selfadjoint operators \break
\phantom{MMMM}commuting with the ${\mathbb Z}_2$--grading $\gamma$; hence
\begin{equation}
L_j \ := \ d\pi (T_j)
\end{equation}
\phantom{MMMM}is an anti-selfadjoint operator commuting with $\gamma$, for all
$j=1,\ldots, n$;

({ii}) \phantom{x}the {\it graded} Lie algebra ${\mathcal G}_{D, \gamma}$, 
    generated by $d\pi ({\mathcal G})$, by $D$ and by arbitrary graded \break
\phantom{MMMM}commutators thereof, satisfies 
\begin{equation}
\left( {\mathcal G}_{D, \gamma} \right)_{\rm even} \ = \
d\,\pi\,({\mathcal G}) 
\end{equation}
\phantom{MMMM}and is finite-dimensional, 
\begin{equation}
{\rm dim}\,\left( {\mathcal G}_{D,\gamma}\right) \ < \ \infty \ .
\end{equation}
\medskip

Let $L_1,\ldots,L_n$, $D_1 := D, D_2,\ldots,D_m$ be a basis of
${\mathcal G}_{D,\gamma}$, where $L_1,\ldots, L_n$ span $({\mathcal
  G}_{D,\gamma})_{\rm even}$ and $D_1,\ldots,D_m$ span $({\mathcal
  G}_{D,\gamma})_{\rm odd}$. Note that the operators $D_1,\ldots,D_m$
are self-adjoint. The structure constants of ${\mathcal G}_{D,\gamma}$ 
\begin{equation}
\left\{ f_{ij}^k \ = \ -\,f_{ji}^k\;, \ g_{ij}^k\;, \ h_{ij}^k \ = \
  h_{ji}^k \right\}
\end{equation}
are real numbers such that
\begin{eqnarray}
&& \left[ L_i, L_j \right] \ = \ f_{ij}^k\;L_k \ ,\nonumber \\
&& \left[ L_i, D_j \right] \ = \ g_{ij}^k\;D_k \ , \\
&&\left\{ D_i, D_j\right\} \ = \ i\,h_{ij}^k\;L_k \ . \nonumber
\end{eqnarray}
The graded Jacobi identities (5.86) yield quadratic relations between
the structure constants.

Our goal, here, is to find the ${\mathcal G}$--invariant ground states
of the operator
\begin{equation}
H \ = \ D^2 \ ,
\end{equation}
i.e., those state vectors $\psi \in {\mathcal H}$ that satisfy
\begin{equation}
D\,\psi \ = 0\;, \ \quad L_j\,\psi \ = \ 0 \ ,
\end{equation}
for all $j=1,\ldots, n$. This problem comes up in string theory [29]
and in $M$(embrane) theory [61]. 
\begin{equation}
\ D_l\,\psi \ = \ 0 \ 
\end{equation}
holds for all $l=1,\ldots,m$, which means that the
${\mathcal G}$--invariant ground states of $H$ are precisely the
${\mathcal G}_{D,\gamma}$--invariant state vectors in ${\mathcal
  H}$. The problem of finding these states can be viewed as a
problem in {\it BRST cohomology}.
\smallskip

Recall that, in Sect.~{\bf 4.2}, the problem to find all ${\mathcal
G}$--invariant states in ${\mathcal H}$ has been formulated as a
problem in BRST cohomology; see eqs.~(4.72), (4.74) and (4.86--89): 
Let $\{ \vartheta^i\}_{i=1}^n$ be a basis of 1-forms dual to
the ``vector fields'' $\{ L_i\}_{i=1}^n$ . Let
\begin{equation}
c^j \ :=\ \vartheta^j \wedge \ , \ \quad b_j \ := \ L_j\rightharpoonup \ .
\end{equation}
Then
\begin{equation}
\left\{ c^i, c^j \right\} \ = \ 
\left\{ b_i, b_j \right\} \ = \ 0, \ \quad
\left\{ c^i, b_j \right\} \ = \ \delta_j^i \ ,
\end{equation}
see (4.71), (4.72). The BRST operator was given by
\begin{equation}
Q_{\rm BRST} \ \equiv \ Q_{d\pi} \ = \ c^j\,L_j\,-\ \frac 1 2 \
f_{ij}^k\;c^i\,c^j\,b_k \ ,
\end{equation}
the ${\mathbb Z}$--grading operator $T$ (``ghost number operator'') by
\begin{equation}
T \ = \ c^j\,b_j \ .
\end{equation}
Then
\begin{equation}
Q_{\rm BRST}^{\ 2 \ } \ = \ 0 \ ,
\end{equation}
and the ${\mathcal G}$--invariant states in ${\mathcal H}$ span \ the
$0^{\rm th}$ cohomology space, $H_{d\pi}^0$, of $Q_{\rm BRST}$.

This theory has a natural extension to a cohomology theory for
{\it graded} Lie algebras with values in a representation. Let $\{
\vartheta^1, \ldots, \vartheta^n, \delta^1,\ldots,\delta^m\}$ be a
basis dual to $\{ L_1, \ldots, L_n, D_1, \ldots, D_m\}$. As above, we
set $c^j = \vartheta^j \wedge$ , $b_j = L_j \rightharpoonup$ , and
define
\begin{equation}
\gamma^j \ = \ \delta^j \otimes_s \ , \ \quad\beta_j \ = 
\ D_j\rightharpoonup \ .
\end{equation}
Then
\begin{equation}
\left[ \gamma^i, \gamma^j\right] \ = \ \left[ \beta_i, \beta_j\right]
\ = \ 0 \ , \quad\ \left[\beta_j, \gamma^i \right] \ = \ \delta_j^i \ ,
\end{equation}
for all $i$ and $j$, and the $\gamma$'s and $\beta$'s commute with the
$c$'s and the $b$'s.

A BRST operator is defined by 
\begin{eqnarray}
Q_{\rm BRST} \ = && c^j\,L_j\;-\;
\frac 1 2 \ f_{ij}^k\;c^i\,c^j\,b_k\nonumber \\
&& -\; g_{ij}^k\,c^i\,\gamma^j\,\beta_k \\
&& + \ \gamma^j\,D_j\;-\;\frac i 2 \
h_{ij}^k\,\gamma^i\,\gamma^j\,b_k\nonumber
\end{eqnarray}
satisfying
\begin{equation}
Q_{\rm BRST}^{\ 2 \ } \ = \ 0 \ ,
\end{equation}
and the ${\mathbb Z}$--grading operator $T$ (``ghost number
operator'') is now given by
\begin{equation}
T \ = \ c^j\,b_j\;+\;\gamma^j\,\beta_j \ .
\end{equation}
Of course, in all these formulas, the summation convention is assumed.

Let ${\mathcal F} \cong  S\bigl(({\mathcal G}_{\rm odd})^*\bigr) \otimes 
\Lambda \bigl(({\mathcal G}_{\rm even})^*\bigr)$
denote the Fock space on which the canonical anti-commutation relations
(5.117) and the canonical commutation relations (5.122) are
represented. This representation on 
${\mathcal F}$ is uniquely characterized by the property that
$T$ defines a positive, self-adjoint operator on ${\mathcal F}$ with 0
as a simple eigenvalue. The corresponding eigenvector, $\varphi_0$, is
called the {\it ``vacuum''}. We define
\begin{equation}
\widetilde{{\mathcal H}} \ := \ {\mathcal H}\;\otimes\;{\mathcal F} \ .
\end{equation}
The space $\widetilde{{\mathcal H}}$ is a ${\mathbb Z}$--graded
complex for $Q_{\rm BRST}$. The $0^{\rm th}$ cohomology space of
$Q_{\rm BRST}$ consists precisely of the ${\mathcal
  G}_{D,\gamma}$--{\it invariant vectors} in ${\mathcal H}$ (tensored
by $\varphi_0$).

Of course, as long as ${\rm dim}({\mathcal G}_{D,\gamma}) < \infty$,
the formalism developed here looks somewhat pompous. But its
usefulness becomes manifest when ${\rm dim}({\mathcal G}_{D,\gamma}) =
\infty$, as is the case e.g.\ in superstring theory~[29].
\smallskip

It is quite clear how this theory can be generalized to
non-commutative geometries described by
$N={n}^\#$ or $N=(n,n)^\#$ supersymmetric spectral data for arbitrary 
$n$, where $n^\#$ denotes $n$ or $\overline{n}$, etc.  We briefly digress
on the example of $N=(1,1)$ (or $N=\overline{(1,1)}$) spectral data,
\[
\left( {\mathcal A}, {\mathcal H}, \dd, \tilde{\gamma}, *\right)
\simeq \left( {\mathcal A}, {\mathcal H}, {\mathcal D}, \gamma,
  \bar{{\mathcal D}}, \bar{\gamma}\right) \ ,
\]           
with ${\mathcal D} = \dd+\dd^*$, $\gamma=*$, $\bar{{\mathcal D}} = i
(\dd-\dd^*)$, $\bar{\gamma} = * \tilde{\gamma}$. Let ${\mathcal G}$ be
a Lie algebra of infinitesimal reparametrizations of $({\mathcal A},
{\mathcal H}, \dd, \tilde{\gamma}, *)$ in the sense of Definition A,
above, but assuming in addition that $L_1,\ldots,L_n$ commute with $\gamma$
and with $\bar{\gamma}$. 
In this situation, a new phenomenon can appear: It may happen that
$d\pi({\mathcal G})$ {\it commutes} with $\dd$. More precisely, let us
assume that, for every $L\in d\pi ({\mathcal G})$, there exists an
element $X\in\Omega_{\dd^*}^1 ({\mathcal A})$, a ``vector field'',
such that
\begin{equation}
L \ = \ \left\{ \dd,\;X\right\} \ ,
\end{equation}
i.e., $L$ is the {\it Lie derivative} associated with $X$. Then
\begin{equation}
\left[ \dd,\,L\right] \ = \ 0 \quad {\rm for \  all} \ L \in d\pi
({\mathcal G}) \ .
\end{equation}
If (5.127) holds for all $L \in d\pi ({\mathcal G})$ we say that
${\mathcal G}$ is a Lie algebra of ``infinitesimal diffeomorphisms'',
or ``vector fields''.

We also assume that the representation $d\pi$ of ${\mathcal G}$ can be
integrated to a unitary representation $\pi$ of a group $G$, with Lie
$G={\mathcal G}$, such that 
\begin{equation}
\pi\,(g)\,a\;\pi\,(g^{-1}) \;\in\;{\mathcal A} \ ,
\end{equation}
for all $g \in G$ and all $a \in {\mathcal A}$ (i.e., $\pi (G)$ acts
by $^*$-automorphisms on ${\mathcal A})$. It then becomes meaningful to
study the $G$--{\it equivariant  cohomology} of the non-commutative
space described by the spectral data $({\mathcal A}, {\mathcal H},
\dd, \tilde{\gamma}, *)$. In the {\it Cartan model}, the
$G$--equivariant cohomology of $({\mathcal A}, {\mathcal H}, \dd,
\tilde{\gamma}, *)$ can be calculated as follows: Let ${\mathcal G}^*$
be the dual of ${\mathcal G}$, with a basis $\{
\gamma^1,\ldots,\gamma^n\}$ dual to the basis $\{ T_1,\ldots,T_n\}$ of
${\mathcal G}$. Let $S({\mathcal G}^*)$ denote the symmetric tensor
algebra over ${\mathcal G}^*$. The algebra $S({\mathcal G}^*) \otimes
\Omega_\dd^{\,^\bullet} ({\mathcal A})$ carries a natural
representation of ${\mathcal G}$ (and of $G$) generated by {\it Lie
  derivatives}: 
\[
L_{T_i}\,\gamma^k \ := \ -\; f_{ij}^k\,\gamma^j, \ \quad\gamma^j \in
{\mathcal G}^* \ ,
\]
where $f_{ij}^k$ are the structure constants of ${\mathcal G}$; and
\[
L_{T_i}\,\alpha \ = \ [ L_i,\,\alpha] \ ,
\]
for all $\alpha \in \Omega_\dd^{\,^\bullet} ({\mathcal A})$,
$i=1,\ldots,n$. One then defines 
\[
{\mathcal L}_i \ := \ L_{T_i}\;\otimes\; \id
\;+\;\id\;\otimes\;L_{T_i} \ .
\]
By $\bigr(S({\mathcal G}^*) \otimes \Omega_\dd^{\,^\bullet} ({\mathcal A})
\bigl)_{\rm inv}$ we denote the $G$-invariant subalgebra of 
$S({\mathcal G}^*) \otimes \Omega_\dd^{\,^\bullet} ({\mathcal A})$, 
consisting of all elements which commute with the ${\mathcal L}_i$. 

The algebra $S({\mathcal G}^*) \otimes \Omega_\dd^{\,^\bullet}
({\mathcal A})$ is represented on the Hilbert space ${\mathcal F}
\otimes {\mathcal H}$, where ${\mathcal F} \cong S({\mathcal G}^*)$
is the symmetric Fock space over ${\mathcal G}^*$. On this Hilbert
space, we introduce the Cartan differential, $\dd_C$, by
\begin{equation}
\dd_C \ := \ \id\;\otimes\;\dd\;-\;\gamma^i\;\otimes\;X_i \ ;
\end{equation}
note that the degree of $\gamma^i$ is $+2$, the one of $X_i$ is $-1$. 
For $\alpha \in S ({\mathcal G}^*) \otimes \Omega_\dd^{\,^\bullet}
({\mathcal A})$, we define 
\begin{equation}
\delta_C\;\alpha \ := \ \left[ \dd_C,\;\alpha\right]_g \ .
\end{equation}
The $G$--equivariant cohomology of $({\mathcal A}, {\mathcal H}, \dd,
\tilde{\gamma}, *)$ is defined by
\begin{equation}
H_{\dd,G}^{\,^\bullet} ({\mathcal A}) \ :=\ H_{\dd_C}^{\,^\bullet}
\left(\left( S \left({\mathcal G}^*\right) \otimes
    \Omega_\dd^{\,^\bullet} ({\mathcal A})\right)_{\rm inv.}\right) \ .
\end{equation}
If supersymmetry is ``unbroken'', in the sense that there exists a
unique vector $\varphi_0 \in {\mathcal F} \otimes {\mathcal H}$ of 
degree 0 which is cyclic and separating for the algebra $S({\mathcal
  G}^*) \otimes \Omega_\dd^{\,^\bullet}({\mathcal A})$ and satisfies 
$\dd\,\phi_0 =0$, then $H_{\dd,G}^{\,^\bullet} ({\mathcal A})$ 
is given by
$H_{\rm de~Rham}^{\,^\bullet} (C(EG)\otimes_G{\mathcal A})$ -- see 
the lectures by J.-L.~Loday for details. 

There is also a BRST model of $G$--equivariant cohomology, where
$S({\mathcal G}^*)$ is replaced by the Weil algebra $W({\mathcal G}) =
S({\mathcal G}^*) \otimes \Lambda {\mathcal G}^*$ and $\dd_C$ by a
differential involving {{\it fermion} creation operators $c^i$
  (besides the {\it bosonic} $\gamma^i$); see e.g.\ [62]. 
\smallskip

This theory can be extended to $N$=$\overline{(2,2)}$ spectral 
data $({\mathcal A}, {\mathcal H}, \partial, {\mathcal T},
\bar{\partial}, \overline{{\mathcal T}}, *)$. 
One then studies e.g.\ anti-holomorphic infinitesimal reparametrizations
generated by operators $L=\{ \bar{\partial},X\}$ -- hence $[ L,
\bar{\partial}]=0$ -- which also satisfy $[L,\partial]=0$. The objective is 
to determine the Dolbeault cohomology of $\bar{\partial}$ equivariant
with respect to a symmetry group generated by operators $L$ with the
properties just described.
\smallskip

Cohomological topological quantum theory includes the study of quantum
theories consisting of the data $({\mathcal A}, {\mathcal H}, Q,
{\mathcal G})$, where $Q$ is an operator satisfying $Q^2=0$ such that
$({\mathcal H}, Q)$ is a ${\mathbb Z}_2$-- or ${\mathbb Z}$--graded
complex (i.e., there is a grading operator $\gamma$ or $T$, as
above), e.g.\ $Q=Q_{\rm BRST}, \ \dd, \bar{\partial},\ldots$ , and
${\mathcal G}$ is a Lie algebra of infinitesimal reparametrizations
represented on ${\mathcal H}$, with the property that, for each $L \in
d\pi ({\mathcal G})$, there exists an odd operator $X$ on ${\mathcal
  H}$ such that $L=\{ Q, X\}$. The object of study is the algebra
$H_{Q,G}^{\,^\bullet}({\mathcal A})$ of cohomology classes and the
dual closed currents as considered in subsection~7) of Sect.~{\bf
  5.2}. Interesting results emerge from the study of the relations
between $H_{\partial, \bar{\partial}, G}^{\,^\bullet} ({\mathcal A})$
and $H_{\partial, \bar{\partial}^*, G}^{\,^\bullet}({\mathcal
  B})$. Among the most interesting ones are those found in
examples where ${\mathcal G}$ is the Witt algebra of infinitesimal
reparametrizations (vector fields) of $S^1$ ($d\pi$ is a projective
representation of ${\mathcal G}$), where one is led to studying ${\rm
  Diff}(S^1)$--equivariant cohomology in the framework of
two-dimensional quantum field theory [63].

\bigskip

The last topic briefly addressed in this section is
``target space supersymmetry''. This is a generalization of the notion of
spectral data admitting Lie algebras of infinitesimal
reparametrizations, as discussed at the beginning of this section. We
start, for example, from $N=n$ supersymmetric spectral data
$({\mathcal A}, {\mathcal H}, D_1,\ldots,D_n, \gamma)$. Let ${\mathcal
  G}$ be a Lie algebra, and $L$ a projective representation of
${\mathcal G}$ on ${\mathcal H}$ with \ ${\rm ad} L \in {\rm Der}({\mathcal
  A})$:
\begin{equation}
\left[\, L_X, L_Y\,\right] \ = \ L_{[X,Y]} \ + \ C_{X,Y} \ ,
\end{equation}
where $C_{X,Y}$ is an operator ${\mathcal G}$--cocycle commuting with
$L({\mathcal G})$ and with $D_1,\ldots,D_n$. We assume that
$D_1,\ldots,D_n$ span a module for ${\mathcal G}$ in the sense that
there are operators
\begin{equation}
\left\{ \lambda_i^{ \ j } (X) \bigm|  i,j =1,\ldots,n,\ X \in {\mathcal
    G}\right\} 
\end{equation}
on ${\mathcal H}$ which {\it commute} with $\{ L_X : X \in {\mathcal
  G}\}$ and with $D_1,\ldots,D_n$ and satisfy 
\begin{equation}
\sum_j \Bigl(
\lambda_i^{ \ j} (Y)\, \lambda_j^{\ k} (X) - \lambda_i^{ \ j} (X)\,
\lambda_j^{ \ k} (Y)\,\Bigr) \ = \ \lambda_i^{ \ k} ([ X,Y]) 
\end{equation}
as well as 
\begin{equation}
[\, L_X, \;D_i\,] \ = \ \lambda_i^{\ j} (X)\;D_j \ ,
\end{equation}
for all $X,Y \in{\mathcal G}$ and $i=1,\ldots,n$. Finally, 
\begin{equation}
\{\, D_i, D_j\, \} \ = \ \triangle_{ij} \ ,
\end{equation}
where $\triangle = (\triangle_{ij})_{i,j=1,\ldots,n}$ \ is a symmetric
$n\times n$ matrix of operators on ${\mathcal
  H}$ with the property that 
\begin{equation}
[\, D_k, \triangle_{ij}\, ] \ = 0 \ ,
\end{equation} 
for all $i,j$ and $k$. The relations (5.137) and (5.138) generalize 
analogous relations in (5.22), Sect.~{\bf5.1}. 

Since $D_1,\ldots,D_n$ commute with the $\lambda_i^{ \ j} (X)$, 
so does $\triangle_{kl}$, for all $k,l$. Equations (5.136) 
and (5.137) are compatible with each other iff
\[
[\, L_X, \triangle_{ij}\,] \ = \ \lambda_i^{ \ l}
(X)\,\triangle_{lj}\;+\;\lambda_j^{ \ k} (X)\,\triangle_{ik} \ ,
\]
but this is a consequence of the graded Jacobi identity (5.86):
\begin{eqnarray}
[L_X, \triangle_{ij}] &=& \left[ L_X, \left\{ D_i, D_j\right\}\right]
\nonumber\\ 
&=& \left\{\left[ L_X, D_i\right],\,D_j\right\}\;+\;\left\{\left[ L_X,
    D_j\right],\,D_i\right\}\nonumber \\
&=& \lambda_i^{ \ l} (X) \left\{ D_l, D_j\right\}\;+\;\lambda_j^{ \ k}
(X) \left\{ D_k, D_i\right\} \nonumber \\
&=& \lambda_i^{ \ l} (X) \,\triangle_{lj}\;+\;\lambda_j^{\ k}
(X)\,\triangle_{ik} \ . \nonumber
\end{eqnarray}
Thus $\{ \triangle_{ij} : i,j=1,\ldots,n\}$ span a module for
${\mathcal G}$ which carries the tensor product of the representation
$\lambda$ and the contragredient representation
$\lambda^\vee$ of ${\mathcal G}$.

Since the operators $D_1,\ldots,D_n$ are self-adjoint,
and assuming that $L$ is unitary in the sense that $L_X^* = - L_X$,
one should require that $\lambda$ is a real representation of
${\mathcal G}$ in the sense that
\begin{equation}
\lambda_i^{ \ j} (X)^*\;=\;\lambda_i^{ \ j}\ ,
\end{equation}
for all $i,j$ and all $X$ in ${\mathcal G}$. 
Furthermore, since $D_i^*=D_i$ and $\{ D_i, D_j\} = \{
D_j,D_i\}$, we have  $\triangle_{ij}^* = \triangle_{ij} =
\triangle_{ji}$ for all $i$ and $j$. This together with 
(5.138) shows that  $\{\, \triangle_{ij}\,\}_{i,j=1,\ldots,n}$ 
is a family of commuting, self-adjoint operators on ${\mathcal H}$ 
which commute with $D_k\,$, $k=1,\ldots,n$. 

We rewrite the matrix $\triangle = \bigl(\triangle_{ij}\bigr)$ of 
commuting operators in the form 
\begin{equation}
\triangle = {1\over n}\,\Bigl(\sum_{j=1}^n \triangle_{jj} \Bigr) \cdot \id
\ + \ \triangle^0\ ,
\end{equation}
where $\sum_{j=1}^n \triangle_{jj}^0 = 0$. We then define the subspace 
${\mathcal H}_0$ of ${\mathcal H}$ to be the eigenspace of the commuting 
operators $\triangle_{ij}^0$ corresponding to the eigenvalue $0$. 
Since the operators $\triangle_{ij}^0$ commute with $D_k$, $k=1,\ldots,n$, 
and with $\triangle_{jj}$, the space ${\mathcal H}_0$ is invariant 
under $D_k$ and $\triangle_{jj}$. (Note, however, that we do {\it not} 
claim that  ${\mathcal H}_0$ is invariant under $L_X$, for all $X\in
{\mathcal G}$.) 
If we restrict the operators $D_k$ to ${\mathcal H}_0$ they satisfy the 
standard $N=n$ supersymmetry algebra, i.e., relations (5.22), with 
$H= {1\over 2n}\,\Bigl(\sum_{j=1}^n \triangle_{jj} \Bigr)\,$.  
\smallskip

The theory sketched here corresponds to what physicists may call
target space supersymmetry. It has a straightforward extension to spectral
data of the form $({\mathcal A}, {\mathcal H}, \{ D_i\}_{i=1}^n, \allowbreak \{
\bar{D}_i\}_{i=1}^n, \tilde{\gamma}, {\mathcal G})$, which is important
when one studies non-real  representations of ${\mathcal G}$ on the
complex vector space spanned by $D_1,\ldots,D_n,\bar{D}_1,\ldots,\bar{D}_n$. 

Target space supersymmetries
appear in the study of superstring theory in the Green-Schwarz
formalism~[29] (which exploits the fact that $SO(8)_v \cong SO(8)_s
\cong SO(8)_{\bar{s}}$), and in $M$(embrane) theory~[61]. The Lie
algebra ${\mathcal G}$ is then supposed to describe some infinitesimal
symmetries of ``target space''. 

For example, ${\mathcal G}$ could be 
the Lie algebra of the group of Lorentz transformations of Minkowski 
space-time. In this case, only the subalgebra of ${\mathcal G}$ 
corresponding to infinitesimal rotations of space leaves the subspace 
${\mathcal H}_0$, introduced above, invariant. The operators $D_k$ are 
then interpreted as target space supersymmetry generators. 

\bigskip\medskip

Our exposition of the general formalism of non-commutative
differential geometry in Section~{\bf 5} has clearly suffered from a
lack of concrete examples and applications to physics. These form 
the subject of the last two sections.

\def \uI {{\displaystyle\mathop{I}_{\widetilde{}}}}
\def \G {{\mbox{\swabfamily{g}}}}
\def \G {{\mathcal Q}}
\def \T {{\hbox{\tt T}}}
\def\klamm#1#2#3{\hskip-#1pt{
\raise-8pt\vbox{\hbox{${}_{\scriptscriptstyle\lfloor}\!
\raise-1.7pt\vbox{
\hbox{\underline{\phantom{#2}}}}\mkern1mu\!
{}_{\scriptscriptstyle\rfloor}$}}}\hskip#3pt}

\setcounter{equation}{0}%

\section{The non-commutative torus}

In order to show how the formalism of non-commutative geometry works
in an explicit example, we discuss one of the simplest non-commutative
spaces: the two-dimensional non-commutative torus ${\mathbb T}_\alpha^2$  
(see [51]).

\vspace{.5cm}

\subsection{Spin geometry ($N=1$)}

To begin with, we describe the $N=1$ data associated to the classical
2-torus ${\mathbb T}_0^2$. By Fourier transformation, the algebra of smooth
functions over ${\mathbb T}_0^2$ is isomorphic to the Schwarz space ${\mathcal
  A}_0 := {\mathcal S}({\mathbb Z}^2)$ over ${\mathbb Z}^2$, endowed
with the (commutative) convolution product:
\begin{equation}
(a\, {\scriptstyle{\bullet}}\, b) (p) \ = \ \sum_{q \in {\mathbb Z}^2} a (q)\; b(p-q)
\end{equation}
where $a, b \in {\mathcal A}_0$ and $p \in {\mathbb Z}^2$. Complex
conjugation 
of functions translates into a ${}^*$--operation: 
\begin{equation}
a^* (p) \ = \ \overline{a (-p)} \ ,\quad \ a\;\in\;{\mathcal A}_0 \ .
\end{equation}
If we choose a spin structure over ${\mathbb T}_0^2$ in such a way that the
spinors are periodic along the elements of a homology basis, then the
associated spinor bundle is a trivial rank 2 vector bundle. With this
choice, the space of square integrable spinors is given by the direct
sum
\begin{equation}
{\mathcal H}_e\;\equiv\;{\mathcal H}\;=\;l^2 ({\mathbb
  Z}^2)\;\oplus\;l^2 ({\mathbb Z}^2) 
\end{equation}
where $l^2 ({\mathbb Z}^2)$ denotes the space of square summable
functions over ${\mathbb Z}^2$. The algebra ${\mathcal A}_0$ acts
diagonally on ${\mathcal H}$ by the convolution product. We choose a
flat metric $(g_{\mu\nu})$ on ${\mathbb T}_0^2$ and we introduce the
corresponding 2-dimensional gamma matrices
\begin{equation}
\left\{ \gamma^\mu, \gamma^\nu \right\} \ = \ -\, 2\,g^{\mu\nu} \ , \quad
\gamma^{\mu*} \ = \ -\, \gamma^\mu \ .
\end{equation}
Then, the Dirac operator $D$ on ${\mathcal H}$ is given by 
\begin{equation}
( D\,\xi) (p) \ = \ i\,p_\mu\, \gamma^\mu\, \xi(p) \ ,\quad \ \xi\;\in\;
{\mathcal H} \ .
\end{equation}
Finally, the ${\mathbb Z}_2$--grading $\sigma$ on ${\mathcal H}$ can
be written as
\begin{equation}
\sigma \ = \ \frac i 2 \ \sqrt{g} \ \varepsilon_{\mu\nu}\;
\gamma^\mu\,\gamma^\nu 
\end{equation}
where $\varepsilon_{\mu\nu}$ denotes the Levi-Civita tensor. The data
$({\mathcal A}_0, {\mathcal H}, D, \sigma)$ are the canonical $N=1$
data associated to the compact spin manifold ${\mathbb T}_0^2$, and it is 
clear that they satisfy all the properties of Definition A in
Sect.~{\bf 5.1}.
\smallskip

The non-commutative torus is obtained by deforming the product of the
algebra ${\mathcal A}_0$. For each $\alpha \in {\mathbb R}$, we define
the algebra ${\mathcal A}_\alpha := {\mathcal S} ({\mathbb Z}^2)$ with
the product 
\begin{equation}
(a\, {\scriptstyle{\bullet}}_\alpha \, b)\;(p) \ = \
\sum_{q\,\in\,{\mathbb Z}^2} a(q)\, b(p-q)\; e^{i\pi\alpha\omega (p,q)}
\end{equation}
where $\omega$ is the integer-valued anti-symmetric bilinear form on
${\mathbb Z}^2 \times {\mathbb Z}^2$ 
\begin{equation}
\omega (p,q)\;=\; p_1 q_2\, -\, p_2 q_1 \ ,\quad \ p, q\,\in\,{\mathbb Z}^2
\ .
\end{equation}
The ${}^*$--operation is defined as before. Alternatively, we could
introduce the algebra ${\mathcal A}_\alpha$ as the unital ${}^*$--algebra
generated by the elements $U$ and $V$ subject to the relations
\begin{equation}
U U^*\;=\;U^*U\;=\:V V^*\;=\;V^*V\;=\;\id\;, \quad \ UV\;=\; e^{- 2\pi
  i\alpha} \, VU \ .
\end{equation}
Having chosen an appropriate closure, the equivalence of the two
descriptions is easily seen if one makes the following
identifications:
\begin{equation}
U (p)\;=\;\delta_{p_{1,1}}\, \delta_{p_{2,0}} \ , \ V
(p)\;=\;\delta_{p_{1,0}}\, \delta_{p_{2,1}} \ .
\end{equation}
If $\alpha$ is a rational number, $\alpha = \frac M N$, where $M$ and
$N$ are co-prime integers, then the, centre $Z ({\mathcal A}_\alpha)$, of
${\mathcal A}_\alpha$ is infinite-dimensional:
\begin{equation}
Z({\mathcal A}_\alpha) \ = \ {\rm span}\left\{ U^{mN} V^{nN} \bigm|
  m, n \in {\mathbb Z} \right\} \ .
\end{equation}
Let ${\rm I}_\alpha$ denote the ideal of ${\mathcal A}_\alpha$ generated by
$Z({\mathcal A}_\alpha) - \id$. Then it is easy to see that the
quotient ${\mathcal A}_\alpha / {\rm I}_\alpha$ is isomorphic, as a unital
${}^*$--algebra, to the full matrix algebra $M_N ({\mathbb C})$.

If $\alpha$ is irrational the centre of ${\mathcal A}_\alpha$ is 
trivial and ${\mathcal A}_\alpha$ 
is of type ${\rm II}_1$, the trace being given by the evaluation
at $p=0$. Unless stated differently, we shall only study the case of
irrational $\alpha$ here, but the finite-dimensional non-commutative 
torus will be used in Sect.~{\bf7.3} below. 

We define the non-commutative 2-torus ${\mathbb T}_\alpha^2$ by its $N=1$ data
$({\mathcal A}_\alpha, {\mathcal H}, D, \sigma)$ where ${\mathcal H},
D$ and $\sigma$ are as in eqs.~(6.3), (6.5) and (6.6), and ${\mathcal
  A}_\alpha$ acts diagonally on ${\mathcal H}$ by the deformed
product, eq.~(6.7). When $\alpha = \frac M N$ is rational, one may
work with the data $({\mathcal A}_\alpha / {\rm I}_\alpha, {\mathbb C}^N
\oplus {\mathbb C}^N, D_\alpha, \sigma)$, where the Dirac operator
$D_\alpha$ is given by
\begin{equation}
D_\alpha \ = \ i\,\gamma^\mu \ \frac{{\rm sin} \left( \frac \pi N\;
    p_\mu\right)}{\frac \pi N } \ .
\end{equation}

\bigskip

\noindent 1) \ub{Differential forms}

\medskip

\noindent  Recall that there is a representation $\pi$ of the algebra of
universal forms $\Omega^{\,^\bullet} ({\mathcal A}_\alpha)$ on ${\mathcal H}$ 
(see Sect.~{\bf 5.1}, subsection 2)). The images of the homogeneous
subspaces of $\Omega^{\,^\bullet} ({\mathcal A}_\alpha)$ under $\pi$
are given by
\begin{eqnarray}
\pi\left( \Omega^0 \left({\mathcal A}_\alpha\right)\right) 
&=& {\mathcal A}_\alpha \quad {\rm (by \ definition)} \\
\pi\left( \Omega^{2 k-1}\left({\mathcal A}_\alpha\right)\right)
&=& \left\{ a_\mu\,\gamma^\mu \bigm| a_\mu\,\in\,{\mathcal A}_\alpha
\right\} \\
\pi\left( \Omega^{2k} \left({\mathcal A}_\alpha\right)\right)
&=& \left\{ a\,+\,b\sigma \bigm| a, b\,\in\,{\mathcal A}_\alpha \right\} 
\end{eqnarray}
for all $k\in{\mathbb Z}_+$. In principle, one should then compute the
kernels $J^n$ of $\pi$ (see eq.~(5.2)), but these are generally huge
and difficult to describe explicitly. To determine the space of
$n$--forms, it is simpler to use the isomorphism (see eq.~(5.3))
\begin{equation}
\Omega_D^n ({\mathcal A}_\alpha)\;=\;\Omega^n ({\mathcal A}_\alpha)
\big/ \left( J^n + \delta J^{n-1}\right) \;\cong\; \pi \left(
  \Omega^n \left( {\mathcal A}_\alpha\right)\right) \big/ \pi (\delta
J^{n-1}) \ .
\end{equation}
The spaces $\pi (\delta J^{n-1})$ are easy to compute, and the result
is
\begin{eqnarray}
\pi\left( \delta J^1\right) &=& {\mathcal A}_\alpha \\
\pi\left( \delta J^{2k}\right)&=& \pi\left( \Omega^{2k+1}\left(
    {\mathcal A}_\alpha\right)\right) \\
\pi\left( \delta J^{2k+1}\right) &=& \pi\left(
  \Omega^{2k+2}\left({\mathcal A}_\alpha\right)\right)
\end{eqnarray}
for all $k\geq 1$. The spaces of $n$--forms are thus given (up to
isomorphism) by
\begin{eqnarray}
\Omega_D^0 ({\mathcal A}_\alpha) &=& {\mathcal A}_\alpha \\
\Omega_D^1 ({\mathcal A}_\alpha) &\cong& \left\{
  a_\mu\,\gamma^\mu\bigm| a_\mu\,\in\,{\mathcal A}_\alpha\right\}\\
\Omega_D^2 ({\mathcal A}_\alpha) &\cong& \left\{ a\,\sigma \bigm|
  a\,\in\,{\mathcal A}_\alpha \right\} \\
\Omega_D^n ({\mathcal A}_\alpha) &=& 0 \quad \ {\rm for}\quad n\;\geq\;3 
\end{eqnarray}
where we have chosen special representatives on the r.s.  Notice that
$\Omega_D^1 ({\mathcal A}_\alpha)$ and $\Omega_D^2({\mathcal
  A}_\alpha)$ are free left ${\mathcal A}_\alpha$--modules of rank 2
and 1, respectively. This reflects the fact that the bundles of 1- and
2-forms over the 2--torus are trivial and of rank 2 and 1, resp.

\bigskip

\noindent 2) \ub{Integration and Hermitian structure over ${\bf
    \Omega_D^1 ({\mathcal A}_\alpha)}$}

\medskip

\noindent It follows from eqs.~(6.13--15) that there is an 
isomorphism $\pi(\Omega^{\,^\bullet} ({\mathcal A}_\alpha)) \cong 
{\mathcal  A}_\alpha \otimes M_2 ({\mathbb C})$. Applying the general
definition of the integral --- see Sect.~{\bf 5.1}, 3) --- to the
non-commutative torus, one finds
\begin{equation}
\bint \omega \ = \ {\rm Tr}_{{\mathbb C}^2} \left( \omega \left(
    0\right)\right) 
\end{equation}
for an arbitrary element $\omega \in \pi (\Omega^{\,^\bullet}
({\mathcal A}_\alpha))\,$. The cyclicity property, Assumption~A in
Sect.~{\bf 5.1}, 3), follows directly from the definition of the product
in ${\mathcal A}_\alpha$ and the cyclicity of the trace on ${\mathbb
  M}_2 ({\mathbb C})$. The kernels $K^n$ of the canonical sesqui-linear
form on $\pi (\Omega^{\,^\bullet} ({\mathcal A}_\alpha))$ --- see
eq.~(5.5) --- coincide with the kernels $J^n$ of $\pi$, and we get for
all $n \in {\mathbb Z}_n$:
\begin{equation}
\tilde{\Omega}^n ({\mathcal A}_\alpha)\;=\;\Omega^n ({\mathcal
  A}_n)\;, \  \quad 
\tilde{\Omega}_D^n ({\mathcal A}_\alpha)\;=\;\Omega_D^n ({\mathcal
  A}_\alpha) \ .
\end{equation}
Note that the equality $K^n=J^n$ holds in all explicit examples of
non-commutative $N=1$ spaces studied so far. It is easy to see that
the canonical representatives $\omega^\bot$ on ${\mathcal H}$ of
differential forms $[\omega] \in \Omega_D^n ({\mathcal A}_\alpha)$,
see eq.~(5.10), coincide with the choices already made in
eqs.~(6.20--23). The canonical Hermitian structure on $\Omega_D^1
({\mathcal A}_\alpha)$ is given by
\begin{equation}
\langle \omega, \eta\rangle_D\;=\;\omega_\mu\,g^{\mu\nu}\,\eta_\nu^*
\in {\mathcal A}_\alpha
\end{equation}
for all $\omega, \eta \,\in\,\Omega_D^1 ({\mathcal A}_\alpha)$. Note
that this is a true Hermitian metric, i.e., it takes values in
${\mathcal A}_\alpha$ and not in the weak closure ${\mathcal
    A}''_\alpha$. Again, this is true in many in other
examples, as well. 

\bigskip

\noindent 3) \ub{Connections on ${\bf \Omega_D^1 ({\mathcal A}_\alpha)}$}

\medskip

\noindent Since $\Omega_D^1 ({\mathcal A}_\alpha)$ is a free left 
${\mathcal  A}_\alpha$--module, it admits a basis which we can choose 
to be $E^\mu := \gamma^\mu$. A connection $\nabla$ on $\Omega_D^1
({\mathcal A}_\alpha)$ is uniquely specified by its coefficients
$\Gamma_{\mu\nu}^\lambda \in {\mathcal A}_\alpha$,
\begin{equation}
\nabla\,E^\mu\;=\;-\,\Gamma_{\nu\lambda}^\mu\,E^\nu \otimes E^\lambda
\in  \Omega_D^1 ({\mathcal A}_\alpha) \otimes_{{\mathcal A}_\alpha}
\Omega_D^1 ({\mathcal A}_\alpha) \ ,
\end{equation}
and these coefficients can be chosen arbitrarily. Note that in the
classical case $(\alpha = 0)$ the basis $E^\mu$ consists of real
1-forms. Thus, we say that the connection $\nabla$ is real if its
coefficients in the basis $E^\mu$ are self-adjoint elements of
${\mathcal A}_\alpha$. A simple computation shows that there is a
unique real, unitary, torsionless connection $\nabla^{L.C.}$ on
$\Omega_D^1 ({\mathcal A}_\alpha)$ given by
\begin{equation}
\nabla^{L.C.}\, E^\mu \ = \ 0 \ .
\end{equation}

\vspace{.5cm}

\subsection{Riemannian geometry ($N=\overline{(1,1)}$)}

In this subsection, we derive a set of $N=(1,1)$ spectral data along
the lines of Sect.~{\bf 5.2}, subsection 5).
Our first task is to find a real structure $J$ on the $N=1$ data
$({\mathcal A}_\alpha, {\mathcal H}, D, \sigma)$. To this end, we
introduce the complex conjugation $\kappa : {\mathcal H} \to {\mathcal
  H}$, $(\kappa \xi)(p) := \bar{\xi} (p) := \overline{\xi(p)}$, as well as the
charge conjugation matrix $C: {\mathcal H} \to {\mathcal H}$ as the
unique (up to a sign) constant matrix such that
\begin{eqnarray}
&C\;\gamma^\mu = -\; \bar{\gamma}^\mu\; C \\
&C\;=\;C^* = C^{-1} \ .
\end{eqnarray}
Then the most natural real structure which satisfies $[\,JaJ^*,b\,]=0$
for all $a,b \in {\mathcal A}_\alpha$ as well as $[\,J,D\,]=0$ 
is simply given by 
\begin{equation}
J := C \kappa\ .
\end{equation}
The right actions of ${\mathcal A}_\alpha$ and
$\Omega_D^1 ({\mathcal A}_\alpha)$ on ${\mathcal H}$ (see~{\bf 5.2.5})
are given as follows
\begin{eqnarray}
\xi\ {\scriptstyle{\bullet}}\ a &\equiv& J a^*\;J^*\xi \;=\;\xi \
{\scriptstyle{\bullet}}_\alpha \ a^\vee \\
\xi \ {\scriptstyle{\bullet}} \ \omega &\equiv& J \omega^* J^* \xi
\;=\; \gamma^\mu \xi \ {\scriptstyle{\bullet}}_\alpha \ \omega_\mu^\vee 
\end{eqnarray}
where $\xi \in {\mathcal H},\ a \in {\mathcal A}_\alpha$,\ $\omega \in
\Omega_D^1 ({\mathcal A}_\alpha)$, $\xi \
{\scriptstyle{\bullet}}_\alpha \ a$ \ denotes the
diagonal right action of $a$ on $\xi$ by the deformed product, and
\[
a^\vee (p) := a(-p)\ .
\] 
Notice
that $(a \ {\scriptstyle{\bullet}}_\alpha \ b)^\vee = a^\vee
{\scriptstyle{\bullet}}_\alpha \  b^\vee$. We
denote by $\displaystyle\mathop{{\mathcal H}}^\circ$ the dense
subspace ${\mathcal 
  S}({\mathbb Z}^2) \oplus {\mathcal S} ({\mathbb Z}^2) \subset
{\mathcal H}$ of smooth spinors. The space
$\displaystyle\mathop{{\mathcal H}}^\circ$  is a
two-dimensional free left ${\mathcal A}_\alpha$--module with 
canonical basis $\{ e_1, e_2\}$. Then any connection $\nabla$ on
$\displaystyle\mathop{{\mathcal H}}^\circ$ 
is uniquely determined by its coefficients $\omega_j^i \in \Omega_D^1
({\mathcal A}_\alpha)$: 
\begin{equation}
\nabla\,e_i \ = \ \omega_i^j \otimes e_j \ = \ \omega_{\mu i}^j\,
\gamma^\mu \otimes e_j\ \, \in\; \Omega_D^1 ({\mathcal A}_\alpha)
\otimes_{{\mathcal A}_\alpha} \displaystyle\mathop{{\mathcal H}}^\circ
\ .
\end{equation}
The ``associated right connection'' $\overline{\nabla}$ is then given by
\begin{equation}
\overline{\nabla}\,e_i = e_j \otimes \bar{\omega}_i^j \ \, \in\;
\displaystyle\mathop{{\mathcal H}}^\circ \otimes_{{\mathcal A}_\alpha}
\Omega_D^1 ({\mathcal A}_\alpha) 
\end{equation}
where 
\begin{equation}
\bar{\omega}_j^i \ = \ -\, C_k^i (\omega_l^k)^*\,C_j^l \ = \ C_k^i
(\omega_{\mu\,l}^k)^*\, C_j^l\,\gamma^\mu \ .
\end{equation}
An arbitrary element in $\displaystyle\mathop{{\mathcal H}}^\circ
\otimes_{{\mathcal A}_\alpha} \displaystyle\mathop{{\mathcal
    H}}^\circ$ can be written as $e_i \otimes a^{ij} e_j$ where
$a^{ij} \in {\mathcal A}_\alpha$. The ``Dirac operators'' ${\mathcal
  D}$ and $\bar{{\mathcal D}}$ on $\displaystyle\mathop{{\mathcal
  H}}^\circ \otimes_{{\mathcal A}_\alpha}
\displaystyle\mathop{{\mathcal H}}^\circ$ associated to the connection
$\nabla$ are given by (see eq.~(5.42))
\begin{eqnarray}
{\mathcal D} \left( e_i \otimes a^{ij}\,e_j\right) &=&
e_i \otimes \left(
  \delta\,a^{ij}\,+\,\bar{\omega}_k^i\,a^{kj}\,+\,a^{ik}\,\omega_k^j\right) \ 
{\scriptstyle{\bullet}} \ e_j \\
\bar{{\mathcal D}} \left( e_i \otimes a^{ij}\,e_j\right) &=&
e_i \ {\scriptstyle{\bullet}} \left(
  \delta\,a^{ij}\,+\,\bar{\omega}_k^i\,a^{kj}\,+\,
  a^{ik}\,\omega_k^j\right) \otimes \sigma\,e_j \ .
\end{eqnarray}
In order to be able to define a scalar product on 
$\displaystyle\mathop{{\mathcal H}}^\circ
\otimes_{{\mathcal A}_\alpha} \displaystyle\mathop{{\mathcal
    H}}^\circ$, we need a Hermitian structure on the {\it right}
module $\displaystyle\mathop{{\mathcal H}}^\circ$, 
denoted by $\langle \cdot, \cdot\rangle$, with values in ${\mathcal
  A}_\alpha$. It is defined by
\begin{equation}
\bint \ \langle \xi, \zeta\rangle\,a \ = \ \left( \xi,\zeta\,a\right)
\quad \forall\, \xi, \zeta \in \displaystyle\mathop{{\mathcal
    H}}^\circ\; , \ \forall \, a \in {\mathcal A}_\alpha \ .
\end{equation}
This Hermitian structure can be written explicitly as 
\begin{equation}
\langle \xi,\zeta\rangle \ = \ \sum_{i=1}^1 \overline{\xi^i} \
{\scriptstyle{\bullet}}_\alpha \ \zeta^{i\vee}\ ,
\end{equation}
and it satisfies
\begin{equation}
\langle \xi\,a, \zeta\,b\rangle \ = \ a^* \,\langle \xi,
\zeta\rangle\, b
\end{equation}
for all $\xi, \zeta \in \displaystyle\mathop{{\mathcal H}}^\circ$ and
$a, b \in {\mathcal A}_\alpha$. Then we define the scalar product on 
$\displaystyle\mathop{{\mathcal H}}^\circ
\otimes_{{\mathcal A}_\alpha} \displaystyle\mathop{{\mathcal
    H}}^\circ$ as (see [5]) 
\begin{equation}
\left( \xi_1 \otimes \xi_2,\;\zeta_1\otimes\zeta_2\right) \ = \ 
\left( \xi_2, \langle \xi_1, \zeta_1\,\rangle\,\zeta_2\right) \ .
\end{equation}
This expression can be written in a more suggestive way if one
introduces a Hermitian structure, denoted
$\langle\cdot,\cdot\rangle_L$, on the {\it left} module
$\displaystyle\mathop{{\mathcal H}}^\circ$: 
\[
\langle \xi, \zeta \rangle_L \ := \ \langle J\,\xi, J\,\zeta \rangle \
.
\]
This Hermitian structure satisfies
\[
\langle a\,\xi, b\,\zeta\rangle_L \ = \ a\,\langle \xi, \zeta\rangle_L\,b^*
\]
for all $a, b \in {\mathcal A}_\alpha$ and $\xi, \zeta \in
\displaystyle\mathop{{\mathcal H}}^\circ$, and the scalar product on 
$\displaystyle\mathop{{\mathcal H}}^\circ
\otimes_{{\mathcal A}_\alpha} \displaystyle\mathop{{\mathcal
    H}}^\circ$
can be written as follows
\[
\left( \xi_1 \otimes \xi_2, \zeta_1 \otimes \zeta_2 \right) \ = \ 
\bint \ \langle \xi_1, \zeta_1\rangle\;\langle \zeta_2, \xi_2\rangle_L
\ . 
\]
A tedious computation shows that the relations
\begin{equation}
{\mathcal D}^* = {\mathcal D}\,,\quad \ \bar{{\mathcal D}}^* = \bar{{\mathcal
    D}}\,,\quad \ \{ {\mathcal D}, \bar{{\mathcal D}} \} = 0\,,\quad \ 
{\mathcal  D}^2 = \bar{{\mathcal D}}^2
\end{equation}
are equivalent to 
\begin{equation}
\nabla\, e_i \ = \ 0 \quad \forall i \ .
\end{equation}
In particular, we see that the original $N=1$ data uniquely determine 
the operators ${\mathcal D}$ and $\bar{{\mathcal D}}$ 
satisfying the $N=(1,1)$ algebra, eq.~(6.43).

One can prove that the ${\mathbb Z}_2$--grading operators $\gamma$ and
$\bar{\gamma}$ (see Sect.~{\bf 5.2}, subsection 1)) are also unique 
(up to a sign):
\begin{equation}
\gamma = \id \otimes \sigma\,, \quad\  \bar{\gamma} = \sigma \otimes \id \ .
\end{equation}
In summary, we see that we get a natural set of $N=(1,1)$ data
$({\mathcal A}_\alpha, {\mathcal H} \otimes_{{\mathcal A}_\alpha}
{\mathcal H}, {\mathcal D}, \gamma, \bar{{\mathcal D}}, \bar{\gamma})$
induced by the original $N=1$ data. Furthermore, there is a unique
operator $T$,
\begin{equation}
T \ = \ \frac{1}{2i} \ g_{\mu\nu}\, \gamma^\mu \,\otimes\,
\gamma^\nu\, \sigma
\end{equation} 
that makes $( {\mathcal A}_\alpha, {\mathcal H} \otimes_{{\mathcal
    A}_\alpha} {\mathcal H}, {\mathcal D}, \gamma, \bar{{\mathcal D}},
\bar{\gamma}, T)$ into a set of $N=\overline{(1,1)}$ data, as defined
at the end of Sect.~{\bf 5.2}, subsection 1). 

\vspace{.5cm}

\subsection{K\"ahler geometry ($N=\overline{(2,2)}$)}

In this subsection, we extend the $N=\overline{(1,1)}$ spectral data
to $N=\overline{(2,2)}$ data. The simplest way to construct this
extension is to determine all anti-selfadjoint operators, collectively
denoted by $I$, that commute with ${\mathcal A}_\alpha, \gamma,
\bar{\gamma}$ and $T$ (see subsection {\bf 5.2.9}). Then one defines
the additional differentials as in eq.~(5.91). The most general
operator $I$ on ${\mathcal H} \otimes_{{\mathcal A}_\alpha} {\mathcal
  H}$ that commutes with all elements of ${\mathcal A}_\alpha$ is of
the form
\begin{equation}
I \ = \ \sum_{\mu,\nu = 0}^3 \gamma^\mu\,\otimes\,\gamma^\nu \
I_{\mu\nu}^R 
\end{equation}
where $I_{\mu\nu}^R$ are elements of ${\mathcal A}_\alpha$ acting on
${\mathcal H} \otimes_{{\mathcal A}_\alpha} {\mathcal H}$ from the
right, and where we have set
\begin{equation}
\gamma^0 \ = \ \id \ , \quad \gamma^3 \ = \ \sigma \ .
\end{equation}
The vanishing of the commutators of $I$ with $\gamma$ and
$\bar{\gamma}$ implies that $I_{\mu\nu}^R=0$ unless $\mu, \nu \in \{
0,3\}$.
The equation $[I,T]=0$ requires $I_{03}^R = I_{30}^R$ and leaves the
coefficients $I_{00}^R$ and $I_{33}^R$ undetermined. Since the
operators $I$ appear only through commutators, their trace part is
irrelevant and we can set $I_{00}^R = 0$. All constraints together
give
\begin{equation}
I\;=\;\left( \sigma\,\otimes\,\id\;+\;\id\,\otimes\,\sigma\right)\,
I_{03}^R \;+\; \left( \sigma\,\otimes\,\sigma\right) \, I_{33}^R 
\end{equation}
where $I_{03}^R$ and $I_{33}^R$ are anti-selfadjoint elements of
${\mathcal A}_\alpha$. We decompose $I$ into two parts
\begin{eqnarray}
I_1 &=& \left( \sigma\,\otimes\,\id \ + \ \id\,\otimes\,\sigma\right)
\; I_{03}^R\\
I_2 &=& \left( \sigma\,\otimes\,\sigma\right)\; I_{33}^R
\end{eqnarray}
and we introduce the new differentials according to eq.~(5.91)
\begin{eqnarray}
d_1 &=& d \ = \ {\mathcal D}\;-\;i\,\bar{{\mathcal D}} \\
d_2 &=& \left[ I_1, d\,\right] \\
d_3 &=& \left[ I_2, d\,\right] \ .
\end{eqnarray}
The nilpotency of $d_2$ and $d_3$ implies that $I_{03}$ and $I_{33}$
are multiples of the identity, and we normalize them as follows
\begin{eqnarray}
I_1 &=& \frac i 2 \ \left( \sigma\,\otimes\,\id \ + \
  \id\,\otimes\,\sigma\right) \ \\
I_2 &=& i\,\left( \sigma\,\otimes\,\sigma\right) \ .
\end{eqnarray}
Comparing eqs.~(6.56) and (6.45) we see that
\begin{equation}
I_2 \ = \ i\,\gamma\,\bar{\gamma}
\end{equation}
and it follows, using eqs.~(6.52) and (6.54), that
\begin{equation}
d_3 \ = \ \left[ I_2, d\right] \ = \ 2\,i\,d\,\gamma\,\bar{\gamma} \ .
\end{equation}
Thus, the differential $d_3$ is a trivial modification of $d$, and we
discard it. It is then easy to verify that $({\mathcal A}_\alpha,
{\mathcal H} \otimes_{{\mathcal A}_\alpha} {\mathcal H}, d_1, d_2,
\gamma, \bar{\gamma}, T, I_1)$ form a set of $N=\overline{(2,2)}$ spectral
data. Furthermore, they are, as we have shown, uniquely determined by
the original $N=\overline{(1,1)}$ data. Therefore, a 
Riemannian non-commutative torus (at irrational deformation parameter
$\alpha$) admits a unique K\"ahler structure.
\smallskip

We have only given the definitions of the spectral data in the
$N=\overline{(1,1)}$ and the $N=\overline{(2,2)}$ setting. As a
straightforward application of the general methods described in
Section~{\bf 5}, we could compute the associated de~Rham resp.~Dolbeault
complexes, as well as the Euler characteristic, the Hirzebruch
signature, or geometrical quantities like curvature. We do not carry
out these calculations here. 
\smallskip

Instead, we emphasize the following feature: For rational
deformation parameter $\alpha = \frac{M}{N}$, the algebra ${\mathcal
  A}_\alpha$ in itself does not specify the geometry of the underlying
non-commutative space. It is only the selection of a specific
$K$--cycle $({\mathcal H}, D)$ that allows us to identify this space
as a deformed torus. In fact, by picking different pairs $({\mathcal
  H}, D)$ for ${\mathcal A}_\alpha = M_N ({\mathbb C})$, one can
obtain the fuzzy two-sphere, and even the fuzzy three-sphere (see
Sec.~{\bf 7.6}).

One might speculate that for irrational $\alpha$ the choice of
$K$--cycles is more restricted. It would be very interesting to
investigate how the geometries for ${\mathcal A}_\alpha, \alpha \notin
{\mathbb Q}$, can be approximated by ``towers of matrix geometries''.

\vspace{1cm}

\setcounter{equation}{0}%
\setcounter{section}{6}

\section{Applications of non-commutative geometry\\
to quantum theories of gravitation}

\noindent In this section we sketch some applications of the tools
described in Sections~{\bf 4--6} to a quantum theory of gravitation yet
to be discovered. We have argued in Section~{\bf 3} that a combination of
quantum theory and general relativity leads to the prediction that
space-time cannot be a classical manifold and that the basic degrees
of freedom of a theory of space-time-matter had better be associated
with extended objects so that space-time uncertainty relations valid
for the {\it location of events} are automatically fulfilled. A
currently popular idea is that those extended 
objects are strings. We therefore sketch some features of string
theory; for a broad exposition of the subject see [29]. However, the
consensus evolves in the direction to say that there are {\it extended
  dynamical objects}, ``branes'', of various dimensions and that,
perhaps, extended objects more fundamental than strings might be
``membranes'' ($M$--theory); see Sect.~{\bf 7.3}. 

\vspace{.5cm}

\subsection{From point-particles to strings}

\noindent Let $M$ be a $d$--dimensional, Lorentzian manifold
interpreted as classical space-time. We consider a point-particle
moving in $M$, as discussed in Section~{\bf 1}. But now we propose to
treat it {\it quantum mechanically}, following Feynman's idea of path
integrals. The action of a relativistic point-particle is given by
(see eq.~(1.6))
\begin{eqnarray}
S_P \left( x, h\right) &:=& \frac {l^{-1}} 2 \ \int\limits_0^1
 g_{\mu\nu}\left( 
x\left(\tau\right)\right) \dot{x}^\mu (\tau) \dot{x}^\nu (\tau)
h (\tau)^{-1/2} \;d \tau \nonumber \\
&\ +& \frac{\mu^2 l^{-1}}{2} \ \int\limits_0^1 h (\tau)^{1/2} \; d \tau \ ,
\end{eqnarray}
where $(g_{\mu\nu})$ is a Lorentzian metric on $M$, $\dot{x} (\tau) :=
\frac{dx(\tau)}{d\tau}$ , $h(\tau) d\tau^2$ is a metric on the unit 
interval $[0,1]$; $\mu^2$ is a positive constant of dimension 
mass$^2$, and $l$ is a constant with the dimension of length. 
Feynman proposed to consider a path integral related to
\begin{equation}
\triangle_F (x,y) \ := \ \int\limits_{ {x(0)\,=\,x } \atop
  {x(1)\,=\,y}} e^{i\,S_P(x,h)} \ {\mathcal D} \, x \;
{\mathcal D}\,h \ ,
\end{equation} 
where, formally, ${\mathcal D} x = \prod_{\tau\in[0,1]}
\frac{d^dx(\tau)}{l^d}$ , ${\mathcal D} h = \prod_{\tau \in [0,1]} dh
(\tau)$ .
Choosing a gauge such that $h (\tau) \equiv T^2, 0 < T < \infty$, and
performing the $x$--integral, one finds that 
\begin{eqnarray}
\triangle_F (x,y) &=& {\rm const}\cdot\ \int\limits_0^\infty dT \left(
  e^{i T l\, \left( \square_g + \mu^2 + i0\right)} \right)\,
(x,y) \nonumber\\
&=& {\rm const}\cdot l^{-1}\, \left( \square_g + \mu^2 + i0\right)^{-1} (x,y) \ .
\nonumber
\end{eqnarray}
This is the {\it Feynman propagator} for a scalar particle 
with mass $\mu$.  For $y^0 > x^0$, $\triangle_F (x,y)$ is a matrix
element of the quantum-mechanical particle propagator from time $x^0$
to time $y^0$. Unfortunately, it is not very meaningful to consider a
single point-particle. First, the principles of local, relativistic
quantum field theory imply that every particle has a twin, the
anti-particle (possibly identical with the particle), and, second,
when the metric $(g_{\mu\nu})$ on $M$ is not static (but $M$ is
asymptotically Minkowskian) then particle creation- and annihilation
processes are observed (for a physically meaningful definition of
particles). So we really must consider a {\it gas of particle
  world-lines}. The partition function, $\Xi$, of 
this gas is obtained by integrating over all configurations of {\it
  closed} world-lines (i.e., loops), each one weighted by $\int {\rm
  exp} (i\,S_P(x,h))\,{\mathcal D} h$.  

According to Symanzik [64], a system of {\it interacting},
relativistic, scalar point-particles can be described in terms of a
gas of world-lines with local soft-core repulsion (``excluded volume
interactions''). This approach has ultimately led to various
non-interaction theorems for scalar quantum field theories (triviality
of $\lambda\varphi^4$ in 
\def \grgl {\hbox{\raise4pt\vbox{\hbox{$\scriptstyle{>}$}}} \hskip-9pt
\hbox{\raise-1pt
\vbox{\hbox{$\scriptscriptstyle{(=)}$}}}}
$d\,\, \grgl \,4\,$)[65]. 

It is clear that the interactions between different point-particles
depend on the way their world-lines are embedded in the classical
space-time background. In other words, the formulation of a local
quantum theory of interacting, relativistic, scalar point-particles
requires a model of classical space-time.

It is not difficult to guess how one might generalize the
Feynman-Symanzik formulation of the quantum theory of relativistic,
scalar point-particles to a quantum theory of relativistic string-like
extended objects. Let us first consider the relativistic mechanics of
a classical string: It sweeps out a {\it world-sheet} $X : \Sigma \to
M$, $\Sigma \ni \xi \mapsto \left( X^\mu (\xi)\right) \in M$, where
$\Sigma $ is a surface equipped with a (Lorentz) metric
$h=(h_{\alpha\beta}(\xi))$. As the equations of motion for $h$ and
$X$, Deser, Zumino and Polyakov [66] have proposed the Euler-Lagrange
equations corresponding to the following action functional (see also
[67]): 
\begin{eqnarray}
&&S_{P,\Sigma}\;(X,h) \ := \ \nonumber \\
&&\qquad \frac{1}{4\pi \alpha'} \int_\Sigma d^2 \xi
\sqrt{|h(\xi)|}\;h^{\alpha\beta} (\xi) \partial_\alpha X^\mu (\xi)
g_{\mu\nu} (X(\xi)) \partial_\beta X^\nu (\xi)\ + \nonumber \\
&&\qquad\ + \ \frac{\Lambda}{4\pi} \int_\Sigma d^2 \xi \sqrt{|h(\xi)|}
\ ,
\end{eqnarray}
where $\alpha'$ and $\Lambda^{-1}$ are constants of dimension 
length$^2$. The solutions to the classical equations of motion are
extremal surfaces in $M$. Actually, the action (7.3) should be
generalized by including two further terms:
\[
S_{P,\Sigma}(X,h)\,\to\,S_{{\rm tot.,}\Sigma}
(X,h)\;:=\;S_{P,\Sigma}(X,h)\,+\,S_\Sigma' (X,h)\,+\,S_\Sigma'' (X,h) \
,  
\]
where
\begin{eqnarray}
&&S_\Sigma' \ := \ \frac{1}{4\pi\alpha'} \int_\Sigma d^2 \xi
\varepsilon^{\alpha\beta} \partial_\alpha X^\mu (\xi) B_{\mu\nu}
(X(\xi)) \partial_\beta X^\nu (\xi) \ ,\\
&& S_\Sigma'' \ := \ \frac{1}{4\pi} \int_\Sigma d^2 \xi
\sqrt{|h(\xi)|} \Phi (X(\xi))\, r \, (\xi) \ .
\end{eqnarray}
Here $\beta \equiv B_{\mu\nu} (x) d x^\mu \wedge dx^\nu$ is a 2-form
on $M$ and $\Phi (x)$, the {\it ``dilaton''}, is a function on $M$;
$r(\xi)$ is the curvature scalar corresponding to the metric
$(h_{\alpha\beta} (\xi))$ on $\Sigma$. The term $S_\Sigma'$ is
proportional to the integral of $\beta$ over the image of $\Sigma$
under the map $X:\Sigma \to M$, $\Sigma \ni \xi \mapsto X(\xi) \in M$,
(which is the integral of the pullback $X^* (\beta)$ over $\Sigma$)
and $S_\Sigma''$ is proportional to the integral of $X^* (\Phi)r$ over
$\Sigma$.
\smallskip

Let us consider a single, relativistic, {\it closed} string
propagating from some initial to some final configuration (at larger
times). Then $\Sigma$ has the topology of a 
cylinder (i.e.\ of a twice punctured sphere). We are actually interested in the
{\it quantum-mechanical} propagation of a relativistic, closed
string. In analogy to Feynman's quantization of the mechanics of
relativistic, scalar point-particles in terms of path integrals (see
(7.2)), one is tempted to guess that the propagator is given by
\begin{equation}
\triangle_F \left( X_i, X_f\right) \ := \ \int e^{i\,S_{{\rm
      tot.,}\Sigma} (X,h)} \ {\mathcal D}_h\,X\,{\mathcal D}\,h \ ,
\end{equation}
where $X_i$ and $X_f$ denote configurations of the string contained in
(co-dimension 1) space-like surfaces $\sigma_i$ and $\sigma_f$,
respectively, embedded in $M$ in such a way that $\sigma_i$ is
e.g.~{\it earlier} than $\sigma_f$ with respect to the causal
orientation of $M$. They provide {\it boundary conditions} for the
functional integral on the r.s.\ of (7.6) at $\partial \Sigma$. 

On the r.s.\ of (7.6), we invoke Fubini's theorem to represent
$\triangle_F (X_i, X_f)$ as 
\begin{eqnarray}
\triangle_F (X_i, X_f) &=& \int {\mathcal D} h \int {\mathcal D}_h X\;
e^{i\,S_{{\rm tot.,}\Sigma} (X,h)} \nonumber\\
&=:& \int {\mathcal D} h\ Z_\Sigma (h) \ .
\end{eqnarray}
Formally, the measure ${\mathcal D}_h X$ is the Riemannian volume form
on the infinite-dimensional {\it Riemannian} manifold of maps $X$ from
$(\Sigma, h)$ to $(M, g)$ and hence depends on $h$ (and on $g$ -- but
$g$ is presently kept fixed). If $\psi$ is a diffeomorphism of $\Sigma$
onto itself then
\begin{equation}
Z_\Sigma (h) \ = \ Z_\Sigma (\psi^* \,h) \ .
\end{equation}
It would seem that eq.~(7.8) holds by construction. However, since the
calculation of $Z_\Sigma (h)$ involves a formal, infinite-dimensional
functional integration, 
one should ask whether ``diffeomorphism (or gravitational) anomalies''
could invalidate (7.8). It turns out that if the field $X(\xi)$, $\xi
= (\sigma, \tau) \in \Sigma$, is a non-chiral field (in the sense that 
left-moving modes of $X$, depending on $\sigma+\tau$, match
right-moving ones, depending on $\sigma-\tau$) then there are no such
anomalies. But if $X$ were {\it chiral} (e.g., purely left-moving)
then gravitational anomalies appear. They can be described as
Lorentz-- and mixed Lorentz--Weyl anomalies and are cancelled by the
ones of a three-dimensional gravitational Chern-Simons action
[77]. (This leads to the prediction that $(M,g)$ should be a
Lorentzian manifold of dimension $26 + n \cdot 24, \ n = 0,1,\ldots$
.)

Eq.~(7.8) implies that $Z_\Sigma (h)$ only depends on the {\it orbit}
$[h]$ of $h$ under the pullback action of the group of
diffeomorphisms of $\Sigma$. Thus the integral (7.7) is ill-defined
before we fix a gauge. On a cylinder $\Sigma$, the orbit $[h]$ of
every metric $h$ contains a conformally flat metric, $e^{\phi (\xi)} \
  {-1 \ 0 \choose \ 0 \ 1}$, \ i.e.,
\begin{equation}
h_{\alpha\beta} (\xi) \ \sim \ e^{\phi (\xi)} \ {-1 \ 0 \choose \quad 0
  \ 1} \ .
\end{equation}
Thus, orbit space is parametrized by the conformal factors $e^{\phi
  (\xi)}$ ($\phi$ is called {\it ``Liouville mode''}), and we can
choose the conformally flat metrics (r.s.\ of (7.9)) as a cross section
in the Riemannian manifold of all metrics on $\Sigma$. (This is what
the physicists call a gauge choice.) {}From (7.9) we conclude that one
can equip $\Sigma$ with a causal structure (a field of light cones),
independently of the choice of $(h_{\alpha\beta})$, and this suggests
that a local (w.r.t. the causal structure on $\Sigma$) ``quantum
theory'' of the metric $(h_{\alpha\beta})$ can be developed (i.e., 
two-dimensional quantum gravity ought to make sense -- recall the
discussion towards the end of Section~{\bf 3}). The functional integral
formulation of this ``quantum theory'' is quite well understood: One
fixes the gauge specified on the r.s.\ of (7.9). We denote
\begin{equation}
\left( \hat{h}_{\alpha\beta}\right) \ := \ {-1 \ 0 \choose \quad 0 \
  1} \ .
\end{equation}
Using the Faddeev-Popov method [68,29], one finds (see [66]) that
\begin{equation}
\triangle_F (X_i, X_f) \ = \ {\rm const}\, \int {\mathcal
  D}_{\hat{h}}\, \phi\; e^{- 26 i\, \Gamma_{\hat{h}} (\phi)} \
Z_\Sigma (e^\phi\,\hat{h}) \ ,
\end{equation}
where
\begin{equation}
\Gamma_{\hat{h}} (\phi) \ = \ \frac{1}{96 \pi}
\int\limits_\Sigma d^2 \xi \sqrt{|\hat{h} (\xi)|} \ \left(
\hat{h}^{\alpha\beta} (\xi)\, \partial_\alpha \phi (\xi)\,
\partial_\beta \phi(\xi)\;+\;4 r_{\hat{h}} (\xi)\,\phi (\xi)\right)\ .
\end{equation} 
Of course, for our gauge choice (7.9), $|\hat{h} (\xi)|=1$ and
$r_{\hat{h}} (\xi) \equiv 0$; but it is useful to know the general
result (7.12) for $\Gamma$, in order to be able to extend these
calculations to more general surfaces $\Sigma$ (where $\hat{h}$ ranges
over some moduli space of conformally inequivalent, non-flat --- and, for
Minkowskian signature, singular --- metrics over which one will have to
integrate; see e.g.~[87]).

Note that (for $\Lambda =0$ on the r.s.\ of (7.3))
\begin{equation}
S_{{\rm tot.,} \Sigma} \bigl( X, \, e^\phi\,\hat{h}\bigr) \ = \
S_{{\rm tot.,} \Sigma} \bigl( X, \, \hat{h}\bigr) \ ,
\end{equation}
i.e., $S_{{\rm tot.,}\Sigma}$ is invariant under {\it Weyl rescaling}.
One might thus expect that $Z_\Sigma (e^\phi \hat{h})= Z_\Sigma
(\hat{h})$. However, this is never true, because ${\mathcal D}_{e^\phi
  \hat{h}} X$ {\it does} depend on $\phi$; one says that
two-dimensional quantum field theories always have a {\it Weyl
  anomaly}. Thus, while we have that 
\[
Z_\Sigma \,( \psi^*\,h) \ = \ Z_\Sigma\, (h) \ ,
\]
for all metrics $h$ and all diffeomorphisms $\psi$ of $\Sigma$ (see
(7.8)), one always finds that
\begin{equation}
Z_\Sigma\, (e^\phi\, h) \ \not\equiv \ Z_\Sigma \,(h) \ .
\end{equation}

One way of trying to give meaning to the $\phi$--integral on the
r.s.\ of eq.~(7.11) is to demand that the $\phi$--dependence of the
integrand be {\it trivial}. Then
\begin{equation}
Z_\Sigma\,( e^\phi\,\hat{h}) \ = \ e^{i\,c\,\Gamma_{\hat{h}} (\phi)} \
Z_\Sigma\,(\hat{h}) \ ,
\end{equation}
with 
\begin{equation}
c \ = \ 26 \ ,
\end{equation}
see e.g.\ [29]. If (7.15) and (7.16) hold, one can omit the 
$\phi$--integration on the r.s.\ of (7.11) (which amounts to 
declaring that ${\rm const}\cdot \int {\mathcal D}_{\hat{h}} \phi = 1$). 
Eqs.~(7.8), (7.15) and (7.16) characterize what one calls (tree-level)
{\it critical bosonic string theory}. The equations (7.8) and (7.15),
for any non-negative value of $c$, mean 
that the two-dimensional quantum field theory defined by the action
$S_{{\rm tot.,}\Sigma} (X, h)$ in eqs.~(7.3--5), for an
arbitrary but fixed choice of $\Sigma$ and $\hat{h}$, should be a
{\it conformal field theory} [69]. A standard argument of quantum field
theory says that
\begin{equation}
(-i)^n\,\frac{\delta^n}{\delta h^{\alpha_1\beta_1}(\xi_1)\ldots \delta
  h^{\alpha_n\beta_n} (\xi_n)}\,{\rm ln}\, Z_\Sigma \,(h)\;=\;
\langle \T\left( T_{\alpha_1\beta_1}(\xi_1)\ldots
  T_{\alpha_n\beta_n}(\xi_n)\right)\rangle_n^c \ ,
\end{equation}
where $\langle \T (\cdot)\rangle_h^c$ denotes the time
$(\tau)$--ordered, connected ``vacuum expectation'' of the field
theory on $(\Sigma, h)$, and $T_{\alpha\beta} (\xi)$ is its {\it
  energy-momentum tensor} at the point $\xi \in \Sigma$. Combining
(7.12), (7.15) and (7.17), and 
setting $T(\xi) = T_{ \ \alpha}^\alpha (\xi) = h^{\alpha\beta} (\xi)
T_{\alpha\beta} (\xi)$ \ (trace of the energy-momentum tensor) we find
that
\begin{eqnarray}
\langle\, T (\xi)\,\rangle_{\hat{h}} &=& -\,i\;
\frac{\delta}{\delta\phi(\xi)} \ {\rm ln}\,Z_\Sigma (e^\phi
\hat{h})\biggm|_{\phi=0} \nonumber\\
&=& \ \frac{c}{24\,\pi} \ r_{\hat{h}} \;(\xi) 
\end{eqnarray}
and
\begin{eqnarray}
\langle \T\left( T(\xi) T(\eta)\right)\rangle_{\hat{h}}^c &=& - \
\frac{\delta^2}{\delta \phi (\xi)\,\delta\phi(\eta)} \ {\rm
  ln}\;Z_\Sigma (e^\phi \hat{h})\biggm|_{\phi=0} \nonumber \\
&=& 0 \ \quad \ {\rm for} \quad \xi \neq \eta \ .
\end{eqnarray}
Eqs.~(7.18) and (7.19) tell us that 
\begin{equation}
T(\xi) \ = \ \frac{c}{24\,\pi} \ r_{\hat{h}} (\xi)\ \id 
\end{equation}
(which vanishes if $\hat{h}$ is flat). Eq.~(7.20) is precisely the
condition for the field theory on $(\Sigma, \hat{h})$ to be {\it
  conformal}. In a renormalization group analysis of Lagrangian field
theory, equation (7.20) can be translated into the condition that the
renormalization group
$\beta$--function {\it vanish}. This condition yields equations for
the tensor fields $g_{\mu\nu}$ (metric), $B_{\mu\nu}$ (2-form) and
$\Phi$ (dilaton) on the (target) space-time manifold $M$. These
equations are generalizations of {\it Einstein's equations} (see
Section~{\bf 1}). They are quite complicated; see [29,70]. When
space-time $M$ is {\it static} then they approximately look as
follows: In local coordinates $X^\mu$ on $M$ with the property that
$g_{0j} (X) = 0$, $g_{00} (X) = -1$, and choosing a gauge such that
$B_{0j} (X) = 0$, $j=1,\ldots,d-1$, 
\begin{eqnarray}
&& R_{ij}\,+\,2\,\nabla_i\,\nabla_j\,\Phi\,-\,\frac 1 4 \
H_{imn}\,H_j^{mn} \ = \ 0 \ ,\nonumber\\
&&-\,\frac 1 2 \ \nabla_m\,H_{ij}^m\,+\,H_{ij}^m\,\partial_m\,\Phi
\phantom{mmmml} = \ 0 \ , \\
&& C^{(d)} \ - \ 26 \phantom{mmmmmmmmmmml} = \  0 \ ,\nonumber
\end{eqnarray}
where $\nabla$ is the Levi-Civita connection on $M$, $R_{ij}$ is the
Ricci tensor, $H_{ijk} = 3 \partial_{[i} \ B_{jk]}$ , 
and
\begin{equation}
C^{(d)}\;=\;d\,-\,\frac 3 2 \ \alpha' \biggl[ r\,-\,\frac{1}{12} \
  H_{ijk}H^{ijk}\,-\,4\,\nabla^j\Phi\,\nabla_j\Phi\,+\,4\,\triangle_g
  \Phi \biggr] \ ,
\end{equation}
where $r$ is the scalar curvature on $M$. Eqs.~(7.21) hold to one-loop
order in the expansion parameter $\alpha'$. (We recall that it is
assumed, here, that $g$, $B$ and $\Phi$ are time-independent.)
Eqs.~(7.21) are the Euler-Lagrange equations corresponding to the
action
\begin{equation}
S^{(d)} (g, B, \Phi) \ = \ \int d^d\,X\,\sqrt{g(X)}\; e^{- 2 \Phi
  (X)}\, \left[ C^{(d)} (X) - 26\right] \ .
\end{equation}
Recalling expression (7.22) for $C^{(d)}$, we observe that this is a
generalization of the {\it Hilbert-Einstein action} with a {\it
  cosmological constant} $\propto d-26$. The vanishing of the
cosmological constant then requires that the dimension $d$ of
space-time should be 26. 

Of course, it is of interest to generalize eqs.~(7.21--23) to {\it
  non-static} space-times; for results see~[70].

Physicists are intrigued by the chain of arguments leading from (7.15)
to (7.23), and they have discovered a number of different ways to
reach these conclusions; see~[29]. They are even more intrigued by the
observation that string theory {\it automatically} describes {\it
  interactions} (scattering) between different strings, and that one
does {\it not} have to talk about (target) space-time $M$ {\it
  explicitly}, in order to describe those interactions (in {\it
  contrast} to point-particle field theory): In order to calculate the
{\it connected part} of a scattering amplitude from $n$ incoming to
$m$ outgoing strings,
one generalizes expression (7.7) to surfaces $\Sigma$ with $n$
positively and $m$ negatively oriented boundary components, and one
sums over all possible topologies (and integrates over the moduli
space of conformal structures) of $\Sigma$. As there is no nice theory
of Lorentzian surfaces of higher genus and with many boundary
components, one performs a {\it Wick rotation}, $\xi \equiv
(\sigma,\tau) \to (\sigma,i\,\tau)$, with the effect that the surfaces
$\Sigma$ become {\it Riemann surfaces}. The different terms in the sum
over topologies are then weighted by factors
\begin{equation}
{\rm exp}\,\left( - {\rm const}\, \langle \Phi\rangle\;
  H(\Sigma)\right) \ ,
\end{equation}
where $H(\Sigma)$ is the number of handles of $\Sigma$ and $\langle
\Phi\rangle$ is some mean value of the dilaton field $\Phi$. Eq.~(7.24)
follows from (7.5) (with $\Phi$ replaced by $\langle \Phi\rangle$) and
the Gauss-Bonnet formula. It is the number \ ${\rm exp}\,(- {\rm
  const.} \langle \Phi \rangle)$ that is a measure for the deformation
parameter, mentioned in Section~{\bf 3}, of the deformation from
classical to quantum space-time geometry.
\smallskip

Of course, it is difficult to calculate the various contributions,
e.g.~to (7.7), in an expansion in the number of handles of $\Sigma$,
and the expansion has been argued to be neither convergent nor Borel
summable [71]. There are good reasons for these problems: First,
critical bosonic string theory is really a sick theory. When one
calculates all the modes of a string propagating in an (approximately)
flat space-time $M$, using (7.11--16), one finds that among these
modes there is a tachyon with {\it negative} \ ${\rm mass}^2 = -
(\alpha')^{-1}$. This is physically unacceptable, but the problem is
cured by replacing the bosonic string by the {\it superstring} and 
performing the GSO
projection -- see e.g.~[29]. But, second, one finds a {\it tower of
  modes} with \ ${\rm mass}^2 = (\alpha')^{-1} n$, 
$n=0,1,2,\ldots$. It is plausible
that $\alpha' \propto l_P^2$, where $l_P$ is the Planck length (see
Section~{\bf 3}). Thus, for large $n$, a string mode has a mass that can
be considerably larger than the Planck mass! Exciting such a mode (and
letting it interact with other strings) ought to produce a {\it major
  perturbation} in the geometry (and, perhaps, the topology) of
space-time $(M,g)$. However, $(M,g)$ is treated as a {\it fixed classical
  background space-time} in string perturbation theory. Thus we are
bound to run into problems with the traditional approach to string
perturbation theory; and the superstring is no better than the bosonic
string, in this respect! 

Let us try to make this a little more precise: Exciting
a string mode in a local region of space-time must perturb space-time
geometry in a neighborhood of that region. This ``back reaction''
can be interpreted as a coherent excitation of massless
modes, such as gravitons, of an {\it arbitrary number} of further
strings\footnote{This {\it infrared problem} is analogous to, but much
  worse than the one familiar from quantum electrodynamics.}. It
really just does not make sense, ultimately, to talk of some finite
number of excited strings propagating through space-time and to
describe them as if they were individual particles in a conventional
quantum field theory on a {\it fixed} space-time manifold
$(M,g)$. Because of graviton emission and absorption --- which should
really be treated non-perturbatively --- the very concept of a single
particle (or of a finite number of particles) does not make sense in a
quantum theory coupling matter to gravitation, and it does not make
sense to treat a single particle as a quantum-mechanical
subsystem. Likewise, it cannot make sense to talk about a finite
number of strings propagating through space-time --- one must search
for a non-perturbative definition of string theory.

Of course, the problem of the gravitational interactions of very
massive string modes should be cured, ultimately, by the feature that
space-time has a quantum structure at very small scales and that very
massive modes cannot be localized in very tiny space-time
regions --- one of the reasons for introducing string theory --- as
described in Section~{\bf 3}, (3.18--21). In fact, it can be argued
[71] that string theory predicts uncertainty relations
of the kind \ $\triangle x \geq \frac{1}{\triangle p} + \alpha'
\triangle p$, which imply (3.18) when $\alpha' \approx l_P^2$.

All we can really hope to learn from the present naive formulation of
string theory is what it might tell us about ``string vacua'', i.e.,
tree-level $(\langle \Phi\rangle \to \infty)$ solutions of string
theory describing some kind of static space-time filled with static
matter fields in which {\it no events} take place (but which might
not be a classical manifold but some {\it non-commutative space} with
the property that ``sub-manifolds'' of certain dimensions, ``branes'',
have fuzzy loci as a consequence of string zero-point oscillations).
\smallskip

Let us briefly return to eq.~(7.11) for the string propagator. Of
course, it may happen that the integrand {\it does} depend on the
field (the Liouville
mode) $\phi$. One then speaks of {\it non-critical string
  theory}. Non-critical string theory is not particularly well
understood. What has been studied in some detail are models leading to
a functional $Z_\Sigma (h)$ that satisfies (7.8) and
\begin{equation}
Z_\Sigma (e^\phi\,\hat{h}) \ \approx \
e^{i\,c\,\Gamma_{\hat{h}}(\phi)} \;Z_\Sigma (\hat{h}) \ ,
\end{equation}
i.e., models which are small perturbations of conformal field
theories, for arbitrary values of $c$. One then appears to find that
either $c \leq 1$ or $c \geq 25$, otherwise the theory is
inconsistent [72]. (However, in [73,74] it is argued that there are other, in
particular discrete values of $c \in (0, 25)$ for which the theory can
be defined). Let us consider a model with $c=25$, and equality in
(7.25), and let us
assume that e.g.
\begin{equation}
Z_\Sigma (h) \ = \ \int {\mathcal D}_h\,X\,e^{i\,S_{{\rm
      tot.,}\Sigma}(X,h)} \ ,
\end{equation}
with $S_{{\rm tot.,}\Sigma} (X,h)$ as in (7.3)--(7.5). Furthermore,
we assume that $(M,g)$ is a twenty-five dimensional {\it Riemannian}
manifold, and that the fields $g_{\mu\nu}$, $B_{\mu\nu}$ and $\Phi$
satisfy eqs.~(7.21) --- more precisely, the equations expressing that
the renormalization group $\beta$--function vanishes. In this
situation, one can identify $M$ with {\it physical space}, space-time
being equal to $N := M\times{\mathbb R}$, coordinate functions on $N$
are given by
\begin{equation}
X^0 \ = \ {\rm const}\cdot\sqrt{\alpha'}\,\phi\ \ \ {\rm and}\ \ \ 
X^\mu,\ \mu=1,\ldots,25, 
\end{equation}
and the metric $(g_{\mu\nu})$ on $N$ is given by
\[
g_{00}\ \equiv\ -\,1\;,\ \  g_{0\mu}\ \equiv\ 0\;,\ \ \mu\;=\;1,\ldots,25,
\]
and $(g_{\mu\nu})$, $\mu,\nu = 1,\ldots,25$, is the metric on
$M$. This interpretation is consistent with eqs.~(7.11,12)
(with $r_{\hat{h}} = 0)$ and (7.15) (for $c=25)$; note that the {\it
  sign} of $g_{00}$ follows from the equation $c-26 = -1$. 
Thus, the Liouville mode $\phi$ appears as the {\it time
  coordinate} on (a static) space-time $N=M\times {\mathbb R}$. We
leave it open to decide whether there is something profound about this
observation; see [72].

Another approach to calculating the Feynman propagator $\triangle_F
(X_i, X_f)$, eqs.~(7.6,7), is to {\it discretize} the surface
$\Sigma$ (e.g.~one replaces $\Sigma$ by the vertices, edges and faces of a
triangulation of $\Sigma$) and to interpret ${\mathcal D} h$ as a sum
over all isomorphism
classes of triangulations of $\Sigma$; see~[75] and refs.~given
there. Finally, in accordance with the general philosophy of these
notes, one can replace $\Sigma$ (and thus $M$) by a {\it
  non-commutative space}, e.g.\ the non-commutative torus 
[76]. These last two approaches offer some chance that one will be
able to sum over different topologies of $\Sigma$.

What has remained conspicuously vague in our discussion is what the
right interpretation of the arguments $X_i, X_f$ in the Feynman
string propagator $\triangle_F (X_i, X_f)$ of eqs.~(7.6,7)
is. In quantum field theory of scalar point-particles, the arguments
$x$ and $y$ of the Feynman propagator $\triangle_F (x,y)$ in eq.~(7.2)
are points in physical space-time, perhaps augmented by internal
degrees of freedom; and, rather than
defining $\triangle_F (x,y)$ by (7.2), it can be defined as the
time-ordered solution of the {\it Schwinger-Dyson equation}
\begin{equation}
\left(\square_g + \mu^2\right)\;\triangle_F\;(x,y) \ = \ {\rm const}\cdot
  l^{-1}\,\delta_y^{(d)} (x)\,;
\end{equation}
$\delta_y^{(d)}$ is the $d$--dimensional $\delta$--function on $M$
concentrated at $y$.

String theory, being intended to be a theory of quantum gravity,
should {\it not} be formulated in a way that refers to any specific
choice of a target space-time $M$ ({\it ``background
  independence''}). Thus, we are actually {\it not} supposed to think of
$X_i$ and $X_f$ as some unparametrized loops embedded in some specific 
target space-time $M$. They really should just represent 
{\it unparametrized  loops}, decorated 
by ``internal degrees of freedom'' intrinsic to the string, but
{\it not} referring to a specific model of target space-time
$M$. Surprisingly, this remark suggests a fairly concrete analogue of
(7.28) as a stringy Schwinger-Dyson equation for
$\triangle_F(X_i,X_f)$. This is the theme of the next section, where
we shall draw on material from Sects.~{\bf 4.2} and {\bf 5.3}.

\vspace{.5cm}

\subsection{A Schwinger-Dyson equation for string Green functions\\
  from reparametrization invariance and world-sheet\\
 supersymmetry}

\noindent 
The Feynman propagator of scalar free field theory on a space-time
$(M,g)$ is a solution of the equation
\[
\left( \square_g+\mu^2\right)\,\triangle_F(x,y) \ = \ {\rm
  const}\cdot l^{-1}\,\delta_y^{(d)}\,(x) \ .
\]
Here $\square_g$, the d'Alembertian on $(M,g)$, is a {\it hyperbolic}
operator. There is no natural Hilbert space to which $\triangle_F$
belongs. Rather, $\triangle_F$ belongs to some space of {\it
  distributions} which is a module for some algebra of hyperbolic
differential operators. Concepts from the theory of (self-adjoint,
normal, \dots) operators on Hilbert 
space are, a priori, a little out of place in attempting to solve
(7.28).

Suppose, however, that $(M,g)$ is a product space
\begin{equation}
(M,g) \ = \ (N,\eta)\;\times\; (L,G),
\end{equation}
where $(N,\eta)$ is a $(d-n)$--dimensional {\it Lorentzian} manifold,
and $(L,G)$ is an $n$--dimensional (e.g.~compact) {\it Riemannian}
manifold; for example, set $d=4$, $n=2$, $N={\mathbb M}^2$ (two-dimensional 
Minkowski space), $L={\rm disk}\subset {\mathbb R}^2$, and think of a wave guide
filled with a scalar field. Then (7.28) can be solved by {\it
  separation of variables}, and we must study the eigenvalues and
eigenfunctions of the Laplace-Beltrami operator $-\triangle_G$ which
defines a positive, self-adjoint operator densely defined
on the Hilbert space $L^2 (L, d{\rm vol}_G)$. Now, this one {\it is}
a problem in the theory of operators on Hilbert space; and once it is
solved, the problem of solving (7.28) is reduced to a hyperbolic
problem on a space of distributions on $(N,\eta )$ --- as we have
learnt in school.

If we are interested in the Feynman propagator for a free field theory
of particles with {\it spin} transforming under the spinor
representation of ${\rm Spin}(d-1,1)$, eq.~(7.28) is replaced by
\begin{equation}
(D+\mu)\,S_F\,(x,y) \ = \ {\rm const}\cdot\delta_y^{(d)} (x) \ ,
\end{equation}
where $D$ is what legitimately is called {\it Dirac operator} (as
opposed to the ``Pauli-Dirac operator'' of Section~{\bf 4}), which is a
hyperbolic differential operator
acting on a space of distributional sections of the spinor bundle over 
$(M,g)$. If $(M,g)$ is of the form (7.29), eq.~(7.30) can be solved by
separation of variables: If $D^N$ denotes the {\it hyperbolic} Dirac
operator acting on distributional sections of the spinor bundle over
$(N,\eta)$ and $D^L$ denotes the {\it elliptic} Pauli-Dirac operator
  acting on smooth sections of the spinor bundle over $(L,G)$, then
\begin{equation}
D \ = \ D^N\,\otimes\, \id \ + \ \gamma\,\otimes\,D^L \ ,
\end{equation}
where $\gamma$ is a ${\mathbb Z}_2$--grading for $D^N$, and $D$ acts
on the tensor product of the two spaces of sections. The solution of
(7.30) involves studying the spectrum and the eigenfunctions of $D^L$,
which is a self-adjoint operator defined on a dense domain
in the Hilbert space ${\mathcal H}_e$ of square-integrable sections
of the spinor bundle over $(L,G)$, as discussed in Section~{\bf
  4}. Again, we encounter a problem in the theory of operators on
Hilbert space. It involves $N=1$ supersymmetric spectral data
$({\mathcal A} = C (L),\, {\mathcal H}_e,\, D^L)$.
\smallskip

We could also study a  free field theory of particles with spin
described by fields which, classically, are differential forms over
$(M,g)$. The calculation of the Feynman propagator then involves {\it
  two} hyperbolic Dirac operators, ${\mathcal D}$ and $\overline{{\mathcal
    D}}$, and, in the situation described in (7.29), this problem
requires the study of $N=(1,1)$ supersymmetric spectral data, 
$({\mathcal A} = C(L),\,{\mathcal H}_{e-p},\, {\mathcal D}^L,\,
\overline{{\mathcal D}}{}^L,\,\gamma,\,\bar\gamma)$, as considered in 
Sects.~{\bf 4} and {\bf 5.2}. Then we have 
\begin{eqnarray}
{\mathcal D} \ &=& \ {\mathcal D}^N \otimes \id \ + \ \gamma \otimes 
{\mathcal D}^L\ , \nonumber\\
\overline{\mathcal D} \ &=& \ \overline{\mathcal D}{}^N \otimes \id \ + \ 
\gamma \otimes \overline{\mathcal D}{}^L \ ,
\end{eqnarray}
where ${\mathcal D}^N,\,\overline{{\mathcal D}}{}^N$ are hyperbolic and 
${\mathcal D}^L,\,\overline{{\mathcal D}}{}^L$ are elliptic. 

An arbitrary Green function $D_F (x,\ldots)$ of this theory
satisfies the equations
\begin{equation}
{\mathcal D}\,D_F(x,\ldots) \ = \ \overline{{\mathcal D}}\,D_F(x,\ldots) \
= \ 0 \ ,
\end{equation}
as long as $x$ does {\it not} coincide with any other argument of
$D_F(x,\ldots)$ and as long as the theory is a theory of {\it free}
fields.

Incidentally, there are Feynman path integral expressions for the
solutions of \break 
eqs.~(7.30,33); they can be inferred from
eqs.~(4.47) and (4.48). Their generalizations to bosonic string theory
have been discussed, in part, in Sect.~{\bf7.2}. We ask: What is the
generalization of eqs.~(7.28), (7.30) and (7.33) to bosonic or
spinning (super) string theory, respectively?

We start with the generalization of (7.28), whose solution should be
the propagator $\triangle_F(X_i, X_f)$ of eq.~(7.7). We first consider
tree-level string theory $(\langle\Phi\rangle \to \infty$ in
(7.24)). According to the discussion following eq.~(7.28), we guess
that $\triangle_F(X := X_i,\ldots)$ belongs to some, as yet
mysterious, ``space of distributions'' which is a module, ${\mathcal
  S}'$, for some, as yet mysterious, ``algebra of hyperbolic
operators''. This algebra must contain an analogue,
$\displaystyle\mathop{\square}_{\widetilde{}}$, of the d'Alembertian
$\square_g$, and one of the
equations satisfied by $\triangle_F(X,\ldots)$ must be 
\begin{equation}
\displaystyle\mathop{\square}_{\widetilde{}}\;\triangle_F \,(X,\ldots)
\ = \ 0 \ ,
\end{equation}
at ``non-coinciding arguments''.

We consider {\it closed} strings. Then the module ${\mathcal S}'$
should carry a (projective) representation of the Witt algebra $W :=
{\rm Der}\,(C\,(S^1))$ of vector fields on $S^1$ (i.e., of
infinitesimal diffeomorphisms of $S^1$), which is interpreted as the
algebra of {\it infinitesimal reparametrizations of the string.} $W$
is an infinite-dimensional Lie algebra, whose complexification has a
basis $\{ l_n\}_{n\in{\mathbb Z}} \ (l_n = i\,e^{i n \sigma}
\frac{d}{d\sigma}, \sigma \in [0,2\pi))$ with structure relations
\begin{equation}
\left[\, l_n, l_m\,\right] \ = \ (n-m)\,l_{n+m} \ .
\end{equation}
This Lie algebra has {\it projective} representations, which 
are representations of a central extension of $W$, called {\it
  Virasoro algebra}, which has a basis $\{ L_n\}_{n\in{\mathbb Z}}$
satisfying the structure relations
\begin{equation}
[L_n, L_m] \ = \ (n-m)\,L_{n+m}\;+\;\frac{c}{12} \
n\,(n^2-1)\,\delta_{n+m,0} \ ,
\end{equation}
where $c$ is the central element. This element is invisible on the
subalgebra $sl_2({\mathbb R})$ of those infinitesimal M\"obius 
transformations that leave the unit circle invariant, with basis 
$\{
L_{-1}, L_0, L_1\}$. 

In quantum field theory, $\triangle_F(x,\ldots)$ is also given by
\begin{equation}
\triangle_F\,(x,\ldots) \ = \ \langle\, \T\,(A(x)\ldots)\rangle \ ,
\end{equation}
where $\langle \T(\ldots)\rangle$ denotes the time-ordered vacuum
expectation value, and $A$, the free scalar field, is an
operator-valued distribution on space-time $M$. 
In analogy to (7.37), one might expect that, in bosonic string theory,
there is a free closed-string field
$\displaystyle\mathop{A}_{\widetilde{}}$ 
which is an operator-valued distribution on the space $M^{S^1}$ of
parametrized loops $X^\mu (\sigma)$ in $M$ ($0 \leq \sigma < 2\pi$) 
such that
\begin{eqnarray}
\triangle_F\,(X_i,X_f) 
&=& {\rm const}\cdot \int {\mathcal D}_{\hat{h}} \phi\; e^{- 26
  i\,\Gamma_{\hat{h}}(\phi)}\, Z_\Sigma (e^\phi \hat{h}) \nonumber\\
&=& {\rm const}\cdot \int\limits_0^\infty dT \left( e^{i T\, 
    (\displaystyle\mathop{{\scriptstyle\square}}_{\widetilde{}} + i
    0)}\right) (X_i, X_f) \\
&=& \langle\, \T\,\big( \displaystyle\mathop{A}_{\widetilde{}}(X_i)
  \displaystyle\mathop{A}_{\widetilde{}}(X_f)\big)\,\rangle^c \
,\nonumber 
\end{eqnarray}
where $X_i = X_i (\sigma)$ and $X_f = X_f(\sigma)$, $0 \leq \sigma <
2\pi$, are loops in $M^{S^1}$. The first equation is (7.11) with the 
surface $\Sigma$ being the cylinder
\begin{eqnarray}
\Sigma &=& \left\{ \xi \equiv (\sigma,\tau) \biggm| 0 \leq \sigma <
  2\pi, \ 0 \leq \tau < \infty \right\} \\
&\sim& \left\{ z \in {\mathbb C} \biggm| 1 \leq |z| < \infty \right\}
\ . \nonumber 
\end{eqnarray}
Furthermore, $\hat{h} = \ { -1 \ 0 \choose \ \; 0 \ 1}$. \ The second
equation
says that $\triangle_F(X_i,X_f)$ is a solution of eq.~(7.34) analogous
to the solution
\begin{equation}
\triangle_F (x,y) \ = \ {\rm const}\cdot \int\limits_0^\infty dT \left(
  e^{i T l(\square_g + \mu^2 + i0)}\right) (x,y)
\end{equation}
of eq.~(7.28). The third equation says that $\triangle_F(X_i, X_f)$ is
the ``time-ordered'' vacuum expectation value of a string field
$\displaystyle\mathop{A}_{\widetilde{}}$. (Thus, one should be able to
express $\triangle_F (X_i, X_f)$ in terms of matrix elements of a {\it
  unitary} quantum-mechanical string propagator.)

Reparametrization invariance should be the statement that, for
``non-coinciding arguments'', 
\begin{eqnarray}
&& \lambda_n \bigg\langle \T\,\bigg( \displaystyle\mathop{A}_{\widetilde{}}
  \left( X^\mu \left( \sigma\right)\right)\ldots\bigg)\bigg\rangle
\klamm{120}{xxxxxxxxx}{50}\nonumber\\
&& \quad\quad = \ \bigg\langle \T\,\bigg(
  \displaystyle\mathop{A}_{\widetilde{}} \left( \left(
      l_n\,X^\mu\right)\left( \sigma\right)\right)\ldots\bigg)
\bigg\rangle  + \ \omega_n\,\bigg\langle \T\,\bigg(
  \displaystyle\mathop{A}_{\widetilde{}} \left( X^\mu
    \left(\sigma\right)\right)\ldots\bigg)\bigg\rangle \
\displaystyle\mathop{=}^{!} \ 0 \ ,
\end{eqnarray}
where the operators $\{ \omega_n \equiv 
\omega (l_n)\}_{n\in {\mathbb Z}}$ are introduced in order to 
allow for projective representations: Eq.~(7.41) is
intended to say that the space ${\mathcal S}'$ be a module for the
Witt algebra $W$ (then $\omega_n=0$ for all $n$), or for a central
extension of $W$ (i.e., for the Virasoro algebra). Thus the operators
$\lambda_n$ representing $l_n$ on ${\mathcal S}'$ should satisfy the
relations
\begin{equation}
\left[ \lambda_n, \lambda_m\right] \ = \ (n-m)\lambda_{n+m} \ + \
\frac{c}{12} \ n (n^2-1) \delta_{n+m,0}
\end{equation}
(and in addition, if the theory is parameter-space parity invariant, $c=0$).

We conclude that $\triangle_F$ is an element of some module ${\mathcal
  S}'$ solving the equations
\begin{equation}
{\displaystyle\mathop{\square}_{\widetilde{}}\tau(X},\ldots)
\klamm{55}{xxxl}{30}\
=\ 0 \ ,\  \quad {\lambda_n \tau(X}, \ldots) \klamm{55}{xxxx}{25}\ = \ 0 \ ,
\end{equation}
for all $n\in{\mathbb Z}$, again at ``non-coinciding''
arguments. According to our previous discussion in Sects.~{\bf 4.2}, 
eqs.~(4.86--89) and {\bf 5.3}, eqs.~(5.116--125), the
problem of solving eqs.~(7.43) can be viewed as a problem in {\it BRST 
  cohomology}. But, in order to find a nilpotent BRST operator 
Q$_{\rm BRST}$ of which $\triangle_F$ will represent a cohomology
class, we must first determine the Lie algebra ${\mathcal
  G}_{\displaystyle\mathop{\scriptscriptstyle{\square}}_{\widetilde{}}}$ generated by $\{\,
\displaystyle\mathop{\square}_{\widetilde{}};\;\lambda_n\,\}$. Logically, 
we do {\it not} seem to have enough data to
find a unique solution for ${\mathcal
  G}{\displaystyle\mathop{\scriptscriptstyle{\square}}_{\widetilde{}}}$ ! But our
discussion between (7.38) and (7.43) suggests a solution: If we define
\[
\triangle_F^{(\tau)} (x,y) \ = \ {\rm const}\cdot \int\limits_\tau^\infty
dT \left( e^{i T l (\square_g + \mu^2 + i0)}\right) (x,y)
\]
then
\[
- il^{-1} \ \frac{d}{d\tau} \ \triangle_F^{(\tau)} (x,y) \biggm|_{\tau=0} \
= \ (\square_g +\mu^2) \triangle_F (x ,y) \ = \ 0
\]
at non-coinciding arguments.
Likewise, we interpret $\triangle_F(X_i,X_f)$ as a {\it time-ordered
  Green function} of the operator
$\displaystyle\mathop{\square}_{\widetilde{}}$, as in eq.~(7.38), and
define
\[
\triangle_F^{(\tau)} (X_i,X_f) \ = \ {\rm const}\cdot
\int\limits_\tau^\infty dT\,\left( e^{i
    T(\displaystyle\mathop{\scriptstyle{\square}}_{\widetilde{}} +
    i0)}\right) (X_i, X_f) \ .
\]
Then we find that
\begin{equation}
-\,i\;\frac{d}{d\tau} \ \triangle_F^{(\tau)} (X_i,X_f)\biggm|_{\tau=0}
\ = \ \displaystyle\mathop{\square}_{\widetilde{}}
\triangle_F^{(\tau)} (X_i, X_f)\biggm|_{\tau=0} \ = \ 0 \ ,
\end{equation}
at ``non-coinciding'' arguments. The path integral representation of
$\triangle_F^{(\tau)} (X_i, X_f)$ involves the cylinder
\begin{eqnarray}
\Sigma_\tau &=& \left\{ \xi = (\sigma,\tau') \bigm| 0 \leq \sigma <
  2\pi, \ \tau \leq \tau' < \infty \right\} \nonumber\\
&\sim& \left\{ x \in {\mathbb C} \bigm| e^\tau \leq |z| < \infty
\right\} \ .
\end{eqnarray}
Thus $\displaystyle\mathop{\square}_{\widetilde{}}$ represents the
generator of the dilatation
\begin{equation}
z \ \longmapsto \ e^{\tau}z \ , \quad \tau\;>\; 0 \ , 
\end{equation}
of the complex plane on the space ${\mathcal S}'$. We already know that
the operators $\lambda_n$, $n\in{\mathbb Z}$, represent complex vector
fields on the circle $\{ e^{i\sigma}\, |\, 0 \leq \sigma < 2\pi\}$ \ as
operators on ${\mathcal S}'$. 
In particular, since $\lambda_0$ represents the generator
of a uniform rotation $e^{i\sigma} \mapsto e^{i(\sigma+\theta)}$, the operator
$\frac 1 2 (\displaystyle\mathop{\square}_{\widetilde{}} + \lambda_0)$ 
generates translations along the light rays $\{ \tau - \sigma = {\rm const.}\}$
\[
\tau+\sigma \longmapsto \tau+\sigma+\theta, \ \quad \tau-\sigma 
\longmapsto\tau-\sigma \ ,
\]
while $\frac 1 2 \,\big(\displaystyle\mathop{\square}_{\widetilde{}}
  - \lambda_0\big)$ represents the generator of  
\[
\tau + \sigma \longmapsto \tau + \sigma \ , \ \quad \tau - \sigma 
\longmapsto\tau - \sigma + \theta \ .
\]
Now we can guess a solution to the problem of determining the Lie
algebra ${\mathcal
  G}_{\displaystyle\mathop{\scriptscriptstyle{\square}}_{\widetilde{}}}$
generated by $\{ \displaystyle\mathop{\square}_{\widetilde{}};
\lambda_n, n\in {\mathbb Z}\}$: There are {\it two} Virasoro algebras,
${\rm Vir}$ and $\overline{\rm Vir}$, with bases $\{
L_n\}_{n\in{\mathbb Z}}$ and $\{\bar{L}_n\}_{n\in{\mathbb Z}}$  such that
\begin{equation}
\displaystyle\mathop{\square}_{\widetilde{}}\;=\;L_0
+\bar{L}_0\;+\;{\rm const.}\,, \ \quad \lambda_n\;=\;L_n - \bar{L}_{-n} \ ,
\end{equation}
and the generators $L_n^\#$ (which denotes $L_n$ or $\bar{L}_n\,$) 
satisfy the relations
\begin{equation}
\left[ L_n^\#, L_m^\#\right] \ = \
(n-m)\,L_{n+m}^\#\,+\,\frac{c^\#}{12}\ n\,(n^2-1)\,\delta_{n+m,0} \ ,
\ \left[ L_n, \bar{L}_m\right]\;=\;0 \ ,
\end{equation}
for all $n,m\in{\mathbb Z}$. The operators $L_n$ and $\bar{L}_n$ represent 
generators of reparametrizations $\tau+\sigma \mapsto f_+(\tau+\sigma)$   
and $\tau - \sigma
\mapsto f_- (\tau-\sigma)$, respectively, on the space ${\mathcal S}'$. 
Thus, suitable combinations of the operators $L_n^\#, n\in{\mathbb Z}$, 
generate the conformal semi-group of maps from $\{ z^\# \biggm| |z^\#|\geq1\}$
into itself. Clearly, the operators $L_0 = \frac 1 2 \
(\displaystyle\mathop{\square}_{\widetilde{}} + \lambda_0 + {\rm
  const.})$, \ $\bar{L}_0 = \frac 1 2 \
(\displaystyle\mathop{\square}_{\widetilde{}} - \lambda_0 + {\rm
  const.})$ and $\{ \lambda_n\}_{n\in{\mathbb Z}}$ as in eq.~(7.47)
provide a representation of ${\rm Vir} \times \overline{\rm Vir}$ on
${\mathcal S}'$.

\smallskip

Thus, in order to solve the equations (7.43), we 
must introduce {\it two} BRST operators, $Q_{\rm BRST}$ and
$\overline{Q}_{\rm BRST}$, whose form is determined by comparing formulas
(4.68), (4.86) and (7.48):
\begin{equation}
Q_{\rm BRST}^\#\ = \ \sum_{n\in{\mathbb Z}}
c_n^\#\,L_{-n}^\#\,-\,\frac 1 2 \;\sum_{n,m\in{\mathbb Z}}
(n-m)\,\hbox{\bf:}\,c_{-n}^\#\,c_{-m}^\#\,b_{n+m}^\#\,\hbox{\bf:}\,- a^\#\,c_0^\# \ ,
\end{equation}
where the double colons denote standard Wick ordering (move operators
with index $n>0$ to the right of operators with index $m<0$, using
anti-commutativity), and $a^\#$ is a constant arising from an
ambiguity in the definition of Wick ordering [29]. The operator $T^\#$
determining the degree of differential forms is given by
\begin{equation}
T^\#\ = \ \frac 1 2 \, \left( c_0\, b_0\, -\, b_0\, c_0\right)\;+
\sum_{n=1}^\infty \left( c_{-n}\, b_n\, -\, b_{-n}\, c_n \right) \ .
\end{equation}
Since ${\rm Vir}$ is an {\it infinite-dimensional} Lie algebra
and because of Wick ordering ambiguities, it is {\it not} automatic
that
\begin{equation}
\left( Q_{\rm BRST}^\#\right)^2 \ = \ 0 \ .
\end{equation}
The condition for eq.~(7.51) to hold turns out to be 
\begin{equation}
c^\# \ = \ 26 \ , \quad a^\# \ = \ 1 \ ,
\end{equation}
see e.g.~[29], and compare to (7.11,12). The solution 
$\triangle_F (X_i, \cdot\,)$ of eqs.~(7.43) must be a {\it cohomology
  class} of the double complex
\[
\left( {\mathcal S}' \otimes \Lambda ( {\rm Vir}^*\,) \otimes
  \Lambda (\overline{\rm Vir}^*\,)\,; \ Q_{\rm BRST}\, , \
  \bar{Q}_{\rm BRST}\right) 
\]
and, upon closer examination (see [29]), it must have degree 
$\left( -\;\frac 1  2\, , \;-\;\frac 1 2 \,\right)$, i.e.\ 
\begin{equation}
{T\,\triangle_F(X_i},X_f)\klamm{68}{xxxxxll}{30}\;=\;
{\overline{T}\,\triangle_F( X_i}, 
X_f)\klamm{68}{xxxxxll}{30}\;=\; -\,\frac 1 2 \ \triangle_F ( X_i, X_f) \ .
\end{equation}
Physicists call the eigenvalues of $T$ and $\overline{T}$ ``ghost
numbers''. It is not very difficult to 
determine the cohomology classes of $Q_{\rm BRST}$ and $\overline{Q}_{\rm
  BRST}$ of ghost number \ $\left( -\,\frac 1 2\, ,-\, \frac 1
  2\right)$, see [29]: As a functional of $X_i$, $\triangle_F (X_i,
\cdot)$ must be a solution of the equations
\[
{c^\#_n\,\tau(X_i},\ldots)\klamm{58}{xxxx}{30}\ = \
{b^\#_{n-1}\,\tau(X_i},\ldots)\klamm{62}{xxxxl}{30} \ = \ 0, 
\quad {\rm for \ all}\ n > 0
\]
and
\[
(L_n^\#-{\delta_{n,0})\,\tau\,(X_i},\ldots) \klamm{95}{xxxxxxxxxxl}{30}\ = \
0, \quad {\rm for \ all} \ n\geq 0 \ ,
\]
where $\tau\,(X_i,\ldots) \in {\mathcal S}' \otimes \Lambda ({\rm Vir}^*)
\otimes \Lambda (\overline{\rm Vir}^{\,*}\,)\,$.  
Likewise, we must have that 
\[
\tau\,(\ldots, {X_f)\,c_n^\#}\klamm{25}{xxl}{1}\  = \ \tau\,(\ldots,
{X_f)\,b_{n+1}^\#} \klamm{35}{xxxl}{10} = \ 0 \quad {\rm for \ all} \ n<0
\]
and
\[
\tau\,(\ldots, {X_f)(L_n^\#}-\delta_{n,0})\klamm{65}{xxx}{40}\ = \ 0, 
\quad {\rm  for \ all} \ n\leq 0 \ .
\]
{}From these equations one can derive that eqs.~(7.43) are satisfied by
$\triangle_F (X_i, X_f)$, at non-coinciding arguments, with
$\displaystyle\mathop{\square}_{\widetilde{}}$ and $\lambda_n$ as in
eqs.~(7.47). If one insists that these equations hold for {\it all}
arguments the solution is {\it not} a string propagator, but it would be 
a {\it  two-string Wightman distribution}. Green functions of {\it
  interacting} string theories are expected to be solutions of {\it
  inhomogeneous versions} of eqs.~(7.43); see [80].
\medskip

The data $({\mathcal S}',
\displaystyle\mathop{\square}_{\widetilde{}}, {\mathcal
  G}_{\displaystyle\mathop{\scriptscriptstyle{\square}}_{\widetilde{}} })$
are analogous to spectral data $({\mathcal A}, {\mathcal H},
\triangle)$ of {\it non-commutative metric spaces}, as described in
point (1) of the introduction to Section~{\bf 5} and generalized in
Sect.~{\bf 5.3}. But there are important differences: The module
${\mathcal S}'$ for ${\mathcal
  G}_{\displaystyle\mathop{\scriptscriptstyle{\square}}_{\widetilde{}} } $
is a space of {\it distributions} and is not equipped with a positive
semi-definite inner product, while ${\mathcal H}$ is a Hilbert space;
the operator $\displaystyle\mathop{\square}_{\widetilde{}}$ is
hyperbolic, while $\triangle$ is elliptic. Moreover, in the data
$({\mathcal S}', \displaystyle\mathop{\square}_{\widetilde{}},
{\mathcal
  G}_{\displaystyle\mathop{\scriptscriptstyle{\square}}_{\widetilde{}} })$,
we have not specified an algebra ${\mathcal A}$, yet, on which the
Witt- or Virasoro algebra acts as an algebra of infinitesimal
reparametrizations. 
\smallskip

At this point, we should recall our brief discussion of {\it
  separation of variables} after eq.~(7.29) and in (7.31,32). 
In analogy to that discussion, we propose the following {\it
  definition}:
\smallskip

We say that, in solving eqs.~(7.43) for the string propagator
$\triangle_F(X_i,X_f)$, one can {\it separate variables} iff
\begin{equation}
\displaystyle\mathop{\square}_{\widetilde{}}\;=\;L_0+\bar{L}_0\;,
\quad \lambda_n\;=\;L_n-\bar{L}_{-n}\,,\quad \ n\in{\mathbb Z} \ ,
\end{equation}
with
\begin{equation}
L_n^\#\;=\;L_n^{\#\,e} \otimes \id\;+\;\id \otimes L_n^{\#\,i}\ 
\end{equation}
(where $L_n^\#$ denotes $L_n$ or $\bar{L}_n$). The sets 
$\{ L_n^{e,i}\}_{n\in{\mathbb Z}}$ and $\{
\bar{L}_n^{e,i}\}_{n\in{\mathbb Z}}$ span commuting Virasoro algebras
${\rm Vir}^{e,i}$ and $\overline{\rm Vir}^{\,e,i}$ with central
charges $c^e, c^i$ and $\bar{c}^e, \bar{c}^i$, respectively, such
that
\begin{equation}
{\rm (i)} \phantom{mm} c^e\;+\;c^i\;=\;26 \ , \quad\ 
\bar{c}^e\;+\;\bar{c}^i\;=\;26 \ , \phantom{ZeichnungZeichnungZeichn}
\end{equation} 
and\\
\phantom{and} \
(ii)  \ the commuting Virasoro algebras ${\rm Vir}^i$ and
$\overline{\rm Vir}^{\,i}$ are {\it unitarily} represented on a {\it
  Hilbert space} ${\mathcal H}^i$, i.e.,
\begin{equation}
\left( L_n^i\right)^* \ = \ L_{-n}^i\;,\quad \ \left( \bar{L}_n^i\right)^*
\ = \ \bar{L}_{-n}^i \ ,
\end{equation}
for all $n\in{\mathbb Z}$, where $^*$ is the adjoint for operators on
the Hilbert space ${\mathcal H}^i$.

One usually also requires that
\begin{equation}
c^i \ = \ \bar{c}^i \ .
\end{equation}

It follows from (7.57) that $c^i, \bar{c}^i \geq \frac 1 2 $ and that
$L_0$ and $\bar{L}_0$ are positive operators on ${\mathcal H}^i$. The
module ${\mathcal S}'$ is then a tensor product, ${\mathcal S}'= 
{\mathcal S}^{e'}\otimes {\mathcal H}^i$, and the solution of (7.43), 
more precisely of the equations
\begin{equation}
{Q_{\rm BRST}^\#\,\tau(X},\ldots)\klamm{68}{xxxxxll}{30}=\;0\,, \quad \ 
{T^\#\,\tau(X},\ldots)\klamm{58}{xxxll}{30}=\;-\;\frac 1 2 \ \tau\,(X,\ldots)
\ ,
\end{equation}
requires the study of the unitary representations of ${\rm Vir}^i$ and
$\overline{\rm Vir}^{\,i}$ on ${\mathcal H}^i$ and, in particular, of
the spectrum and the eigenvectors of $L_0$ and $\bar{L}_0$.
\smallskip

The data $\left( {\mathcal H}^i, \{ L_n^{i\#}\}_{n\in{\mathbb
      Z}}\right)$ could come from a {\it unitary conformal field
  theory}, as discussed in the next section. In this case, the
mathematical problem to be studied is to understand in how far a
unitary conformal field theory determines a (generally
non-commutative) Riemannian space $(L,G)$ describing the {\it geometry
  of ``internal degrees of freedom''} of a tree-level string
theory. This is the problem addressed in refs.~[78,24]. Thus,
apparently, unitary conformal field theories take the place of the
spectral data
\[
\left( {\mathcal A}\;:=\;C(L), \ {\mathcal H}\;:=\; L^2 \left( L,d {\rm
      vol}_G\right)\, , \ \triangle\;=\;-\,\triangle_G\right) \ ,
\]
which appear in the solution of the Schwinger-Dyson equation (7.28) in
the situation described in (7.29).

If the Virasoro algebras ${\rm Vir}^e$ and $\overline{\rm Vir}^{\,e}$
describe string propagation in an ``external'' Minkowski space
$(N,\eta)$ and if the string theory is non-chiral (left- and right
moving sectors are isomorphic) then eqs.~(7.43) and (7.59) imply that the 
{\it  mass} $m$ of a string mode is given by the formula
\begin{equation}
m^2 \ = \ h\;+\;\bar{h}\;+\;n\;-\;2 \ ,
\end{equation}
where $h^\# \in {\rm spec}\, L_0^{i\#}$ and $n=0,1,2,\ldots$ ;
the contribution $-2$ on the r.s.\ of (7.60) comes from the equation
$a^\# =1$, see (7.49,52). Unfortunately, 
because of this $-2$, it could happen that
$m^2<0$, i.e., a tachyon appears. This problem is eliminated in {\it
  superstring theory}. In fact, the analysis of tree-level bosonic
string theory just presented can be extended to superstring theory. 
\medskip

A string theory with $N=1$ supersymmetric data $({\mathcal S}',
\displaystyle\mathop{D}_{\widetilde{}}, {\mathcal
  G}_{\displaystyle\mathop{\scriptstyle{D}}_{\widetilde{}}})$,
generalizing the free field theory of particles with spin as
considered in eq.~(7.30) is {\it heterotic string theory}. We refer
the reader to [29] and references given there for details. On the {\it
  Ramond sector} of this theory, one identifies the generalized Dirac
operator $\displaystyle\mathop{D}_{\widetilde{}}$ with e.g.~a
left-moving Ramond generator $G_0$, 
and sets
\begin{equation}
L_0 \ := \ \displaystyle\mathop{D}_{\widetilde{}}\,^2 \ + \ \frac{c}{24}
\ ,
\end{equation}
for some constant $c$ which will turn out to be the central charge of
a super-Virasoro algebra. If $\{ \lambda_n\}_{n\in{\mathbb Z}}$ are
the operators representing reparametrizations of the parameter
space $S^1$ of a closed string on the module ${\mathcal S}'$ (see
eq.~(7.41)), one sets, for arbitrary $n\in{\mathbb Z}$, 
\begin{eqnarray}
&& [\, \lambda_n, \displaystyle\mathop{D}_{\widetilde{}}\,] \ \equiv\
[\,\lambda_n,G_0\,]\;=:\;\frac n 2 \ G_n\ ,\phantom{ZeichnungZeichnung}\nonumber\\
\hskip-4cm{\rm and} \phantom{Zeichnung}&& \\
&& \{ G_n, G_m\} \;=:\; 2\,L_{n+m}\;+\;\frac c 3 \, \left(n^2-\frac 1 4\right)
\,\delta_{n+m,0} \ . \nonumber
\end{eqnarray}
One then demands (or, under suitable hypotheses, {\it proves}) that
the operators $\{ G_n, L_n\}_{n\in{\mathbb Z}}$ obey the additional 
commutation relations
\begin{eqnarray}
&&\left[\, L_n, L_m\,\right] \ = \ \left(n-m\right)\,L_{n+m} \ + \
\frac{c}{12} \ n\left( n^2-1\right)\,\delta_{n+m,0} \ , \nonumber\\
&&\left[\, L_n, G_m\,\right] \ = \ \left( \frac n 2 \ -m\right)\,G_{n+m} \ , 
\end{eqnarray}
which together with (7.62) define the {\it super-Virasoro algebra}\ $\,s{\rm
Vir}$. Finally one introduces
\[
\bar{L}_n \ = \ L_{-n} \, - \, \lambda_{-n} \ ,
\]
and it follows from (7.42) and (7.63) that $\{ \bar{L}_n\}$ generate a
second Virasoro algebra $\overline{\rm Vir}$ with some central
charge $\bar{c}$. Thus
\begin{equation}
{\mathcal G}_{\displaystyle\mathop{\scriptstyle{D}}_{\widetilde{}}} \
= \  s{\rm Vir} \ \times \ \overline{\rm Vir} \ .
\end{equation}

The propagator $S_F (X_i, X_f)$ of tree-level heterotic string
theory is a solution of the equations 
\begin{equation}
{L_n\,\tau(X},\ldots) \klamm{58}{xxxll}{30} = \ 0\,, \quad
{G_n\,\tau(X},\ldots) \klamm{58}{xxxll}{30} = \ 0\,, \quad
{\bar{L}_n\,\tau(X},\ldots) \klamm{58}{xxxll}{30} = \ 0 \ ,
\end{equation}
for all $n\in{\mathbb Z}$, at ``non-coinciding arguments''. As
outlined in Sect.~{\bf 5.3}, the problem of solving eqs.~(7.65) can be
reformulated as a problem in BRST {\it cohomology}: Let \ ${\rm Vir}
\equiv s{\rm Vir}_{\rm even}$ be the Virasoro subalgebra contained
in $s{\rm Vir}$ spanned by $\{ L_n\}_{n\in{\mathbb Z}}$, and let
$s{\rm Vir}_{\rm odd}$ be the subspace of $s{\rm Vir}$ spanned by
$\{ G_n\}_{n\in{\mathbb Z}}$. We consider the module
\begin{equation}
\underline{{\mathcal S}}' \ := \ {\mathcal S}' \otimes \Lambda\bigl(
 ( s{\rm Vir}_{\rm even})^*\bigr) \otimes S\bigl(( s{\rm Vir}_{\rm
    odd})^*\bigr) \otimes \Lambda\bigl(\overline{\rm Vir}^{\,*}\bigr) \ .
\end{equation}
Here $\Lambda\bigl((s{\rm Vir}_{\rm even})^*\bigr)$ and 
$\Lambda \bigl(\overline{\rm  Vir}^{\,*}\bigr)$ 
are {\it anti-symmetric} Fock spaces carrying the Fock
  representation of the canonical anti-commutation relations
\begin{equation}
\left\{ c_n^\#, c_m^\#\right\} \ = \ \left\{ b_n^\#, b_m^\#\right\} \ = \
0\,,\quad \ \left\{ c_n^\#, b_m^\#\right\} \ = \ \delta_{n+m,0} \ ,
\end{equation}
and $S\bigl((s{\rm Vir}_{\rm odd})^*\bigr)$ is a {\it symmetric Fock space}
carrying the Fock representation of the canonical commutation
relations
\begin{equation}
\left[ \gamma_n, \gamma_m\right] \ = \ \left[
  \beta_n,\beta_m\right]\,=\, 0\,,\quad \ \left[
  \beta_n,\gamma_m\right]\,=\,\delta_{n+m,0} \ . 
\end{equation}
Furthermore, $c_n,b_n$ anti-commute with $\bar{c}_m, \bar{b}_m$, and
$\gamma_n,\beta_n$ commute with $c_m, \bar{c}_m, b_m$ and
$\bar{b}_m$. We define the BRST operator
\begin{eqnarray}
Q_{\rm BRST} &=& \sum_n c_n\,L_{-n} - \frac 1 2 \sum_{n,m}
(n-m)\,\hbox{\bf:}\,c_{-n}\,c_{-m}\,b_{n+m}\;\hbox{\bf:}\; -\;a\,c_0\nonumber\\
&& + \sum_{n,m} \left( \frac 3 2
  \;n+m\right)\,\hbox{\bf:}\,c_{-n}\,\beta_{-m}\,\gamma_{n+m}\;\hbox{\bf:} \\
&& +\; \sum_n \gamma^n\,G_{-n} \;-\; \sum_{n,m}
\gamma_{-n}\,\gamma_{-m}\,b_{n+m} \ , \nonumber
\end{eqnarray}
see eq.~(5.123). An operator $\overline{Q}_{\rm BRST}$ is defined as in
(7.49). Then one can prove (see [29] and references given there) that
\begin{eqnarray}
&&Q_{\rm BRST}^{\ 2} \ \vphantom{\sum}= \ 0 \ \Longleftrightarrow \ 
c \ = \ 15 \quad {\rm and} \ \quad a \ =  \ 0\ ,\quad\quad \quad \nonumber \\
&&\bar{Q}_{\rm BRST}^{\ 2} \  \vphantom{\sum^x}= \ 0 \ \Longleftrightarrow \ \bar{c} \ = \
26\quad {\rm and} \quad \ \bar{a} \ = \ 1\ ,\quad\quad\quad
\end{eqnarray}
\smallskip

\noindent
see (7.52). The space $\underline{{\mathcal S}}'$ of eq.\ (7.66) is a
${\mathbb Z} \times {\mathbb Z}$ graded double complex for $\left(
  Q_{\rm BRST}, \overline{Q}_{\rm BRST}\right)$, and the string propagator
$S_F (X_i, X_f)$ can be characterized as a {\it cohomology class} of
$\left( Q_{\rm BRST}, \bar{Q}_{\rm BRST}\right)$ of 
``ghost number'' $\left( - \frac 1 2 , - \frac 1 2 \right)\,$; compare 
also to the discussion following eq.~(7.52). 

It is well known [29] that heterotic string theory has a second
sector, the {\it Neveu-Schwarz sector}. There, the spectral
data have the form $\left( {\mathcal S}_{NS}', Q, Q^+, {\mathcal
    G}_{Q,Q^+}\right)$ with
\begin{equation}
\left\{ Q, Q^+\right\} \ =: \ 2\,L_0 \ ,
\end{equation}
and ${\mathcal G}$ is still a Virasoro algebra providing a projective
representation of infinitesimal reparametrizations of $S^1$. One sets
$Q =: G_{1/2\,}$, $Q^+ =: G_{-1/2}\,$, and defines
\begin{eqnarray}
&& \left[ \lambda_n, G_{1/2}\right] \ =: \ \frac{n-1}{2} \ G_{n+1/2} \
, \ n \in {\mathbb Z} \ ,\nonumber \\
&& \left\{ G_n, G_m\right\} \ =: \ 2\,L_{n+m}\;+\;\frac c 3 \ \left(
  n^2 - \frac 1 4 \right)\,\delta_{n+m,0} \ .
\end{eqnarray}
The algebra generated by $\{\,G_{n+1/2}\;,L_n\,\}_{n\in{\mathbb Z}}$ 
is again characterized by the relations  given in (7.62,63),
but the operators $G_n$ now have labels $n\in{\mathbb Z} + \frac 1
2$ .

In the Ramond sector, with ${\mathcal S}' =: {\mathcal S}_R'$, the
left-moving fermionic string modes have {\it periodic} boundary
conditions on parameter space $S^1$, while in the Neveu-Schwarz
sector, with ${\mathcal S}' =: {\mathcal S}_{NS}'\,$, they have {\it
  anti-periodic} boundary conditions on $S^1$. In the context of 
string theory, the existence of the
Neveu-Schwarz sector and the disappearance of tachyons from the
spectrum of modes of heterotic string theory follow from the condition
that the amplitude for closed string propagation along a toroidal
world-sheet be {\it modular invariant}. 
\smallskip

{}From our discussion in (7.31) and (7.54--58) 
one can guess how to define a notion of separation of variables in 
heterotic string theory. Separation of variables leads to the
consideration of spectral data $\left( {\mathcal H}^i, \{
  L_n^{i\#}\}_{n\in{\mathbb Z}}\right)$ defining a unitary conformal
field theory. But, in contrast to purely bosonic string theory, the
conformal field theories encountered in the study of heterotic string
theory typically have {\it supersymmetries}: In addition to the
Virasoro generators $L_n^i, \bar{L}_n^i, n\in{\mathbb Z}$, there are
Ramond (and Neveu-Schwarz) generators $G_n^i, \bar{G}_n^i$ with
$n\in{\mathbb Z}$ (resp.\ $n\in{\mathbb Z} + 1/2$), and
unitarity is the constraint that {$(G_n^{i\#})^* = G_{-n}^{i\#}$} on
the Hilbert space ${\mathcal H}^i$. 
The formula corresponding to eq.~(7.60) is
\begin{equation}
m^2 \ = \ h+\bar{h}+n -1, \quad n \ = \ 0,1,2,\ldots \ , 
\end{equation}
where $h^\#$ is an eigenvalue of $L_0^{i\#}$. Thus, in order to
identify the {\it massless} string modes corresponding to particles
like gravitons, gluons, photons, light fermions, we must study the
eigenspaces of $L_0^i$ and $\bar{L}_0^i$ corresponding to the
eigenvalue $0$ and ${1\over2}$.
\smallskip

In Sect.~{\bf 4.1} we have studied Pauli's non-relativistic quantum
theory of an electron and a positron. {\it Heterotic string theory is
  the stringy analogue of Pauli's electron}. The role of the
electromagnetic U(1)--connection $A$ in the quantum theory of Pauli's
electron is  
played by string modes transforming under a gauge group $G=SO(32)$ or
$E_8 \times E_8$; see [29]. The gauge symmetry appears in the study of
the right-moving modes (with infinitesimal reparametrizations
represented by the generators $\bar{L}_n$) which
contain a Kac-Moody current algebra at level 1 based on the group
$G$. The analogue of Pauli's positron is a heterotic string theory
with reversed roles of left- and right-moving modes. In Sect.~{\bf
  4.1}, we also studied the quantum theory of bound states,
positronium, of an electron and a positron. This provided us with
spectral data $({\mathcal A}, {\mathcal H}_{e-p}, {\mathcal D},
\overline{{\mathcal D}})$ displaying $N=(1,1)$ supersymmetry,  see
eqs.~(4.30--35), from which de~Rham-Hodge theory and
Riemannian geometry of a classical manifold could be reconstructed. A
more general analysis of the passage from $N=1$ supersymmetric
spectral data (electron and positron -- spin geometry) to $N=(1,1)$
supersymmetric spectral data (positronium -- Riemannian geometry) has
been sketched in subsection 5) of Sect.~{\bf 5.2}. That analysis
suggests that it should be possible to construct closed string theories with
$N=(1,1)$ supersymmetric data $\bigl( {\mathcal S}_R',
  \displaystyle\mathop{{\mathcal D}}_{\widetilde{}},
  \overline{\displaystyle\mathop{{\mathcal D}}_{\widetilde{}}}, {\mathcal
    G}_{\displaystyle\mathop{\scriptstyle{{\mathcal
          D}}}_{\widetilde{}}
    \overline{\displaystyle\mathop{\scriptstyle{{\mathcal D}}}}}\bigr)$ 
 (and with corresponding Neveu-Schwarz data)
from two copies of heterotic string theory with $N=1$ supersymmetric 
data. 

There are two such theories with $N=(1,1)$ supersymmetry,
the {\it type IIA} and the {\it type IIB} string theories,
distinguished from each other by different combinations of left- and
right-moving modes describing chiral fermions; see [29] and
refs.~given there. In these string theories,
\begin{equation}
{\mathcal G}_{\displaystyle\mathop{\scriptstyle{{\mathcal
        D}}}_{\widetilde{}},
    \overline{\displaystyle\mathop{\scriptstyle{{\mathcal
            D}}}_{\widetilde{}}}} \ = \ s{\rm Vir}\;\times\; s\overline{\rm
    Vir} \ ,
\end{equation}
where the super-Virasoro algebras have generators $\{ L_n^\#,
G_n^\#\,\}_{n\in{\mathbb Z}}$ that satisfy the  relations (7.62,63), with
$\displaystyle\mathop{{\mathcal D}}_{\widetilde{}} = G_0$ and
$\bar{\displaystyle\mathop{{\mathcal D}}_{\widetilde{}}} =
\bar{G}_0$. 

We are interested in calculating tree-level string Green functions
$D_F (X,\ldots)$ which are solutions of the equations
\[
\lambda_n\,\tau(X,\ldots) \klamm{58}{xxxll}{30} = \ 0\;,\  \quad
\displaystyle\mathop{{\mathcal
      D}}_{\widetilde{}}{}^\#\,\tau(X, \ldots) \klamm{62}{xxxxx}{33}\ = \ 0 
\]
(at ``non-coinciding'' arguments). Under suitable
hypotheses, these equations imply the following more precise ones:
\begin{equation}
L_n^\#\,\tau(X,\ldots)\klamm{55}{xxxll}{30}\;=\;0\;, \quad
G_m^\#\,\tau(X,\ldots)\klamm{55}{xxxll}{30}\;=\;0
\end{equation}
for all $n\in{\mathbb Z}$ , $m\in {\mathbb Z} \left( + \frac 1
  2\right)$ and where $\tau \in {\mathcal S}_R'$ (resp.\ $\tau \in {\mathcal
  S}_{NS}'$).

It is of interest to study solutions of a system of weaker
equations. We define a differential $\dd$ by setting
\[
\dd \ := \ G_0 \ - \ i\,\bar{G}_0 \ .
\]
The operators $\dd$ and $\dd^* := G_0 + i\,\bar{G}_0$ can be
interpreted as the exterior derivative and its adjoint of a centrally
extended $N=(1,1)$ supersymmetry algebra; see subsection 8) of
Sect.~{\bf 5.2}. In addition, we define
\[
\dd_n \ := \ [\,\lambda_n, \dd\,] \ = \ G_n\,-\,i\,\bar{G}_{-n} \ .
\]
In type-II string theories, the central charges $c$ and $\bar{c}$
of the left- and the right-moving super-Virasoro algebras coincide
$(c=\bar{c}=15)$. Identifying $\lambda_n$ with $L_n-\bar{L}_{-n}$ as in 
eq.~(7.47), it follows that the $\lambda_n$ satisfy the Witt algebra 
\[
[\,\lambda_n, \lambda_m\,] \ = \ (n-m)\,\lambda_{n+m} \ ,
\]
see eq.~(7.35); furthermore,
\[
[ \,\lambda_n, \dd_m\,] \ = \ \left( \frac n 2 \;- m\right)\;\dd_{n+m} \ ,
\]
and
\begin{equation}
\{ \dd_n, \dd_m\} \ = \ 2\,\lambda_{n+m} \ .
\end{equation}
These commutation relations define the ``super-Witt algebra''.

One may argue that equations (7.75) for the Green functions $D_F
(X,\ldots)$ of \break
type-II string theory are really more restrictive than
they ought to be. The correct general equations for the Green
functions $D_F (X,\ldots)$ with Ramond-Ramond boundary conditions (at
``non-coinciding arguments'') are the {\it weaker} equations
\begin{equation}
\lambda_n\,\tau(X,\ldots) \klamm{55}{xxxll}{30}\ = \ 0\,, \quad
\dd_n\,\tau(X,\ldots) \klamm{55}{xxxll}{30}\ = \ 0 \ ,
\end{equation}
for all $n\in{\mathbb Z}$. It follows from the structure relations of
the super-Witt algebra that solutions of eqs.~(7.77) are {\it
  cohomology classes} of the operator $\dd$, which is {\it nilpotent}
on the subspace of the module ${\mathcal S}_R'$ annihilated by $\{
\lambda_n\}_{n\in{\mathbb Z}}$. Among solutions of eqs.~(7.77) one
would expect to find ones describing type-II string ``solitons'' --- 
see also [85].
\smallskip

In attempting to solve eqs.~(7.75) for the (tree-level) Green
functions of type-II string theory by {\it separation of variables},
one is led to studying $N=(1,1)$ supersymmetric spectral data
\begin{equation}
\left( {\mathcal H}^i, \left\{ L_n^{i\,\#}\right\}_{n\in{\mathbb
      Z}}\;,\;\left\{ G_m^{i\,\#}\right\}_{m\in{\mathbb Z}\;(+
    1/2)}\right) 
\end{equation}
derived from an ``internal'' $N=(1,1)$ supersymmetric, unitary 
conformal field theory, as briefly studied in Sect.~{\bf 7.4}. 
The mass formula (7.73) continues to hold, with $h$ and $\bar{h}$ 
eigenvalues of $L_0^i$ and $\bar{L}_0^i$, respectively. It suggests 
that, in low-energy physics, essentially only the
eigenstates of $L_0^i$ and $\bar{L}_0^i$ corresponding to the
eigenvalues 0 and ${1\over2}$ are important; see also Sect.~{\bf 7.5}. 
\medskip
\eject

\noindent The analogy 
\begin{eqnarray}
{\rm electron, \ positron} &\longleftrightarrow& {\rm heterotic \
  string \ theory} \nonumber \\
{\rm positronium\phantom{electr}} &\longleftrightarrow& {\rm type \ IIA \
    and \ IIB \ superstring \ theory} \nonumber
\end{eqnarray}
suggests that, just as positronium can be realized as a bound state of
an electron and a positron, type IIA and IIB superstrings can be
realized as ``bound states'' of heterotic strings. This calls for the
study of interacting string theory.

Perturbative string scattering amplitudes can be calculated, in the
operator formalism studied in this section, with the help of the {\it
  Krichever-Novikov} generalizations [79] of the Virasoro and super
Virasoro algebras for Riemann surfaces of higher genus. But we shall
not get into this fairly technical subject. String perturbation theory
will not be adequate for the study of non-perturbative phenomena, such
as string theory solitons and bound states. Our best bet for
getting a first look at such phenomena is the theory of $D$--branes, 
see e.g.\ the comments in the lectures by R.~Dijkgraaf and B.~Greene, 
and in particular [85]. But the problem remains to
find a {\it genuinely non-perturbative formulation of interacting
  string theory}; see [80] for some attempts in this direction. Our
discussion in Sections~{\bf 3} and {\bf 7.1} indicates where one of
the key problems may lie: Space-time $(N,\eta)$ and internal space
$(L,G)$ (see eq.~(7.29)) will, according to the ideas of 
Section~{\bf 3}, ultimately be deformed to non-commutative spaces. 
The study of this deformation calls for a non-perturbative formulation
of string theory involving summing over string world-sheets of {\it
  arbitrary topology,} with an arbitrary number of {\it punctures.}
This sum appears to be ill-behaved [71] --- see also Part~I of [75] and
refs.~given there. It is not hard to guess why one runs into
problems: Due to the towers of Planck-scale modes of perturbative
string theory, whose recoil on space-time geometry is not properly
taken into account, the bounds (3.22) and (3.23)  on the number of
events and the dimension of local ``algebras of observables'',
respectively, described in Section~{\bf 3} 
 are {\it violated} by
perturbative string theory (which, at its outset, treats target- and
parameter space as classical). One way out of these difficulties might
be to {\it deform} the {\it parameter spaces} of string world-sheets
from classical to non-commutative Riemann surfaces, as envisaged in
[76]. One is entitled to expect that it is easier to sum over ``all
non-commutative Riemann surfaces'' than to integrate over the entire
moduli space of all classical Riemann surfaces. (But the program
alluded to, here, is still in its infancy.)

\vspace{.5cm}

\subsection{Some remarks on $M$(atrix) models}

\noindent 
A proposal inspired by $M$--theory [88] currently attracting
attention is to trade strings for higher-dimensional extended objects,
in particular membranes, with {\it non-commutative parameter
  spaces}. This proposal originates in the thesis of J.~Hoppe [81] and
in subsequent work of de~Wit, Hoppe, Nicolai and others [61], which
has recently been reinterpreted and extended by Banks, Fischler,
Shenker and Susskind [82]. In this approach, parameter space
supersymmetry is replaced by ``target space supersymmetry'', see
Sect.~{\bf 5.3}.

Let us choose a non-commutative 2-torus as a parameter space (see
Section~{\bf 6}). A basis in the ``algebra of functions'' on the
non-commutative 2-torus is given by the $N\times N$ matrices
\begin{equation}
T_{\underline{p}}^{(N)} \ := \ \frac{i}{4\pi}\;\frac N M \
q^{1/2\ p_1\,p_2}\;U^{p_1}\,V^{p_2} \ ,
\end{equation}
where $q=e^{4\pi\,i\,M/N}$ for two co-prime integers $M$ and $N$, 
$p_1,p_2 \in \{\,- \frac{N-1}{2}, -
\frac{N-3}{2},$ $\ldots, \frac{N-1}{2}\,\}$ and   
\[
U \ = \ \left( \begin{array}{llll}
1&&&\\
 &q&&0\\
&&\ddots&\\
&0&&q^{N-1}
\end{array}\right) \ , \quad
V \ = \ \left( \begin{array}{llll}
0 &1&&0\\
&\ddots&\ddots&\\
&&\ddots&1\\
1&&&0
\end{array} \right) \ ; 
\]
therefore $UV=q^{-1}VU$. One checks that 
\begin{equation}
\left[ T_{\underline{p}}^{(N)},\,
  T_{\underline{q}}^{(N)}\right] \ = \ \frac{N}{2\pi M}\;{\rm sin}\left(
    \frac{2\pi M}{N} \left( p_1q_2-p_2q_1\right)\right)
\,T_{\underline{p}+\underline{q}\,({\rm mod}\,N)}^{(N)}\ .
\end{equation}  
In the limit $N\to\infty$, $\frac M N \to 0$, these commutation
relations approach the relations
\[
\left[
  T_{\underline{p}}^{(\infty)},\,T_{\underline{q}}^{(\infty)}\right]_P
\ = \ \left( p_1 q_2- p_2q_1\right)\;
T_{\underline{p}+\underline{q}}^{(\infty)} 
\]
defining the Lie algebra of functions on the 2-torus with respect to
the obvious Poisson bracket. Integration over the non-commutative
torus is given by the {\it normalized trace} on $N\times N$
matrices. The $N\times N$ matrices $\Bigl\{
  T_{\underline{p}}^{(N)}\Bigr\}$ span $gl(N,{\mathbb C})$
. In a unitary representation of the commutation relations (7.80) one
has that
\[
\left( T_{\underline{p}}^{(N)} \right)^* \ = \
\,-\,T_{-\underline{p}}^{(N)} \ \quad {\rm and}\quad\ {\rm tr}\left(
  T_{\underline{p}}^{(N)}\right) \ = \ 0 
\]
for $\underline{p} \neq 0$. Anti-selfadjoint combinations of these
generators \ ${\rm span}\, su (N)$. In fact, $su (N)$ is the {\it
  algebra of ``infinitesimal reparametrizations''} of the
non-commutative 2-torus described by (7.79,80). The ``algebra of
functions'' on this non-commutative torus, $M_N ({\mathbb
  C})$, is denoted by ${\mathcal A}^{(N)}$.
\smallskip

In the light-cone gauge (which appears to be incompatible with Poincar\'e
covariance), a (classical) membrane model with parameter space given
by the non-commutative 2-torus ${\mathbb T}^2_{(N)}$ with 
data $({\mathcal A}^{(N)}, {\mathbb C}^N, {\rm tr}\,(\cdot))$, and a
target space corresponding to an eleven-dimensional, non-commutative 
Minkowski space $\widetilde{{\mathbb M}}^{11}$, is described by
9-tuples \break
$\{ X^1,\ldots,X^9\}$ of matrices $X^j \in {\mathcal A}^{(N)}
, \, j=1,\ldots,9$, and a \ 32-component Majorana \break
spinor $\Theta$ satisfying
$\Gamma^+\Theta = 0\,$, where $\Gamma^+$
is the \ 32$\times$32 \ Dirac matrix corresponding to the 
one-form $d(x^0-x^{10})$; the 16 non-zero, independent components
$\theta_\alpha$ of $\Theta$ are $N\times N$ matrices of 
Grassmann generators. 

One may view $\widetilde{{\mathbb M}}^{11}$ as a non-commutative space
described by a non-abelian ``algebra of functions'' ${\mathcal
  A}^{(\infty)}$ which is an infinite-dimensional $C^*$--algebra with
a trace ${\rm tr}(\cdot)$. One might expect that an unquantized
membrane with parameter space ${\mathbb T}^2_{(N)}$ embedded in
$\widetilde{{\mathbb M}}^{11}$ can be described as a ${}^*$--homomorphism
from ${\mathcal A}^{(\infty)}$ to ${\mathcal A}^{(N)}$ (in analogy to
the description of a classical sub-manifold embedded in a manifold).
However, the right idea appears to be to describe an unquantized
membrane as an embedding of ${\mathcal A}^{(N)}$ into ${\mathcal
  A}^{(\infty)}$. 

Fixing the center of mass coordinates of an unquantized membrane
amounts to imposing the conditions \ ${\rm
  tr}\,(X^j)={\rm tr}\,(\theta_\alpha)=0$ for all $j=1,\ldots,9$ and 
$\alpha=1,\ldots,16$. 
We may then expand $X^j$ and $\theta_\alpha$ in a basis $\{ T_A\}$, 
$A=1,\ldots,\;\frac{N(N+1)}{2} - 1$, of $su(N)$:  
\begin{equation}
X^j\ =\ i\,X^{jA}\,T_A\, , \quad \theta_\alpha \ = \
\theta_\alpha^A\;T_A \ ,
\end{equation}
where
\begin{equation}
{\rm tr}\,(T_A\,T_B) \ = \ -\,\delta_{AB}\;,\quad \ [T_A,\,T_B] \ = \
f_{AB}^C\,T_C \ ;
\end{equation}
$\{ f_{AB}^C\}$ are structure constants of $su(N)$, and the summation
convention is imposed. The coefficients $X^{jA}$ are real variables,
$\theta_\alpha^A$ are Grassmann variables, and $T_A^* = - T_A$ (so
that the $X^j$ are Hermitian).

Canonical quantization of the model proceeds as usual: One introduces
variables $P_j^B$ canonically conjugate to $X^{iA}$ and imposes the
commutation relations
\begin{eqnarray}
&& \left[\, X^{jA},\,P_k^B\,\right] \ = \ i\,\delta_k^j\,\delta^{AB} \ , \
\quad\left\{ \theta_\alpha^A,\,\theta_\beta^B\right\} \ = \
-\,\delta_{\alpha\beta}\,\delta^{AB} \ , \nonumber \\
&& \left[\, X^{jA},\,\theta_\alpha^B\,\right] \ = \ \left[\,
  P_j^A,\,\theta_\alpha^B\,\right] \ = \ 0 \ .
\end{eqnarray}
These commutation relations have an irreducible $^*$--representation
on a Hilbert space ${\mathcal H}^{(N)}$, which one interprets as the
space of state vectors of a quantized membrane or, perhaps more
appropriately, of $N$ {\it 0--branes} [82,85]. Let $\{
\gamma^i\}_{i=1}^8$ be 16$\times$16 symmetric Dirac matrices
generating the Clifford algebra $Cl({\mathbb R}^8)$, and $\gamma^9 :=
\gamma^1\cdots\gamma^8$. One defines self-adjoint super-charges on 
${\mathcal H}^{(N)}$ by 
\begin{equation}
D_\alpha\ := \ \sum_{A,\beta} \Bigl(
  P_j^A\,\left(\gamma^j\right)_\alpha^\beta\;+\;\frac 1 2 \
  f_{BC}^A\,X^{iB}\,X^{jC}\,\left[ \gamma_i,
    \gamma_j\right]_\alpha^\beta\Bigr) \,\theta_\beta^A 
\end{equation}
The Hilbert space ${\mathcal H}^{(N)}$ carries unitary representations
of ${\rm Spin}(9)$ and of ${\rm SU}(N)$:  ${\rm SO}(9)$ is a global
symmetry group of the target space $\widetilde{{\mathbb M}}^{11}$, and
${\rm SU} (N)$ is the group of reparametrizations of parameter space
${\mathbb T}^2_{(N)}$. The Clifford generators $\theta_\alpha^A$ and
the super-charges $D_\alpha$ transform as spinors under the adjoint
action of the representation of ${\rm Spin}(9)$ on ${\mathcal
  H}^{(N)}$; the generators $X^{jA}, \; P_j^A$ transform as vectors
under ${\rm Spin}(9)$ and in the adjoint representation under ${\rm
  SU}(N)$; the generators $\theta_\alpha^A$ transform in 
the adjoint representation of ${\rm SU}(N)$, and the super-charges
$D_\alpha$ are ${\rm SU}(N)$--invariant.

Next one computes the anti-commutators $\{ D_\alpha, D_\beta\}$ and
finds that 
\begin{equation}
\left\{ D_\alpha, D_\beta\right\} \ = \
2\,\delta_{\alpha\beta}\,H^{(N)}\;+\;
2\,X^{jA}\,(\gamma_j)_{\alpha\beta}\,L_A \ ,
\end{equation}
where $H^{(N)}$ is the light-cone Hamiltonian, i.e., (classically) 
the generator of translations along the light rays $x^0+x^{10} = 
{\rm const.,}$ and $L_A = i\,L(T_A)$ are self-adjoint generators of 
the unitary representation of ${\rm SU}(N)$ on ${\mathcal H}^{(N)}$ 
--- compare to the structure described in eq.\ (5.137).  The
space of {\it physical} state vectors of the theory is the subspace
${\mathcal H}_0^{(N)}$ of {\it reparametrization-invariant}, i.e.,
${\rm SU}(N)$--{\it invariant} vectors in ${\mathcal H}^{(N)}$. On
this subspace, the relations (7.85) reduce to
\begin{equation}
\left\{ D_\alpha, D_\beta\right\} \ = \
2\,\delta_{\alpha\beta}\,H^{(N)} 
\end{equation}
so that $\left( {\mathcal A}_0, {\mathcal H}_0^{(N)}, \left\{
    D_\alpha\right\}_{\alpha=1}^{16}\right)$ \ are $N=16$ {\it
  supersymmetric spectral data}, where ${\mathcal A}_0$ is the largest
$^*$--subalgebra of $B({\mathcal H}_0^{(N)})$ with the property that
for every $a\in{\mathcal A}_0$, $[D_\alpha, a]$ is a bounded operator
on ${\mathcal H}_0^{(N)}$ .

The light-cone Hamiltonian $H_0^{(N)} := H^{(N)}\Bigm|_{{\mathcal
    H}_0^{(N)}}$ is clearly a positive, self-adjoint operator on
${\mathcal H}_0^{(N)}$. Its spectrum covers the half-axis
$[0,\infty)$; see [61]. One expects that the spectrum of $H_0^{(N)}$
is purely absolutely continuous, except for a possible
finitely-degenerate eigenvalue at 0; see [83] for some preliminary
results.
\smallskip

The model discussed here fits nicely into the general framework
considered in Sect.~{\bf 5.3}. It is quite 
clear that ``physically relevant'' results can only be expected to
emerge in a limiting regime as $N\to\infty$ (with $\frac M N$ \
in (7.79,80) approaching 0 or an irrational number). Lots of
conjectures about such limiting regimes have recently been discussed;
see e.g.~[82,84].

Attempts to interpret these models as a formulation of some sort of
non-perturbative quantum gravity appear slightly premature: Global
symmetries of target space-time should not enter a formulation of
quantum gravity, and the ``light-cone gauge'' is not a meaningful 
concept, in general. In this respect, perturbative string theory is at
a much more advanced stage. Yet, some of the problems arising in the 
analysis of the matrix models considered above are, of course, 
interesting, at least mathematically. 

A generalization of these matrix models in the form of dimensionally
reduced super Yang-Mills theories appears in the study of $D$--branes
in superstring theory [85]. The gauge group is ${\rm U}(N)$, where $N$
is the number of $D$--branes. An action functional for $N$ parallel
$D$--branes of dimension $p<10$ can be obtained using the {\it
  Connes-Lott construction} [50]. One starts from the algebra
\[
{\mathcal A}^{(N,p)}\ :=\ C^\infty\,(M) \otimes M_N\,({\mathbb C})\ ,
\]
where $M$ is a $(p+1)$--dimensional manifold parametrizing the
world-volume of a $D$--brane, and considers $N=1$ supersymmetric
spectral data 
\begin{equation}
\left( {\mathcal A}^{(N,p)},\, {\mathcal H}^{(N,p)},\,D^{(N,p)}
\right) \ ,
\end{equation}
where ${\mathcal H}^{(N,p)}$ is the Hilbert space of square-integrable
spinors on $M$ with values in $M_N({\mathbb C})$, and the Dirac operator 
is given by 
\begin{equation}
D^{(N,p)} \ = \ D_M\;+\;\sum_{j\,=\,p+1}^9 \;\gamma^j\,X_j \ \quad
{\rm with}\ \quad
D_M \ = \ \sum_{\mu=0}^p\;\gamma^\mu\,\nabla_\mu \ ;
\end{equation}
$\nabla$ is the Levi-Civita connection on $M$, the matrices
$\gamma^0,\ldots,\gamma^9$ are 32$\times$32 Dirac matrices, and
$X_{p+1},\ldots,X_9$ are commuting $N\times N$ matrices describing the
transversal coordinates of $N$ \ $D$--branes in the ground state
configuration. The form (7.88) of the Dirac operator is derived from
(open) superstring theory [85]. A (low-energy effective) action
functional for $N$ parallel, $p$--dimensional $D$--branes can be
obtained from (7.87,88) e.g.\ by following the constructions in [50,89].
\smallskip

Of course, it is presumably not correct to describe the world-volumes
of $D$--branes as {\it classical} manifolds. Our proposal is to
replace them by {\it non-commutative spaces} described by spectral
data $\left( {\mathcal B}_p \otimes M_N ({\mathbb C}),
  {\mathcal H}, D\right)$, where ${\mathcal B}_p$ is a non-abelian
``algebra of functions'' on the world-volume. {}From these data one can
construct Yang-Mills(-Higgs) action functionals as in [5,50].
If ${\mathcal B}_p$ is a finite-dimensional matrix algebra it is not
difficult to quantize the systems described by these action
functionals using functional integrals. Reasons why the world-volumes
of $D$--branes might typically be non-commutative spaces will become
apparent in the next section.

\vfil\eject

\subsection{Two-dimensional conformal field theories}

In Sect.~{\bf 7.2}, we have observed that unitary (super-)conformal
field theories play a fundamental role in the study of string theory
vacua when one is able to ``separate variables'', see
(7.54--58,78). They ought to describe the geometry of
``internal spaces'', denoted ($L,G$) in (7.29). This idea has
motivated a program initiated in [78,24], and stimulated, in part, by
the work in [15,16,5]: to reconstruct loop space and target
space geometries from algebraic data provided by super-conformal field
theories. The observation is that, in general, those geometries are
{\it non-commutative geometries}, in the sense of Connes [5].
To develop this theme would require more
room than is left. We refer the reader to [24,78,90,107] for
various technical aspects of this program, but hasten to add that much
more technical work remains to be done. 

\bigskip

\noindent 1) \ub{Recap of two-dimensional, local quantum field theory}

\medskip

\noindent Parameter space-time is chosen to be a two-dimensional cylinder
$\Sigma$ with
coordinates $(\sigma,\tau)$, $0 \leq \sigma < 2\pi$, $\tau \in
{\mathbb R}$, equipped with a Lorentz metric \ ${- 1 \ 0 \choose \ 0 \
  1}$ . 
We consider a local, relativistic quantum field theory on $\Sigma$,
see [91], with a Hilbert space ${\mathcal H}$ of physical state
vectors carrying a continuous, unitary representation of the group of
translations 
on $\Sigma$ with infinitesimal generators $H$ ($\tau$--translations)
and $P$ ($\sigma$--translations) such that
\begin{equation}
H \;\pm\;P\;\geq\;0 
\end{equation}
(spectrum condition). It is also assumed that ${\mathcal H}$ contains
a vector $\Omega$, the ``{\it vacuum} vector'', with the property
that $(H\pm P) \Omega = 0$. The vacuum vector is assumed to be a
cyclic vector for a $^*$--algebra generated by {\it local field
  operators} $\{ \varphi_I (\xi)\}_{I\in{\mathcal J}}$ , $\xi =
(\sigma,\tau) \in \Sigma$. Local {\it bosonic} field operators satisfy
{\it locality} in the form
\begin{equation}
\left[ \varphi_I (\xi), \;\varphi_J (\eta)\right] \ = \ 0 \ ,
\end{equation}
for all $I, J$ in ${\mathcal J}$, whenever $(\xi - \eta)^2 < 0$ 
(i.e., whenever $\xi$ and $\eta$ are space-like separated). The fields
$\varphi_I (\xi)$ are operator-valued tempered distributions  
with the usual properties described in [91]. The fields $\{ \varphi_I
(f)\,\bigm|\, I \in {\mathcal J},\, f \in {\mathcal S} (\Sigma)$ , 
${\rm supp}\,f \subset
{\mathcal O}\}$ where ${\mathcal O}$ is a contractible open region
in $\Sigma$ (specifically a contractible ``diamond''), form a
$^*$--algebra ${\mathcal A} ({\mathcal O})$ of unbounded operators
defined on an invariant domain ${\mathcal D}$ dense in ${\mathcal
  H}$. By (7.89) and (7.90), the vacuum $\Omega$ is a {\it cyclic} and
{\it separating} vector for ${\mathcal A}({\mathcal O})$, [91]. 

We assume that, among the local bosonic fields  $\varphi_I (\xi)$
of the theory, there is a field $T_{\mu\nu}
(\xi)$, {\it the energy-momentum tensor} of the theory, such that
\begin{equation}
H \ =  \int\limits_{\tau = {\rm const.}} T_{00}
(\sigma,\tau)\,d\sigma\, , \ \quad P  \ =  \int\limits_{\tau={\rm const.}}
T_{01} (\sigma,\tau)\,d \sigma \ .
\end{equation} 
Wightman's reconstruction theorem [91] asserts that the entire
structure of a local relativistic quantum field theory is encoded in
its Wightman distributions 
\begin{equation}
W_{I_1\ldots I_n} \left( \xi_1, \ldots,\xi_n\right) \ := \ \bigg\langle
\Omega,\prod_{j=1}^N\,\varphi_{I_j} \,\left(
  \xi_j\right)\,\Omega\bigg\rangle \ .
\end{equation}
By (7.89), these distributions are boundary values of functions
$W_{I_1\ldots I_n} (\zeta_1,\ldots,\zeta_n)$ analytic in $\zeta_1,
\ldots,\zeta_n$ on the domain
\[
\left\{ \left( \zeta_1,\ldots,\zeta_n\right) \in {\mathbb C}^{2n} \biggm|
 {\rm Im}\left( \zeta_{j+1} - \zeta_j\right) \in V_+\right\} \ ,
\]
where $V_+ = \{ (\sigma,\tau) \bigm| \tau > |\sigma|\}$ is the forward
light cone, and by (7.90) the domain of analyticity can be extended
to the ``permuted forward tube'', ${\rm Im}( \zeta_{\pi (j+1)}
      - \zeta_{\pi(j)}) \,\in\,V_+$, $ \pi \,\in\,S_n$ ,
  which contains the {\it Euclidean region}
\begin{equation}
\left\{ \left( \zeta_1,\ldots,\zeta_n\right) \;\bigm|\; \zeta_j = \left(
    \sigma_j, i\,\tau_j\right)\,, \ \left( \sigma_j, \tau_j\right) \in
  {\mathbb R}^2\right\} \ .
\end{equation}
One defines the {\it Schwinger functions} by
\begin{equation}
S_{I_1\ldots I_n} \left( \xi_1,\ \ldots, \xi_n\right) \ := \ W_{I_1
  \ldots I_n} \left(\left( \sigma_1, i\,\tau_1\right), \ldots,
  \left(\sigma_n, i\,\tau_n\right)\right) \ .
\end{equation}
The key result concerning Schwinger functions is 
the Osterwalder-Schrader reconstruction theorem [92]. Defining
$\phi_I (\sigma,\tau) := e^{- \tau H} \varphi_I (\sigma, 0) e^{\tau
  H}$ ,  a dense set of vectors in ${\mathcal H}$ is spanned by
\begin{equation}
\psi\;:=\;\prod_{j=1}^n\, \phi_{I_j}\left( \sigma_j,
  \tau_j\right)\,\Omega \ ,\quad \ 0 < \tau_1 <\tau_2 < \ldots < \tau_n \ ,
\end{equation}
with $I_1,\ldots, I_n \in {\mathcal J}$ and $n=0,1,2,\ldots$ . The
scalar products $\langle \psi,\psi'\rangle$, with $\psi$ and $\psi'$
as in (7.95) can then be expressed in terms of the Schwinger functions
introduced in (7.94); see [92].

It is sometimes convenient (``radial quantization'') to introduce the
variables
\begin{equation}
z\;=\;e^{-\tau\,+\,i\,\sigma} \ , \quad \bar{z}\;=\; e^{-\tau\,-\,i\,\sigma}
\ , 
\end{equation}
with $(\sigma,\tau) \in {\mathbb C}^2$. The Euclidean region (7.93)
corresponds to $\bar{z}_j = z_j^*$ ($\equiv$ complex conjugate of
$z_j$) for $j=1,\ldots, n$. 

\bigskip

\noindent 2) \ub{Conformal field theory} [99]

\medskip
\noindent A relativistic quantum field theory is {\it M\"obius-invariant}
if there are positive numbers {(conformal
weights)} $h_I,\, \bar{h}_I, I\in{\mathcal J}$, such that the forms
\begin{equation}
W_{I_1\ldots I_n}\,\left( z_1, \bar{z}_1, \ldots, z_n,
  \bar{z}_n\right) \prod_{j=1}^n (d\,z_j)^{h_{I_j}}\,
(d\,\bar{z}_j)^{\bar{h}_{I_j}} 
\end{equation}
are invariant under M\"obius transformations
\begin{equation}
z_j\;\longmapsto\; \frac{a\,z_j + b}{c\,z_j + d} \ , \ \quad
\bar{z}_j\;\longmapsto\; \frac{a^*\,\bar{z}_j + b^*}{c^*\,\bar{z}_j +
  d^*} \ , \quad\ j=1,\ldots, n \ ,
\end{equation}
for \ ${a \ b \choose c \ d} \ \in {\rm SL}(2,{\mathbb C})$, for arbitrary
$I_1,\ldots,I_n$ in ${\mathcal J}$, and for all $n$, and if the
generators of the virtual representation [100] of the M\"obius group
on ${\mathcal H}$ can be expressed in terms of Fourier modes
\[
\frac{1}{2\pi} \int\limits_0^{2\pi} T_{\mu\nu}
(\sigma,0)\,e^{i\,n\,\sigma}\; d\,\sigma \ , \quad\ n = 0,\,\pm 1 \ , 
\]
of the energy-momentum tensor.

A theorem due to L\"uscher and Mack [93] says that if a local,
relativistic quantum field 
theory on $\Sigma$ is M\"obius-invariant in the sense just described,
then it is a {\it conformal field theory}, i.e.\ $(u_\pm := \tau \pm
\sigma)$
\begin{equation}
T_{\mu\nu}(\xi) \ = \left( \begin{array}{ll}
\quad 0 \quad\quad T_{++}\,(u_+) \\
T_{--}\,(u_-) \quad\quad 0 \\
\end{array} \right) \ ,
\end{equation}
with $T(\xi)\equiv{\rm tr}\,\bigl[\,T_{\mu\nu}(\xi)\,\bigr]=2T_{+-}(\xi)=0\,$ and 
\[
\frac{\partial}{\partial u_-} \ T_{++} \ = \ \frac{\partial}{\partial
  u_+} \ T_{--} \ = \ 0 \ ,
\]
and the Fourier modes
\begin{eqnarray}
L_n &=& \frac{1}{2\pi} \int\limits_0^{2\pi} T_{++} (u_+)\ e^{inu_+} \
du_+ \ , \nonumber \\
\bar{L}_n &=& \frac{1}{2\pi} \int\limits_0^{2\pi} T_{--} (u_-) \
e^{inu_-} \ du_- \ ,
\end{eqnarray}
$n\in{\mathbb Z}$, span two commuting Virasoro algebras, ${\rm Vir}$
and $\overline{{\rm Vir}}$, with structure relations as in eq.~(7.36). 
The energy-momentum tensor is a conformal tensor of
dimension~2. Recalling that $z=e^{-\tau+i\,\sigma}$, $\bar{z} =
e^{-\tau-i\,\sigma}$  in the Euclidean region
$\{ \xi = (\sigma, i\,\tau) \bigm| 0 \leq \sigma < 2\pi\ , \tau > 0 \}$, 
this motivates us to define
\begin{equation}
T(z) \ := \ z^{-2} \,T_{++} (\tau + i\,\sigma) \; , \; \overline{T}
(\bar{z}) \ := \ \bar{z}^{\,-2} \, T_{--} \,(\tau - i\,\sigma) \ .
\end{equation}
Then
\begin{eqnarray}
L_n &=& \frac{1}{2 \pi i} \oint\limits_{|z|=1} \ z^{n+1} \;T\,(z)\;d z \ ,
\nonumber\\
\bar{L}_n &=& \frac{1}{2\pi i} \oint\limits_{|\bar{z}|=1} \bar{z}^{n+1}
\;\overline{T}\,(\bar z)\;d\bar z \ .
\end{eqnarray}
Under somewhat stronger hypotheses, one can prove that, in a conformal
field theory, the domain of analyticity of the Wightman functions \
$W_{I_1\ldots I_n}\,(z_1, \bar{z}_1,\ldots,z_n.\bar{z}_n)$ \ is given
by
\begin{equation}
M_n \ \times \ \overline{M}_n \ ,
\end{equation}
where $M_n$ is the universal covering of $\left\{ z_1, \ldots,
  z_n \bigm| z_i \neq z_j \ {\rm for} \ i \neq   j \right\}$, and analogously 
for $\overline{M}_n\,$; \ see [97].
\smallskip

In conformal field theory, one should attempt 
to find {\it all} \ local fields $\psi^{(K)} (u_\pm)$ , $K \in {\mathcal
  K}_\pm$ \ for some index sets ${\mathcal K}_\pm$, with the
property that $\frac{\partial}{\partial u_\mp} \; \psi^{(K)} (u_\pm) =
0\,$, i.e., all local {\it chiral} fields (depending only on {\it
  one} of the two light-cone coordinates). Obviously, $T_{++}$ and
$T_{--}$ are examples of local chiral fields of conformal dimension
2. Local chiral fields  are completely determined by their Fourier
modes
\begin{equation}
\psi_n^{(K)} \ = \ \frac{1}{2\pi} \int\limits_0^{2\pi} \psi^{(K)}
\,(u_+) \ e^{inu_+} \ du_+ \ ,
\end{equation}
and, if $\psi^{(K)}$ is a conformal tensor of weight $(h_K,0)$ independent
of $u_-$ we find that
\begin{equation}
\psi_n^{(K)} \ = \ \frac{1}{2\pi i} \oint\limits_{|z|=1} z^{n+h_K-1} \
\psi^{(K)} (z) \;dz \ .
\end{equation}
A similar equation holds for the Fourier modes $\bar{\psi}_n^{(K)}$ of 
a chiral field $\psi^{(K)} (u_-)$, $K\in{\mathcal K}_-$, which is 
assumed to be a conformal tensor of weight $(0, \bar{h}_K)$. These
definitions make sense provided $h_K \in {\mathbb N}$, i.e., for
{\it local, chiral Bose fields}. They can be extended to local, chiral
{\it Fermi fields} \ $\psi^{(K)} (z)$ with $h_K + \frac 1 2 \in
{\mathbb N}$, after one has chosen a spin structure on the circle $0
\leq \sigma < 2 \pi$, i.e., either {\it periodic} (``Ramond'') 
or {\it anti-periodic} (``Neveu-Schwarz'') boundary
conditions. For periodic boundary conditions, the Fourier modes are
labeled by integers; for anti-periodic boundary conditions, they are
labeled by half-integers.

A chiral field $\psi^{(K)}$ is a Bose field iff
\begin{equation}
\left[ \psi^{(K)}\,(u_+)\;,\; \psi^{(K')}\,(u_+')\right] \ = \ 0 
\end{equation}  
for $u_+ \neq u_+'$, where $\psi^{(K')}$ is an arbitrary chiral Bose-
or Fermi field; then $h_K \in {\mathbb N}$. Chiral
fields $\psi^{(K)}$ and $\psi^{(K')}$ are Fermi fields iff
\begin{equation}
\left\{ \psi^{(K)} \,(u_+) \;,\; \psi^{(K')}\,(u_+')\right\} \ = \ 0 
\end{equation}
for $u_+ \neq u_+'$ ; then $h_K + \frac 1 2\; , h_{K'} + \frac 1 2
\; \in {\mathbb N}$. 

We shall always assume that there is an involution $^+$ : ${\mathcal
  K}_\pm \to {\mathcal K}_\pm$ \ such that
\begin{equation}
\left( \psi_n^{(K)}\right)^* \ = \ \psi_{-n}^{(K^+)} 
\end{equation}
or, equivalently, $\psi^{(K)} (u_+)^* = \psi^{(K^+)} (u_+)$ .

The {\it chiral algebra} ${\mathcal E}$ of a conformal field theory
on the cylinder $\Sigma$ is the {\it unital \break
$^*$--algebra} of
(generally unbounded) operators on ${\mathcal H}$ generated by $\id$
and $\{ \psi_n^{(K)} \bigm| n \in {\mathbb Z}$, $K\in {\mathcal K}_+$,
$h_K \in {\mathbb N}\}$, with $\psi_n^{(K)}$ as in (7.104,105).
The {\it anti-chiral algebra} $\bar{{\mathcal E}}$ is
defined similarly. The conformal field theory is {\it left--right
  symmetric} iff
\begin{equation}
{\mathcal E} \ \cong \ \bar{{\mathcal E}} \ ,
\end{equation}
i.e., iff ${\mathcal E}$ and $\bar{{\mathcal E}}$ are $^*$--isomorphic
$^*$--algebras.

One can define ${\mathbb Z}_2$--graded, {\it extended (anti-)chiral
  algebras} ${\mathcal C}_\alpha, \bar{{\mathcal C}}_\alpha$ with 
$\alpha$ = Ramond or $\alpha$ = Neveu-Schwarz, by including in
their definition the Fourier modes of {\it chiral Fermi fields} with
periodic or anti-periodic boundary conditions, respectively. Then
${\mathcal E}^\#$ is the even part of ${\mathcal C}_\alpha^\#$. The
algebra ${\mathcal C}_\alpha$ is the universal enveloping algebra of a
{\it graded Lie algebra} iff its generators obey relations
\begin{equation}
\left[\, \psi_n^{(K)}\,,\, \psi_m^{(K')}\,\right]_g \ = \
f_{K''}^{KK'}(n,m)\;\psi_{n+m}^{(K'')}\;+\; g^{KK'}(n)\;
\delta_{n+m,0} 
\end{equation}
for some structure constants $f_{K''}^{KK'}(n,m)$ and ``central
elements'' $g^{KK'}(n)$.

By applying the calculus of residues to 
local chiral fields one finds that a chiral algebra ${\mathcal E}$
can be equipped with a family $\{ \triangle_z \bigm| z \in {\mathbb
  C}^*\} $ of {\it co-products} $\triangle_z : {\mathcal E} \to
{\mathcal E} \otimes {\mathcal E}$, defined by
\[
\triangle_z \left( \psi_n^{(K)}\right) \ = \ \delta_z \left(
  \psi^{(K)}\right)_n \otimes \id + \id \otimes \psi_n^{(K)} \ ,
\]
where
\begin{eqnarray}
&&\delta_z \left( \psi^{(K)}\right)_n \ := \ \sum_{m=0}^\infty \left(
\begin{array}{c} n+h_K-1\\ m \end{array} \right) \ z^{n-m+h_K-1} 
\psi_{m-h_K+1}^{(K)} \ \quad{\rm for}\ n>-h_K\;\ ,\nonumber\\
&&\hskip-1cm{\rm and} \\
&&\delta_z \left( \psi^{(K)}\right)_n \ := \ \sum_{m=0}^\infty 
(-1)^m \left(
\begin{array}{c} m-n-h_K\\ m \end{array} \right) \ z^{n-m+h_K-1} \;
\psi_{m-h_K+1}^{(K)} \ \quad{\rm for}\ n\leq -h_K\;\ ,\nonumber 
\end{eqnarray}
see e.g.\ [96]. These co-products can obviously be extended to maps
\begin{eqnarray}
&&\triangle_z \ : \ {\mathcal C}_{NS} \ \longrightarrow \ {\mathcal C}_{NS} 
\otimes {\mathcal C}_{NS} \nonumber \\
&& \qquad\quad  {\mathcal C}_R \ \;  \longmapsto \ {\mathcal C}_{NS} 
\otimes {\mathcal C}_R \ . 
\end{eqnarray}
Clearly $\delta_z (\id) = 0$, so that $\triangle_z (\id) = \id \otimes \id $ .
\smallskip

Conformal field theory can be viewed as the representation theory of a
pair of a chiral algebra ${\mathcal E}$ and an anti-chiral algebra
$\bar{{\mathcal E}}$ which are always assumed to contain $U({\rm
  Vir})$, $U(\overline{\rm Vir})$, respectively --- i.e., $T^\# (z^\#)$
is among the generators of ${\mathcal E}^\#$. A unitary
$^*$--representa\-tion $j$ of ${\mathcal E}^\#$ is a $^*$--homomorphism
from ${\mathcal E}^\#$ to a $^*$--algebra of densely defined,
unbounded operators on a Hilbert space $\h_j$ such that
\begin{equation}
j \left( \psi_{\; n}^{\# (K)}\right)^* \ = \ j\,\left(\psi_{\; -n}^{\#
    (K^+)}\right) \ .
\end{equation}
Because of the relativistic spectrum condition (7.89) we only consider
{\it ``positive-energy representations'':} These are representations
$j$ of ${\mathcal E}^\#$ with the property that $j(L_0^\#)$ is a
positive self-adjoint operator 
\begin{equation}
j\,(L_0^\#) \ \geq \ 0 \ .
\end{equation}
Note that, formally, $j(L_0^\#) = \frac 1 2  (H\pm P) \bigg|_{\h_j}$ ,
which according to (7.89) must be positive operators. 

It follows from  (7.102) and (7.105) that 
\begin{equation}
\left[\, L_0^\#, \psi_n^{\#(K)}\,\right] \ = \ -\, n \psi_{\; n}^{\# (K)}
, \ \quad n \in {\mathbb Z}\ ( +\, 1/2 ) \ ,
\end{equation}
hence ${\mathcal E}$ and $\bar{{\mathcal E}}$ are ${\mathbb
  Z}$--graded algebras. Suppose that $\chi$ is a vector in $\h_j$
with the property that 
\begin{equation}
\big\langle \chi, j (L_0^\#)\,\chi\big\rangle \ \leq \ \left( h_j^\# +
  \varepsilon \right) \langle \chi, \chi \rangle \ , 
\end{equation}
where $h_j^\# := {\rm inf \ spec}\, j (L_0^\#)$ and $\varepsilon <
\frac 1 2 $ . Combining (7.115) and (7.116), we conclude that 
\begin{equation}
j\,\left( \psi_n^{\#(K)}\right)\,\chi \ = \ 0\  \quad {\rm for \
  all} \ n > 0 \ .
\end{equation}
A vector $\chi$ satisfying (7.117) is called a {\it ``highest weight
  vector''.} If the representation space $\h_j$ is separable it
follows from (7.114--117) that it can be decomposed into a
direct integral of unitary ``highest weight'' modules for ${\mathcal
  E}^\#$. 
A {\it vacuum representation} $e$ of an (anti-)chiral algebra
${\mathcal E}^\#$ is an irreducible, unitary positive-energy
representation of ${\mathcal E}^\#$ on a Hilbert space $\h_e^\#$
containing a highest weight vector $\Omega$ (the vacuum), i.e., 
\begin{equation}
e \left( \psi_n^{\#(K)}\right)\,\Omega \ = \ 0 \quad {\rm for \ all}
\ n > 0 \  
\end{equation}
for all $K \in {\mathcal K}_\pm$ which moreover is {\it M\"obius-invariant},
i.e., 
\begin{equation}
e\,(L_{\pm 1})\,\Omega \ = \ e\,(L_0)\,\Omega \ = \ 0 \ .
\end{equation}
In the following, we usually omit the symbol $e$.

A chiral algebra ${\mathcal E}^\#$ is called {\it rational} iff it
only has a {\it finite} number of irreducible unitary positive-energy
representations $j$, including a {\it unique} vacuum representation
$e$. Examples are the universal enveloping algebras of the Virasoro
algebras with central charge $c = 1 - \frac{1}{p(p+1)}\, , \,
p=1,2,\ldots\,$,  and of simply-laced Kac-Moody algebras at
integer level.

Thanks to the existence of the co-products $\triangle_z$ defined in
(7.111), the irreducible unitary positive-energy representations of a
rational (anti-)chiral algebra ${\mathcal E}^\#$ form the irreducible
objects of a semi-simple, rigid, braided $C^*$--tensor category
${\mathcal T}$, with sub-objects of direct sums (see e.g.~[98]): 
The vacuum representation $e$ plays the role of the unit in ${\mathcal
  T}$, as one derives without difficulty from (7.111). Given two
unitary positive-energy representations $j$ and $k$ of ${\mathcal
  E}^\#$, one defines their tensor product 
\begin{equation}
j \otimes_z k \ := \ (j \otimes k) \;\circ\;\triangle_z \ ,
\end{equation} 
which is independent of $z$ up to isomorphism. Then we have 
\begin{equation}
e \otimes_z j \ \cong \ j \otimes_z e\;\cong\; j \ .
\end{equation}
Moreover, given an irreducible unitary positive-energy representation
$j$, one can show (in the proper setting:\ see [101,102,103,12]) that
there exists a unique irreducible unitary positive-energy
representation $j^\vee$, the representation {\it conjugate} to $j$, 
such that $j \otimes_z j^\vee \cong j^\vee \otimes_z j$ 
contains $e$ as a sub-representation precisely once.

Let $\uI$ denote the finite set of all irreducible, unitary
positive-energy representation of a rational chiral algebra ${\mathcal
  E}$. The {\it fusion rule algebra} is the abelian algebra generated
by $\{ N_{jk}^i\}$, where $N_{jk}^i$ is the  multiplicity of $i
\in \uI$ in the tensor product representation $j \otimes_z k
\cong k \otimes_z j$, for $j$ and $k$ in $\uI$. Then
\begin{equation}
\sum_l N_{il}^m\,N_{jk}^l \ = \ \sum_l N_{ij}^l\,N_{lk}^m\;, \ \quad
N_{ij}^k\;=\;N_{ji}^k\;=\; N_{ik^\vee}^{j^\vee} \ .
\end{equation}
Given three representations $i, j$ and $k$ in $\uI$, there exist
intertwiners
\begin{equation}
V_i^\alpha \left( \chi_j, z\right)_k \ : \ \h_k \to \h_i \ ,
\end{equation}
$\chi_j \in \h_j$ , $z \in {\mathbb C}$, $\alpha = 1, \ldots,
N_{jk}^i$, such that 
\begin{equation}
i \left( \psi_{\; n}^{(K)}\right) \, V_i^\alpha\left( \chi_j,
z\right)_k \, - \, V_i^\alpha\left( \chi_j, z\right)_k\,k \left(
\psi_n^{(K)}\right)\
= \ V_i^\alpha\left(\delta_z\left(
    \psi^{(K)}\right)_n\chi_j, z\right)_k \ ,
\end{equation}
for every generator $\psi_n^{(K)}$ of ${\mathcal E}$, with $\delta_z$
as in (7.111). 

The intertwiners $V_i^\alpha (\chi_j, z)_k$ are called {\it chiral
  vertex operators} and obey braid commutation relations and fusion
equations involving braiding  matrices $R^\pm
[i,j,l,m]_{k\gamma\delta}^{n\alpha\beta}$ \ which are solutions of the
celebrated polynomial equations, see [94,95,96].

In ``radial quantization'', the product
\[
V_i^\alpha\left( \chi_j,z_2\right)_k \; V_k^\beta\left(\chi_l,
  z_1\right)_m 
\] 
is a well-defined operator from $\h_m$ to $\h_i$  provided
$|z_1| < |z_2|$ . Matrix elements of this product have an analytic
continuation $(z_1, z_2)$ along the paths $\gamma^+$ and $\gamma^-$,
with

\phantom{xx}
\[
\gamma^+ \ \longleftrightarrow\ \  z_2\,\bullet\quad\qquad\qquad\bullet\,z_1\ ,
\ \quad\quad
 \gamma^- \ \longleftrightarrow\ \ z_2\,\bullet\quad\qquad\qquad\bullet\,z_1\ ,
\] 
\vskip-1.30cm
\hbox{\hskip4.33cm\epsfbox{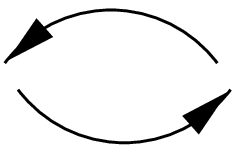}\hskip4.28cm\epsfbox{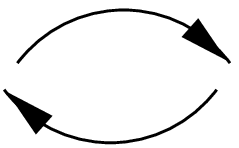}}
\bigskip

\noindent and
\begin{eqnarray}
&&\!\!\!\!\!\!\!\left[\, V_i^\alpha (\chi_j, z_1)_k\, V_k^\beta (\chi_l,
  z_2)_m\,\right]_{\gamma^\pm}\phantom{xxxx} \nonumber \\
&& \qquad\quad = \sum R^\pm \left[
  i,j,l,m\right]_{k\,\gamma\,\delta}^{n\,\alpha\,\beta}\, V_i^\gamma
(\chi_l, z_2)_n\, V_n^\delta (\chi_j, z_1)_m \ .
\end{eqnarray}
Furthermore,
\begin{eqnarray}
&& V_i^\alpha (\chi_j,z_1)_k\, V_k^\beta(\chi_l, z_2)_m \phantom{xxxx}\nonumber \\
&& \qquad\quad = \vphantom{\sum^K}\sum F [i,j,l,m]_{k\,
\gamma\,\delta}^{n\,\alpha\,\beta}\;
V_i^\gamma \left( V_n^\delta \left(\chi_j, z_1-z_2\right)_l\, \chi_l,
  z_2\right)_m \ ,
\end{eqnarray}
where $F [i,j,l,m]_{k\,\gamma\,\delta}^{n\,\alpha\,\beta}$ are the 
fusing matrices; see [94,95,96].

A chiral vertex operator $V_i^\alpha (\chi_j, z)_k$ is called {\it
  primary}
iff $\chi_j \in \h_j$ is a highest weight vector for ${\mathcal
  E}$.
It is not hard to show [99,97] that a primary chiral vertex operator
satisfies the differential equations 
\begin{eqnarray}
&&i (L_n)\, V_i^\alpha (\chi_j, z)_k - V_i^\alpha (\chi_j, z)_k\, k
(L_n) \phantom{xxxxxxxxxxxxxxxxx}\nonumber\\
&&\qquad\quad\ \ \  = \left[ z^{n+1}\, \frac{d}{dz}\,+\,z^n (n+1)
  \,h_j\right]\; V_i^\alpha (\chi_j, z)_k \ ,
\end{eqnarray}
for all $n \in {\mathbb Z}$, where $h_j$ is the eigenvalue of $j(L_0)$ \ (the
``highest weight'') corresponding to the eigenvector $\chi_j$.

{}From the chiral vertex operators of a chiral algebra ${\mathcal E}$
and an anti-chiral algebra $\bar{{\mathcal E}}$ one can attempt to
construct {\it local fields} $\varphi_{\chi_j \otimes
  \chi_{\bar{j}}}^\alpha (z,\bar{z})$ by setting
\begin{equation}
\varphi_{\chi_j \otimes \chi_{\bar{j}}}^\alpha \,(z,\bar{z}) \ := \
\sum D\left[ i,k; \bar{i},
  \bar{k}\right]_{j\bar{j}\beta\gamma}^\alpha\; 
V_i^\beta (\chi_j, z)_k \otimes
V_{\bar{i}}^\gamma (\chi_{\bar{j}}, \bar{z})_{\bar{k}} \ ; 
\end{equation}
the $D$'s are complex ``sewing coefficients''. Here \
$V_{i^\#}^\alpha ( \ldots )_{k^\#} \bigm|_{\,\h_{l^\#}} = 0$ if
$k^\# \neq l^\#$. These local fields are operator-valued distributions
from the Hilbert space 
\begin{equation}
{\mathcal H} \ = \ \bigoplus_{(k,\bar{k}) \in \Pi}
\ \h_k \otimes \h_{\bar{k}} \otimes {\mathbb C}^{\, n (k,\bar{k})} 
\end{equation}
to itself, where $\Pi$ is a subset of the product $\uI \times
\bar{\uI}$ of irreducible unitary positive-energy representations of
${\mathcal E}$ and $\bar{{\mathcal E}}$, determined by
the set of non-zero sewing coefficients $D$, and ${\mathbb C}^{\, n
  (k,\bar{k})}$ is a ``multiplicity space'' corresponding to the index
$\alpha = 1,\ldots, n(k,\bar k)$ which labels different left-right 
sewings. 

{\it Locality} is the constraint that $\varphi_{\chi_j
  \otimes \chi_{\bar{j}}}^\alpha$ and
$\varphi_{\chi_k\otimes\chi_{\bar{k}}}^\beta$ {\it commute} whenever
their arguments, $(\sigma,\tau)$ and $(\sigma',\tau')$, are space-like
separated. This constraint yields over-determined algebraic equations
for the sewing coefficients $D$ in terms of matrix elements of the
braid matrices $R^\pm$, see e.g.~[97]. Examples of solutions of these
equations can be found in [104,105] (and in the refs.~given there).
Note that, by (7.124),
\begin{equation}
\left[\, \psi_n^{(K)}\, ,
  \;\varphi_{\chi_j\otimes\chi_{\bar{j}}}^\alpha\,(z,\bar{z})\,\right] \
= \ 
 \varphi_{\delta_z (\psi^{(K)})_n\,\chi_j
  \otimes \chi_{\bar{j}}}^\alpha\,(z,\bar{z}) \ .
\end{equation}
This equation can be understood by applying eq.~(7.105) and Cauchy's
theorem to the l.s., and this calculation has originally motivated the
definition of the co-products in (7.111). 

{}From eqs.~(7.125,126) and (7.128) one can derive the so-called operator
product expansion (OPE) of two local conformal fields,  see [99]: 
Let $\chi_{i^\#} \in \h_{i^\#}$ be
an eigenvector of $i^\# (L_0^\#)$ corresponding to an 
eigenvalue $h_{i^\#} \geq 0$  for $i^\# = j^\#, k^\#, l^\#$. There
are invariant tensors $C\left( \chi_j, \chi_{\bar{j}}, \alpha \bigm|
  \chi_k, \chi_{\bar{k}}, \beta \bigm| \chi_l, \chi_{\bar{l}},
  \gamma\right)$ such that
\begin{eqnarray}
&& \hskip-.7cm\varphi_{\chi_j \otimes \chi_{\bar{j}}}^\alpha\,(z,\bar{z}) \;
\varphi_{\chi_k \otimes \chi_{\bar{k}}}^\beta\, (w, \bar{w}) \nonumber
\\
&&\phantom{Zeichn} = \ \sum C\left( \chi_j, \chi_{\bar{j}}, \alpha
  \bigm| \chi_k, \chi_{\bar{k}}, \beta \bigm| \chi_l, \chi_{\bar{l}},
  \gamma \right) \;\left( z-w\right)^{-\,h_j - h_k +h_l} \\
&&\phantom{Zeichnungg}\quad\quad \times\; \left( \bar{z} - \bar{w}\right)^{-
  h_{\bar{j}} - h_{\bar{k}} + h_{\bar{l}}}\; \varphi_{\chi_l \otimes
  \chi_{\bar{l}}}^\gamma\, \left( w,\bar{w}\right) \ , \nonumber
\end{eqnarray}
where the sum extends over a complete, orthonormal set of vectors
$\chi_{l^\#} \in \h_{l^\#}$ , over all $l$ and $\bar{l}$ and all
$\gamma$ .
The coefficients $C$ on the r.s.\ of (7.131) can be expressed in terms
of the fusing matrices $F$, the sewing coefficients $D$ and matrix
elements of chiral vertex operators.

Local commutativity of $\varphi_{\chi_l \otimes \chi_{\bar{l}}}^\gamma
\, (w,\bar{w})$ implies that $h_l-h_{\bar{l}} \in {\mathbb Z}$. It
follows from general results of local, relativistic quantum field
theory that the operators
\begin{equation}
\varphi_{\chi_j \otimes \chi_{\bar{j}}}^\alpha \, (\sigma, f) \ := \
\int\limits_{-\infty}^\infty d\tau\; f (\tau)\; \varphi_{\chi_j
  \otimes \chi_{\bar{j}}}^\alpha \, (\sigma, \tau) \; , 
\end{equation}
where $(\sigma, \tau) \in \Sigma$  and $f$  is an arbitrary Schwartz
space test function, are densely defined operators on ${\mathcal
  H}$ \  \footnote{``Smearing'' $\varphi_{\chi_j \otimes
    \chi_{\bar{j}}}^\alpha (\sigma, \tau)$ in $\sigma$, for fixed
  $\tau$, does usually {\it not} yield a well-defined
  operator!}. They generate a unital $^*$--algebra ${\mathcal F}$ of
{\it ``functions on quantized phase space over loop space''} --- compare
to eq.~(2.9). 
\smallskip

The representation-theoretic approach to local conformal field theory
outlined in this subsection 
(see [99,94,95,96,97] and refs.~given there) can be translated into
the general framework of algebraic quantum field theory [101,102,103,12,98], 
where one works with $^*$--algebras and von~Neumann algebras of
bounded operators. This offers considerable advantages in rendering
the analysis mathematically rigorous but makes the theory more
abstract. There is no room to review the algebraic approach in these
notes.

\bigskip

\noindent 3) \ub{A dictionary between conformal field theory and Lie group
  theory}
\medskip

\noindent We consider a compact, semi-simple Lie group $G$ with Lie algebra
$\hbox{{\tt g}}$ as in Sect.~{\bf 4.2}. Let ${\mathcal R} = {\mathcal
  R}_G$ denote the list of all irreducible representations of $G$, and
let ${\mathcal H}_G = L^2 (G, dg)$, where $dg$ is the Haar measure on
$G$. By the Peter-Weyl theorem, 
\begin{equation}
{\mathcal H}_G \ = \ \bigoplus_{I\,\in\,{\mathcal
    R}_G} \, W_I \otimes W_{I^\vee} \ , 
\end{equation}
\vbox{\noindent where $W_I$ is the representation space for the representation $I$; 
see (4.79). Let $\hbox{{\tt g}}_L$ (resp.\ 
$\hbox{{\tt g}}_R$) denote the Lie algebra of left (resp.\ right) invariant 
vector fields on $G$.

Comparing the summary of group representation theory presented in
Sect.~{\bf 4.2} with the review of two-dimensional conformal field
theory in the last subsection, we arrive at the following dictionary .
\def\vp{$\vphantom{\displaystyle\sum^{}}$}
\bigskip
\begin{center}
\begin{tabular}{l|ll}
$\vphantom{\displaystyle\sum_{-}}$Lie group theory\phantom{ZeichnungZeich} 
&& Conformal field
theory \\ 
\hline
\small
&& \\
compact, semi-simple &&two-dimensional (rational) \\
Lie group $G$ &&conformal field theory $\G$\\
\vp\qquad ${\mathcal H}_G$ , eq.~(7.133) && \qquad ${\mathcal H}$ , eq.~(7.129) \\
\vp\qquad\qquad$U (\hbox{{\tt g}}_L)$ && \qquad\qquad${\mathcal E}$ \\
\vp\qquad\qquad$U (\hbox{{\tt g}}_R)$ && \qquad\qquad$\bar{{\mathcal E}}$ \\
\vp\qquad\quad ${\mathcal R}_L \cong{\mathcal R}$  && \qquad\qquad$\uI$ \\
\vp\qquad\quad${\mathcal R}_R \cong{\mathcal R}^\vee\, \cong \, {\mathcal R}$ &&
\qquad\qquad$\bar{\uI}$\\ 
\vp$\triangle\,:\,U(\hbox{{\tt g}}) \to U(\hbox{{\tt g}})\otimes U(\hbox{{\tt
g}})$ 
&& \quad$\triangle_z$  \ \  as in (7.111) \\
$\hbox{{\tt g}} \ni X \mapsto X \otimes \id + \id \otimes X$ && \\
\vp\qquad$N_{IJ}^K$\ \   as in (4.80) && \quad$N_{ij}^k$\ \   as in (7.122) \\
\vp\quad $V^\alpha (I, J|K)$ , eq.~(4.81) &&\quad $V_i^\alpha (\chi_j, z)_k$ ,
eq.~(7.123) \\\qquad
&&\\
\vp\quad$C (I, J|k)$ , eq.~(4.82) && matrix elements of 
          $\varphi_{\chi_j \otimes \chi_{\bar{j}}}^\alpha(z, \bar{z})$ , \\
&&see eq.~(7.128);\   OPE, eq.~(7.131) \\
$C(G)$ \ (algebra of conti- &&formal ``algebra''\vp\ of local fields \\
nuous functions on $G$) &&$\varphi_{\chi_j \otimes
    \chi_{\bar{j}}}^\alpha (\sigma, 0) $ \\
\vp Laplace--Beltrami &&$H = L_0 + \bar{L}_0$  (Hamilton \\
operator $\triangle_G$ on \ ${\mathcal H}_G$ && operator on
${\mathcal H}$)\\
\vp algebra ${\mathcal F}_\hbar$  of ``functions  && algebra
${\mathcal F}$ defined below \\
on quantized phase space'' &&eq.~(7.132) \\
over $G$, see eq.~(2.9) && \\
\vp spectral data $\left( {\mathcal F}_\hbar , {\mathcal H}_G , \triangle_G
\right)$ 
&&spectral data 
$\left( {\mathcal F}, {\mathcal H}, H = L_0 + \bar{L}_0\right) $\\
\vp Hilbert space ${\mathcal H}_{e-p}$ of && state space of an $N=(1,1)$ \\
square-integrable diffe- && superconformal extension of $\G\,$,\\
rential forms on $G$ && see Sect.\ {\bf7.6} below\\
\vp Operators ${\mathcal D}, \overline{{\mathcal D}}$ defined && Ramond
generators $G_0, \overline{G}_0$ \\
in eq.~(4.76)  &&of Sect.\ {\bf7.6} below\\ 
&&\\ \hline
\end{tabular}
\normalsize
\end{center}
}

Since the parameter space-time of a conformal field theory $\G$ is the
cylinder $\Sigma$, it is plausible that the spectral data
$({\mathcal F}, {\mathcal H}, H)$ of $\G$ describe the
(non-commutative) geometry of some loop space $M^{S^1}$, where, for
a {\it rational} conformal field theory, ``target space'' $M$ can be
expected to be some {\it compact} (non-commutative) space. In other
words, a two-dimensional conformal field theory is 
always a conformal, non-linear $\sigma$--model of maps $X:\Sigma \to
M$, but the target space $M$ might, in general, be a {\it
  non-commutative space} in the sense of Connes [5]. This is the
claim advanced and partially substantiated in [78,24]. 

In fact, the local fields
$\varphi_{\chi_j\otimes\chi_{\bar{j}}}^\alpha(\sigma, \tau = 0)$
--- which, unfortunately, are not well-defined --- could be interpreted as
``functions on loop space $M^{S^1}\,$'' in a natural way. To see
this, we consider a compact, smooth classical manifold, $M$. When
equipped with Tychonov's topology, loop space $M^{S^1}$ is a compact
Hausdorff space. The Stone-Weierstrass theorem then says that any set
of continuous functions on $M^{S^1}$ that separate points in $M^{S^1}$
(i.e., that distinguish two arbitrary loops $X_1$ and $X_2$ in $M$) is
{\it total} (i.e., spans a dense set in $C(M^{S^1})$). 
Clearly, $C(M)$ separates points of $M$. Let $\varphi$ be an element
of $C(M)$. Then $\varphi$ determines a continuous function
$\varphi_\sigma$ on $M^{S^1}$ defined by
\[
\varphi_\sigma (X) \ := \ \varphi \left( X\,\left(\sigma\right)\right)
\]
$\forall X \in M^{S^1}$ (i.e., $X : S^1 \to M$). The set
\[
\left\{ \varphi_\sigma \bigm| \varphi \in C (M)\; , \ 0 \leq \sigma <
  2 \pi \right\}
\]
separates points in $M^{S^1}$ and hence is total in $M^{S^1}$. 

It is tempting to identify the local fields
$\varphi_{\chi_j\otimes\chi_{\bar{j}}}^\alpha (\sigma, 0)$ with the
functions $\varphi_\sigma$, for a suitable choice of the target space
$M$. This interpretation would be particularly natural in the examples
of the two-dimensional Wess-Zumino-Witten models [41,42]: Let $G$ be a 
simply-laced compact Lie group. Let $\widehat{\hbox{{\tt g}}}_k$ 
be the corresponding Kac-Moody current algebra at positive integer level $k$. 
The WZW model based on the group $G$ at level
$k$ is defined by setting
\begin{equation}
{\mathcal E} \ = \ \bar{{\mathcal E}} \ = \ U
(\widehat{\hbox{{\tt g}}}_k) \ .
\end{equation}
In this model, the {\it primary} local fields 
$\varphi_{\chi_j\otimes\chi_{\bar{j}}} (\sigma,\tau)$ are given by
\begin{equation}
\varphi_{\chi_j\otimes\chi_{\bar{j}}} (\sigma, \tau) \ = \ j\,\left(
  g\,(\sigma, \tau)\right)_{\alpha \bar{\beta}} \ ,
\end{equation}
where $j$ is an {\it ``integrable''} representation of $G$ (see
e.g.\ [24]), $g(\cdot, \tau)$ denotes a loop in $G$, i.e., an element
of $G^{S^1}$, and $j(g)_{\alpha\bar{\beta}}$ denotes the
$\alpha\bar{\beta}$ matrix element of $j(g)$.
\smallskip

\noindent This example is interesting in two respects:

\noindent (1) In general, only the {\it smeared} field operators
\begin{equation}
\int\limits_{- \infty}^\infty d\tau\; f (\tau)\;j (g (\sigma,
\tau))
\end{equation}
are well-defined operators on the Hilbert space ${\mathcal H} =
{\mathcal H}_{G,k}$ of the model, where $f$ is an arbitrary Schwartz
space test function (see (7.132)). Thus, only the algebra ${\mathcal
  F}$ of ``functions on quantized phase space over $G^{S^1}$ '' makes
sense, rather than the analogue of $C(G^{S^1})$. 

\smallskip

\noindent (2) Given $G$ and $k <\infty$, the list $\uI$ of {\it ``integrable''
  representations} of $G$, which correspond to the irreducible
unitary positive-energy representations of ${\mathcal E} \cong U
(\widehat{\hbox{{\tt g}}}_k)$, is {\it finite}. Thus,
\begin{equation}
\left\{ j(g)_{\alpha\bar{\beta}} \bigm| j \in \uI \right\} \ \cong \
\bigoplus_{j\,\in\,\uI} W_j \otimes W_{j^\vee} 
\end{equation}
is not nearly dense in $C(G)$.

We conclude that if the WZW model based on the group $G$ at level
$k<\infty$ describes the geometry of some loop space $M^{S^1}$ then $M$ 
{\it cannot} be the group manifold of $G$. It turns out (see [24,106,107]) 
that $M$ can be interpreted as a non-commutative space
corresponding to a quantum deformation of $G$, described by the
property that the ``algebra of functions on $M$'' is the ``algebra of
functions on a quantum group'' corresponding to $G$, with deformation
parameter
\[
q \ = \ {\rm exp}\, i\pi\,/\,(k+g^\vee) 
\]
where $g^\vee$ is the dual Coxeter number of $\hbox{{\tt g}}$.

The target space geometry of the WZW model (with $G=SU(2)$) has been
studied in some detail in [24,106,107]. We shall outline the results
in the next subsection. 

\vspace{0.5cm}

\subsection{Reconstruction of (non-commutative) target spaces from
  conformal field theory}

\noindent In the last subsection, we have argued that 
two-dimensional conformal field theory describes loop space
geometry. We have encountered the technical problem that the local
fields $\varphi_{\chi_j \otimes \chi_{\bar{j}}}^\alpha (\sigma, \tau =
0)$ , which, formally, can be identified with functions on loop space
$M^{S^1}$ of some target space $M$, are not well-defined operators on
the Hilbert space ${\mathcal H}$ of state vectors of the conformal
field theory. However, the field operators $\varphi_{\chi_j \otimes
  \chi_{\bar{j}}}^\alpha (\sigma, f)$ {\it smeared in parameter-time}
$\tau$, defined in (7.132), {\it are} well-defined operators on
${\mathcal H}$ and generate a {\it non-abelian} $^*$--algebra 
${\mathcal F}$ of ``functions on quantized phase space'' of
$M^{S^1}$. 

Unfortunately, there are no local fields
$\varphi_{\chi_j\otimes\chi_{\bar{j}}}^\alpha (\sigma, \tau)$ that are
{\it independent} of $\sigma$; i.e., a priori, there
are no candidates for ``functions on constant loops'' from which one
could construct an algebra of ``continuous functions'' on target space
$M$. The algebra ${\mathcal F}$ does {\it not} contain any
$^*$--subalgebra that could be interpreted as the algebra of
continuous functions on $M$; worse: there does {\it not} appear to
exist any non-trivial $^*$--homomorphism from ${\mathcal F}$ to some
$C^*$--algebra that could be interpreted as the algebra of continuous
functions on $M$. (Note that e.g.\ in the $\lambda \varphi^4$ theory, where 
$\varphi$ is a real-valued scalar field in $1+1$ dimensions, one can define 
the commutative $^*$--algebra generated by all operators $\varphi(f, \tau=0)$, 
$f$ an arbitrary test function, which describes loop space over 
${\mathbb R}$. In a conformal field theory, it is, in general, not possible to 
multiply fields smeared out with test functions $f(\sigma)$, at fixed $\tau$, 
simply because the scaling dimensions of the fields are too large.)

Thus, in order to reconstruct the target space $M$
from spectral data $({\mathcal F}, {\mathcal H}, H)$ of some conformal
field theory $\G$, one needs some new ideas.
In the following, we briefly review such ideas, as proposed in 
[78,24,106,107].

\medskip

\noindent{\bf(1)} We start by identifying {\it vector fields} on $M$. {}From
eq.~(7.115) in  subsection 2), it is 
clear that the (anti-)chiral algebras ${\mathcal E} (\bar{{\mathcal
    E}})$ of conformal field theories $\G$ are always
${\mathbb Z}$--graded. The grading operator is the generator $L_0$ of
${\rm Vir} \subseteq {\mathcal E}$. Let $\{ \psi_n^{(K)} \bigm| n \in
{\mathbb Z}, \; K \in {\mathcal K}_+\}$ be a system of {\it $W$-algebra
generators} of ${\mathcal E}$ with conformal weights $h_K \geq
0$; i.e., ${\mathcal E}$ is spanned linearly by the $\psi^{(K)}$ and
their normal ordered products (see e.g.\ [111]). 

For example, if ${\mathcal E}$ is the universal enveloping algebra of
some Kac-Moody current algebra based on a semi-simple Lie algebra
$\hbox{{\tt g}}$ then
\begin{equation}
\psi_n^{(K)} \ = \ J_n^A\, , \ A=1,\ldots, \, {\rm dim} \ \hbox{{\tt g}}
\ ,
\end{equation}
where $J_n^A$ are the modes of chiral currents, $J^A (\sigma + \tau)$,
of conformal weight $h_A = 1$.

By eq.~(7.115),
\[
\left[ L_0, \psi_n^{(K)} \right] \ = \ -\,n\,\psi_n^{(K)}\,,\ \quad n \in
{\mathbb Z} \ .
\]
We define ${\mathcal E}^{(0)}$ to be the $^*$--subalgebra of
${\mathcal E}$ generated by 
\begin{equation}
\left\{ \psi_n^{(K)} \bigm| n \in {\mathbb Z}\, , \ |n| \leq h_k -1\,
  , \ K \in {\mathcal K}_+\right\} \ .
\end{equation}
Similarly, $\bar{{\mathcal E}}^{(0)}$ is the $^*$--algebra generated
by 
\begin{equation}
\left\{ \bar{\psi}_n^{(K)} \bigm| n\in{\mathbb Z}\,, \ |n| \leq
  \bar{h}_K -1\,, \ K\in{\mathcal K}_- \right\} \ .
\end{equation}
It is tempting to interpret the generators (7.139,140) as {\it
  vector fields} on the target space, $M=M_{\G}$, of the conformal
field theory $\G$: If $\G$ is a WZW model based on the Lie algebra
$\hbox{{\tt g}}$ of a compact semi-simple Lie group $G$ then $h_K =
\bar{h}_{\bar{K}} = 1$, for all $K\in{\mathcal K}_+$ and $\bar{K} \in
{\mathcal K}_-$, and the generators (7.139) and (7.140) can indeed be
identified with {\it left-} and {\it right-invariant vector fields} on
$G$, respectively.

An important observation is that the co-products $\triangle_z$ map
${\mathcal E}^{(0)}$ to ${\mathcal E}^{(0)} \otimes {\mathcal
  E}^{(0)}$, as is implied by eq.~(7.111). In particular, if
${\mathcal E} = U(\widehat{\hbox{{\tt g}}}_k)$ then 
${\mathcal E}^{(0)} = U (\hbox{{\tt g}})$, and the co-products
$\triangle_z\bigg|_{{\mathcal E}^{(0)}}$ coincide with the usual
co-product of $U(\hbox{{\tt g}})$.

\medskip

\noindent{\bf(2)} Given a highest weight representation $j \in \uI$ of ${\mathcal
  E}$, let $\h_{j,0}$ be the subspace of the representation space
$\h_j$ spanned by all highest weight vectors in $\h_j$ (in the sense
of eq.~(7.117)). We define $\h_j^{(0)}$ to be the closure of the
subspace of $\h_j$ spanned by
\begin{equation}
\left\{ a \chi\; \bigm|\; \chi \in \h_{j,0} \ , 
\ a \in {\mathcal E}^{(0)}
\right\} \ .
\end{equation}
In the example of a WZW model based on a compact, semi-simple Lie
group $G$ with Lie algebra $\hbox{{\tt g}}, \uI = \bar{\uI}$ consists of
all integrable representations of $\hbox{{\tt g}}$, ${\mathcal E}^{(0)}
\cong U (\hbox{{\tt g}})$, and $\h_j^{(0)}$ is the finite-dimensional
representation space for the representation $j \in \uI$ of {\tt g}.

\medskip

\noindent{\bf(3)} We are now prepared to define an ``algebra of continuous
functions'' ${\mathcal A}_{\G}$ on the target space $M_{\G}$ of a
rational conformal field theory $\G$. There are (at least) the following two
ways of defining ${\mathcal A}_{\G}$, which, in the example of WZW models, are
 expected to be equivalent (and related to quantum group theory; see [106]).

\smallskip

\noindent{\bf(3.1)} In the first approach, we take ${\mathcal A}_{\G}$ 
to be ``generated'' (in a sense made precise below) by {\it primary local fields} 
$\varphi_{\chi_j\otimes \chi_{\bar{j}}}^\alpha (\sigma, \tau)$ of $\G$ 
as defined below eq.~(7.129), which are labeled by  certain pairs
of representations $(j,\bar{j})$ of ${\mathcal E} \times \bar{{\mathcal E}}$ 
ranging over some subset $\Pi \subseteq \uI \times \bar{\uI}$. 
In restricting to primary fields, we again draw inspiration from the 
example of WZW models, or, put differently, from string theory on 
group manifolds: In this example, the primary fields can be regarded as 
functions on the (deformed) group manifold or as functions of ``centre of mass'' 
coordinates of the moving string, while descendant fields would describe 
excited string oscillations around the centre of mass. 

Moreover, the example of a string moving in a toroidal target tells us 
that it might be appropriate to restrict the ``generators'' of 
${\mathcal A}_{\G}$ further by only using primary fields 
$\varphi_{\chi_j\otimes \chi_{\bar{j}}}^\alpha$
with $(j,\bar{j}) \in \Pi^{(0)}$, where $\Pi^{(0)}$ is a subset of $\Pi$ 
containing the label $(j_0,\bar j_0)$ of the field of lowest non-trivial 
scaling dimension $d_0 = h_{j_0} +h_{\bar j_0}\,$ such that 
$h_{j_0}=h_{\bar j_0}\,$, together with those of all primaries 
that arise from repeated OPE of this field with itself. (More 
precisely, we require that $\Pi^{(0)}$ is a sub-ring of the fusion 
ring.) If, in the case with toroidal targets, we were to include 
{\it all} primary fields as ``generators'' of ${\mathcal A}_{\G}$, 
we would obtain functions depending not only on the usual Fourier
(``momentum'') modes, but also on the winding number. This would, 
make it impossible to distinguish the torus from its $T$-dual.

To give a precise meaning to the algebra ${\mathcal A}_{\G}$, we 
define a subspace $\widehat{{\mathcal H}}_{\G}^{(0)}$ of the Hilbert 
space ${\mathcal H}_{\G}$ of $\G$ (defined in eq.~(7.129)) by setting 
\begin{equation}
\widehat{{\mathcal H}}_{\G}^{(0)} \ = \
\bigoplus_{(k,\bar{k}), \in \Pi^{(0)}} \, \h_k^{(0)}
\otimes \h_{\bar{k}}^{(0)} \otimes {\mathbb C}^{n (k,\bar{k})} \ ,
\end{equation}
where $\h_{k^\#}^{(0)}$ is as in (7.141). Obviously, $\widehat{{\mathcal
  H}}_\G^{(0)}$ is {\it invariant} under ${\mathcal E}^{(0)} \otimes
\bar{{\mathcal E}}^{(0)}$ and under the Virasoro generators
$H = L_0 + \bar{L}_0\,,\ P  =  L_0 - \bar{L}_0$, therefore we can 
restrict to the zero-momentum subspace 
\begin{equation}
{\mathcal H}_{\G}^{(0)} := \bigl\{\,\chi \in \widehat{{\mathcal
  H}}_\G^{(0)}\;\big\vert \; P\chi = 0 \, \bigr\}\ .
\end{equation}
With each pair $(\chi_k, \chi_{\bar{k}})$ of vectors in $\h_k^{(0)}
\times \h_{\bar{k}}^{(0)}\,$, for $(k,\bar{k}) \in \Pi^{(0)}$, and each
local field $\varphi_{\chi_k \otimes \chi_{\bar{k}}}^\beta (z, \bar{z})$
of the conformal field theory $\G$, we associate an element
$\Phi_{\chi_k \otimes \chi_{\bar{k}}}^\beta$ of the algebra ${\mathcal
  A}_\G$, which 
we define in terms of its {\it matrix elements} between a total set of
vectors in ${\mathcal H}_\G^{(0)}$: For $i=j$ or $l$, $\chi_i \otimes
\chi_{\bar{i}} \otimes u_\alpha \in \h_i^{(0)} \otimes
\h_{\bar{i}}^{(0)} \otimes {\mathbb C}^{n (i,\bar{i})} \subset
{\mathcal H}_\G^{(0)}, (i,\bar{i}) \in \Pi^{(0)}$ , we define
\begin{eqnarray}
&&\big\langle \chi_j \otimes \chi_{\bar{j}} \otimes u_\alpha \bigm|
\Phi_{\chi_k \otimes \chi_{\bar{k}}}^\beta \bigm| \chi_l \otimes
\chi_{\bar{l}} \otimes u_\gamma \big\rangle \nonumber\\
&&\qquad := \ C\left( \chi_j, \chi_{\bar{j}}, \alpha \bigm| \chi_k,
  \chi_{\bar{k}}, \beta \bigm| \chi_l, \chi_{\bar{l}}, \gamma \right)
\\
&&\qquad\, = \ \big\langle \chi_j \otimes \chi_{\bar{j}} \otimes
u_\alpha\, , \ \varphi_{\chi_k \otimes \chi_{\bar{k}}}^\beta (1,1)
\left( \chi_l \otimes \chi_{\bar{l}} \otimes
  \gamma\right)\big\rangle_{{\mathcal H}_\G} \ , \nonumber
\end{eqnarray}
where the $C$'s are the coefficients in the operator product expansion
(7.131). This is the proposal made in ref.~[24]. Because the local
fields $\varphi_{\chi_k\otimes\chi_{\bar{k}}}^\beta (z,\bar{z})$ form
a $^*$--algebra, the algebra ${\mathcal A}_\G$ generated by \ $\{
\Phi_{\chi_k \otimes \chi_{\bar{k}}}^\beta \bigm| \chi_k \otimes
\chi_{\bar{k}} \otimes u_\beta \in {\mathcal H}_\G^{(0)} \}$ \ is a
$^*$--{\it algebra of operators represented on} ${\mathcal
  H}_\G^{(0)}$. If ${\mathcal H}_\G^{(0)}$ is finite-dimensional, an
assumption that holds e.g.\ for the 
WZW models (at finite level), then the operators $\Phi_{\chi_k \otimes
  \chi_{\bar{k}}}^\beta$ are {\it bounded} and ${\mathcal A}_\G$ is a
direct sum of full matrix algebras. 

It is crucial to observe that, by definition (7.144),
\begin{equation}
\left[\, \psi_n^{(K)}\, , \, \Phi_{\chi_k \otimes \chi_{\bar{k}}}^\beta
\,\right] \ = \ \Phi_{k \left( \delta_1 \left(\psi^{(K)}\right)_n\right)
  \chi_k \otimes \chi_{\bar{k}}}^\beta 
\end{equation}
for all generators $\psi_n^{(K)}$ of ${\mathcal E}^{(0)}$, with
$\delta_1 (\psi^{(K)})_n \in  {\mathcal E}^{(0)}$, and that
\begin{equation}
\left[\, \bar{\psi}_n^{(K)}\,,\,
  \Phi_{\chi_k\otimes\chi_{\bar{k}}}^\beta \,\right] \ = \
\Phi_{\chi_k\otimes\bar{k}\,\left( \delta_1\,\left(
      \bar{\psi}^{(K)}\right)_n\right)\,\chi_{\bar{k}}}^\beta \ ,
\end{equation}
for all generators $\bar{\psi}_n^{(K)}$ of $\bar{{\mathcal E}}^{(0)}$,
with $\delta_1 (\bar{\psi}^{(K)})_n \in \bar{{\mathcal E}}^{(0)}$ .
This follows quite easily from (7.139--142,144) and 
the definition (7.111) of the co-products $\triangle_z$. In
particular, in the example of a WZW model based on a compact,
semi-simple Lie group $G$ with Lie algebra $\hbox{{\tt g}}$, we have
that
\begin{equation}
\left[\, J^A\,,\,\Phi_{\chi_k\otimes\chi_{\bar{k}}}^\beta\,\right] \ = \
\Phi_{k(J^A)\chi_k\otimes\chi_{\bar{k}}}^\beta \ ,
\end{equation}
where $\{ J^A \equiv J_{n=0}^A\}$ is a basis of $\hbox{{\tt g}}$, and
\begin{equation}
\left[\,
  \bar{J}^{\,A}\,,\,\Phi_{\chi_k\otimes\chi_{\bar{k}}}^\beta\,\right] \
= \ \Phi_{\chi_k \otimes\bar{k}(\bar{J}^{\,A})\chi_{\bar{k}}}^\beta \
.  
\end{equation}
We concentrate on the WZW model with diagonal modular invariant; then 
$\Pi^{(0)}$ is itself given by the diagonal in $\uI \times
\bar{\uI}$, i.e., $k\cong\bar{k}$ for all $(k,\bar{k}) \in
\Pi^{(0)}$, and $\beta$ has only a single value (and hence can be
omitted). Eqs.~(7.147) and (7.148) are analogous to the two
intertwining relations stated in Sect.~{\bf 4.2}, below
eq.~(4.82). When the level of the WZW model tends to $+\infty$ (i.e.,
in the {\it ``classical limit''}), eqs.~(7.147) and (7.148) become
{\it equivalent} to those intertwining relations!

In accordance with our discussion in Section~{\bf 2}, below eq.~(2.9),
an algebra ${\mathcal F}_\G^{(0)}$ of ``functions on quantized phase
space''
over target space $M_\G$ can be defined as the {$^*$--algebra} of
operators on ${\mathcal H}^{(0)}$ generated by ${\mathcal A}_\G,
\bar{{\mathcal E}}^{(0)}$ and the Hamiltonian $H$.

The metric non-commutative geometry of the target space $M_\G$ is thus
encoded in the spectral data
\begin{equation}
\left( {\mathcal A}_\G, {\mathcal H}^{(0)}, H\right) \quad {\rm and}
\quad \left( {\mathcal F}_\G^{(0)}, {\mathcal H}^{(0)}, H \right) \ .
\end{equation}
In the example of the SU(2)--WZW model at integer level $k$, the 
non-commutative Riemannian geometry of the {\it ``fuzzy
  three-sphere''} described by $({\mathcal A}_\G^{(0)}, {\mathcal
  H}^{(0)}, H)$ has been studied quite explicitly in [107].

\medskip

\noindent{\bf(3.2)} There is an alternative definition of the algebra ${\mathcal
  A}_\G$ of ``functions on $M_\G$'' studied in [106]: It is based on 
the idea that the target space will, to a large extent, be determined 
by its (quantum) symmetries. One defines
${\mathcal A}_\G$ to be the (generally {\it non-associative})
$^*$--algebra generated by operators
\[
\left\{ \phi_{\chi_k\otimes\chi_{\bar{k}}}^\beta \bigm| \chi_k \in
  \h_k^{(0)},\,\chi_{\bar{k}} \in \h_{\bar{k}}^{(0)},\, (k,\bar{k})
  \in \Pi^{(0)},\, \alpha = 1, \ldots, n \,(k,\bar{k}) \right\}
\]
with {\it multiplication table} given by
\begin{equation}
\phi_{\chi_j \otimes \chi_{\bar{j}}}^\alpha
\;\star\;\phi_{\chi_k\otimes\bar{\chi}_k}^\beta \ = \ \sum C\left(
  \chi_j, \chi_{\bar{j}}, \alpha \bigm| \chi_k, \chi_{\bar{k}}, \beta
  \bigm| \chi_l, \chi_{\bar{l}}, \gamma \right)\,
\phi_{\chi_l\otimes\chi_{\bar{l}}}^\gamma \ .
\end{equation}
This approach may help to clarify the connections of non-commutative
target space geometry to {\it quantum group theory}. But we shall not
pursue it here.
\medskip

In order to study the cohomology of non-commutative target spaces and
their Riemannian or complex non-commutative geometries, we should
study {\it supersymmetric extensions} of conformal field theory. This
is the subject of the following last subsection.
\medskip

In the example of the SU(2)--WZW model at level $k$, the coefficients
$C(\cdot|\cdot|\cdot)$ in eqs.~(7.130), (7.144), (7.150) have been
calculated explicitly: Let $\{ \chi_s^i\}$ be an orthonormal basis in
the representation space $W_s$ of SU(2) of spin \ $s \leq \frac k 2$
. Then
\begin{equation}
C\,\left( \chi_{s_1}^i, \chi_{s_1}^{\bar{i}} \bigm| \chi_{s_2}^j,
  \chi_{s_2}^{\bar{j}} \bigm| \chi_{s_3}^l, \chi_{s_3}^{\bar{l}}
\right) 
=   C^k\left( s_1, s_2, s_3\right) \, C_{\,jli}\left(
  s_2,s_3\bigm| s_1\right)_{\bar{j}\,\bar{l}\,\bar{i}} \ ,
\end{equation}
where the tensors $C_{\,jli}(s_2,s_3|s_1)_{\bar{j}\,\bar{l}\,\bar{i}}$
(proportional to squares of Clebsch-Gordan matrices) are defined in (4.82),
and the coefficients $C^k(s_1,s_2,s_3)$ enforce the
$\widehat{su}(2)_k$ fusion rules; explicit expressions for
$C^k(s_1,s_2,s_3)$ may be found in [24] and refs.~given there. It is
shown in [107] that, in this example, the algebra ${\mathcal
  A}_{\rm{SU(2)-WZW}}$ of ``functions'' over $M$ is a full matrix
algebra; the same is true for the algebra ${\mathcal
  F}_{\rm{SU(2)-WZW}}$ of ``functions on quantized phase space''
over~$M$.
In [106] some steps are undertaken to reconstruct the conformal field
theory $\G$ from the data $({\mathcal A}_\G, {\mathcal H}^{(0)}, H)$
together with ${\mathcal E}^{(0)}, \bar{{\mathcal E}}^{(0)}$, in the
example of WZW models. This program makes contact with the theory of
lattice Kac-Moody algebras.
\medskip

Clearly, the program sketched in this subsection to reconstruct the
(generally non-commutative) target spaces of conformal field theories
{\it remains tentative} and must be tested in some interesting
examples. First steps in this direction have been taken in 
[24,107,106]. Examples that are reasonably well understood involve 
chiral algebras obtained from Kac-Moody current algebras or from 
coset constructions based on Kac-Moody algebras [24]. Whereas 
WZW models based on compact semi-simple Lie groups describe 
{\it non-commutative} targets, 
those  built on direct products of U(1)--current algebras yield 
target spaces which are {\it tori}. In these examples, {\it
  dual tori} are identified, thanks to our choice of the set
$\Pi^{(0)}$ in the definition of ${\mathcal A}_\G$. They are the
simplest examples for $T$--{\it duality}. Moreover, the U(1)--models also 
provide simple examples of {\it mirror symmetry} [110]; see [24].

Superconformal field theories of considerable interest in string 
theory would be the Gepner models, whose target spaces are expected to
correspond to (non-commutative deformations of) Calabi-Yau spaces. They
remain to be understood more precisely.

In attempting to reconstruct target spaces of conformal field theories
one finds that the definition of the algebra ${\mathcal A}_\G$ of
``functions on target space'' usually involves considerable
arbitrariness. This arbitrariness is at the origin of $T$--duality and of
mirror symmetry. The latter is related to an arbitrariness in the definition
of the {\it degree} of field operators of $N=(2,2)$ superconformal
field theories (cf.\ the next section) and to the fact that, usually, 
there are several options for choosing ${\mathcal A}_\G$ as a subalgebra 
of an algebra ${\mathcal F}_\G^{(0)}$ of ``functions on quantized 
phase space'', for a given conformal field theory $\G$. 
These issues deserve further study.

Let us add an observation on the nature of the ``full'' target space-time
of a string theory reconstructed by the scheme outlined above: After 
separation of variables, as in (7.29,55), and identifying the algebra of 
functions on ``internal space'' 
with the algebra ${\mathcal A}_\G$ --- which typically is a finite-dimensional
matrix algebra --- we are led to targets described by algebras of the form 
$C^{\infty}(M^4) \otimes {\mathcal A}_\G$ which resemble the space-times 
underlying the Connes-Lott construction of the Standard Model! 
\vspace{0.5cm}

\subsection{Superconformal field theories, and the topology of target
  spaces}

\noindent In this subsection, we consider conformal field theories
whose chiral and anti-chiral algebras ${\mathcal E}$ and $\bar{\mathcal E}$
have ${\mathbb  Z}_2$--graded extensions 
${\mathcal C}_R^\#$ and ${\mathcal  C}_{NS}^\#$, 
as discussed in subsection~2) after eq.~(7.109),
i.e., ${\mathcal C}_{\,_\bullet}$ and $\bar{{\mathcal C}}_{\,_\bullet}$ 
contain fermionic (odd) generators besides the  bosonic ones.
We demand that ${\mathcal C}_{\,_\bullet}^\#$ contain  a super-Virasoro 
algebra. Examples are the supersymmetric WZW models  (see [24] 
and refs.~given there). Superconformal field theories realize 
the mathematical structure discussed in Sect.~{\bf 5.3}.

There are three supersymmetric extensions of the Virasoro algebra that 
are important
in the study of superstring vacua: the $N=1,2$ and $4$ super-Virasoro
algebras. 

\medskip

\noindent 1) \ub{The $N=1$ super-Virasoro algebra} [108]
\medskip

\noindent It has generators $\{ L_n\}_{n\in{\mathbb Z}}$ and $\{
G_r\}_{r\in{\mathbb Z}(+\frac 1 2)}$ satisfying the commutation
relations
\begin{eqnarray}
&& \left[\, L_n, L_m\,\right] \ = \ \left(n-m\right)
L_{n+m}\;+\;\frac{c}{12}\;n \left( n^2-1\right) \delta_{n+m,0} \ ,
\nonumber \\
&& \left[\, L_n, G_r\,\right] \ = \ \left( \frac n 2 \ - \ r\right)
G_{n+r} \ , \\
&& \left\{\, G_r, G_s\,\right\} \ = \ 2\,L_{r+s}\ + \ \frac c 3 \ \left(
  r^2\,-\,\frac 1 4 \right) \delta_{r+s,0} \ . \nonumber
\end{eqnarray}
On the Ramond sector (periodic b.c.), the indices $r$ of the ``Ramond
generators'' $G_r$ range over ${\mathbb Z}$, while, on the
Neveu-Schwarz sector (anti-periodic b.c.), the indices $r$ of the
``Neveu-Schwarz generators'' $G_r$ range over ${\mathbb Z} + \frac 1 2
$.

In a unitary representation, we have that
\begin{equation}
L_n^* \ = \ L_{-n} \ , \ G_r^* \ = \ G_{-r} \ .
\end{equation}
An $N=(1,1)$ supersymmetric conformal field theory
has the property that both ${\mathcal C}_{\,_\bullet}$ and $\bar{{\mathcal
    C}}_{\,_\bullet}$ contain an $N=1$ super-Virasoro algebra (and we
shall assume, for simplicity, that ${\mathcal C}_{\,_\bullet} \cong
\bar{{\mathcal C}}_{\,_\bullet}$). Then the operators 
\begin{equation}
{\mathcal D}\ := \ G_0\ ,\ \quad\overline{{\mathcal D}}\ :=\ \overline{G}_0
\end{equation}
play the role of the two Pauli-Dirac operators in $N=(1,1)$
supersymmetric spectral data, and
\begin{equation}
\dd\ :=\ G_0-i\,\overline{G}_0\,,\quad\ \dd^*\ :=\ G_0+i\,\overline{G}_0 
\end{equation}
can be interpreted as exterior derivative and its adjoint in spectral
data with {\it centrally extended} $N=(1,1)$ supersymmetry, as studied
in subsection 8) of Sect.~{\bf 5.2}.
\smallskip

\noindent We define
\begin{eqnarray}
&&\phantom{ZeichnungZeichnung} \lambda_n \ := \ L_n - \bar{L}_{-n} \ ,
\nonumber\\
&&\hskip-.7cm {\rm and} \\
&&\phantom{ZeichnungZeichnung} \d_n \ := \ \frac 2 n \ \left[
  \lambda_n, \dd \right] \ = \ G_n - i\,\bar{G}_{-n} \
. \phantom{ZeichnungZeichnungZeich} \nonumber
\end{eqnarray}
If the central charges $c$ and $\bar{c}$ of the two Virasoro
algebras in ${\mathcal C}_R$ and $\bar{{\mathcal C}}_R$ coincide, as
assumed above, then we have that
\begin{eqnarray}
&& \left[\, \lambda_n, \lambda_m\,\right] \ = \ (n-m)\,\lambda_{n+m}
\qquad {\rm (Witt \ algebra)} \ ,\phantom{ZeichnungZeichnung} \nonumber\\
&& \left[ \,\lambda_n, \dd_m\,\right] \ = \ \left( \frac n 2 \ -
  m\right)\,\dd_{n+m} \ , \nonumber\\
\noalign{\leftline{\rm and}}
&&\left\{ \dd_n,\,\dd_m\right\} \ = \ 2\,\lambda_n \ . \nonumber
\end{eqnarray}
These commutation relations are the structure relations of the {\it
  super-Witt algebra}, compare to Sect.~{\bf 5.3}, eqs.~(7.77,.78).

If ${\mathcal F}_\G$ denotes the field algebra (algebra of ``functions
on quantized phase space'') of an $N=(1,1)$ superconformal field
theory $\G$, as defined below eq.~(7.132) of subsection~2), then the
spectral data of $\G$ on the {\it Ramond sector} are given by
\begin{equation}
\left( {\mathcal F}_\G\,,\ {\mathcal H}_{\rm Ramond}\,,\, {\mathcal
    D}\,,\ \bar{{\mathcal D}}\,, \ \gamma\,,\ \bar{\gamma}\,\right) \ ,
\end{equation}
where $\gamma^\# = (-1)^{F^\#}$ and $F^\#$ counts the number of left
(resp.\ right) moving fermions. The (anti-)chiral algebras ${\mathcal
  E}^\#$ of $\G$ play the role of reparametrization
symmetries. $N=(1,1)$--superconformal field theories are perfect
examples for the mathematical structure described in subsection~8) of
Sect.~{\bf 5.2} (see also Sect.~{\bf 5.3}): There, we have 
reviewed the topological information encoded
in the spectral data (7.157). For example, index theory for a
superconformal field theory is concerned with the calculation of the
following elliptic genera: the {\it Euler characteristics}
\[\chi \ = \ {\rm tr}_{{\mathcal H}_{\rm Ramond}} \left( \gamma
  \otimes \bar{\gamma}\; e^{i(\tau\,{\mathcal D}^2 -
    \bar{\tau}\,\bar{{\mathcal D}}^2)} \right) \ ,
\]
(which is independent of $\tau, \bar{\tau}$) and the {\it signature genus}
\[
\Phi (\sigma) \ = \ {\rm tr}_{{\mathcal H}_{\rm Ramond}} \left( \left(
    \id \otimes \bar{\gamma}\right)\; e^{i\left( \tau\,{\mathcal D}^2
      - \bar{\tau}\,\bar{{\mathcal D}}^2\right)} \right) \ .
\]
This genus and the $\hat{A}$ genus are modular forms; for details 
see [109] and refs.~given there. See also eqs.~(5.76,77).
\smallskip

The de Rham-Hodge theory has been outlined in subsections 7) and 8) of
Sect.~{\bf 5.2}. The {\it Ramond ground states} of $\G$, i.e., the
highest weight vectors $\psi \in {\mathcal H}_{\rm Ramond}$ satisfying
\begin{equation}
{\mathcal D}\,\psi \ = \ \bar{{\mathcal D}}\,\psi \ = \ (L_0 -
\bar{L}_0)\,\psi \ = \ 0 \ ,
\end{equation}
can be interpreted as {\it harmonic forms} on the {\it target space}
$M_\G$ of $\G$; see also [15,16] --- unless supersymmetry is {\it
  spontaneously broken}, as described e.g.\ in Sect.~{\bf 4.2},
eqs.~(4.76,77), and in subsection 7) of Sect.~{\bf 5.2}, below
eq.~(5.60). Spontaneous supersymmetry breaking is encountered e.g.\ in
the study of the $N=(1,1)$ supersymmetric WZW models; see [24] and
refs.~given there. 

When supersymmetry is spontaneously broken there are no Ramond ground
states and the cohomology of the complex of {\it vector forms}, see
subsection 7) of Sect.~{\bf 5.2}, eqs.~(5.48--55), is {\it
  trivial}. But this does {\it not} imply that the cohomology of the
complexes $\Omega_\dd^{\,^\bullet} ({\mathcal F}_\G)$ ,
$\Omega_{{\mathcal D}}^{\,^\bullet} ({\mathcal F}_\G)$,
$\Omega_\dd^{\,^\bullet} ({\mathcal F}_\G^{(0)})$ \ is
trivial, too, where ${\mathcal F}_\G$ is the $^*$--algebra defined below
eq.~(7.132), and ${\mathcal F}_\G^{(0)}$ is the $^*$--algebra of
``functions on quantized phase space over target space $M_\G$'', as
defined above eq.~(7.149). The differential $^*$--algebras
$\Omega_{{\mathcal D}}^{\,^\bullet} ({\mathcal F}_\G)$ and
$\Omega_{{\mathcal D}}^{\,^\bullet} ({\mathcal F}_\G^{(0)})$ are
defined as in subsection 2) of Sect.~{\bf 5.1}. (We are using here
that the ``small'' Ramond Hilbert space ${\mathcal H}_{\rm
  Ramond}^{(0)}$,  whose definition can 
be inferred from eqs.~(7.141,142), is {\it invariant} under
${\mathcal F}_\G^{(0)}$ and under ${\mathcal D}$ (and $\bar{{\mathcal
    D}}$); see also [24].) The differential ${}^{\natural}$--algebras
$\Omega_\dd^{\,^\bullet} ({\mathcal F}_\G)$ and
$\Omega_\dd^{\,^\bullet} ({\mathcal F}_\G^{(0)})$ are defined as in
eqs.~(5.34--36) of Sect.~{\bf 5.2}. The cohomology
rings $H_\dd^{\,^\bullet} ({\mathcal F}_\G)$, $H_\dd^{\,^\bullet}
({\mathcal F}_\G^{(0)})$ are defined as in eqs.~(5.58,59) of
Sect.~{\bf 5.2}. 
\smallskip

{\it Superconformal field theories and their ``target space geometry''
  fit perfectly into the framework developed in Section}~{\bf 5}! In
particular, the $S^1$--equivariant cohomology of
$\Omega_\dd^{\,^\bullet} ({\mathcal F}_\G)$ is determined according
to the theory outlined in subsection~8) of Sect.~{\bf 5.2}. 
\medskip

In order to illustrate the general theory, we summarize results for
the example where $\G$ is the 
supersymmetric SU(2)--WZW model at level $k=1,2,\ldots$ . Proofs for
the results stated here can be found in [107].

\smallskip

\noindent (i) The ``small'' Ramond Hilbert space, ${\mathcal H}^{(0)} =
{\mathcal H}_{\rm Ramond}^{(0)}$, is given by
\begin{equation}
{\mathcal H}^{(0)} \ = \ {\mathcal H}_{\rm bos.}^{(0)} \ \otimes \
F^{(0)} \ ,
\end{equation}
where 
\begin{equation}
{\mathcal H}_{\rm bos.}^{(0)} \ := \
\bigoplus_{s\,\leq\,\frac k 2}\ W_s\otimes W_s \ ,
\end{equation}
and $s\equiv s^\vee = 0, \frac 1 2 ,\, 1, \ldots,\, \frac k 2 $ \ is
the spin of the representation of SU(2) on $W_s$; furthermore, $F^{(0)}$
is the representation space for the unique irreducible representation
of the Clifford algebra $Cl({\mathbb R}^6)$ with six self-adjoint
generators $\{
\psi^A, \bar{\psi}^A\}_{A=1}^3$ \ satisfying
\[
\left\{ \psi^A, \psi^B\right\} \ = \ \delta^{AB}\,,\quad\left\{
  \bar{\psi}^A, \bar{\psi}^B \right\} \ = \ \delta^{AB}\,, \quad \left\{
  \psi^A, \bar{\psi}^B \right\} \ = \ 0\ . 
\]
If the direct sum on the r.s.\ of (7.160) were {\it unrestricted} 
${\mathcal H}^{(0)}$ would be the Hilbert space ${\mathcal H}_{e-p}$
of square-integrable differential forms on ${\rm SU(2)}\simeq S^3$;
see Sect.~{\bf 4.2}.

\smallskip

\noindent (ii) The algebra ${\mathcal A}^{(0)} = {\mathcal
  A}_{\widehat{su}(2)_k}^{(0)}$, which coincides with ${\mathcal F}^{(0)}$\ 
in this example, turns out to be a full matrix algebra,
\begin{equation}
{\mathcal A}^{(0)} \ ={\mathcal F}^{(0)}\ \cong \ {\rm End}\,\left( 
{\mathcal H}_{\rm  bos.}^{(0)}\right)   \ .
\end{equation}

\noindent (iii) The Pauli-Dirac operators ${\mathcal D} :=
G_0\bigg|_{{\mathcal 
   H}^{(0)}}$ and $\overline{{\mathcal D}} := \overline{G}_0\bigg|_{{\mathcal
        H}^{(0)}}$ are given by
\begin{eqnarray}
&& {\mathcal D} \ = \ \psi^A\,\Bigl( J_A\,-\,\frac{i}{12}\;
  \varepsilon_{ABC}\,\psi^B\,\psi^C\Bigr) \ , \nonumber\\
&& \overline{{\mathcal D}} \ = \ \bar{\psi}^A\,\Bigl(
  \bar{J}_A\,-\,\frac{i}{12}\;
  \varepsilon_{ABC}\,\bar{\psi}^B\,\bar{\psi}^C\Bigr) \ ,
\end{eqnarray}
where $J_A^\#$ is the 0--mode of the current $J_A^\#(z)$ generating
the left- (right-) Kac-Moody algebra $\widehat{su}(2)_k$ , whose
enveloping algebra ${\mathcal E} (\bar{{\mathcal E}})$ 
is the (anti-)chiral algebra of the theory. Formulas (7.162) are a
special case of eqs.~(4.76), Sect.~{\bf 4.2}: $i\Gamma^j\to\psi^A\,,\ 
i \bar{\Gamma}^j\to \bar{\psi}^A\,,\  \break
i T_j\to J_A \,,\ i \bar{T}_j\to \bar{J}_A \,,\ f_{ijk} \to\varepsilon_{ABC}\,$ .

Thus, the $N=(1,1)$ spectral data
\begin{equation}
\left( {\mathcal A}^{(0)}\,,\,{\mathcal H}^{(0)}\,,\,{\mathcal
    D}\,,\,\bar{{\mathcal D}} \right)
\end{equation}
describe the non-commutative geometry of the target space 
$M_{\widehat{su}(2)_k}$, which is the ``fuzzy three-sphere'' $S_k^3$.

\smallskip

(iv) The differential $^*$--algebras $\Omega_{{\mathcal
    D}}^{\,^\bullet} ({\mathcal A}^{(0)}) \cong
\Omega_{\bar{{\mathcal D}}}^{\,^\bullet} ({\mathcal A}^{(0)})$
considered in Sects.~{\bf 5.1, 5.2} turn out to be given by
\begin{equation}
\Omega_{{\mathcal D}}^{\,^\bullet} ({\mathcal A}^{(0)}) \ = \
\bigoplus_{n=0}^3 \ \Omega_{{\mathcal D}}^n\,
({\mathcal A}^{(0)}) \ , 
\end{equation}
where $\Omega_{{\mathcal D}}^n ({\mathcal A}^{(0)})$, $n=0,\ldots,3,$
are {\it free} ${\mathcal A}^{(0)}$--modules:
$\Omega_{{\mathcal D}}^0 ({\mathcal A}^{(0)}) = {\mathcal
  A}^{(0)}$ and 
\begin{eqnarray}
&& \Omega_{{\mathcal D}}^1\,({\mathcal A}^{(0)}) \ {\rm has \
  dimension \ 3, \ with \ basis} \ \{ \id \otimes \psi^A\}_{A=1}^3 \ ,
\nonumber\\
&& \Omega_{{\mathcal D}}^2\,({\mathcal A}^{(0)}) \ {\rm has \
  dimension \ 3, \ with \ basis} \ \{\id \otimes \psi^A \psi^B\}_{A<B} \
, \\
&& \Omega_{{\mathcal D}}^3\,({\mathcal A}^{(0)}) \ {\rm has \
  dimension \ 1, \ with \ basis} \ \{ \id \otimes \psi^1 \psi^2
\psi^3\} \ . \nonumber
\end{eqnarray}
Thus, every element $\alpha \in \Omega_{{\mathcal D}}^{\,^\bullet}
({\mathcal A}^{(0)})$ can be represented uniquely as
\[
\alpha \ = \ \alpha_0 \otimes \id + \alpha_{1,A} \otimes \psi^A +
\alpha_{2,A} \otimes \psi^{A+1} \psi^{A+2} + \alpha_3 \otimes \psi^1
\psi^2 \psi^3 \ ,
\]
where the coefficients $\alpha_{n}, \alpha_{n,A}$ are elements of ${\mathcal
  A}^{(0)}$. Integration of forms is given by
\begin{equation}
\bint \ \alpha \ = \ {\rm tr}_{{\mathcal H}_{\rm bos.}^{(0)}} 
(\alpha_0) \ ,
\end{equation}
and the metric (Hermitian structure) on $\Omega_{{\mathcal
    D}}^{\,^\bullet} ({\mathcal A}^{(0)})$ by
\begin{equation}
\langle \alpha,\,\beta\rangle \ = \ \alpha_0\,\beta_0^*\,+\,\frac 1 2 \
\alpha_{1,A}\, \beta_{1,A}^*\,+\,\frac 1 4 \
\alpha_{2,A}\,\beta_{2,A}^*\,+\,\frac 1 8 \ \alpha_3\,\beta_3^*\ .
\end{equation}
Following Sect.~{\bf 5.1}, one can equip the ``cotangent bundle''
$\Omega_{{\mathcal D}}^1 ({\mathcal A}^{(0)})$ with (left- or right-)
connections, $\nabla$, and calculate their Riemann-, Ricci- and scalar
curvature; see~[107].

Next, we report on the cohomology groups of the fuzzy three-sphere
following Connes' definition of cohomology rings, which is suitable
for $N=1$ spectral data as considered in Sect.~{\bf 5.1}. We define
\[
{\mathcal A}_{\; R}^{(0)} \ := \
\bigoplus_{s\,\leq\,\frac k 2} \ \left(\id\,\bigg|_{W_s}
\otimes {\rm End}\;(W_s) \right)\ ;
\]
(compare with eq.~(7.160)). A lengthy calculation (see [107]) shows
that
\begin{eqnarray}
&& \phantom{ZeichnungZeichnung} H^0\,({\mathcal A}^{(0)}) \ \cong \
H^3\,({\mathcal A}^{(0)}) \ \cong \ {\mathcal A}_{\;R}^{(0)} \ ,
\phantom{ZeichnungZeichnungZeich} \nonumber\\
&&\hskip-.8cm{\rm and} \\
&& \phantom{ZeichnungZeichnung} H^1\,({\mathcal A}^{(0)}) \ = \
H^2\,({\mathcal A}^{(0)}) \ = \ \{ 0\} \ . \nonumber
\end{eqnarray}
These results support the interpretation of target spaces of
SU(2)--WZW models as ``fuzzy three-spheres''.

One can view the target spaces of WZW models based on a compact,
semi-simple group $G$ (at finite level $k=1,2,3,\ldots$) as examples
of non-commutative {\it Riemannian} spaces and describe them in terms of
$N=(1,1)$ spectral data 
\begin{equation}
\left( {\mathcal A}^{(0)}\,,\,{\mathcal H}^{(0)}\,,\,
  \dd\,,\,\dd^*\,,\,\tilde{\gamma}\,,\,*\right) \ .
\end{equation}
One chooses the differential $\dd$ to be given by the BRST operator
corresponding to the representation of $G$ on ${\mathcal H}^{(0)}$, as
in eq.~(4.86) of Sect.~{\bf 4.2}; the generators $T_j$ in eq.~(4.86)
are defined in terms of the zero-modes of the left- and/or
right-moving Kac-Moody currents of the model. In the example of the
SU(2)--WZW-models at level $k$, the ``de Rham cohomology
groups'' (see Sect.~{\bf 5.2}) determined by the BRST operators have the form 
\[
H^0({\mathcal A}^{(0)})\ \cong \ H^3({\mathcal A}^{(0)}) \ \neq \
\{0\}\, , \quad  H^1({\mathcal A}^{(0)}) \ = \ H^2({\mathcal A}^{(0)}) \ =
\ \{0\} \ .
\]
For further details see [107,24,106].

\medskip

\noindent 2) \ub{$N=2$ and $N=4$ supersymmetry; mirror symmetry}
\medskip

\noindent In the study of superstring vacua exhibiting space-time 
supersymmetry one is led to consider superconformal field theories 
with higher (world sheet) 
supersymmetries, in particular, with 2 or 4 supersymmetries in each
chiral sector; see [29]. Properties of such conformal field theories
can be derived from the representation theory of $N=2$ or $N=4$
super-Virasoro algebras.

The $N=2$ super-Virasoro algebra has generators $\{ L_n, G_{n+a}^\pm ,
J_n\}_{n\in{\mathbb Z}}$ , with $a=0$ on the Ramond sector (periodic
boundary conditions), and $a=\frac 1 2$ \ on the Neveu-Schwarz sector
(anti-periodic boundary conditions). They satisfy the commutation
relations 
\begin{eqnarray}
&&(i) \qquad \ \left[ L_n, L_m\right]\ = \ (n-m) L_{n+m} + \frac{c}{12}
\ n (n^2-1) \,\delta_{n+m,0} \ , \nonumber \\
&&(ii)\qquad  \left[ L_n, J_m\right] \ = \ -\; m\;J_{n+m} \ ,
\nonumber \\
&&(iii) \quad \ \;\left[ J_n, J_m\right] \ = \ \frac c 3 \
n\;\delta_{n+m,0} \ , \nonumber \\
&&(iv)\qquad\!\!\left[ L_n, G_{m+a}^\pm \right] \ = \ \left( \frac n 2 \
  -\;(m+a)\right)\,G_{n+m+a}^\pm \\
&&(v) \qquad\! \left[ J_n, G_{m+a}^\pm \right] \ = \ \pm\;G_{n+m+a}^\pm
\nonumber \\
&&(vi) \quad \ \left\{ G_{n+a}^+ , G_{m+a}^-\right\}  =  2\,
L_{n+m}+\left( n-m+2a\right) J_{n+m}+ \frac c 3 \left(
  (n+a)^2-\frac 1 4 \right)\delta_{n+m,0}\nonumber \\ 
&&(vii)\quad \left\{ G_{n+a}^+\,,\,G_{m+a}^+\right\} \ = \ \left\{
  G_{n+a}^-\,,\,G_{m+a}^-\right\} \ = \ 0 \ . \nonumber
\end{eqnarray}
In a unitary representation of the $N=2$ super-Virasoro algebra,
\begin{equation}
L_n^*\;=\;L_{-n}\,,\quad \ J_n^*\;=\;J_{-n}\,,\quad  \ 
\left( G_{n+a}^+\right)^*\;=\;G_{-n-a}^- \ .
\end{equation}
Relations (7.170) and (7.171) show that the generators $G_0^+$ and
$G_0^-$ of an $N=2$ super-Virasoro algebra correspond to the
operators 
$\partial = D_1 - i\,D_2$  and $\partial^* = D_1 +
i\,D_2$ of $N=2$ spectral data, as discussed 
in subsection 8) of Sect.~{\bf 5.1}, eqs.~(5.24, 5.25). 
The zero mode $J_0$ of the current $J$ corresponds to
the ${\mathbb Z}$--grading operator ${\mathcal T}$ of
eq.~(5.26). Thus, the $N=2$ super-Virasoro algebra is related to
spectral data describing {\it K\"ahler geometry}. The Virasoro
subalgebra with generators $\{ L_n\}_{n\in{\mathbb Z}}$ plays the
r\^ole of a Lie algebra ${\mathcal G}$ of {\it infinitesimal
  reparametrizations}, as discussed at the beginning of Sect.~{\bf
  5.3}; see Definition A of Sect.~{\bf 5.3}.

A {\it local, unitary}, ``left-right symmetric'' {\it superconformal
  field theory} with $N=(2,2)$ supersymmetry on the Ramond sector
${\mathcal H}_R$ is given in terms of spectral data
\begin{equation}
\left( {\mathcal F}, {\mathcal H}_R, \partial, \partial^*,
  \overline{\partial}, \overline{\partial}^{\, *}, {\mathcal T},
  \overline{{\mathcal T}}, {\mathcal G}, \overline{{\mathcal G}} \right)  
\end{equation}
where ${\mathcal F}$ is a $^*$--algebra of operators on ${\mathcal
  H}_R$ constructed from local bosonic fields of the theory, $\partial
= G_0^+$, $\partial^* =  G_0^-$ are the zero-modes of the left-moving
Ramond generators, $\overline{\partial} = \overline{G}_0^{\,\pm}$,
$\overline{\partial}^{\,*} = \overline{G}_0^{\,\mp}$ --- see below 
for the choice of sign --- are the zero modes of the
right-moving Ramond generators, ${\mathcal T} = J_0$ is the grading
operator of the left-movers, while $\overline{{\mathcal T}} = \pm
\overline{J}_0$ is the grading operator of the right-movers (but the
spectra of $J_0$ and $\overline{J}_0$ are {\it not}, in general, 
contained in the integers), \ ${\mathcal G}$ and
$\overline{{\mathcal G}}$ are Virasoro algebras associated with left
and right movers,
respectively. The graded Lie algebra ${\mathcal
  G}_{\partial,\partial^*,{\mathcal T}}$ generated by $\partial,
\partial^*, {\mathcal T}$ and ${\mathcal G}$ is the $N=2$
super-Virasoro algebra described in (7.170) and (7.171); likewise, the
graded Lie algebra $\overline{{\mathcal
    G}}_{\overline{\partial},\overline{\partial}{}^*,\overline{{\mathcal T}}}$
generated by $\overline{\partial}, \overline{\partial}{}^{*},
\overline{{\mathcal T}}$ and $\overline{{\mathcal G}}$ is another copy
of the $N=2$ super-Virasoro algebra, and usually the central charges
$c$ and $\bar{c}$ coincide. Elements of ${\mathcal G}_{\partial,
  \partial^*, {\mathcal T}}$ graded-commute with elements of
$\overline{{\mathcal G}}_{\overline{\partial},
  \overline{\partial}{}^{*}, \overline{{\mathcal T}}}$ . This implies,
in particular, that $\{ \partial, \overline{\partial}{}^{\#}\} = 0\,$, 
i.e., part of the K\"ahler conditions are satisfied automatically. 

By eq.~(7.170), (iii) the generators $\{ J_n^\#\}_{n\in{\mathbb Z}}$
form a U(1) current algebra. Setting
\[
G_{n+a}^\# \ := \ \frac{1}{\sqrt{2}} \ \left( G_{n+a}^{\#\,+} +
  G_{n+a}^{\#\,-} \right) \ ,
\]
one finds that $\{ L_n^\#, G_{n+a}^\# \}_{n\in{\mathbb Z}}$ generate an
$N=1$ super-Virasoro algebra, as described in eq.~(7.152) above. 

In the identifications following eq.~(7.172), we have indicated the
possibility of two choices of sign in the right moving sector. Indeed,
the automorphism 
\begin{eqnarray}
&& G_{n+a}^\pm \ \longmapsto \ G_{n+a}'^\pm \ := \ G_{n+a}^\pm\,, 
\quad\ J_n \ \longmapsto \ J_n' \ := \ J_n , \nonumber \\
&&\overline{G}{}_{n+a}^{\,\pm} \ \longmapsto \ \overline{G}{}_{n+a}'^\pm \ :=
\ \overline{G}{}_{n+a}^{\,\mp}\,, \quad\ \overline{J}_n \ \longmapsto
\overline{J}{}'_n \ := \ -\,\overline{J}{}_n 
\end{eqnarray}
describes the {\it mirror map}, which is a symmetry of the conformal
field theory; see [110,112].
\smallskip

{}From the spectral data (7.172) of an $N=(2,2)$ superconformal field
theory ${\mathcal Q}$ one can attempt to reconstruct {\it target spaces} 
$M_{\mathcal Q}$ and $M_{{\mathcal Q}'}$ by passing from (7.172) to 
spectral data
\begin{eqnarray}
&&\left( {\mathcal F}_{{\mathcal Q}^{(0)}}, 
{\mathcal H}_{{\mathcal Q}^{(0)}},
\partial_0,  \partial_0^*, \overline{\partial}_0, 
\overline{\partial}{}_0^{\,*}, {\mathcal T}_0,
  \overline{{\mathcal T}}_0 \right) \ , \nonumber\\
{\rm or}\phantom{ZeichnungZeichnung} && \phantom{Zeichnung}\\
&& \left( {\mathcal F}_{\mathcal Q}'^{(0)} , 
{\mathcal H}_{\mathcal Q}'^{(0)},
    \partial_0', \partial_0'^{*}, \overline{\partial}{}_0',
    \overline{\partial}{}{}_0'^{*}, {\mathcal T}_0',
    \overline{{\mathcal T}}{}'_0 \right) \ ,
  \phantom{ZeichnungZeichnung} \nonumber
\end{eqnarray}
following the constructions in Sect.~{\bf7.5} and subsection 1), above: The
construction of the algebras ${\mathcal F}_{\mathcal Q}^{(0)}$ and ${\mathcal
  F}_{\mathcal Q}'^{(0)}$, represented on ``small Ramond spaces'' ${\mathcal
H}_{\mathcal Q}^{(0)}$, ${{\mathcal H}}_{\mathcal Q}'^{(0)}$, respectively, (see [24]) 
involves selecting suitable subrings of local bosonic fields of grade
(charge) = $(0,0)$ indexed by sets $\Pi^{(0)}$ and $\Pi'^{(0)}$,
respectively, of pairs of representations of the chiral algebras, as
described in Sects.~{\bf7.4} and {\bf7.5}. The Dolbeault operators $\partial_0,
\partial_0^*, \overline{\partial}_0, \overline{\partial}{}_0^{*}$ are
obtained by restricting $G_0^+, G_0^-, \overline{G}{}_0^{+}$ and
$\overline{G}{}_0^{-}$ to ${\mathcal H}_{\mathcal Q}^{(0)}$, and ${\mathcal T}_0,
\overline{{\mathcal T}}_0$ by restricting $J_0$ and $\overline{J}_0$
to ${\mathcal H}_{\mathcal Q}^{(0)}$; analogously, the operators $\partial_0',
\partial_0^{'*}, \overline{\partial}{}'_0,
\overline{\partial'}_0^{\,*}$ are obtained by restricting $G_0^+,
G_0^-, \overline{G}{}_0^{-}$ and $\overline{G}{}_0^{+}$ to ${\mathcal
  H}_{\mathcal Q}'^{(0)}$, respectively, and ${\mathcal T}_0', \overline{\mathcal
    T}{}_0'$ by restricting $J_0$ and $- \overline{J}_0$ to
${{\mathcal H}}_{\mathcal Q}'^{(0)}$, respectively. Thus, the two sets of data in
(7.174) are 
interchanged by the mirror map (7.173), and $M_{\mathcal Q}$ and $M_{\mathcal Q}'$ form a
{\it mirror pair} of (non-commutative) {\it K\"ahler spaces}. Of
course, the details of the construction of ${\mathcal F}_{\mathcal Q}^{(0)}$ and
${\mathcal H}_{\mathcal Q}^{(0)}$ or ${\mathcal F}_{\mathcal Q}'^{(0)}$ 
and ${\mathcal H}_{\mathcal Q}'^{(0)}$ 
depend on the superconformal field theory under consideration; a
satisfactory, general (model-independent) construction remains to be
found. Some simple examples are described in [24].

In spite of the fact that the detailed procedure to reconstruct the (generally 
{\it non-commutative}) target spaces $M_{\mathcal Q}$ and $M_{\mathcal Q}'$ is
not in general known, at present, a remarkable piece of general theory about 
$M_{\mathcal Q}$ and $M_{\mathcal Q}'$ {\it is known}: The theory of {\it
  chiral-chiral} and {\it chiral-antichiral rings} [112]. In
subsection~9) of Sect.~{\bf 5.2}, eq.~(5.93) and below, we have defined
an algebra
\begin{equation}
\Omega_{\partial,\overline{\partial}}^{\,^{\bullet,\bullet}} \
({\mathcal A}) \ = \ \bigoplus_{p,q} \
\Omega_{\partial, \overline{\partial}}^{p,q} \ ({\mathcal A}) 
\end{equation}
of Dolbeault forms which is a bi-graded bi-differential
algebra. Because $\partial^2 = \overline{\partial}{}^{\,2} = 0$, 
$\Omega_{\partial,\overline{\partial}}^{\,^{\bullet,\bullet}} $ is a
bi-graded complex with respect to graded commutation by $\partial$ and
by $\overline{\partial}$.

Setting ${\mathcal A} := {\mathcal F}_{\mathcal Q}^{(0)}$ , $\partial :=
\partial_0$ , $\overline{\partial} := \overline{\partial}_0$ , as in
(7.174), we obtain the differential algebra
$\Omega_{\partial_0,\overline{\partial}_0}^{\,^{\bullet,\bullet}} (
  {\mathcal F}_Q^{(0)} )$ of Dolbeault forms on $M_{\mathcal Q}$ . The
choice ${\mathcal A} := {\mathcal F}_{\mathcal Q}'^{(0)}$ , $\partial :=
\partial_0'$ , $\overline{\partial} := \overline{\partial}{}'_0$ yields
the differential algebra
$\Omega_{\partial_0',\overline{\partial}{}'_0}^{\,^{\bullet,\bullet}}
( {\mathcal F}_{\mathcal Q}'^{(0)})$ of 
Dolbeault forms on the mirror target $M_{\mathcal Q}'$.
\smallskip

Actually, the correct {\it general definition} of an algebra
$\Omega_{\partial_0,\overline{\partial}_0}^{\,^{\bullet,\bullet}}(
{\mathcal Q}) =\displaystyle\mathop{\oplus}_{p,q} 
\Omega_{\partial_0,  \overline{\partial}_0}^{p,q} 
({\mathcal Q})$ of Dolbeault forms of an $N=(2,2)$
superconformal field theory ${\mathcal Q}$ is to demand that 
$\Omega_{\partial_0,  \overline{\partial}_0}^{p,q} ({\mathcal Q})$ 
contain {\it all} functionals $\varphi^{p,q}$ of fields of ${\mathcal Q}$ 
with the properties that
$\varphi^{p,q}$ leaves ${\mathcal H}_{\mathcal Q}^{(0)}$ invariant and
\begin{equation}
\left[\,J_0, \varphi^{p,q}\,\right] \ = \ p\,\varphi^{p,q}\, , \ \quad
\left[\,
  \overline{J}_0, \varphi^{p,q}\,\right] \ = \ q\,\varphi^{p,q} \ .
\end{equation}
In general, the charges (grades) $p$ and $q$ are {\it not}
integers. However, if $c = \overline{c} = 3 n$, $n = 1,2,3,\ldots\,,$
and some additional properties are satisfied, 
$p$ and $q$ turn out to be integers. In this case, a correct
choice of the algebra ${\mathcal F}_{\mathcal Q}^{(0)}$ is one for which 
\begin{equation}
\Omega_{\partial_0,\overline{\partial}_0}^{\,^{\bullet,\bullet}}
({\mathcal Q})
\ = \ \Omega_{\partial_0,\overline{\partial}_0}^{\,^{\bullet,\bullet}}
( {\mathcal F}_{\mathcal Q}^{(0)}) \ .
\end{equation}
An algebra
$\Omega_{\partial_0',\overline{\partial}{}_0'}^{\,^{\bullet,\bullet}}
({\mathcal Q})$ is defined similarly, and if $c = \overline{c} = 3 n$, 
one must attempt to choose ${{\mathcal F}}_{\mathcal Q}'^{(0)}$
such that
\begin{equation}
\Omega_{\partial_0',\overline{\partial}{}_0'}^{\,^{\bullet,\bullet}}
({\mathcal Q})
\ = \ \Omega_{\partial_0',\overline{\partial}{}_0'}^{\,^{\bullet,\bullet}}
( {{\mathcal F}}_{\mathcal Q}'^{(0)}) \ .
\end{equation}
Eqs.~(7.177) and (7.178) are crucial consistency conditions.
\smallskip

The operators $\partial_0$ and $\overline{\partial}_0$ act on 
$\Omega_{\partial_0,\overline{\partial}_0}^{\,^{\bullet,\bullet}} \
({\mathcal Q})$ by graded commutation. One can then attempt to determine the
cohomology groups $H_{\partial_0,\overline{\partial}_0}^{p,q} ({\mathcal Q})$. It
turns out that $H_{\partial_0,\overline{\partial}_0}^{p,q} ({\mathcal Q})$
contains ``harmonic forms'' $\varphi_{\ \alpha}^{p,q}$ , $\alpha =
1,2,3,\ldots,$ for all $p,q,$ which are in a one-to-one correspondence
to {\it chiral-chiral, primary states} in the Neveu-Schwarz sector of
$Q$. A state $\big| \varphi_{\ \alpha}^{p,q}\rangle$ \ in the
Neveu-Schwarz sector is chiral-chiral iff
\begin{equation}
G_{- 1/2}^+ \ \big| \varphi_{\ \alpha}^{p,q}\rangle \ = \
\overline{G}{}_{- 1/2}^{+} \ \big| \varphi_{\ \alpha}^{p,q} \rangle \
= \ 0
\end{equation}
and primary iff it is a highest weight vector for the $N=2$
super-Virasoro algebras. It then follows that $h=\frac p 2$ ,
$\overline{h} = \frac q 2$ , where $h$ and $\overline{h}$ are the
conformal weights of \ $\big| \varphi_{\ \alpha}^{p,q} \rangle$. One
can show that $h \leq \frac c 6$ , $\overline{h} \leq
\frac{\overline{c}}{6}$ .

It turns out (see [113]) that chiral-chiral primary operators $\{
\varphi_{\ \alpha}^{p,q} \}$ form a ring, the {\it chiral-chiral ring}
$H_{\partial_0, \overline{\partial}_0}^{\,^\bullet} ({\mathcal Q})$, which, in
examples, can often be determined explicitly. If $c = \overline{c} = 3
n$ for some positive integer $n$, and assuming that (7.177) holds, then
$H_{\partial_0, \overline{\partial}_0}^{\,^\bullet} ({\mathcal Q})$ is what one
might interpret as the {\it Dolbeault cohomology ring} of $M_{\mathcal Q}$. 

Analogous results hold when $\partial_0, \overline{\partial}_0$ and 
$\Omega_{\partial_0,\overline{\partial}_0}^{\,^{\bullet,\bullet}} 
({\mathcal Q})$ are replaced by $\partial_0', \overline{\partial}{}'_0$ and
$\Omega_{\partial_0',\overline{\partial}{}'_0}^{\,^{\bullet,\bullet}}
({\mathcal Q})$, respectively. One then arrives at the {\it chiral-antichiral
  ring} $H_{\partial'_0, \overline{\partial}{}'_0}^{\,^\bullet} 
({\mathcal Q})$
describing the Dolbeault cohomology ring of $M_{\mathcal Q}'$.
\smallskip

The ring structure of $H_{\partial_0,
  \overline{\partial}_0}^{\,^\bullet} ({\mathcal Q})$ and $H_{\partial_0',
  \overline{\partial}{}'_0}^{\,^\bullet} ({\mathcal Q})$ is, in general, 
not that of cohomology rings of classical manifolds, but of certain
deformations of such rings, although the dimensions of the spaces of
harmonic forms of definite U(1) charge may coincide with the Hodge
numbers of a classical Calabi-Yau space. 

One can verify that the theory of chiral-chiral and chiral-antichiral
rings [112] fits into the general cohomology theory of complex
non-commutative geometry, as outlined in Sect.~{\bf 5.2} and in [18]. 

For further material on $N=(2,2)$ superconformal field theories see
the lectures by B.~Greene.
\medskip

In the study of string vacua with internal target spaces described by
hyper-K\"ahler manifolds one also encounters $N=(4,4)$ superconformal
field theories. They are based on the representation theory of two
copies of the $N=4$ super-Virasoro algebra with generators $\{ L_n,
G_r^{A\pm}, T_m^I \}$ satisfying the commutation relations 
\begin{eqnarray}
&&(i)\qquad \left[\, L_n,L_m\,\right] \ = \ (n-m)\,L_{n+m} + \frac{c}{12} \
n (n^2-1)\,\delta_{n+m,0} \nonumber \\
&&(ii)\quad\ \ \left[\, L_n, T_m^I \,\right] \ = \ -\,m\,T_{n+m}^I \nonumber\\
&&(iii)\quad\ \left[\, T_n^I, T_m^J\,\right] \ = \
i\,\varepsilon^{IJK}\,T_{n+m}^K + \delta^{IJ}
\frac{c}{12}\,n\,\delta_{n+m,0} \nonumber\\
&&(iv)\quad\ \, \left[\, L_n, G_r^{A\pm}\,\right] \ = \ \left( \frac n 2
  \ -\,r\right) \,G_{n+r}^{A\pm}\nonumber\\
&&(v)\quad\ \ \left[\, T_n^I, G_r^{A+}\,\right] \ = \ \frac 1 2 \ 
\overline{(\sigma^I)^{AB}}\; G_{n+r}^{B+} \ , \nonumber\\
&&\phantom{(v)}\quad\ \  \left[\, T_n^I, G_r^{A-}\,\right] \ = \ -\, \frac 1 2
 \,(\sigma^I)^{AB}\;G_{n+r}^{B-} \nonumber\\
&&(vi)\quad\ \left\{ G_r^{A+}, G_s^{B-}\right\} \ = \
2\,\delta^{AB}\,L_{r+s} + 2\,(r-s) (\sigma^I)^{AB}\, T_{r+s}^I
+ \frac c 3 \left( r^2-\frac 1 4
\right)\,\delta^{AB}\,\delta_{r+s,0} \nonumber\\
&&(vii)\quad \left\{ G_r^{A+}, G_s^{B+}\right\} \ = \ \left\{
  G_r^{A-}, G_s^{B-}\right\} \ = \ 0 \ .\nonumber
\end{eqnarray}

\noindent Here $A, B \in \{ 1,2\}, \ I, J, K \in \{ 1,2,3\}, \ r,
s\in {\mathbb Z}$ (Ramond) or $r, s \in {\mathbb Z} + \frac 1 2 $ \
(Neveu-Schwarz), and $\sigma^I, \ I=1,2,3,$ are the 2$\times$2 Pauli
matrices. In a unitary representation, one has that
\[
L_n^* = L_{-n}\,, \quad\ \left( T_m^I\right)^* \ = \ T_{-m}^I\,, \
\quad\left(G_r^{A\pm}\right)^* \ = \ G_{-r}^{A\mp} \ .
\]
The operators $\{ T_n^I\}_{n\in{\mathbb Z}}$ are the Fourier modes of
an $SU(2)$--current $T$ generating an $\widehat{su}(2)$--Kac-Moody
algebra at level $k=\frac c 6$. The operators $(G_r^{1\pm},
G_r^{2\pm})$ form SU(2)--doublets; thus SU(2) is a {\it ``vertical
  symmetry''} of the $N=4$ super-Virasoro algebra in the sense
explained in subsection 9) of Sect.~{\bf 5.2}. It corresponds to the
vertical SU(2) symmetry generated by the holomorphic symplectic form
and the holomorphic ${\mathbb Z}$--grading on the space of holomorphic
differential forms on a hyper-K\"ahler manifold.
\medskip

Unfortunately, we cannot enter into a more detailed discussion of the
mathematically fascinating world of the (non-commutative) 
target- and loop space geometry of superconformal field theories. We
refer the reader to [110,112,24] and the references given there for
examples. But we hope that we have made the point that a combination
of $N=2$ (and $N=4$) superconformal field theory with the methods of
non-commutative K\"ahler and hyper-K\"ahler geometry, as described in
Section {\bf 5} and in [18], provides a natural conceptual framework
for the study of topics such as {\it mirror symmetry}, {\it topology 
changes}, {\it supersymmetric cycles}, etc.

\vspace{.6cm}

\newpage

\section{Conclusions}

In these notes we have attempted to review some physical foundations
and some conceivably useful mathematical methods that may guide a way
towards a ``quantum theory of space-time-matter'', yet to be
discovered. We have argued (Section~{\bf 3}) that in such a theory,
space, time and matter loose their individuality and that classical
space-time is an approximate notion that is only appropriate for the
description of some asymptotic regimes of a fundamental quantum theory
of space-time-matter. The intrinsic geometry of space-time-matter is
expected to be {\it non-commutative}. This feature can best be 
taken into account by trying to conceive the fundamental theory as a
{\it theory of extended objects}. For such a theory to have
geometrical content, it is natural to require that its solutions
exhibit {\it supersymmetry}, i.e., take the form of supersymmetric
quantum theories. Key examples of supersymmetric quantum theories are
Pauli's quantum theories of a non-relativistic electron with spin, of
its twin, the non-relativistic positron, and of positronium (i.e., of
a bound state of an electron and a positron), as described in
Section~{\bf 4}. Pauli's quantum theory of non-relativistic particles
with spin neatly encodes the classical differential topology and geometry
of Riemannian manifolds and suggests natural generalizations of
classical differential topology and geometry, called non-commutative
geometry, as described in Connes' book [5] and in Section~{\bf 5} and
[18]. We have discussed some examples of non-commutative geometrical
spaces in Section~{\bf 6} (non-commutative torus) and Sects.~{\bf 7.5} 
and {\bf 7.6} (e.g.\ the ``fuzzy 3-sphere'').
\smallskip

First quantized, tree-level superstring theory, as briefly described
in Sects.~{\bf 7.1} and {\bf 7.2}, is a very sophisticated analogue of
Pauli's quantum theory of non-relativistic, spinning particles. It
encodes the topology and geometry of a certain class of loop spaces
over generally {\it non-commutative} geometrical spaces describing
physical space-times. The supersymmetry algebras, more precisely the
superconformal field theories, describing superstring vacua provide
key tools to explore the geometry of those loop spaces. Unfortunately,
space-times described by the vacua of first quantized, tree-level
superstring theory are {\it static}.

For purposes of physics, the present formulation of first quantized
superstring theory is ultimately inadequate in that it is an
intrinsically perturbative approach towards understanding the
presumably intrinsically non-perturbative quantum dynamics of
space-time-matter. It does not appear to enable one to properly
describe the dynamical degrees of freedom of non-static,
non-commutative quantum space-times. 

In order to overcome the shortcomings of first quantized superstring
theory, one is tempted to search for ``second quantized'' theories. In
passing from first quantized to second quantized theories, one appears
to trade parameter space supersymmetry for target space supersymmetry,
and one should worry that one may loose ``back ground
independence''. Some preliminary ideas about  second quantized,
non-perturbative formulations of a quantum theory of space-time-matter
have been reviewed in Sect.~{\bf 7.3} (``matrix models'').
They have the positive features that parameter- and target space are
treated as non-commutative spaces and that they appear to incorporate
some of the general principles reviewed in Section~{\bf 3}. But they
have the negative feature that they are based on too rigid a notion of
target space (involving {\it global} symmetries) and that their
very formulation requires choosing a light-cone gauge, so far. In how
far superstring theory emerges from matrix models in a limiting regime
is only partially understood.
\smallskip

All theories alluded to in Section~{\bf 7} have the {\it common feature}
that they yield {\it supersymmetric spectral data} (of the kind
studied in Section~{\bf 5}) which enable one to construct
non-commutative geometric spaces. While the geometric spaces
constructed from the spectral data of vacua of first quantized
superstring theory have a more or less direct relationship with
space-time, the geometric spaces constructed from spectral data
provided by second quantized theories are spaces describing, in
principle, all dynamical degrees of freedom (of ``space-time-matter'';
in a sense they are the configuration- or {\it quantized phase spaces}
of ``space-time-matter''), and it is not clear, yet, how one may 
extract from them geometrical features of physical space-time. 

Yet, the common features of the theories described in Section~{\bf 7}
may encourage us to propose the following 

\begin{quote}
{\bf ``Geometrization Principle''}. A fundamental theory of
space-time-matter has solutions yielding supersymmetric spectral data
analogous to those described in Section~{\bf 5} from which models of
non-commutative space-time can be reconstructed.
\end{quote}

\noindent It is likely that the way to finding a satisfactory 
non-perturbative formulation of a fundamental quantum theory of 
space-time-matter remains long and steep, resembling an ascent 
to Mount Everest rather than to Mont Blanc.

\vfill
\newpage

%
\def\spa{\hskip10pt}\def\spb{\hskip11pt}

\noindent {\large{\bf References}}

\begin{description} \parskip=-1pt
\item[[1$\!\!\!$]]\spb\, D.\ Christodoulou, S.\ Klainerman, ``The global non-linear
  stability of the Minkowski space'', Princeton University Press 1993.
\item[[2$\!\!\!$]]\spb\, M.\ Kac, ``Can one hear the shape of a drum?'',
  Amer.~Math.~Monthly {\bf 73} (1966) 1--23.
\item[[3$\!\!\!$]]\spb\, B.\ Colbois, G.\ Courtois, ``A note on the first non-zero
  eigenvalue of the Laplacian acting on $p$-forms'', Manuscripta
  Math.~{\bf 68} (1990) 143--160;\\
   J.\ Cheeger, ``A lower bound for the smallest eigenvalue of the
  Laplacian'' in ``Problems in analysis (Papers dedicated to Salomon
  Bochner, 1969)'', Princeton University Press 1970, pp.\ 195--199.
\item[[4$\!\!\!$]]\spb\, O.\ Bratteli, D.W.\ Robinson, ``Operator Algebras and
  Quantum Statistical Mechanics 1'', Springer Verlag 1987.
\item[[5$\!\!\!$]]\spb\, A.\ Connes, ``Noncommutative Geometry'', Academic Press
  1994.
\item[[6$\!\!\!$]]\spb\, J.\ Fr\"ohlich, P.\ Pfeifer, ``Generalized time-energy
  uncertainty relations and \break
bounds on life times of resonances'',
  Rev.~Mod.~Phys.~{\bf 67} (1995) 759--779.
\item[[7$\!\!\!$]]\spb\, S.\ Doplicher, K.\ Fredenhagen, J.E.\ Roberts, ``The 
  quantum
  structure of space-time at the Planck scale and quantum fields'',
  Commun.~Math.~Phys.~{\bf 172} (1995) 187--220.
\vskip-0.5cm
\item[[8$\!\!\!$]]\spb\, J.\ Fr\"ohlich, ``The non-commutative geometry of
  two-dimensional supersymmetric conformal field theory'', in PASCOS,
  Proc.~of the Fourth Intl.~Symp.~on Particles, Strings and Cosmology,
  K.C.\ Wali (ed.), World Scientific 1995.
\item[[9$\!\!\!$]]\spb\, S.W.\ Hawking, ``Black hole explosions?'', Nature~{\bf
    248} (1974) 30--31;\\
``Particle creation by black holes'', Commun.~Math.~Phys.~{\bf 43}
(1975) 199--220.
\item[[10$\!\!\!$]]\spa I.D.\ Novikov, V.P.\ Frolov, ``Physics of Black
  Holes'', Kluwer 1989.
\item[[11$\!\!\!$]]\spa K.\ Schoutens, H.\ Verlinde, E.\ Verlinde, ``Quantum black
  hole evaporation'', Phys.\ Rev.~D~{\bf 48} (1993) 2670--2685.
\item[[12$\!\!\!$]]\spa F.\ Gabbiani, J.\ Fr\"ohlich, ``Operator algebras and
  conformal field theory'', Commun.~Math.~Phys.~{\bf 155} (1993) 
  569--640.
\item[[13$\!\!\!$]]\spa D.\ Buchholz, E.H.\ Wichmann, ``Causal independence and the
  energy-level density of states in local quantum field theory'',
  Commun.~Math.~Phys.~{\bf 106} (1986) 321--344.
\item[[14$\!\!\!$]]\spa G.\ 't Hooft, ``Dimensional reduction in quantum gravity'',
  in Salamfest 1993, gr-qc/9310026;\\
L.\ Susskind, ``The world as a hologram'', J.~Math.~Phys.~{\bf 36}
(1995) 6377--6396.
\item[[15$\!\!\!$]]\spa E.\ Witten, ``Constraints on supersymmetry breaking'',
  Nucl.~Phys.~B~{\bf 202} (1982) 253--316.
\item[[16$\!\!\!$]]\spa E.\ Witten, ``Supersymmetry and Morse theory'',
  J.~Diff.~Geom.~{\bf 17} (1982) 661--692.
\item[[17$\!\!\!$]]\spa N.\ Berline, E.\ Getzler, M.\ Vergne, ``Heat Kernels and
  Dirac Operators'', Springer Verlag 1992.
\item[[18$\!\!\!$]]\spa J.\ Fr\"ohlich, O.\ Grandjean, A.\ Recknagel,
  ``Supersymmetric quantum theory and (non-commutative) differential
  geometry'', hep-th/9612205, to be published in Commun.\ Math.\ Phys.
\item[[19$\!\!\!$]]\spa J.\ Fr\"ohlich, O.\ Grandjean, A.\ Recknagel, ``Supersymmetry
  and non-commutative geometry'', to be published in the proceedings
  of the Carg\`ese 1996 Summer School on ``Quantum Fields and Quantum
  Space-Time''. 
\item[[20$\!\!\!$]]\spa D.\ Salamon, ``Spin Geometry and Seiberg-Witten
  Invariants'', preprint 1995, to be published by Birkh\"auser
  Verlag. 
\item[[21$\!\!\!$]]\spa J.\ Fr\"ohlich, E.H.\ Lieb, M.\ Loss,
  ``Stability of Coulomb 
  systems with magnetic fields~I'', Commun.~Math.~Phys.~{\bf 104}
  (1986) 251--270;\\
E.H.\ Lieb, M.\ Loss, ``Stability of Coulomb systems with magnetic
fields II'', Commun.~Math.~Phys.~{\bf 104} (1986) 271--282.
\item[[22$\!\!\!$]]\spa L.\ Alvarez-Gaum\'e, ``Supersymmetry and the Atiyah-Singer
  index theorem'', Commun.~Math.~Phys.~{\bf 90} (1983) 161--173.
\item[[23$\!\!\!$]]\spa E.\ Nelson, ``Feynman integrals and the Schr\"odinger
  equation'', J.~Math.~Phys.~{\bf 5} (1964) 332--343.
\item[[24$\!\!\!$]]\spa J.\ Fr\"ohlich, K.\ Gaw\c edzki, ``Conformal field theory
  and the geometry of strings'', CRM Proceedings and Lecture Notes
  Vol.~{\bf 7} (AMS Publications) (1994) 57--97.
\item[[25$\!\!\!$]]\spa J.\ Feldman, D.\ Lehmann, H.\ Kn\"orrer, E.\ Trubowitz, 
   ``Fermi liquids in two space dimensions'', in: ``Constructive Physics'', 
    V.\ Rivasseau (ed.), Lecture Notes in Phys\-ics {\bf 446}, Springer Verlag 1995;\\
  T.\ Chen, J.\ Fr\"ohlich, M.\ Seifert, ``Renormalization
  group methods: Landau-Fermi liquid and BCS superconductor'', in: Les
  Houches 1994: Fluctuating Geometries in Statistical Mechanics and
  Field Theory, F.\ David, P.\ Ginsparg, J.\ Zinn-Justin (eds.), Elsevier
  Science B.V.\ 1996, pp.\ 913--970. 
\item[[26$\!\!\!$]]\spa A.L.\ Besse, ``Einstein Manifolds'', Springer Verlag 1987.
\item[[27$\!\!\!$]]\spa D.D.\ Joyce, ``Compact hyper-complex and quaternionic
  manifolds'', J.~Diff.~Geom.\ {\bf 35} (1992) 743--762;\\
``Manifolds with many complex structures'',
Q.J.~Math.~Oxf.~II.~Ser.~{\bf 46} (1995) 169--184.
\item[[28$\!\!\!$]]\spa B.\ de Wit, A.K.\ Tollst\'en, H.\ Nicolai, ``Locally
  supersymmetric $D=3$ non-linear sigma models'', Nucl.~Phys.~B~{\bf
    392} (1993) 3--38.
\item[[29$\!\!\!$]]\spa M.B.\ Green, J.H.\ Schwarz, E.\ Witten, ``Superstring Theory
  I'', Cambridge University Press 1987.
\item[[30$\!\!\!$]]\spa C.\ Chevalley, S.\ Eilenberg, ``Cohomology of Lie groups and
  Lie algebras'', Trans.\ Amer.~Math.~Soc.~{\bf 63} (1948) 85--124.
\item[[31$\!\!\!$]]\spa J.\ Fr\"ohlich, ``Classical and quantum statistical
  mechanics in one and two dimensions: two-component Yukawa- and
  Coulomb systems'', Commun.~Math.~Phys.~{\bf 47} (1976) 233--268.
\item[[32$\!\!\!$]]\spa C.\ Fefferman, ``Stability of Coulomb systems in a magnetic
  field'', Proc.~Natl.~Acad.\ Sci.~{\bf 92} (1995) 5006--5007;\\
E.H.\ Lieb, M.\ Loss, J.P.\ Solovej, ``Stability of matter in magnetic
fields'', Phys.~Rev.\ Lett.~{\bf 75} (1995) 985--989;\\
E.H.\ Lieb, ``The stability of matter: from atoms to stars'', Springer
1997, 2nd Ed.
\item[[33$\!\!\!$]]\spa M.\ Loss, H.-T.\ Yau, ``Stability of Coulomb systems with
  magnetic fields III'', Commun.~Math.~Phys.~{\bf 104} (1986) 
  283--290.
\item[[34$\!\!\!$]]\spa N.\ Seiberg, E.\ Witten, ``Electric-magnetic duality,
  monopole condensation, and confinement in $N=2$ supersymmetric
  Yang-Mills theory'', Nucl.~Phys.~B~{\bf 426} (1994) 19--52, Erratum
  ibid.~B~{\bf 430} (1994) 485--486;\\
``Monopoles, duality, and chiral symmetry breaking in $N=2$
supersymmetric QCD'', Nucl.~Phys.~B~{\bf 431} (1994) 484--550;\\
E.\ Witten, ``Monopoles on 4-manifolds'', IAS preprint, 29 pp.,
hep-th/9411102.
\item[[35$\!\!\!$]]\spa J.\ Anandan, ``Electromagnetic effects in the quantum
  interference of dipoles'', Phys.\ Lett.\ A {\bf 138} (1989) 
  347--352;\\
J.\ Fr\"ohlich, U.\ Studer, ``Gauge invariance and  current algebra in
non-relativistic many-body theory'', Rev.~Mod.~Phys.~{\bf 65} (1993) 
733--802.\ 
\item[[36$\!\!\!$]]\spa M.\ Baake, P.\ Reinicke, V.\ Rittenberg, ``Fierz identities
  for real Clifford algebras and the number of supercharges'',
  J.~Math.~Phys.~{\bf 26} (1985) 1070--1071;\\
M.\ Claudson, M.B.\ Halpern, ``Supersymmetric ground state wave
functions'', Nucl.\ Phys.~B~{\bf 250} (1985) 689--715.
\item[[37$\!\!\!$]]\spa V.\ Bach, J.\ Fr\"ohlich, I.M.\ Sigal, ``Mathematical theory
  of non-relativistic matter and radiation'', Lett.~Math.~Phys.~{\bf
    34} (1995) 183--201.
\item[[38$\!\!\!$]]\spa L.\ Alvarez-Gaum\'e, D.Z.\ Freedman, ``Geometrical structure
  and ultraviolet finiteness in the supersymmetric $\sigma$-model'',
  Commun.~Math.~Phys.~{\bf 80} (1981) 443--451;\\
  N.J.\ Hitchin, A.\ Karlhede, U.\ Lindstrom, M.\ Rocek, 
   ``Hyperk\"ahler metrics and supersymmetry'', 
   Commun.\ Math.\ Phys.\ {\bf108} (1987) 535-589; \\
E.\ Bergshoeff, E.\ Szegin, H.\ Nishino, ``(8,0) Locally supersymmetric
$\sigma$-models with conformal invariance in two dimensions'',
Phys.~Lett.~B~{\bf 186} (1987) 167--172;\\
H.\ Nicolai, ``The integrability of $N=16$ supergravity'',
Phys.~Lett.~B~{\bf 194} (1987) 402--407;\\
B.\ de Wit, P.\ van Nieuwenhuizen, ``Rigidly and locally supersymmetric
two-dimen\-sional non-linear $\sigma$-models with torsion'',
Nucl.~Phys.~B~{\bf 312} (1989) 58--94.
\item[[39$\!\!\!$]]\spa O.\ Augenstein, ``Supersymmetry, non-commutative geometry
  and model building'', Diploma Thesis ETH Z\"urich, March 1996.\ 
\item[[40$\!\!\!$]]\spa J.\ Wess, J.\ Bagger, ``Supersymmetry and supergravity'',
  Princeton University Press 1983;\\
P.\ West, ``Introduction to supersymmetry and supergravity'', World
Scientific 1986.
\item[[41$\!\!\!$]]\spa E.\ Witten, ``Non-abelian bosonization in two dimensions'',
  Commun.~Math.~Phys.\ {\bf 92} (1984) 455--472.
\item[[42$\!\!\!$]]\spa D.\ Gepner, E.\ Witten, ``String theory on group
  manifolds'', Nucl.~Phys.~B~{\bf 278} (1986) 493--549;\\
G.\ Felder, K.\ Gaw\c edzki, A.\ Kupiainen, ``Spectra of
Wess-Zumino-Witten models with arbitrary simple groups'',
Commun.~Math.~Phys.~{\bf 117} (1988) 127--158;\\
A.\ Pressley, G.\ Segal, ``Loop groups'', Clarendon Press 1986.\ 
\item[[43$\!\!\!$]]\spa A.\ Connes, ``Noncommutative differential geometry'', 
   Inst.\ Hautes \'Etudes Sci.\ Publ.\ Math.\ {\bf62} (1985) 257-360.\ 
\item[[44$\!\!\!$]]\spa A.\ Connes, M.\ Karoubi, ``Caract\`ere multiplicatif d'un
  module de Fredholm'', $K$-Theory~{\bf 2} (1988) 431--463.
\item[[45$\!\!\!$]]\spa A.H.\ Chamseddine, G.\ Felder, J.\ Fr\"ohlich, ``Gravity in
  non-commutative geometry'', Commun.~Math.~Phys.~{\bf 155} (1993) 
  205--217.
\item[[46$\!\!\!$]]\spa A.\ Connes, ``The action functional in non-commutative
  geometry'', Commun.\ Math.\ Phys.~{\bf 117} (1988) 673--683.
\item[[47$\!\!\!$]]\spa R.G.\ Swan, ``Vector bundles and projective modules'',
  Trans.~Amer.~Math.~Soc.\ {\bf 105} (1962) 264--277.
\item[[48$\!\!\!$]]\spa N.\ Jacobson, ``Basic Algebra II'', W.H.\ Freeman and
  Company 1985.
\item[[49$\!\!\!$]]\spa A.H.\ Chamseddine, J.\ Fr\"ohlich, O.\ Grandjean, ``The
  gravitational sector in the Connes-Lott formulation of the standard
  model'', J.~Math.~Phys.~{\bf 36} (1995) 6255--6275.
\item[[50$\!\!\!$]]\spa A.\ Connes, J.\ Lott, ``Particle models and non-commutative
  geometry'', Nucl.~Phys.\ Proc.\ Suppl.\ B~{\bf 18} (1990) 29--47; \\
Proceedings of 1991 Carg\`ese Summer School, ed.~by J.\ Fr\"ohlich,
G.\ 't Hooft, A.\ Jaffe, G.\ Mack, P.K.\ Mitter, and R.\ Stora, Plenum 1992.
\item[[51$\!\!\!$]]\spa A.\ Connes, ``$C^*$-alg\`ebres et g\'eom\'etrie
  diff\'erentielle'', C.R.\ Acad.~Sci.~Paris S\'er.~A-B, {\bf 290}
  (1980) 599--604;\\
M.\ Rieffel, ``Non-commutative tori -- a case study of non-commutative
differentiable manifolds'' Contemp.~Math.~{\bf 105} (1990) 191--211.
\item[[52$\!\!\!$]]\spa A.\ Jaffe, A.\ Lesniewski, K.\ Osterwalder, ``Quantum
  $K$-theory I: The Chern character'', Commun.~Math.~Phys.~{\bf 118}
  (1988) 1--14;\\
 ``On super-KMS functionals and entire cyclic
  cohomology'', $K$-theory~{\bf 2} (1989) 675--682;\\
A.\ Jaffe, K.\ Osterwalder, ``Ward identities for non-commutative
geometry'', Commun.~Math.~Phys.~{\bf 132} (1990) 119--130.
\item[[53$\!\!\!$]]\spa C.\ Itzykson, J.-B.\ Zuber, ``Quantum Field Theory'',
  McGraw-Hill 1980.
\item[[54$\!\!\!$]]\spa A.\ Connes, ``Reality and non-commutative geometry'',
  J.~Math.~Phys.~{\bf 36} (1995) 6194--6231.
\item[[55$\!\!\!$]]\spa J.\ Fr\"ohlich, O.\ Grandjean, A.\ Recknagel, unpublished
  notes.
\item[[56$\!\!\!$]]\spa M.F.\ Atiyah, R.\ Bott, ``A Lefshetz fixed
  point formula for 
  elliptic differential operators'', Bull.~Am.~Math.~Soc.~{\bf 12}
  (1966) 245;\\
``A Lefshetz fixed point formula for elliptic complexes I, II'',
Ann.~Math.~{\bf 86} (1967) 374; Ann.~Math.~{\bf 88} (1968) 451.
\item[[57$\!\!\!$]]\spa R.\ Haag, J.\ Lopuszanski, M.\ Sohnius, ``All possible
  generators of supersymmetries of the $S$-matrix'',
  Nucl.~Phys.~B~{\bf 88} (1975) 257--274.
\item[[58$\!\!\!$]]\spa R.\ Haag, ``Local Quantum Physics'', Springer Verlag
1992.
\item[[59$\!\!\!$]]\spa R.\ Longo, ``Index of subfactors and statistics of quantum
  fields I, II'', Commun.~Math.\ Phys.~{\bf 126} (1989) 217--247;\\
S.\ Doplicher, J.E.\ Roberts, ``Compact Lie groups associated with
endomorphisms of $C^*$-algebras'', Bull.~Am.~Math.~Soc.~{\bf 11}
(1984) 333--338;\\
``A new duality theory for compact groups'', Invent.~math.~{\bf 98}
(1989) 157--218;\\
``Endomorphisms of $C^*$-algebras, cross products and duality for
compact groups'', Ann.~Math.~{\bf 130} (1989) 75--119;\\
``Why there is a field algebra with a compact gauge group describing
the superselection structure in particle physics'',
Commun.~Math.~Phys.~{\bf 131} (1990) 51--107;\\
J.\ Fr\"ohlich, C.\ King ``Two-dimensional conformal field theory and
three-dimensional topology'', Int.~J.~of Mod.~Phys.~A~{\bf 20} (1989) 
5321--5399;\\
see also [95], [101], [102], [103].
\item[[60$\!\!\!$]]\spa K.\ Kodaira, ``Complex manifolds and
  deformation of complex 
  structures'', Springer 1986.
\item[[61$\!\!\!$]]\spa J.\ Hoppe, ``Quantum theory of a massless relativistic 
   surface and a two-dimensional boundstate problem'',  Ph.D.\ Thesis, MIT 1982;\\
  ``Quantum theory of a relativistic surface'', in ``Constraint's Theory and
  Relativistic Dynamics'', G.\ Longhi, L.~Lusanna (eds.), Proc.\ Florence
  1986, World Scientific; \\
 B.\ de Wit, J.\ Hoppe, H.\ Nicolai, ``On the quantum mechanics of
supermembranes'', Nucl.~Phys.~B~{\bf 305} [FS23] (1988) 545--581;\\
B.\ de Wit, M.\ L\"uscher, H.\ Nicolai, ``The supermembrane is
unstable'', Nucl.~Phys.\ B~{\bf 320} (1989) 135--159;\\
 see also [82], [85].
\item[[62$\!\!\!$]]\spa S.\ Cordes, G.\ Moore, S.\ Ramgoolam, ``Lectures on 2D
  Yang-Mills theory, equivariant cohomology and topological field theories'', 
  in ``Fluctuating geometries in statistical mechanics and field
  theory: Les Houches 1994'', F.\ David, P.\ Ginsparg, J.\ Zinn-Justin (eds.), 
  North-Holland 1996.
\item[[63$\!\!\!$]]\spa E.\ Witten, ``Topological sigma models'',
  Commun.~Math.~Phys.~{\bf 118} (1988) 411-419; \\
``Topological Gravity'', Phys.~Lett.~{\bf 206} B (1988) 601--606.
\item[[64$\!\!\!$]]\spa K.\ Symanzik, ``Euclidean quantum field theory'', 
  in ``Local Quantum Theory'', proceedings of the International School of Physics
  ``Enrico Fermi'' Varenna 1968, Academic Press 1969.
\item[[65$\!\!\!$]]\spa M.\ Aizenman, ``Proof of the triviality of $\varphi_d^4$
  field theory and some mean-field features of Ising models for
  $d>4$'', Phys.~Rev.~Lett.~{\bf 47} (1981) 1--4, 886 (E);\\
  ``Geometric analysis of $\varphi^4$ fields and Ising models I,II'', Commun.\
   Math.\ Phys.\ {\bf86} (1982) 1--48;\\
   J.\ Fr\"ohlich, ``On the triviality of $\lambda\varphi_4^4$ theories
   and the approach to the critical point in $d\,\, \grgl \,4\,$
   dimensions'', Nucl.~Phys.~B~{\bf 200} [F54] (1982) 281--296;\\
   R.\ Fernandez, J.\ Fr\"ohlich, A.D.\ Sokal, ``Random Walks, Critical
   Phenomena, and Triviality in Quantum Field Theory'', Texts and
    Monographs in Physics, Springer 1992.
\item[[66$\!\!\!$]]\spa S.\ Deser, B.\ Zumino, ``A complete action for the spinning
  string'', Phys.~Lett.~{\bf 65} B (1976) 369--373;\\
A.M.\ Polyakov, ``Quantum geometry of bosonic strings'',
Phys.~Lett.~{\bf 103} B (1981) 207--210; 
``Quantum geometry of fermionic strings'', ibid.\ {\bf 103} B
(1981) 211--213.
\item[[67$\!\!\!$]]\spa T.\ Goto, ``Relativistic quantum mechanics of
  one-dimensional mechanical continuum and subsidiary condition of
  dual resonance model'', Prog.~Theor.~Phys.~{\bf 46} (1971) 
  1560--1569;\\
Y.\ Nambu, Lectures at the Copenhagen Summer Symposium, 1970
(unpublished).
\item[[68$\!\!\!$]]\spa L.D.\ Faddeev, V.N.\ Popov, ``Feynman Diagrams for the
  Yang-Mills Field'', Phys.\ Lett.~{\bf 25} B (1967) 29--30.
\item[[69$\!\!\!$]]\spa K.\ Gaw\c edzki, ``Lectures on conformal field
  theory'', preprint IHES-P-97-2.
\item[[70$\!\!\!$]]\spa D.\ Friedan, ``Non-linear models in $2+\varepsilon$
  dimensions'', Ph.D.~thesis (1980) published in Ann.~Phys.~{\bf 163}
  (1985) 318--419; Phys.~Rev.~Lett.~{\bf 45} (1980) 1057--1060;\\
C.G.\ Callan, D.\ Friedan, E.\ Martinec, M.J.\ Perry, ``Strings in
background fields'', Nucl.~Phys.~B~{\bf 262} (1985) 593--609;\\
C.\ Schmidhuber, A.A.\ Tseytlin, ``On string cosmology and the RG flow
in 2-d field theory'', Nucl.~Phys.~B~{\bf 426} (1994) 187--202.
\item[[71$\!\!\!$]]\spa D.J.\ Gross, P.F.\ Mende, ``String theory beyond the Planck
  scale'', Nucl.~Phys.\ B~{\bf 303} (1988) 407--454.
\item[[72$\!\!\!$]]\spa V.G.\ Knizhnik, A.M.\ Polyakov, A.B.\ Zamolodchikov,
  ``Fractal structure of 2-d quantum gravity'',
  Mod.~Phys.~Lett.~A~{\bf 3} (1988) 819--826;\\
F.\ David, ``Conformal field theories coupled to 2-d gravity in the 
conformal gauge'', Mod.~Phys.~Lett.~A~{\bf 3} (1988) 
1651--1660;\\
J.\ Distler, H.\ Kawai, ``Conformal field theory and 2d quantum
gravity'', Nucl.~Phys.\ B~{\bf 321} (1989) 509--527;\\
C.\ Schmidhuber, ``Exactly marginal operators and running coupling
constant in 2D gravity'', Nucl.~Phys.~B~{\bf 404} (1993) 342--358.
\item[[73$\!\!\!$]]\spa A.H.\ Chamseddine, ``A solution to two-dimensional quantum
  gravity.\ Non-critical strings'', Phys.~Lett.~B~{\bf 256} (1991)
  379--386;\\
``A study of non-critical string in arbitrary dimensions'',
Nucl.~Phys.~B~{\bf 368} (1992) 98--120.\ 
\item[[74$\!\!\!$]]\spa J.L.\ Gervais, A.\ Neveu, ``The dual string spectrum in Polyakov's
quantization'', Nucl.~Phys.~B~{\bf 199} (1982) 59--76.
\item[[75$\!\!\!$]]\spa see section I.7 in the last reference in [65].
\item[[76$\!\!\!$]]\spa Research Group in $M\cup\Phi$, ETH Z\"urich, unpublished.
\item[[77$\!\!\!$]]\spa A.H.\ Chamseddine, J.\ Fr\"ohlich, ``Two-dimensional
  Lorentz-Weyl anomaly and gravitational Chern-Simons theory'',
  Commun.~Math.~Phys.~{\bf 147} (1992) 549--562.
\item[[78$\!\!\!$]]\spa A.H.\ Chamseddine, J.\ Fr\"ohlich, ``Some elements of
  Connes' non-commutative geometry, and space-time geometry'', in
  ``Chen Ning Yang, a Great Physicist of the Twentieth Century'',
  C.S.\ Liu and S.-T.\ Yau (eds.), International Press 
  1995, pp.\ 10--34.\ 
\item[[79$\!\!\!$]]\spa I.M.\ Krichever, S.P.\ Novikov, ``Virasoro-Gelfand-Fuks type
  algebras, Riemann surfaces, operator's theory of closed strings'',
  J.~Geom.~Phys.~{\bf 5} (1988) 631--661.
\item[[80$\!\!\!$]]\spa See the references given in M.B.\ Green, J.H.\ 
  Schwarz, E.~Witten, ``Superstring theory II'', Cambridge University
  Press 1987, pp.\ 586--590.\ 
\item[[81$\!\!\!$]]\spa see the first two references in [61].
\item[[82$\!\!\!$]]\spa T.\ Banks, W.\ Fischler, S.H.\ Shenker, L.\ Susskind,
  ``$M$-theory as a matrix model: a conjecture'', Phys.~Rev.~D~{\bf 55}
  (1997) 5112--5128.
\item[[83$\!\!\!$]]\spa J.\ Fr\"ohlich, J.\ Hoppe, ``On zero-mass ground states in
  super-membrane matrix models'', preprint ETH-TH/96-53,
  hep-th/9701119.
\item[[84$\!\!\!$]]\spa T.\ Banks, N.\ Seiberg, S.\ Shenker, ``Branes from
  matrices'', hep-th/9612157.
\item[[85$\!\!\!$]]\spa J.\ Polchinski, ``Dirichlet branes and Ramond-Ramond
  charges'', Phys.~Rev.~Lett.\ {\bf 75} (1995) 4724--4727;\\
``TASI Lectures on D-Branes'', hep-th/9611050;\\
E.\ Witten, ``Bound states of strings and $p$-branes'',
Nucl.~Phys.~B~{\bf 460} (1996) 335--350.
\item[[86$\!\!\!$]]\spa Research Group in $M\cup\Phi$, ETH Z\"urich, unpublished.
\item[[87$\!\!\!$]]\spa E.\ D'Hoker, D.H.\ Phong, ``The geometry of string
  perturbation theory'', Rev.~Mod.\ Phys.~{\bf 60} (1988) 917--1065.
\item[[88$\!\!\!$]]\spa E.\ Witten, ``String theory dynamics in various
  dimensions'', Nucl.~Phys.~B~{\bf 443} (1995) 85--126.
\item[[89$\!\!\!$]]\spa A.H.\ Chamseddine, G.\ Felder, J.\ Fr\"ohlich, ``Grand
  unification in non-commutative geometry'', Nucl.~Phys.~B~{\bf 395}
  (1993) 672--698;\\
A.H.\ Chamseddine, J.\ Fr\"ohlich, ``SO(10)-Unification in
non-commutative geometry'', Phys.~Rev.~D~{\bf 59} (1994) 
2893--2907;\\
A.H.\ Chamseddine, ``Connection between space-time supersymmetry and
non-comm\-utative geometry'', Phys.~Lett.~B~{\bf 332} (1994) 349--357.
\item[[90$\!\!\!$]]\spa A.H.\ Chamseddine, J.\ Fr\"ohlich, ``The Chern-Simons action
  in non-commutative geometry'', J.~Math.~Phys.~{\bf 35} (1994) 
  5195--5218.
\item[[91$\!\!\!$]]\spa R.F.\ Streater, A.S.\ Wightman, ``PCT, Spin and Statistics,
  and All That'', Addison-Wesley 1964;\\
R.\ Jost, ``The General Theory of Quantized Fields'', Lectures in
Applied Mathematics, American Mathematical Society 1965.
\item[[92$\!\!\!$]]\spa K.\ Osterwalder, R.\ Schrader, ``Axioms for Euclidean
  Green's functions I, II'', Commun.~Math.~Phys.~{\bf 31} (1973) 
  83--112; {\bf 42} (1975) 281--305;\\
  V.\ Glaser, ``On the equivalence of the Euclidean and Wightman formulation 
  of field theory'', Commun.~Math.~Phys.~{\bf 37} (1974) 257--272.
\item[[93$\!\!\!$]]\spa M.\ L\"uscher, G.\ Mack, ``The energy-momentum tensor 
  of critical quantum field theory in $1+1$ dimensions'', 
  Hamburg 1976, unpublished; \\
  P.\ Furlan, G.M.\ Sotkov, I.T.\ Todorov, ``Two-dimensional conformal 
  quantum field theory'', Riv.\ Nuovo Cim.\ {\bf12} (1989) 1--202.
\item[[94$\!\!\!$]]\spa J.\ Fr\"ohlich, ``Statistics of fields, the Yang-Baxter
equation, and the theory of knots and links'', in ``Non-perturbative
Quantum Field Theory'', G.\ 't Hooft et al.~(eds.),  Plenum
Press 1988;\\
G.\ Felder, J.\ Fr\"ohlich, G.\ Keller, ``On the structure of unitary
conformal field theory I, II'', Commun.~Math.~Phys.~{\bf 124} (1989)
417--463; {\bf 130} (1990) 1--49.
\item[[95$\!\!\!$]]\spa K.H.\ Rehren, B.\ Schroer, ``Einstein causality
  and Art in 
  braids'', Nucl.~Phys.~B~{\bf 312} (1989) 715--750.
\item[[96$\!\!\!$]]\spa G.\ Moore, N.\ Seiberg, ``Classical and quantum conformal field
theory'', Commun.\ Math.~Phys.~{\bf 123} (1989) 177--254;\\
``Polynomial equations for rational conformal field theories'',
Phys.~Lett.~B~{\bf 212} (1988) 451--460.
\item[[97$\!\!\!$]]\spa see J.\ Fr\"ohlich, C.\ King cited in [59].
\item[[98$\!\!\!$]]\spa J.\ Fr\"ohlich, T.\ Kerler, ``Quantum Groups, Quantum
  Categories and Quantum Field Theory'', Lecture Notes in Mathematics  
  {\bf1542}, Springer Verlag 1993.
\item[[99$\!\!\!$]]\spa A.A.\ Belavin, A.M.\ Polyakov, A.B.\ Zamolodchikov,
  ``Infinite conformal symmetry in two-dimensional quantum field
  theory'', Nucl.~Phys.~B~{\bf 241} (1984) 333--380.
\item[[100$\!\!\!$]]\ J.\ Fr\"ohlich, K.\ Osterwalder, E.\ Seiler, ``On virtual
  representations of symmetric spaces and their analytic
  continuation'', Ann.~Math.~{\bf 118} (1983) 461--489.
\item[[101$\!\!\!$]] \ S.\ Doplicher, R.\ Haag, J.E.\ Roberts, ``Fields,
  observables and gauge transformations I, II'',
  Commun.~Math.~Phys.~{\bf 13} (1969) 1--23; {\bf
    15} (1969) 173--200; \\
  ``Local observables and particle statisticsI, II'', Commun.~Math.\ 
  Phys.~{\bf 23} (1971) 199--230; {\bf 35} (1974) 49--85.
\item[[102$\!\!\!$]]\ K.\ Fredenhagen, K.H.\ Rehren, B.\ Schroer, ``Superselection
  sectors with braid group statistics and exchange algebras I, II'',
  Commun.~Math.~Phys.~{\bf 125} (1989) 201--226;
  Rev.~Math.~Phys.~(Special Issue) (1992) 111--154.\ 
\item[[103$\!\!\!$]]\ J.\ Fr\"ohlich, F.\ Gabbiani, ``Braid statistics in local
  quantum theory'', Rev.~Math.\ Phys.~{\bf 2} (1990) 251--353.
\item[[104$\!\!\!$]]\ A.\ Cappelli, C.\ Itzykson, J.B.\ Zuber, ``The ADE
  classification of minimal and $A_1^{(1)}$ conformal invariant
  theories'', Commun.~Math.~Phys.~{\bf 113} (1987) 1--26.
\item[[105$\!\!\!$]]\ G.\ Felder, J.\ Fr\"ohlich, G.\ Keller, ``Braid matrices and
  structure constants for minimal conformal models'',
  Commun.~Math.~Phys.~{\bf 124} (1989) 647--664.
\item[[106$\!\!\!$]]\ A.Yu.~Alekseev, L.\ Faddeev, J.\ Fr\"ohlich, V.\ Schomerus,
 ``Representation theory of lattice current algebras'', q-alg/9604017, 
  Commun.\ Math.\ Phys., in press;\\
A.Yu.~Alekseev, J.\ Fr\"ohlich, V.\ Schomerus, in preparation.
\item[[107$\!\!\!$]]\ O.\ Grandjean, Ph.D.\  thesis, ETH Z\"urich, June 1997.\ 
\item[[108$\!\!\!$]]\ H.\ Eichenherr, ``Minimal operator algebras in
  superconformal quantum field theory'', Phys.~Lett.~{\bf 151}~B
  (1985) 26--30.
\item[[109$\!\!\!$]]\ ``Elliptic curves, modular forms, and Fermat's last
  theorem'', J.\ Coates, S.T.\ Yau (eds.), International Press 1995.
\item[[110$\!\!\!$]]\ B.R.\ Greene, M.R.\ Plesser, ``Duality in Calabi-Yau 
  moduli spaces'', Nucl.\ Phys.\ B {\bf338} (1990) 15--37;\\
  P.\ Candelas, X.C.\ de la Ossa, P.S.\ Green, L.\ Parkes, ``An exactly 
  soluble superconformal theory from a mirror pair of Calabi-Yau manifolds'', 
  Phys.\ Lett.\ B {\bf258} (1991) 118-126;\\
  ``Essays on Mirror Manifolds'', S.-T.\ Yau (ed.),  International Press 1992.
\item[[111$\!\!\!$]]\ W.\ Nahm, ``Chiral algebras of two-dimensional chiral
   field theories and their normal ordered products'', in ``Recent
    developments in  conformal field  theories'', Proc.\ Trieste Oktober 1989;\\
    R.\ Blumenhagen, M.\ Flohr, A.\ Kliem, W.\ Nahm, A.\ Recknagel, R.\ Varnhagen,
    ``W-algebras with two and three generators'', Nucl.\ Phys.\ B {\bf 361}
    (1991) 255--289.
\item[[112$\!\!\!$]]\ W.\ Lerche, C.\ Vafa, N.P.\ Warner, ``Chiral rings 
in $N=2$ superconformal theories'', Nucl.\ Phys.\ B {\bf324} (1989) 427--485.\ 
\end{description}

\end{document}